\documentclass[11pt,a4paper,titlepage]{article}

\usepackage[a4paper,margin=30mm,headheight=14pt]{geometry}
\usepackage[T1]{fontenc}

\usepackage{amsmath,amssymb,amsfonts,mathtools}
\usepackage{bm}
\usepackage{physics}
\usepackage[version=4]{mhchem}
\usepackage{libertine}
\usepackage[libertine]{newtxmath}
\usepackage{microtype}

\usepackage{graphicx}
\usepackage{capt-of}
\usepackage{xcolor}
\usepackage{booktabs}
\usepackage{tabularx}
\usepackage{longtable}
\usepackage{array}
\usepackage{makecell}
\usepackage{colortbl}
\usepackage[most]{tcolorbox}
\usepackage{pifont}
\usepackage{tikz,pgfplots}
\usetikzlibrary{arrows,arrows.meta,positioning,shapes.geometric}
\usetikzlibrary{decorations.pathmorphing}
\pgfplotsset{compat=1.18}
\usepackage{algorithmicx,algorithm,algpseudocode}

\usepackage{titlesec}
\usepackage{fancyhdr}
\usepackage{stackengine}
\usepackage{scalerel}
\usepackage[numbers,sort&compress]{natbib}
\usepackage{hyperref}

\graphicspath{{figures/}}
\newcommand{\hfsketchcredit}{Sketch: AI-assisted adaptation of the credited portrait.}

\definecolor{hfNavy}{HTML}{17324D}
\definecolor{hfBlue}{HTML}{1769AA}
\definecolor{hfTeal}{HTML}{007C83}
\definecolor{hfRed}{HTML}{B23A48}
\definecolor{hfOrange}{HTML}{D96C22}
\definecolor{hfPurple}{HTML}{9A7300}
\definecolor{hfViolet}{HTML}{775DA6}
\definecolor{hfInk}{HTML}{202A33}
\definecolor{hfPale}{HTML}{EDF2F6}
\definecolor{hfGreen}{HTML}{00685C}
\definecolor{hfGreenDark}{HTML}{00574D}
\definecolor{hfGreenLight}{HTML}{F2F8F6}

\newcommand{\hftablestyle}{\small\renewcommand{\arraystretch}{1.00}}
\newcommand{\hftablehead}[1]{\textcolor{hfGreenDark}{\bfseries\boldmath #1}}
\newcommand{\hftablesection}[2]{%
  \rowcolor{hfGreenLight}%
  \multicolumn{#1}{@{}l@{}}{\strut\hftablehead{#2}}\\*}

\tcbset{hfworked/.style={
  enhanced,breakable,sharp corners,
  boxrule=0.5pt,boxsep=0pt,left=2mm,right=2mm,top=2mm,bottom=2mm,
  grow to left by=2.3mm,grow to right by=2.3mm,
  before skip=12pt,after skip=12pt,
  fonttitle=\small\bfseries,coltitle=white,
  lefttitle=2mm,righttitle=2mm,toptitle=1.5mm,bottomtitle=1.5mm}}
\newcounter{hfexercisenum}
\newtcolorbox[use counter=hfexercisenum]{hfexercise}[2][]{hfworked,
  colframe=hfViolet!80!black,colback=hfViolet!3,
  colbacktitle=hfViolet!80!black,
  title={Exercise~\thetcbcounter: #2},
  title after break={Exercise~\thetcbcounter\ (continued): #2},#1}
\newtcolorbox{hfsolution}[1]{hfworked,
  colframe=hfGreenDark,colback=hfGreenLight,colbacktitle=hfGreenDark,
  title={Solution~\thehfexercisenum: #1},
  title after break={Solution~\thehfexercisenum\ (continued): #1},
  before upper={\allowdisplaybreaks[1]}}
\newtcolorbox{hfabstract}{hfworked,
  colframe=hfPurple,colback=hfPurple!4,colbacktitle=hfPurple,
  title={Abstract \textbar\ ESQC 2026}}
\newtcolorbox{hftheorem}[1]{hfworked,
  colframe=hfBlue,colback=hfBlue!3,colbacktitle=hfBlue,
  title={#1},title after break={#1 (continued)}}
\newtcolorbox[use counter=algorithm]{hfscfsummary}[1][]{
  enhanced,unbreakable,float=tbp,
  colframe=hfGreenDark,colback=white,colbacktitle=hfGreenDark,
  boxrule=0.7pt,arc=2mm,boxsep=0pt,
  left=4mm,right=4mm,top=3mm,bottom=3mm,
  fonttitle=\bfseries,fontupper=\small,coltitle=white,
  lefttitle=4mm,righttitle=4mm,toptitle=2mm,bottomtitle=2mm,
  title={Algorithm~\thetcbcounter\enspace\textbar\ The RHF self-consistent-field procedure},
  #1}
\newtcolorbox{hfhistorical}[1]{hfworked,unbreakable,
  colframe=hfOrange!75!black,colback=hfOrange!3,
  colbacktitle=hfOrange!75!black,fontupper=\small,
  title={Historical perspective: #1}}
\newcommand{\hfbiography}[4]{%
  \begin{hfhistorical}{#1}
  \phantomsection
  \addcontentsline{loh}{subsection}{#1}
  \if\relax\detokenize{#2}\relax
    #3\par
  \else
    \noindent\begin{minipage}[t]{0.17\linewidth}
      \vspace{0pt}\centering #2
    \end{minipage}\hfill
    \begin{minipage}[t]{0.79\linewidth}
      \vspace{0pt}#3
    \end{minipage}\par
  \fi
  \smallskip{\scriptsize\color{hfInk!75}#4\par}
  \end{hfhistorical}}
\makeatletter
\newcommand{\listofhistoricalperspectives}{%
  \phantomsection
  \section*{Historical perspectives}
  \addcontentsline{toc}{section}{Historical perspectives}
  \@starttoc{loh}}
\makeatother
\newcommand{\hfbiophoto}[1]{%
  \includegraphics[width=\linewidth,height=27mm,keepaspectratio]{#1}}
\newcommand{\hfbioperson}[3]{%
  \noindent\begin{minipage}[t]{0.17\linewidth}
    \vspace{0pt}\centering\hfbiophoto{#2}
  \end{minipage}\hfill
  \begin{minipage}[t]{0.79\linewidth}
    \vspace{0pt}\textbf{#1.} #3
  \end{minipage}\par}

\newcommand{\hfBioHartree}{%
\hfbiography{Douglas Rayner Hartree (1897--1958)}
{\hfbiophoto{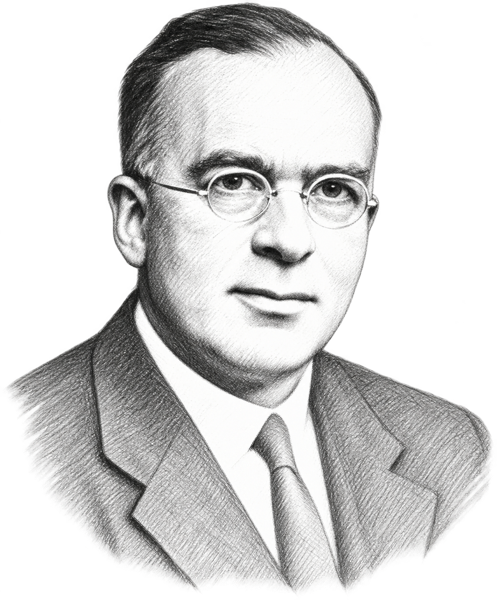}}
{An English mathematician and physicist, Hartree worked at Manchester before
returning to Cambridge as Plummer Professor of Mathematical Physics in 1946.
His 1928 self-consistent-field method made atomic structure accessible through
iterative numerical solutions of one-electron equations \citep{Hartree1928}.
He also helped develop mechanical and electronic computing: for Hartree,
theoretical approximations and the means of actually solving their equations
were inseparable concerns.}
{Biography: \href{https://mathshistory.st-andrews.ac.uk/Biographies/Hartree/}{University of St Andrews, MacTutor}.
Source portrait: unknown photographer, c.~1940;
\href{https://commons.wikimedia.org/wiki/File:Douglas_hartree.jpg}{Centre for Computing History / Wikimedia Commons}, listed as public domain.
\hfsketchcredit}}

\newcommand{\hfBioWilliamHartree}{%
\hfbiography{William Hartree (1870--1943)}
{\hfbiophoto{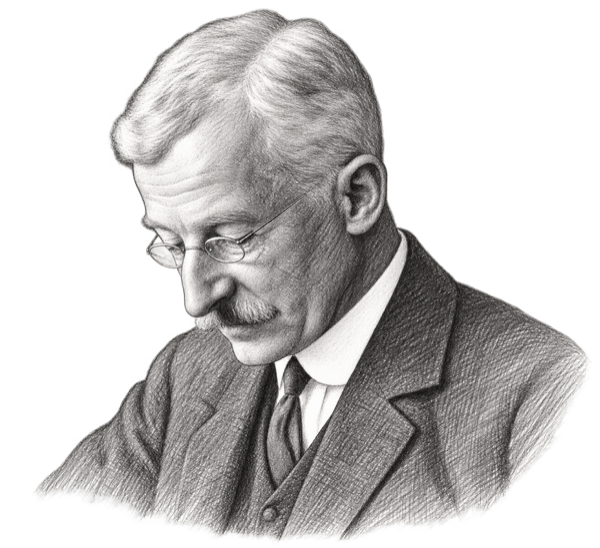}}
{An English engineer and numerical researcher, William Hartree taught at
Cambridge and later collaborated with the physiologist A.~V.~Hill. He was
Douglas Hartree's father and an important collaborator in turning atomic
self-consistent-field theory into numerical results. Their joint 1935
beryllium study included exchange explicitly \citep{HartreeHartree1935}.
His contribution illustrates a less visible part of the method's history:
obtaining accurate atomic orbitals required sustained numerical work as
well as the formulation of the equations.}
{Biography: \href{https://doi.org/10.1038/152154a0}{A.~V.~Hill, Nature obituary, 1943};
\href{https://mathshistory.st-andrews.ac.uk/Biographies/Hartree/}{MacTutor Hartree biography}.
Source photograph: \href{https://sliderulemuseum.org/Yearbooks.shtml}{International Slide Rule Museum archive}, late 1930s;
photographer unspecified. \textbf{Draft reproduction: permission pending.}
\hfsketchcredit}}

\newcommand{\hfBioSwirles}{%
\hfbiography{Bertha Swirles, Lady Jeffreys (1903--1999)}
{\hfbiophoto{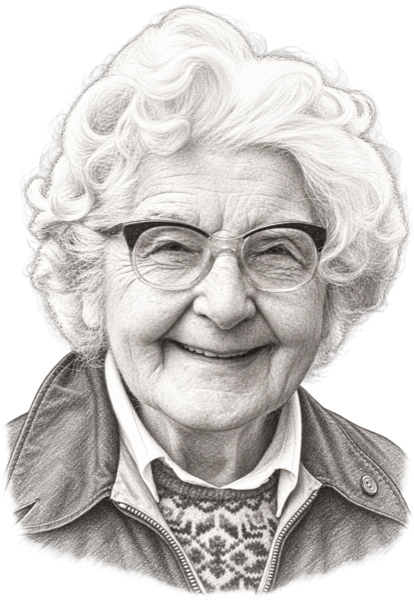}}
{A British mathematician trained at Girton College, Cambridge, Swirles
worked on quantum theory before returning to Girton as a Fellow in 1938.
Her 1935 relativistic self-consistent-field treatment extended the atomic
orbital framework beyond the non-relativistic approximation
\citep{Swirles1935}. With Douglas and William Hartree, she also studied
exchange and the superposition of configurations in a 1939 paper on oxygen
\citep{HartreeHartreeSwirles1939}. These were distinct extensions of the
ordinary single-determinant theory developed here. Her subsequent teaching
and the book \emph{Methods of Mathematical Physics}, written with Harold
Jeffreys, influenced generations of mathematical physicists.}
{Biography: \href{https://mathshistory.st-andrews.ac.uk/Biographies/Jeffreys_Bertha/}{University of St Andrews, MacTutor}.
Source portrait: Dr John Davies, \href{https://www.girton.cam.ac.uk/sites/default/files/2020-10/CoB10YearBooklet.pdf}{Girton College, Celebrating Our Benefactors booklet, 2017 entry}; photograph date unspecified.
\textbf{Draft reproduction: permission pending.}
\hfsketchcredit}}

\newcommand{\hfBioDirac}{%
\hfbiography{Paul Adrien Maurice Dirac (1902--1984)}
{\hfbiophoto{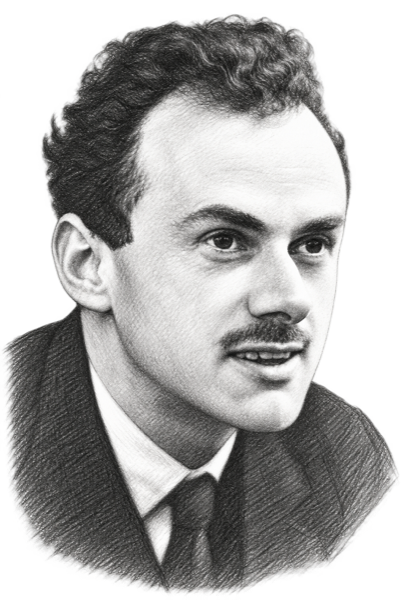}}
{Born in Bristol, Dirac trained in engineering and mathematics before
becoming one of the founders of quantum mechanics at Cambridge. His 1930
paper on exchange expressed the exchange-corrected self-consistent-field
theory in terms of a one-particle density matrix and explicitly examined
invariance under changes of occupied orbitals \citep{Dirac1930}. This
formulation makes the occupied subspace, rather than a particular set of
orbitals spanning it, the central object. It is distinct from the local
exchange approximation for the Thomas atom also developed in that paper.}
{Biography: \href{https://mathshistory.st-andrews.ac.uk/Biographies/Dirac/}{University of St Andrews, MacTutor}.
Source portrait: Nobel Foundation, 1933,
\href{https://commons.wikimedia.org/wiki/File:Paul_Dirac,_1933.jpg}{via Wikimedia Commons, listed as public domain}.
\hfsketchcredit}}

\newcommand{\hfBioMcWeeny}{%
\hfbiography{Roy McWeeny (1924--2021)}
{\hfbiophoto{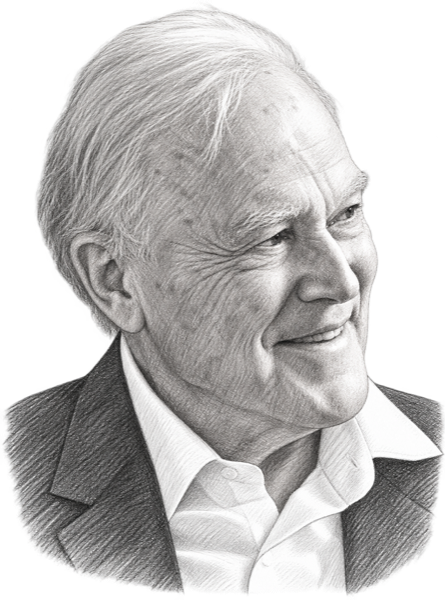}}
{A British theoretical chemist, McWeeny earned his doctorate with Charles
Coulson at Oxford and worked in Newcastle, Keele, and Sheffield before
moving to Pisa in 1982. His work made reduced density matrices a powerful
language for electronic structure. His 1960 account \citep{McWeeny1960}
brought together their use in orbital theories, connecting electron
densities, energies, and the constraints associated with a determinantal
state. His wider contributions included valence-bond theory,
multiconfigurational SCF methods, and the response of molecules to external
fields.}
{Biography: \href{https://www.iaqms.org/obituary/McWeenyObituary.pdf}{Brian Sutcliffe and Michael Robb, IAQMS obituary, 2021}.
Source portrait: \href{https://www.iaqms.org/deceased/mcweeny.php}{IAQMS archive}; photographer and date unspecified.
\textbf{Draft reproduction: permission pending.}
\hfsketchcredit}}

\newcommand{\hfBioPaldus}{%
\hfbiography{Josef Paldus (1935--2023)}
{\hfbiophoto{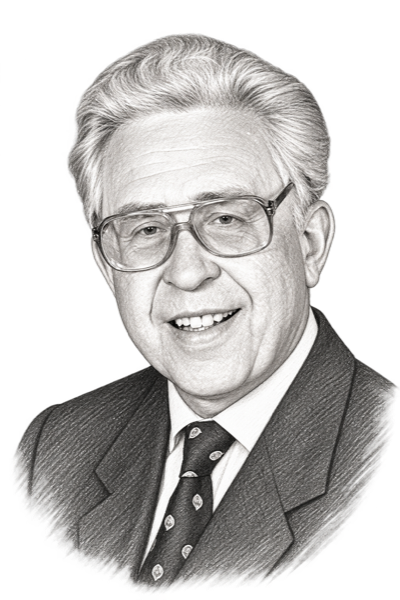}}
{A Czech-born Canadian theoretical chemist, Paldus trained in Prague and
joined the University of Waterloo in 1968. With Ji\v{r}\'{i}
\v{C}\'{i}\v{z}ek, he developed the analysis of Hartree--Fock stability,
including the distinction between singlet and triplet instabilities in
closed-shell systems \citep{CizekPaldus1967}. This work clarified why a
self-consistent solution need not be an energy minimum. Their collaboration
also helped establish coupled-cluster theory in molecular quantum chemistry.
Paldus developed unitary-group methods for treating spin-adapted
many-electron states \citep{Paldus1974}, bringing algebraic structure to
configuration-interaction calculations. His later account with Jason
Stuber \citep{StuberPaldus2003} places the symmetry-broken determinants
discussed here in a systematic framework. His work thus connects the
classification of independent-particle states with the treatment of
electron correlation beyond a single determinant.}
{Biography: \href{https://uwaterloo.ca/applied-mathematics/news/memoriam-josef-paldus}{University of Waterloo, 2023 obituary};
\href{https://www.iaqms.org/deceased/paldus.php}{IAQMS profile}.
Source portrait: IAQMS archive; photographer and date unspecified.
\textbf{Draft reproduction: permission pending.}
\hfsketchcredit}}

\newcommand{\hfBioBernardi}{%
\hfbiography{Fernando Bernardi}{}
{An Italian theoretical organic chemist at the University of Bologna,
Bernardi studied with Boys in Cambridge in 1969--1971. Their joint 1970
paper introduced the counterpoise procedure discussed here
\citep{BoysBernardi1970}, making the comparison of fragment basis spaces an
explicit part of intermolecular-energy calculations. His later research
connected electronic-structure methods with organic reactivity and
photochemistry, including work with Massimo Olivucci and Michael Robb on
potential-energy-surface crossings. The Boys--Bernardi name acknowledges
both contributors to a correction that remains central to the practical
use of atom-centered basis sets.}
{Biography: author profile in \href{https://doi.org/10.1039/CS9962500321}{Bernardi, Olivucci, and Robb, Chemical Society Reviews, 1996, p.~321}.}}

\newcommand{\hfBioFock}{%
\hfbiography{Vladimir Aleksandrovich Fock (1898--1974)}
{\hfbiophoto{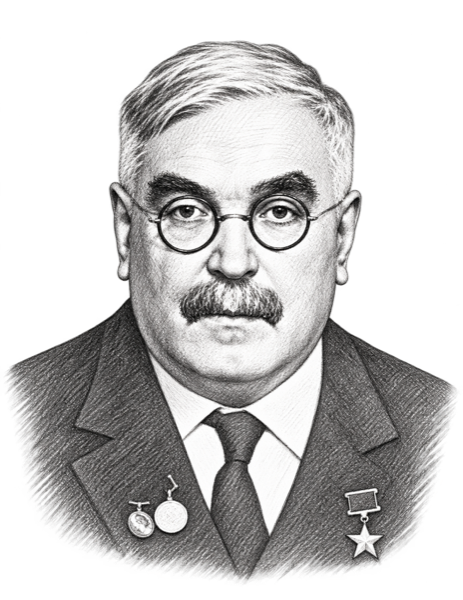}}
{Fock was a Russian theoretical physicist whose career was closely associated
with the university in his native St Petersburg, later called Leningrad.
In 1930 he incorporated exchange into the self-consistent-field equations by
using an antisymmetric many-electron wavefunction \citep{Fock1930}.
His work ranged from wave mechanics and gravitation to quantum field theory;
the Fock-space construction carries his name as well.}
{Biography: \href{https://www.iaqms.org/deceased/fock.php}{International Academy of Quantum Molecular Science (IAQMS)} and
\href{https://mathshistory.st-andrews.ac.uk/Biographies/Fock/}{MacTutor}.
Source portrait: \href{https://www.iaqms.org/deceased/fock.php}{IAQMS archive}; photographer and date unspecified.
\textbf{Draft reproduction: permission pending.}
\hfsketchcredit}}

\newcommand{\hfBioBornOppenheimer}{%
\hfbiography{Born and Oppenheimer}{}
{\hfbioperson{Max Born (1882--1970)}{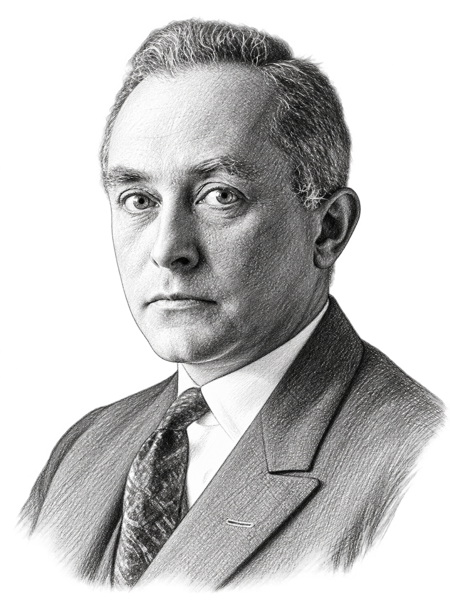}
{A German-born physicist, Born helped establish quantum mechanics at
G\"ottingen and later taught at Edinburgh. His statistical interpretation of
the wavefunction brought him a share of the 1954 Nobel Prize in Physics.
His work with Oppenheimer addressed a different foundational question:
how to organize the electronic and nuclear motions of a molecule.}
\smallskip
\hfbioperson{J.~Robert Oppenheimer (1904--1967)}{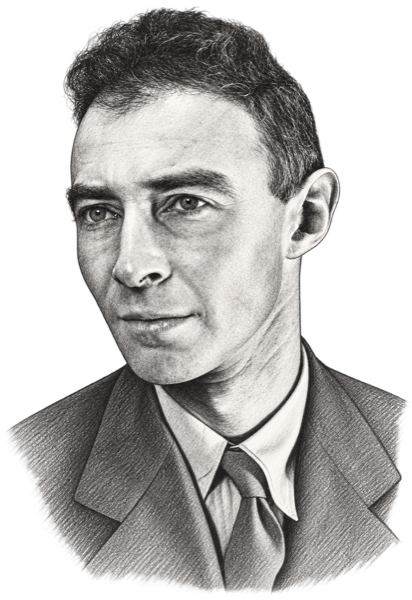}
{The American theoretical physicist completed his doctorate at G\"ottingen
in 1927, the year of their joint molecular paper \citep{BornOppenheimer1927}.
He later taught at Berkeley and Caltech and directed the Institute for
Advanced Study. The approximation bearing their names supplies the
electronic-structure framework used here; it is distinct from the subsequent
Hartree--Fock approximation.}}
{Biographies: \href{https://www.nobelprize.org/prizes/physics/1954/born/facts/}{Nobel Foundation};
\href{https://www.ias.edu/oppenheimer-legacy}{Institute for Advanced Study}.
Source portraits: Born, unknown photographer, \href{https://commons.wikimedia.org/wiki/File:Max_Born.jpg}{via Wikimedia Commons, listed as public domain};
Oppenheimer, 1944, US National Archives,
\href{https://commons.wikimedia.org/wiki/File:Oppenheimer_(cropped).jpg}{record 558579, crop by Artem.G, tonal restoration by MyCatIsAChonk}, public domain.
\hfsketchcredit}}

\newcommand{\hfBioMulliken}{%
\hfbiography{Robert Sanderson Mulliken (1896--1986)}
{\hfbiophoto{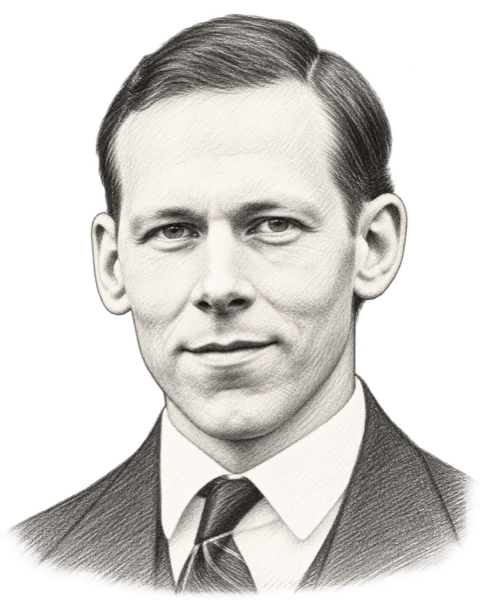}}
{An American physicist and chemist based at the University of Chicago,
Mulliken helped develop the molecular-orbital description of chemical bonding
and spectroscopy. He received the 1966 Nobel Prize in Chemistry for this
work. His population analysis \citep{Mulliken1955} connects the orbital
coefficients and overlap matrix to a chemical interpretation of the electron
distribution, while remaining dependent on the chosen partitioning convention.}
{Biography: \href{https://www.nasonline.org/wp-content/uploads/2024/06/mulliken-robert.pdf}{R.~Stephen Berry, NAS memoir};
\href{https://www.nobelprize.org/prizes/chemistry/1966/mulliken/facts/}{Nobel Foundation}.
Source portrait: Chicago, 1929, Friedrich Hund estate / Gerhard Hund (GFHund),
\href{https://commons.wikimedia.org/wiki/File:Mulliken,Robert_1929_Chicago.jpg}{via Wikimedia Commons},
\href{https://creativecommons.org/licenses/by/3.0/}{CC BY 3.0}.
\hfsketchcredit}}

\newcommand{\hfBioCoulsonFischer}{%
\hfbiography{Coulson and Fischer-Hjalmars}{}
{\hfbioperson{Charles Alfred Coulson (1910--1974)}{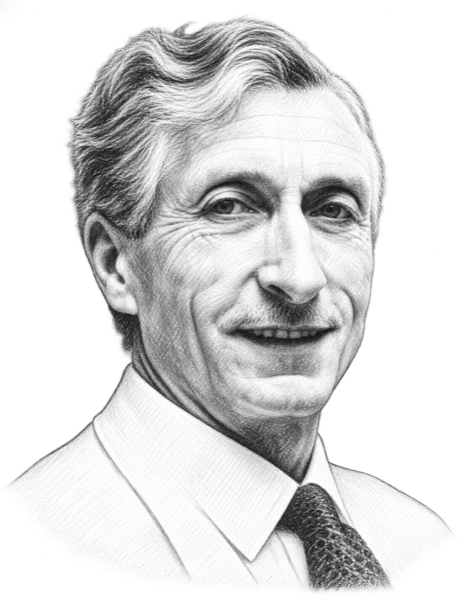}
{A British theoretical chemist and mathematician, Coulson held positions
in London and Oxford and made molecular-orbital theory a powerful language
for discussing chemical bonding. His research and teaching emphasized
physical understanding alongside calculation. His collaboration with Fischer
examined the limitations of simple orbital descriptions of the hydrogen molecule.}
\smallskip
\hfbioperson{Inga Fischer-Hjalmars (1918--2008)}{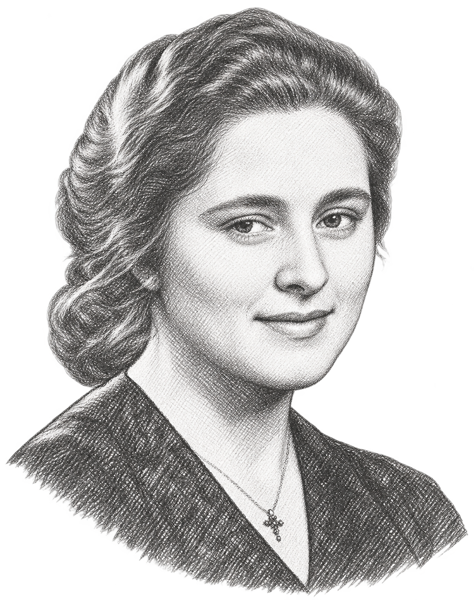}
{Initially trained in pharmacy, Fischer-Hjalmars became a pioneer of Swedish
quantum chemistry and succeeded Oskar Klein as professor of theoretical
physics at Stockholm in 1963. She published the 1949 hydrogen-molecule
paper as Inga Fischer \citep{CoulsonFischer1949}. The Coulson--Fischer point
discussed here preserves her earlier surname in the terminology of
Hartree--Fock symmetry breaking.}}
{Biographies: \href{https://www.iaqms.org/deceased/coulson.php}{IAQMS Coulson profile};
\href{https://www.iaqms.org/obituary/IngaFish-Hjalmar.pdf}{Bj\"orn Roos, IAQMS obituary}.
Source portraits: Coulson, \href{https://www.iaqms.org/deceased/coulson.php}{IAQMS archive}, photographer/date unspecified,
\textbf{draft reproduction: permission pending};
Fischer-Hjalmars, unknown photographer, Svenskt Portr\"attarkiv,
\href{https://commons.wikimedia.org/wiki/File:Inga_Fischer_SPA_(cropped).jpg}{existing Commons crop},
\href{https://creativecommons.org/licenses/by/4.0/}{CC BY 4.0}.
\hfsketchcredit}}

\newcommand{\hfBioPopleNesbet}{%
\hfbiography{Pople and Nesbet}{}
{\hfbioperson{John Anthony Pople (1925--2004)}{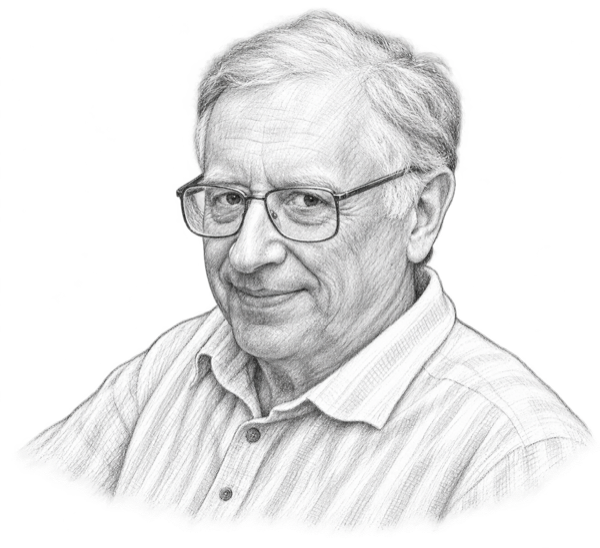}
{A British theoretical chemist, Pople developed both semiempirical and
ab initio molecular-orbital methods. His work helped establish reproducible
computational procedures and widely used Gaussian basis-set families.
He shared the 1998 Nobel Prize in Chemistry for the development of
computational methods in quantum chemistry.}
\smallskip
\hfbioperson{Robert K.~Nesbet (b.~1930)}{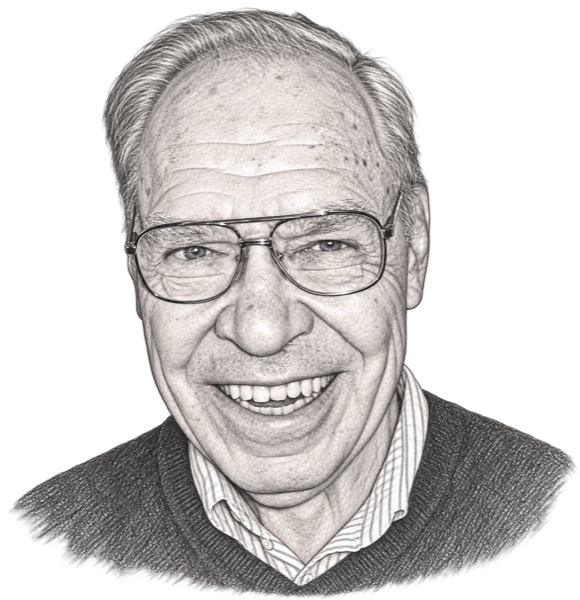}
{Nesbet earned his Cambridge doctorate under Boys in 1954, using early
digital computers for molecular calculations. He later worked with Slater
at MIT and joined IBM in 1962. His joint paper with Pople on self-consistent
orbitals for radicals \citep{PopleNesbet1954} underlies the coupled
spin-dependent matrix equations developed in this subsection.}}
{Biographies: \href{https://www.iaqms.org/deceased/pople.php}{IAQMS Pople profile};
\href{https://bitsavers.org/pdf/ibm/IBM_Journal_of_Research_and_Development/421/nesbet.pdf}{Nesbet's IBM author biography};
\href{https://sciencefeatured.com/2024/11/27/conformal-theory-of-gravitation-and-cosmic-expansion-new-light-on-dark-matter/}{Nesbet author profile}.
Source portraits: Pople, IAQMS archive; Nesbet, Science Featured author profile.
Photographers/dates unspecified. \textbf{Draft reproductions: permissions pending.}
\hfsketchcredit}}

\newcommand{\hfBioBerthier}{%
\hfbiography{Gaston Berthier (1923--2009)}
{\hfbiophoto{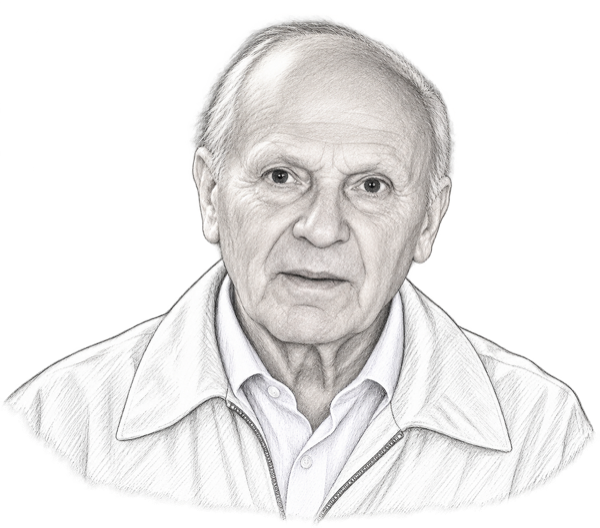}}
{A French theoretical chemist, Berthier was a pioneer of quantum chemistry
in France, working with the Pullman group and at the \'{E}cole normale
sup\'erieure. In 1954 he developed an open-shell self-consistent-field
treatment allowing electrons of opposite spin to occupy different spatial
orbitals \citep{Berthier1954}, independently of Pople and Nesbet
\citep{PopleNesbet1954}. His contribution is recognized here in the name
\emph{Pople--Nesbet--Berthier equations}. With Bernard L\'evy, he later
developed a generalized Brillouin theorem for multiconfigurational SCF
wavefunctions. Alongside his research, he trained generations of theoretical
chemists and helped build the French and wider Latin-language
quantum-chemistry community.}
{Biography: \href{https://exobiologie.fr/deces-de-gaston-berthier/}{Yves Ellinger, Soci\'et\'e Fran\c{c}aise d'Exobiologie obituary, 2010}.
Source portrait: Fabriceparis,
\href{https://commons.wikimedia.org/wiki/File:Photo-GB-150px.jpg}{via Wikimedia Commons},
\href{https://creativecommons.org/licenses/by-sa/3.0/}{CC BY-SA 3.0}; photograph date unspecified.
\hfsketchcredit}}

\newcommand{\hfBioPulay}{%
\hfbiography{P\'eter Pulay (b.~1941)}
{\hfbiophoto{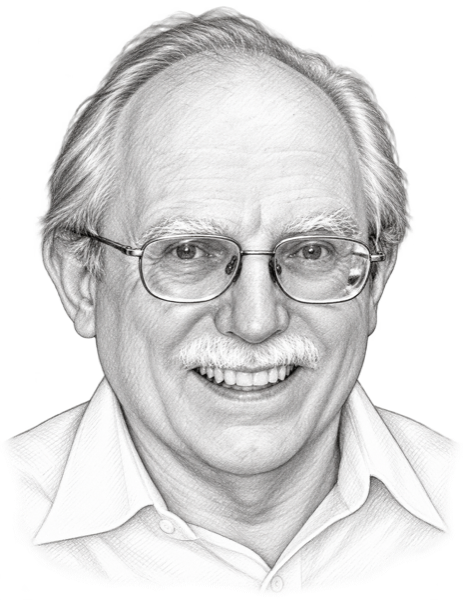}}
{Born in Hungary, Pulay developed methods that made molecular electronic
structure more practical, including analytic energy gradients and efficient
geometry optimization. His DIIS procedure \citep{Pulay1980,Pulay1982}
accelerates self-consistency by combining information from earlier iterations.
His career at the University of Arkansas also encompassed vibrational force
fields and local correlation, linking algorithmic developments with
calculations of molecular structure and properties.}
{Biography and source portrait: \href{https://www.iaqms.org/members/pulay.php}{IAQMS profile}.
Source portrait photographer/date unspecified. \textbf{Draft reproduction: permission pending.}
\hfsketchcredit}}

\newcommand{\hfBioMollerPlesset}{%
\hfbiography{M{\o}ller and Plesset}{}
{\hfbioperson{Christian M{\o}ller (1904--1980)}{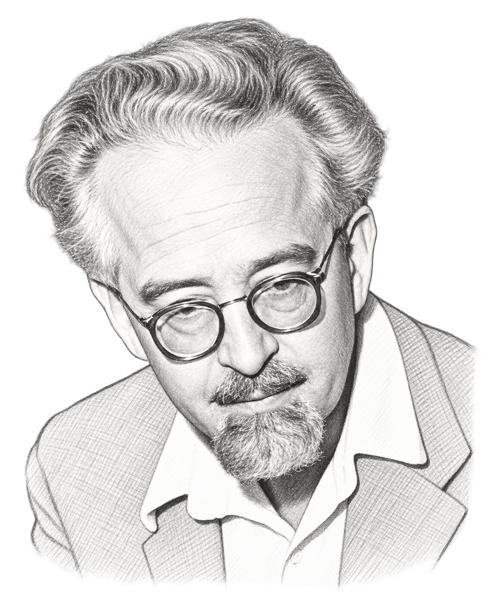}
{A Danish theoretical physicist closely associated with Copenhagen and
Niels Bohr's institute, M{\o}ller worked on relativistic quantum theory as well
as many-electron systems. His teaching and research helped shape Danish
theoretical physics. The perturbation treatment developed with Plesset in
1934 used a Hartree--Fock reference \citep{MollerPlesset1934}.}
\smallskip
\hfbioperson{Milton S.~Plesset (1908--1991)}{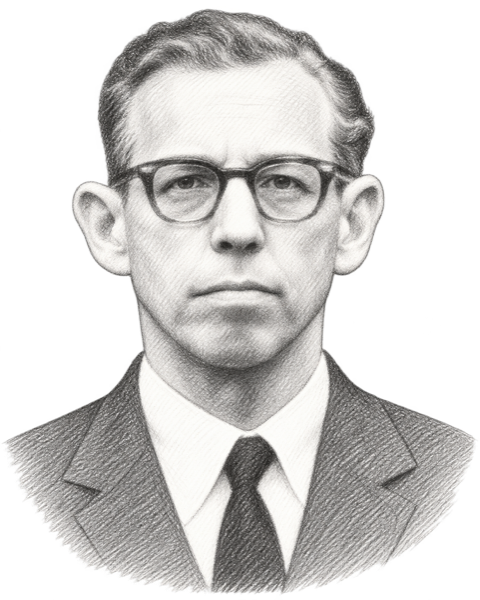}
{An American physicist, Plesset received his doctorate at Yale and spent
1933--1934 at Bohr's institute after postdoctoral work at Caltech.
His later Caltech research centered on applied mechanics, hydrodynamics,
and cavitation. His name therefore appears both in electronic-structure
perturbation theory and, through the Rayleigh--Plesset equation, in bubble dynamics.}}
{Biographies: \href{https://biografiskleksikon.lex.dk/Christian_M\%C3\%B8ller}{Mogens Pihl, Danish Biographical Dictionary};
\href{https://oralhistories.library.caltech.edu/127/}{Caltech oral history}.
Source portraits: M{\o}ller, Rochester, 1957, AIP Emilio Segr\`e Visual Archives, Marshak Collection,
\href{https://commons.wikimedia.org/wiki/File:Christian_Moller.jpg}{existing Commons detail},
\textbf{draft reproduction: permission pending};
Plesset, Copenhagen, 1963, Friedrich Hund estate / Gerhard Hund (GFHund),
\href{https://commons.wikimedia.org/wiki/File:Plesset,Milton_1963_Kopenhagen.jpg}{via Commons},
\href{https://creativecommons.org/licenses/by/3.0/}{CC BY 3.0}.
\hfsketchcredit}}

\newcommand{\hfBioHuzinaga}{%
\hfbiography{Sigeru Huzinaga}
{\hfbiophoto{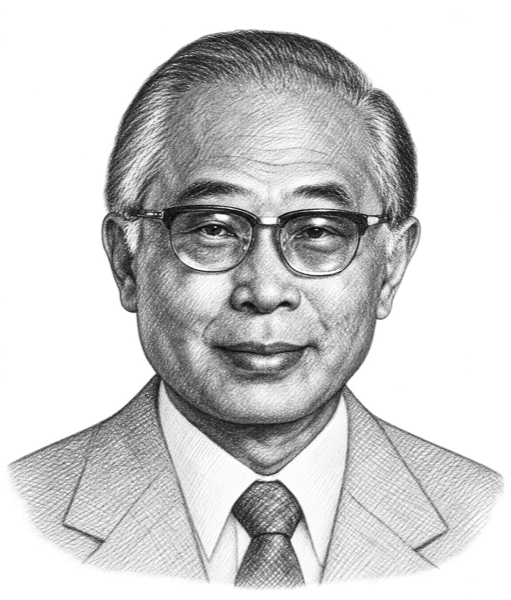}}
{A Japanese quantum chemist and professor at the University of Alberta,
Huzinaga helped turn Gaussian basis functions into systematic tools for
molecular calculations. His 1965 atomic Gaussian expansions
\citep{Huzinaga1965} provided an influential foundation for practical basis
construction. His work connected the choice and contraction of atomic
functions with the accuracy and computational demands of molecular
electronic-structure calculations.}
{Biography: \href{https://www.registrar.ualberta.ca/calendar-archive/calendar-2015-2016/University-Staff/Emeriti/251.html}{University of Alberta emeriti record};
\href{https://www.molsci.jp/wp-content/uploads/A0002.pdf}{Huzinaga's account of computational chemistry, 2007}.
Source portrait: \href{https://yamadazaidan.jp/wordpress/wp-content/uploads/2021/09/S60_Vol.1_18.pdf}{Yamada Science Foundation report, printed p.~16 (detail)};
photographer/date unspecified. \textbf{Draft reproduction: permission pending.}
\hfsketchcredit}}

\newcommand{\hfBioDunning}{%
\hfbiography{Thom H.~Dunning Jr. (b.~1943)}
{\hfbiophoto{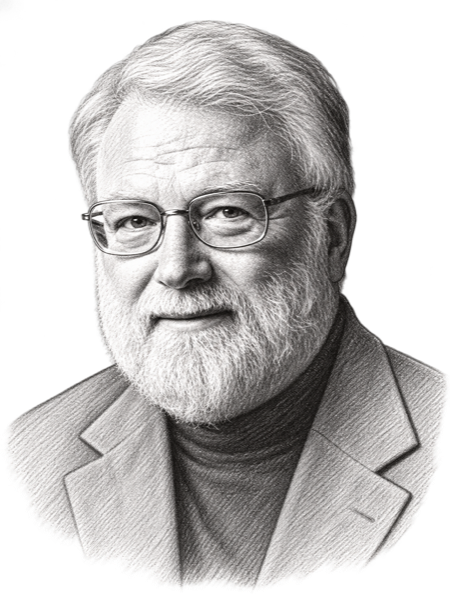}}
{An American theoretical chemist trained at Caltech, Dunning developed
Gaussian basis sets first for Hartree--Fock calculations and subsequently
for correlated wavefunctions. His correlation-consistent construction
\citep{Dunning1989} organizes basis functions into balanced hierarchical
sequences. This made systematic convergence toward the complete-basis limit
a practical strategy, linking basis design to the reliable assessment of
electronic-structure methods.}
{Biography and source portrait: \href{https://www.iaqms.org/members/dunning.php}{IAQMS profile}.
Source portrait photographer/date unspecified. \textbf{Draft reproduction: permission pending.}
\hfsketchcredit}}

\newcommand{\hfBioPauli}{%
\hfbiography{Wolfgang Pauli (1900--1958)}
{\hfbiophoto{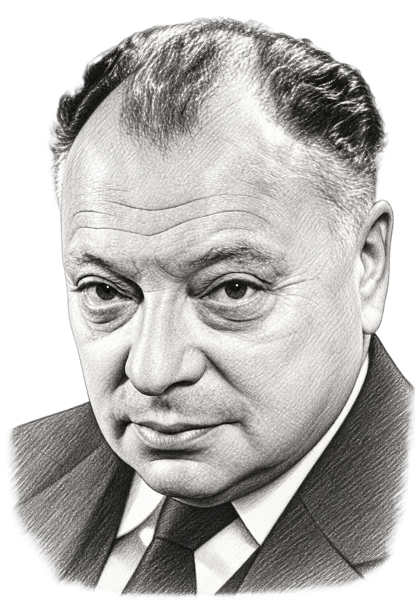}}
{Born in Vienna, Pauli became professor of theoretical physics at ETH
Z\"urich in 1928. His exclusion principle, proposed in 1925, was decisive
in understanding atomic electronic configurations \citep{Pauli1925} and
earned him the 1945 Nobel Prize in Physics. In the many-electron formulation
used here, exclusion follows from the antisymmetry of the electronic
wavefunction rather than from an additional rule imposed on its orbitals.}
{Biography: \href{https://archives.web.cern.ch/pauli_archive/biography/}{CERN Pauli Archive};
\href{https://www.nobelprize.org/prizes/physics/1945/pauli/facts/}{Nobel Foundation}.
Source portrait: Bettina Katzenstein / ETH-Bibliothek Z\"urich, Bildarchiv,
\href{https://commons.wikimedia.org/wiki/File:Wolfgang_Pauli_ETH-Bib_Portr_01042_(cropped).jpg}{Portr 01042, crop by Xabier},
\href{https://creativecommons.org/licenses/by-sa/3.0/}{CC BY-SA 3.0}.
\hfsketchcredit}}

\newcommand{\hfBioSlater}{%
\hfbiography{John Clarke Slater (1900--1976)}
{\hfbiophoto{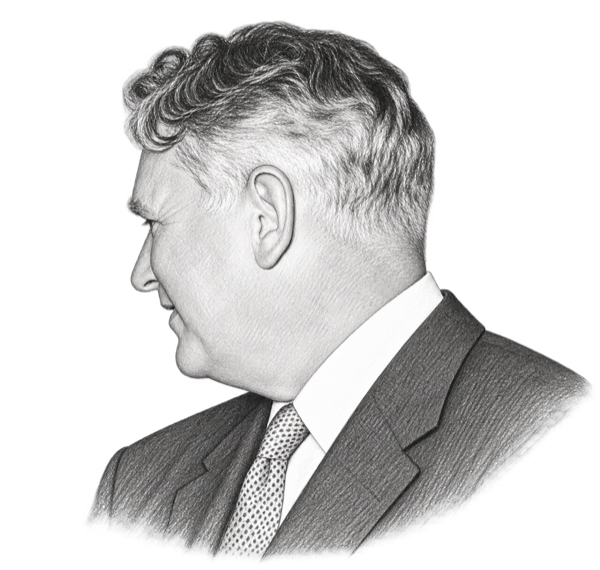}}
{An American physicist with a long career at MIT and later at the University
of Florida, Slater helped shape the quantum theory of atoms, molecules, and
solids. His 1929 determinantal wavefunctions provide the antisymmetric
building blocks used throughout this chapter \citep{Slater1929}.
His analytic atomic orbitals and screening rules also connect this
many-electron construction with the Slater-type basis functions introduced
in the second part \citep{Slater1930Orbitals}.}
{Biography: \href{https://www.nasonline.org/wp-content/uploads/2024/06/slater-john.pdf}{P.~M.~Morse, NAS memoir}.
Source portrait detail: Larry C.~Huston, 1967; identified on the right in the
\href{https://photoarchive.lib.uchicago.edu/db.xqy?one=apf1-01019.xml}{archive description}.
University of Chicago Photographic Archive, apf1-01019, Hanna Holborn Gray Special Collections Research Center, University of Chicago Library.
\href{https://photoarchive.lib.uchicago.edu/rights.html}{Educational/scholarly reuse; commercial permission required}.
\hfsketchcredit}}

\newcommand{\hfBioCondon}{%
\hfbiography{Edward Uhler Condon (1902--1974)}
{\hfbiophoto{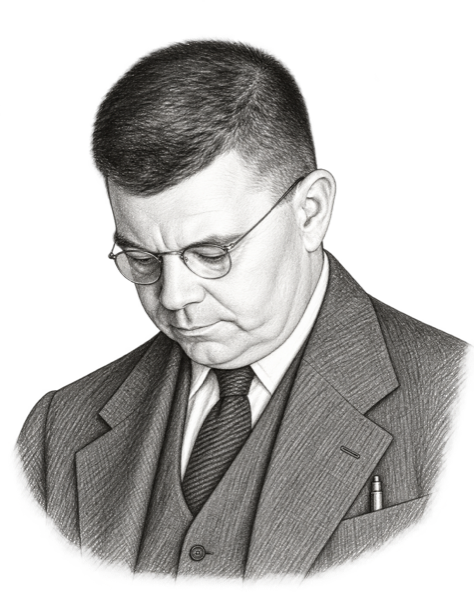}}
{Condon was an American theoretical physicist whose work connected the new
quantum mechanics with atomic and molecular spectroscopy. His treatment of
complex spectra contributed to the determinant matrix-element rules now
named after Slater and Condon \citep{Condon1930}. He worked at Princeton and
Westinghouse before directing the US National Bureau of Standards from 1945
to 1951, and later held academic appointments at Washington University and
the University of Colorado.}
{Biography: \href{https://history.aip.org/catalog/icos/1368.html}{American Institute of Physics, Condon papers};
\href{https://www.nist.gov/history/directors-national-bureau-standards-1901-1988-and-national-institute-standards-and}{NIST historical record}.
Source portrait: National Bureau of Standards, 1947,
\href{https://commons.wikimedia.org/wiki/File:Edward_U._Condon_in_1947.jpg}{via Wikimedia Commons}; US federal-government photograph, public domain.
\hfsketchcredit}}

\newcommand{\hfBioBrillouin}{%
\hfbiography{L\'eon Nicolas Brillouin (1889--1969)}
{\hfbiophoto{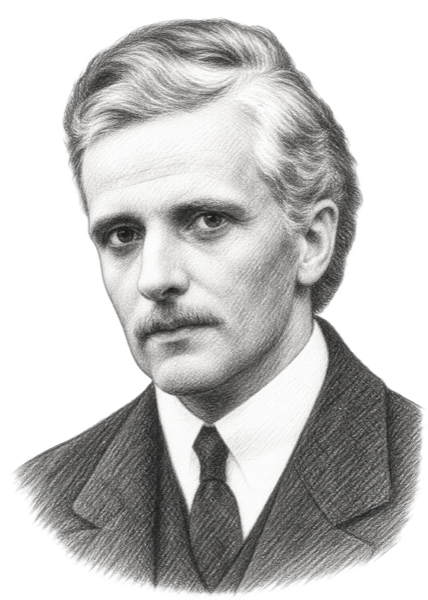}}
{A French theoretical physicist, Brillouin held chairs in Paris before
continuing his career in the United States. His research encompassed wave
propagation, the electronic structure of solids, and the relation between
information and thermodynamics. His work on self-consistent fields led to
the stationarity result discussed here \citep{Brillouin1932}; his name is also
associated with Brillouin zones and the semiclassical WKB approximation.}
{Biography: \href{https://www.nasonline.org/wp-content/uploads/2024/06/brillouin-lon-n.pdf}{L.~Hilleth Thomas, NAS memoir}.
Source portrait: Benjamin Couprie, Institut International de Physique Solvay,
\href{https://commons.wikimedia.org/wiki/File:L\%C3\%A9on_Brillouin_Solvay_conference_1927.jpg}{1927 Solvay photograph, crop by Setreset}; public domain.
\hfsketchcredit}}

\newcommand{\hfBioKoopmans}{%
\hfbiography{Tjalling Charles Koopmans (1910--1985)}
{\hfbiophoto{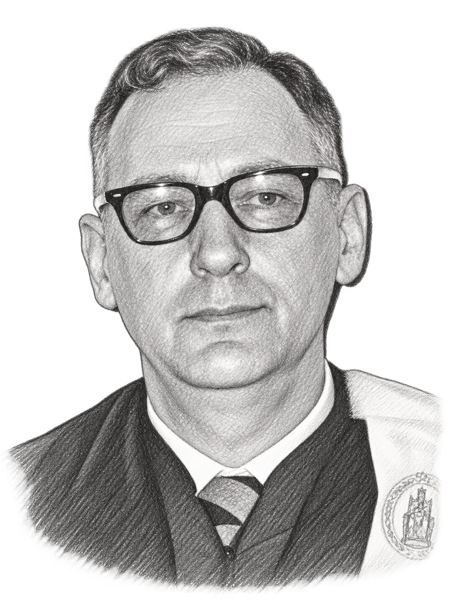}}
{Koopmans was born in the Netherlands and initially trained in mathematics
and physics. His 1934 paper related electron-removal energies to
Hartree--Fock orbital energies under the frozen-orbital approximation
\citep{Koopmans1934}. He subsequently turned to economics and joined Yale
in 1955. His later work on the allocation of resources brought him a share
of the 1975 Nobel Memorial Prize in Economic Sciences---an unusual career
link between electronic-structure theory and mathematical economics.}
{Biography: \href{https://cowles.yale.edu/people/tjalling-c-koopmans}{Yale University, Cowles Foundation}.
Source portrait: Eric Koch / Anefo, Nationaal Archief,
\href{https://commons.wikimedia.org/wiki/File:TjallingKoopmans1967_(cropped).jpg}{Leuven, 1967, crop by Wabbuh},
\href{https://creativecommons.org/publicdomain/zero/1.0/}{CC0 1.0}.
\hfsketchcredit}}

\newcommand{\hfBioRoothaan}{%
\hfbiography{Clemens C.~J.~Roothaan (1918--2019)}
{\hfbiophoto{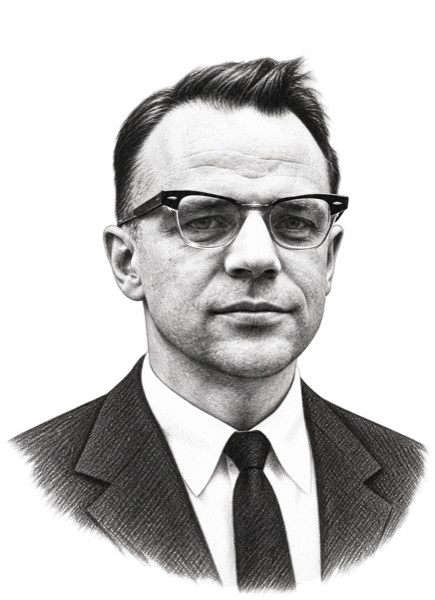}}
{Born in Nijmegen in the Netherlands, Roothaan earned his doctorate at the
University of Chicago under Robert Mulliken in 1950. His 1951 formulation
of the molecular self-consistent-field problem in a finite basis
\citep{Roothaan1951}, developed independently of Hall's, helped turn
Hartree--Fock theory into a practical computational method. He remained
closely involved in scientific computing, directing Chicago's Computation
Center from 1962 to 1968.}
{Biography: \href{https://emeritifaculty.uchicago.edu/news-announcements/article/clemens-c.j.-roothaan-1918-2019}{University of Chicago obituary}.
Source portrait: University of Chicago Photographic Archive,
\href{https://photoarchive.lib.uchicago.edu/db.xqy?one=apf1-12856.xml}{apf1-12856}, Hanna Holborn Gray Special Collections Research Center, University of Chicago Library.
\href{https://photoarchive.lib.uchicago.edu/rights.html}{Educational/scholarly reuse; commercial permission required}.
\hfsketchcredit}}

\newcommand{\hfBioHall}{%
\hfbiography{George Garfield Hall (1925--2018)}
{\hfbiophoto{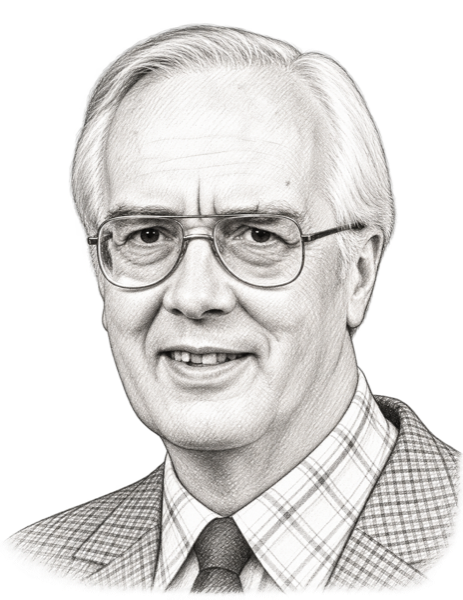}}
{Born in Belfast, Hall studied at Cambridge under John Lennard-Jones.
His 1951 treatment of molecular orbitals gave an independent derivation
of the matrix self-consistent-field equations now known as the
Roothaan--Hall equations \citep{Hall1951}. He became professor of applied
mathematics at Nottingham in 1962. Alongside quantum chemistry, his career
included substantial work in mathematical education, illustrating the
mathematical breadth behind the molecular-orbital methods.}
{Biography: \href{https://www.iaqms.org/obituary/GeorgeGarfieldHall.pdf}{IAQMS obituary}.
Source portrait: \href{https://www.iaqms.org/deceased/hall.php}{IAQMS archive}; photographer and date unspecified.
\textbf{Draft reproduction: permission pending.}
\hfsketchcredit}}

\newcommand{\hfBioLowdin}{%
\hfbiography{Per-Olov L\"owdin (1916--2000)}
{\hfbiophoto{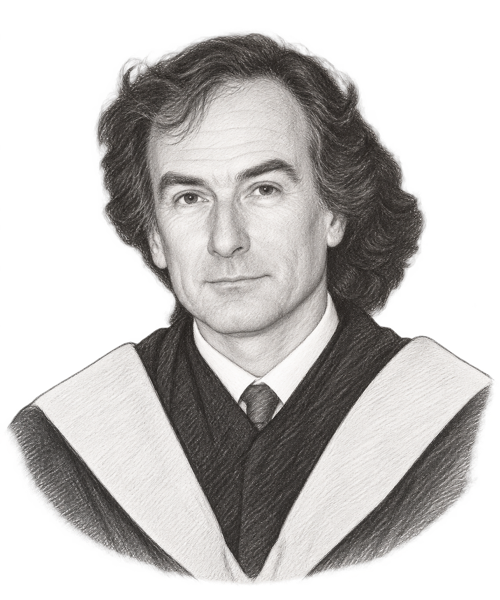}}
{A Swedish theoretical scientist, L\"owdin built influential quantum-chemistry
schools at Uppsala and the University of Florida. His 1950 study of
nonorthogonal atomic orbitals established the symmetric orthogonalization
used here \citep{Lowdin1950}. His research emphasized the mathematical
foundations of quantum chemistry, while his international summer and winter
schools and the Sanibel symposia helped train successive generations of
electronic-structure researchers.}
{Biography: \href{https://www.iaqms.org/deceased/lowdin.php}{IAQMS profile}.
Source portrait: Sorbonne Universit\'e, unknown photographer, 25 November 1977,
\href{https://patrimoine.sorbonne-universite.fr/idurl/1/8392}{honorary-doctorate ceremony (detail)},
\href{https://www.etalab.gouv.fr/licence-ouverte-open-licence}{Licence Ouverte};
\href{https://commons.wikimedia.org/wiki/File:Per-Olov_L\%C3\%B6wdin.jpg}{via Wikimedia Commons}.
\hfsketchcredit}}

\newcommand{\hfBioBoys}{%
\hfbiography{Samuel Francis Boys (1911--1972)}
{\hfbiophoto{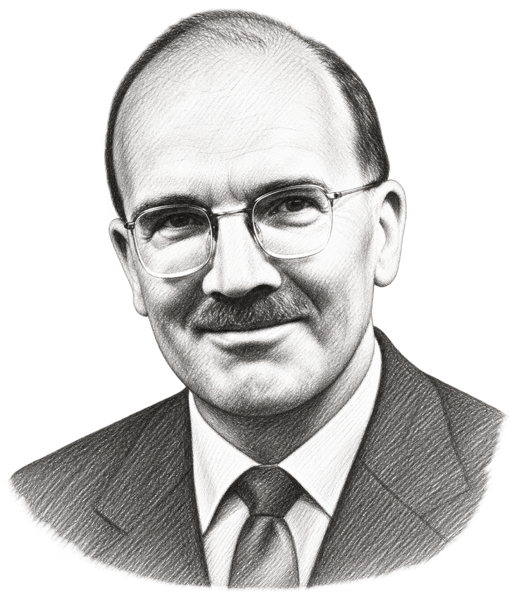}}
{A British theoretical chemist at Cambridge, Boys made Gaussian functions
central to practical molecular calculations through his 1950 work on
electronic wavefunctions \citep{Boys1950}. Efficient multicenter integrals
were a major part of their appeal. His later contributions included orbital
localization and, with Fernando Bernardi, the counterpoise correction for
basis-set superposition error \citep{FosterBoys1960,BoysBernardi1970}.
He was elected a Fellow of the Royal Society in 1972.}
{Biography: \href{https://www.iaqms.org/deceased/boys.php}{IAQMS profile};
\href{https://makingscience.royalsociety.org/people/na1119/samuel-francis-boys}{Royal Society record}.
Source portrait: \href{https://www.iaqms.org/deceased/boys.php}{IAQMS archive}; photographer and date unspecified.
\textbf{Draft reproduction: permission pending.}
\hfsketchcredit}}

\newcommand{\hftheoremportrait}[4]{%
  \noindent\begin{minipage}[t]{0.16\linewidth}
    \vspace{0pt}\centering
    \includegraphics[width=0.85\linewidth,height=24mm,keepaspectratio]{#1}\\[2pt]
    {\scriptsize #2}
  \end{minipage}\hfill
  \begin{minipage}[t]{0.80\linewidth}
    \vspace{0pt}#3
  \end{minipage}\par\smallskip
  {\scriptsize\color{hfInk!75}#4\par}\smallskip}

\hypersetup{
  colorlinks=true,
  linkcolor=hfTeal,
  filecolor=hfPurple,
  urlcolor=hfTeal,
  citecolor=hfPurple,
  pdftitle={Many Electrons, One Determinant: Hartree--Fock Theory, Basis Sets, and Molecular Integrals},
  pdfauthor={Pierre-Fran\c{c}ois Loos}
}
\newcommand{\doi}[1]{doi:\ \href{https://doi.org/#1}{\nolinkurl{#1}}}

\titleformat{\section}
  {\normalfont\Large\bfseries\color{hfGreenDark}}
  {\thesection}{0.75em}{}
\titleformat{\subsection}
  {\normalfont\large\bfseries\color{hfNavy}}
  {\thesubsection}{0.75em}{}
\titleformat{\subsubsection}
  {\normalfont\normalsize\bfseries\color{hfNavy}}
  {\thesubsubsection}{0.75em}{}

\fancypagestyle{plain}{
  \fancyhf{}
  \fancyfoot[C]{\thepage}
  
}

\numberwithin{equation}{section}

\newcommand{\Ec}{E_\text{c}}
\newcommand{\EHF}{E_\text{HF}}

\newcommand{\ba}{\boldsymbol{a}}
\newcommand{\bb}{\boldsymbol{b}}
\newcommand{\bc}{\boldsymbol{c}}
\newcommand{\bd}{\boldsymbol{d}}
\newcommand{\be}{\boldsymbol{e}}
\newcommand{\bff}{\boldsymbol{f}}

\newcommand{\bk}{\boldsymbol{k}}

\newcommand{\br}{\boldsymbol{r}}
\newcommand{\bs}{\boldsymbol{s}}
\newcommand{\bx}{\boldsymbol{x}}
\newcommand{\bA}{\boldsymbol{A}}
\newcommand{\bB}{\boldsymbol{B}}
\newcommand{\bC}{\boldsymbol{C}}
\newcommand{\bD}{\boldsymbol{D}}

\newcommand{\bF}{\boldsymbol{F}}
\newcommand{\bG}{\boldsymbol{G}}
\newcommand{\bH}{\boldsymbol{H}}
\newcommand{\bI}{\boldsymbol{1}}
\newcommand{\bJ}{\boldsymbol{J}}
\newcommand{\bK}{\boldsymbol{K}}
\newcommand{\bM}{\boldsymbol{M}}
\newcommand{\bO}{\boldsymbol{0}}
\newcommand{\bP}{\boldsymbol{P}}
\newcommand{\bQ}{\boldsymbol{Q}}
\newcommand{\bR}{\boldsymbol{R}}
\newcommand{\bS}{\boldsymbol{S}}

\newcommand{\bU}{\boldsymbol{U}}
\newcommand{\bV}{\boldsymbol{V}}
\newcommand{\bX}{\boldsymbol{X}}
\newcommand{\beps}{\boldsymbol{\epsilon}}

\newcommand{\hh}{\Hat{h}}
\newcommand{\hf}{\Hat{f}}

\newcommand{\chA}{\Hat{\mathcal{A}}}
\newcommand{\chO}{\Hat{\mathcal{O}}}
\newcommand{\chJ}{\Hat{\mathcal{J}}}
\newcommand{\chK}{\Hat{\mathcal{K}}}
\newcommand{\chH}{\Hat{\mathcal{H}}}
\newcommand{\chT}{\Hat{\mathcal{T}}}
\newcommand{\chV}{\Hat{\mathcal{V}}}
\newcommand{\chP}{\Hat{\mathcal{P}}}
\newcommand{\chS}{\Hat{\mathcal{S}}}

\newcommand{\cE}{\mathcal{E}}
\newcommand{\cJ}{\mathcal{J}}
\newcommand{\cK}{\mathcal{K}}

\newcommand{\purple}[1]{\textcolor{hfPurple}{#1}}
\newcommand{\red}[1]{\textcolor{hfRed}{#1}}
\newcommand{\orange}[1]{\textcolor{hfOrange}{#1}}
\newcommand{\green}[1]{\textcolor{hfGreen}{#1}}
\newcommand{\blue}[1]{\textcolor{hfBlue}{#1}}
\newcommand{\violet}[1]{\textcolor{hfViolet}{#1}}

\DeclarePairedDelimiter{\sbra}{[}{|}
\DeclarePairedDelimiter{\sket}{|}{]}

\DeclarePairedDelimiterX{\sbraket}[2]{[}{]}{%
  #1\,\delimsize|\,#2%
}

\DeclarePairedDelimiterX{\smel}[3]{[}{]}{%
  #1\,\delimsize|\,#2\,\delimsize|\,#3%
}

\DeclarePairedDelimiter{\rbra}{(}{|}

\DeclarePairedDelimiterX{\rbraket}[2]{(}{)}{%
  #1\,\delimsize|\,#2%
}

\DeclarePairedDelimiterX{\rmel}[3]{(}{)}{%
  #1\,\delimsize|\,#2\,\delimsize|\,#3%
}

\newcommand{\ii}{\mathrm{i}}
\newcommand{\cre}[1]{\Hat{a}_{#1}^\dag}
\newcommand{\ani}[1]{\Hat{a}_{#1}}

\newcommand{\HtwoOrbitalPlot}[1]{%
  \begin{tikzpicture}
    \begin{axis}[
      width=\linewidth,
      height=0.44\textheight,
      xmin=-3,xmax=3,
      ymin=-0.05,ymax=1.48,
      axis x line=middle,
      axis y line=none,
      xtick=\empty,
      ytick=\empty,
      clip=false,
      samples=120,
      domain=-3:3,
    ]
      \addplot[hfBlue,very thick]
        {cos(#1)*(exp(-1.7*(x+1.15)^2)+exp(-1.7*(x-1.15)^2))/sqrt(2)
        +sin(#1)*(exp(-1.7*(x+1.15)^2)-exp(-1.7*(x-1.15)^2))/sqrt(2)};
      \addplot[hfRed,very thick,dashed]
        {cos(#1)*(exp(-1.7*(x+1.15)^2)+exp(-1.7*(x-1.15)^2))/sqrt(2)
        -sin(#1)*(exp(-1.7*(x+1.15)^2)-exp(-1.7*(x-1.15)^2))/sqrt(2)};
      \addplot[only marks,mark=*,mark size=1.7pt,hfInk]
        coordinates {(-1.15,0) (1.15,0)};
      \node[below,font=\scriptsize] at (axis cs:-1.15,0) {$A$};
      \node[below,font=\scriptsize] at (axis cs: 1.15,0) {$B$};
      \path (axis cs:-3,-0.05) rectangle (axis cs:3,1.05);
    \end{axis}
  \end{tikzpicture}%
}

\tikzset{snake it/.style={
  decoration={snake,amplitude=.4mm,segment length=2mm},decorate
}}

\algnewcommand\algorithmicassert{\texttt{assert}}
\algnewcommand\Assert[1]{\State \algorithmicassert(#1)}

\title{\color{hfGreenDark}Many Electrons, One Determinant\\[0.5em]
  \large Hartree--Fock Theory, Basis Sets, and Molecular Integrals}
\author{
	Pierre-Fran\c{c}ois (Titou) Loos
	\bigskip
	\\
	\small Laboratoire de Chimie et Physique Quantiques (UMR 5626), \\
	\small Universit\'e de Toulouse, CNRS, UPS, Toulouse, France
}
\date{European Summerschool in Quantum Chemistry (ESQC) 2026}
\renewcommand{\maketitle}{
\begin{titlepage}
  \thispagestyle{empty}
  \begin{tikzpicture}[remember picture,overlay]
    \fill[hfGreenDark]
      (current page.north west) rectangle
      ([xshift=9mm]current page.south west);

    \node[anchor=north west,inner sep=0pt,text=hfGreenDark]
      at ([xshift=30mm,yshift=-27mm]current page.north west)
      {\sffamily\small\bfseries QUANTUM CHEMISTRY};
    \node[anchor=north east,inner sep=0pt,align=right,text=hfGreenDark]
      at ([xshift=-30mm,yshift=-24mm]current page.north east)
      {\sffamily\bfseries\fontsize{22}{24}\selectfont ESQC 2026};
    \draw[hfGreenDark,line width=0.7pt]
      ([xshift=30mm,yshift=-47mm]current page.north west) --
      ([xshift=-30mm,yshift=-47mm]current page.north east);

    \node[anchor=north west,inner sep=0pt,text=hfGreenDark]
      at ([xshift=30mm,yshift=-64mm]current page.north west)
      {\begin{minipage}{150mm}
        \raggedright
        {\fontsize{38}{43}\selectfont\bfseries
          Many Electrons,\\One Determinant\par}
        \vspace{7mm}
        {\fontsize{23}{29}\selectfont
          Hartree--Fock Theory,\\Basis Sets, and Molecular Integrals\par}
      \end{minipage}};

    \node[anchor=center,inner sep=0pt]
      at ([xshift=105mm,yshift=-199mm]current page.north west)
      {\includegraphics[width=190mm]{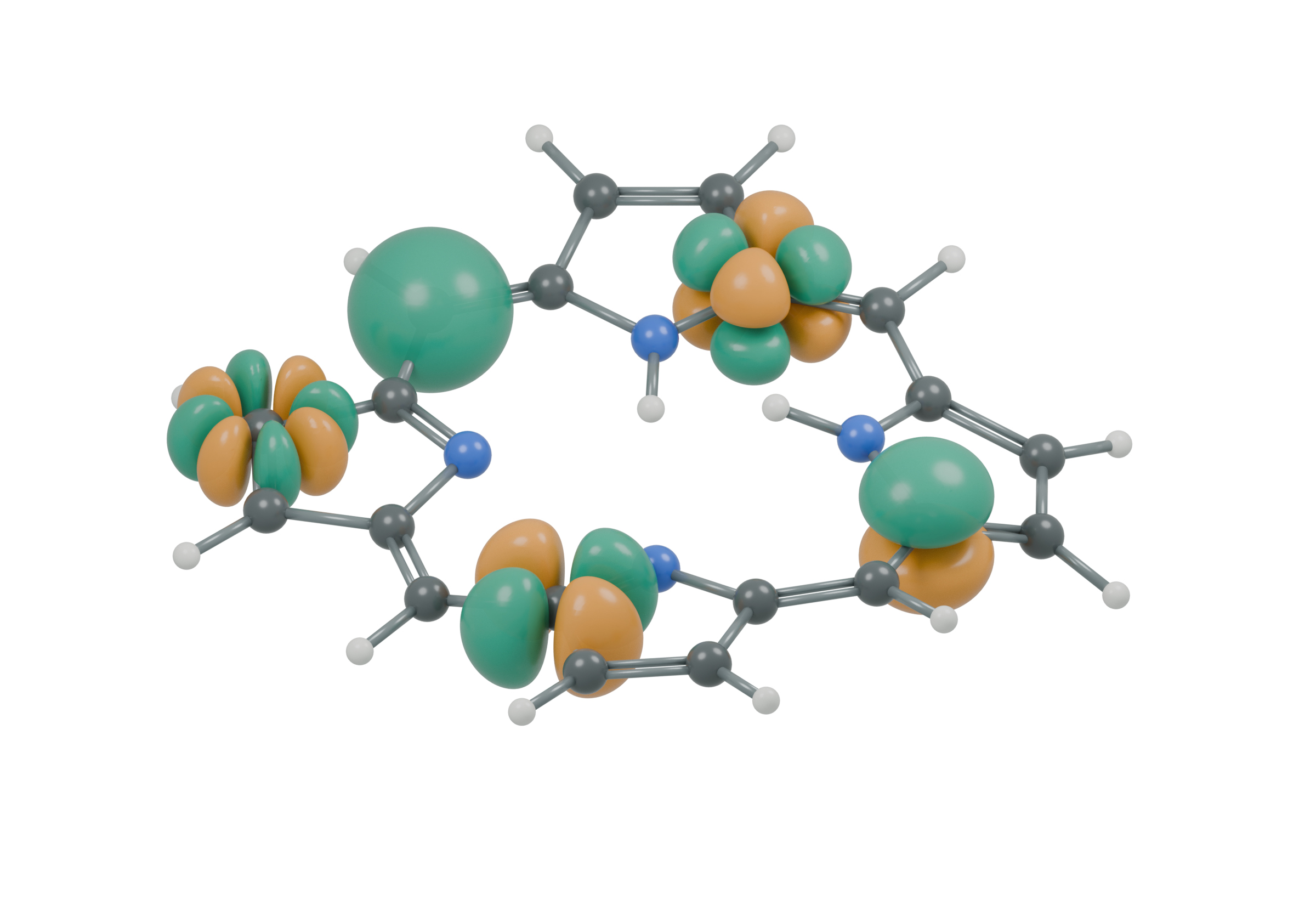}};

    \node[anchor=north west,inner sep=0pt,text=hfGreenDark]
      at ([xshift=30mm,yshift=-243mm]current page.north west)
      {\begin{minipage}{150mm}
        \raggedright
        {\large\bfseries Pierre-Fran\c{c}ois (Titou) Loos\par}
        \vspace{3mm}
        {\small\color{hfInk}
          Laboratoire de Chimie et Physique Quantiques (UMR 5626)\\
          Universit\'e de Toulouse, CNRS, UPS, Toulouse, France\par}
      \end{minipage}};
    \draw[hfGreen!35,line width=0.5pt]
      ([xshift=30mm,yshift=22mm]current page.south west) --
      ([xshift=-30mm,yshift=22mm]current page.south east);
    \node[anchor=south west,inner sep=0pt,text=hfGreenDark]
      at ([xshift=30mm,yshift=14mm]current page.south west)
      {\sffamily\footnotesize European Summer School in Quantum Chemistry (ESQC)};
  \end{tikzpicture}
  \null
\end{titlepage}

}

\begin{document}

\hypersetup{pageanchor=false}
\maketitle
\pagenumbering{roman}
\hypersetup{pageanchor=true}
\begin{hfabstract}
This chapter accompanies the lectures on Hartree--Fock theory and on basis
sets and molecular integrals at the \href{https://www.esqc.org}{European Summer School of Quantum
Chemistry (ESQC) 2026}. Intended for doctoral students, it develops the
single-determinant description of electronic structure from antisymmetry and
the variational principle to the orbital equations and their finite-basis
matrix representation. Detailed derivations connect the energy functional,
the Fock operator, excited determinants, and the interpretation of orbital
energies. Restricted, unrestricted, and generalized formulations show how
the choice of variational space affects spin symmetry, symmetry breaking,
and the limitations of the Hartree--Fock approximation.

The computational development covers orthogonalization, self-consistent-field
iteration, convergence acceleration, overlap-based occupation selection,
and orbital analysis, distinguishing orbital stationarity from stability.
A connection to second-order M{\o}ller--Plesset perturbation theory explains
the correlation-energy expression and the cost of transforming
electron-repulsion integrals from atomic to molecular orbitals.

The basis-set and integral development examines radial and angular
flexibility, contraction, basis-set incompleteness, and superposition error.
The Gaussian product theorem leads to one- and two-electron integrals,
recurrence relations, integral screening, and low-rank representations.
Worked examples and exercises with detailed solutions connect these
derivations to practical calculations. Historical perspectives and explicit
notation conventions provide context for the approximations and algorithms
that underpin molecular electronic-structure theory.
\end{hfabstract}
\clearpage
\tableofcontents
\clearpage
\listofhistoricalperspectives
\clearpage
\phantomsection
\section*{Notation and conventions}
\addcontentsline{toc}{section}{Notation and conventions}

Table~\ref{tab:notation} summarizes the chapter's symbols and acronyms.
Atomic units are used unless stated otherwise. Read each
orbital sum with its spin-orbital or spatial-orbital convention, and each
density matrix with its occupation normalization; local reuses are identified
by topic. Acronyms and abbreviations are collected alphabetically at the end
of the table.

\begingroup
\hftablestyle
\setlength{\LTleft}{0pt}
\setlength{\LTright}{0pt}
\begin{longtable}{@{}p{0.26\linewidth}p{\dimexpr0.74\linewidth-2\tabcolsep\relax}@{}}
  \caption{Symbols and acronyms, with formulation-specific local conventions.}
  \label{tab:notation}\\
  \toprule
  \hftablehead{Symbol or acronym} &
  \hftablehead{Meaning and convention}\\
  \midrule
  \endfirsthead
  \caption[]{Symbols, acronyms, and local conventions (continued).}\\
  \toprule
  \hftablehead{Symbol or acronym} &
  \hftablehead{Meaning and convention}\\
  \midrule
  \endhead
  \midrule
  \multicolumn{2}{r@{}}{\footnotesize Continued on the next page}\\
  \endfoot
  \bottomrule
  \endlastfoot

  \rowcolor{hfGreenLight}
  \multicolumn{2}{@{}l@{}}{\strut\textcolor{hfGreenDark}{\textbf{Particles, coordinates, and orbital indices}}}\\*
  $N$, $K$ & Electron number; spatial basis size ($2K$ spin orbitals).\\
  $N^\alpha$, $N^\beta$ & Collinear spin populations, with
  $N^\alpha+N^\beta=N$. Superscript spin labels are used throughout;
  $N^\sigma$ denotes either population for $\sigma=\alpha,\beta$.\\
  $O$, $V$ & Closed-shell MP2 spatial counts: $O=N/2$ occupied and
  $V=K-O$ virtual orbitals.\\
  $\br$, $\bx$, $\omega$ & Spatial coordinate, combined coordinate
  $\bx=(\omega,\br)$, and spin coordinate. Integration over $\bx$
  includes the spin degree of freedom.\\
  $A,B$, $Z_A$, $\bR_A$ & Nuclear labels, charges, and positions;
  $M$ counts nuclei.\\
  $\phi_\mu$, $\psi_p$, $\chi_p$ & Spatial AO, spatial MO, and spin orbital.
  $\chi_\mu^\text{AO}$ is a spin-AO basis function in the general AO exercises.\\
  $\Bar{\psi}_p$ & Shorthand for the $\beta$ spin partner, not conjugation
  (denoted by a star).\\
  $\mu,\nu,\lambda,\sigma$ & AO indices, normally running over $K$ spatial
  functions. In the general spin-orbital AO Fock/energy exercises, the
  basis label temporarily includes spin.\\
  $p,q,r,s$ & General spatial- or spin-orbital indices, as specified locally.\\
  $i,j,k,l$; $a,b,c,d$ & Occupied and virtual indices, relative to the chosen
  reference. Spin-orbital sums contain $N$ occupied functions; closed-shell
  spatial sums contain $N/2$.\\
  $\alpha$, $\beta$; $S$, $M_S$ & The two spin functions; total-spin and
  spin-projection quantum numbers. $\chS^2$ and $\chS_z$ are the associated
  operators.\\

  \rowcolor{hfGreenLight}
  \multicolumn{2}{@{}l@{}}{\strut\textcolor{hfGreenDark}{\textbf{Operators, states, and energies}}}\\*
  $\chH_\text{elec}$, $\hh$, $\hf$ & Electronic Hamiltonian, one-electron
  core Hamiltonian, and density-dependent Fock operator.\\
  $\chJ_i$, $\chK_i$ & Coulomb and exchange operators generated by occupied
  orbital $i$. Their spatial versions follow the spin integration.\\
  $\Psi_\text{HF}$, $\Psi_0$ & HF determinant; the same determinant when
  used as the reference for substitutions or perturbation theory.\\
  $\chA$, $\chP_\pi$ & Antisymmetrizer weighted by $1/\sqrt{N!}$ and
  electron permutation operator. This $\chA$ is not an idempotent
  $1/N!$-weighted projector; Eq.~\eqref{eq:hf-determinant}.\\
  $\Psi_i^a$, $\Psi_{ij}^{ab}$ & Singly and doubly substituted determinants.
  Their phases follow the stated ordering of creation and annihilation
  operators, $\cre{p}$ and $\ani{p}$.\\
  $\epsilon_{ji}$, $\beps$ & Hermitian occupied-space orthonormality
  multipliers before canonicalization; diagonalized by an occupied-space
  unitary transformation.\\
  $\epsilon_p$, $\beps$ & Canonical orbital energy and diagonal matrix of
  orbital energies. The matrix spans $K$ spatial orbitals in RHF, or $2K$
  spinors in GHF; $\beps^\sigma$ denotes one UHF spin sector.
  $\beps_\text{o}$ and $\beps_\text{v}$ denote occupied and virtual blocks.\\
  $\EHF$, $U_I$ & Electronic HF energy; PES including nuclear repulsion.
  H$_2$ formulas specify $1/R$. $E_\text{H}$ means the Hartree contribution
  in the stationarity exercise or atomic H energy in the dissociation example.\\
  $\mathrm{IP}$, $\mathrm{EA}$ & Removal energy $E^{N-1}-E^N$; attachment
  energy $E^N-E^{N+1}$. Koopmans comparisons fix geometry and basis.\\
  $\widetilde E$, $E_\text{corr}^M$ & A tilde on an ionic energy denotes
  frozen orbitals in the Koopmans discussion; $E_\text{corr}^M$ is the exact
  minus optimized HF energy for electron number $M$.\\
  $\chH^{(0)}$, $\chV$ & MP zeroth-order Hamiltonian $\sum_i\hf(i)$ and
  perturbation; Eq.~\eqref{eq:mp-partition}.\\
  $\Ec^{(2)}$, $\Delta_{ij}^{ab}$ & MP2 correlation energy and denominator
  $\epsilon_i+\epsilon_j-\epsilon_a-\epsilon_b$, negative for the gapped
  Aufbau reference.\\

  \rowcolor{hfGreenLight}
  \multicolumn{2}{@{}l@{}}{\strut\textcolor{hfGreenDark}{\textbf{Integral brackets and complex conjugation}}}\\*
  $\braket{\chi_i}{\chi_j}$ & Ordinary one-particle overlap; orthonormal
  spin orbitals give $\delta_{ij}$.\\
  $\braket{ij}{kl}$ & Physicists' Coulomb integral: electron 1 carries
  $i,k$ and electron 2 carries $j,l$; Eq.~\eqref{eq:eri-physicists}.\\
  $\rbraket{ij}{kl}$ & Chemists' Coulomb integral: electron 1 carries
  $i,j$ and electron 2 carries $k,l$. Thus
  $\braket{ij}{kl}=\rbraket{ik}{jl}$; Eq.~\eqref{eq:eri-chemists}.\\
  $\mel{ij}{}{kl}$ & Antisymmetrized integral
  $\braket{ij}{kl}-\braket{ij}{lk}$;
  Eq.~\eqref{eq:antisymmetrized-eri}.\\
  $h_{ij}$, $\cJ_{ij}$, $\cK_{ij}$ & Spin-orbital core, Coulomb, and
  exchange matrix elements. Spatial versions use the indicated spatial
  orbitals and spin-summation factors.\\
  $*$, $\dag$, $\cdot$ & Complex conjugation, Hermitian adjoint, and explicit
  matrix multiplication. Bra arguments are conjugated on forming the bra.
  Eightfold ERI symmetry requires real orbitals.\\

  \rowcolor{hfGreenLight}
  \multicolumn{2}{@{}l@{}}{\strut\textcolor{hfGreenDark}{\textbf{AO matrices and density normalizations}}}\\*
  $C_{\mu p}$, $\bC$, $\bC_\text{o}$ & MO coefficient, coefficient matrix,
  and its occupied columns. The spatial expansion is
  $\psi_p=\sum_\mu C_{\mu p}\phi_\mu$.\\
  $\bS$, $\bX$ & AO overlap metric and orthogonalizer, satisfying
  $\bX^\dag\cdot\bS\cdot\bX=\bI$ in the retained space.\\
  $\bH$, $\bF$, $\bG[\bP]$ & AO core-Hamiltonian, Fock, and two-electron
  matrices, with $\bF=\bH+\bG[\bP]$, including in GHF.\\
  $f_{pq}$, $\bff$ & Fock matrix elements and blocks in the MO basis;
  distinct from the AO representation $F_{\mu\nu}$ and $\bF$.\\
  $\bJ$, $\bK$ & Coulomb and exchange matrices, with local density/spin
  prefactors; distinct from orbital operators $\chJ_i$ and $\chK_i$.\\
  $\bP$ (RHF) & Spin-summed spatial density,
  $\bP=2\bC_\text{o}\cdot\bC_\text{o}^\dag$ and
  $\Tr(\bP\cdot\bS)=N$. A spin-orbital density counts each occupied
  spin orbital once instead.\\
  $\bP^\alpha$, $\bP^\beta$ (UHF) & Separate single-spin densities,
  $\Tr(\bP^\sigma\cdot\bS)=N^\sigma$ for
  $\sigma=\alpha,\beta$.\\
  $\bP^\text{T}$, $\bP^\text{S}$ (UHF) & Charge and collinear spin-density
  matrices: the sum and difference of $\bP^\alpha$ and $\bP^\beta$.\\
  $\boldsymbol{\mathcal P}$ (GHF) & Full $2K\times2K$ spinor density, with
  blocks $P_{\mu\nu}^{\sigma\sigma'}=\sum_i C_{\mu i}^\sigma
  C_{\nu i}^{\sigma' *}$. Spin is the outer block index.\\
  $\bP$, $\bM_x$, $\bM_y$, $\bM_z$ (GHF) & In the Pauli decomposition,
  $\bP=\tfrac12(\bP^{\alpha\alpha}+\bP^{\beta\beta})$ is
  \emph{half} the charge-density matrix; $\bM_k$ are the magnetization blocks.
  This $\bP$ is not the spin-summed RHF/UHF density.\\
  $\boldsymbol{\mathcal C}$, $\boldsymbol{\mathcal S}$ & Stacked spinor
  coefficients and block-diagonal overlap metric in GHF;
  $\Tr(\boldsymbol{\mathcal P}\cdot\boldsymbol{\mathcal S})=N$.\\
  $\rho(\br)$, $\rho^\sigma(\br)$ & Real-space total and single-spin
  electron densities, integrating to $N$ and $N^\sigma$, respectively.\\

  \rowcolor{hfGreenLight}
  \multicolumn{2}{@{}l@{}}{\strut\textcolor{hfGreenDark}{\textbf{Iteration, orbital transformations, and numerical parameters}}}\\*
  $\bU$, $\kappa_{ai}$ & Unitary orbital transformation and infinitesimal
  occupied-virtual rotation parameter. Occupied-only rotations preserve
  the determinant's physical state, up to a phase.\\
  $\be^{(n)}$, $\tau_\text{SCF}$ & Commutator residual at SCF iteration $n$
  and its convergence threshold. A zero residual establishes stationarity,
  not stability.\\
  $m$, $c_i$, $B_{ij}$ (DIIS) & History size, real weights
  ($\sum_i c_i=1$), and residual inner products. Here $i,j$ label stored
  iterations, not occupied orbitals.\\
  $w_p$ (MOM) & Overlap weight used to select the occupied space relative
  to the chosen reference. The MOM matrix $\boldsymbol{O}$ contains reference-to-new
  overlaps.\\
  $I$, $J$ (AO-to-MO) & Temporary four-index arrays in the successive
  one-index transformations; not identity or Coulomb operators.\\

  \rowcolor{hfGreenLight}
  \multicolumn{2}{@{}l@{}}{\strut\textcolor{hfGreenDark}{\textbf{Basis functions and Gaussian integrals}}}\\*
  $\bk$, $\Omega$, $E_\text{cut}$ & Plane-wave vector, simulation-cell
  volume, and kinetic-energy cutoff.\\
  $\alpha,\beta,\gamma,\delta$ & Positive primitive Gaussian exponents
  in the integral sections, rather than spin labels. The combined exponents
  are $\zeta=\alpha+\beta$ and $\eta=\gamma+\delta$.\\
  $\zeta$, $n$, $\ell$, $m$ (STO/GTO) & In the radial/angular comparison,
  $\zeta$ is an STO exponent, $n$ labels the radial form, and $\ell,m$ are
  angular-momentum labels. This use of $\zeta$ precedes its Gaussian-pair use.\\
  $\ba=(a_x,a_y,a_z)$ & Cartesian polynomial powers, with total angular
  momentum $a=a_x+a_y+a_z$; not a virtual orbital. The same convention
  applies to $\bb,\bc,\bd$.\\
  $\sket{\ba}$, $\ket{\ba}$ & Unnormalized primitive and contracted
  Cartesian Gaussian. Square and round integral brackets distinguish
  primitive and contracted Gaussian integrals in this context.\\
  $D_k$, $K_a$ & Contraction coefficient and number of primitives in
  contraction $a$. Primitive normalization factors are absorbed consistently;
  the local upper limit $K$ in a contraction also counts primitives.\\
  $\bA,\bB,\bC,\bD$ & Gaussian centers, not coefficient matrices or
  Cartesian angular-momentum vectors.\\
  $\bP$, $\bQ$, $K_{\zeta,\bP}$ & Gaussian product centers and a
  separation-dependent product prefactor. Here $\bP$ is a position, not a density matrix.\\
  $\vartheta$, $T$, $F_m(T)$ & Reduced pair exponent
  $\vartheta=\zeta\eta/(\zeta+\eta)$, dimensionless Boys argument
  $T=\vartheta\abs{\bP-\bQ}^2$, and Boys function of order $m$;
  Section~\ref{sec:boys-functions}.\\
  $\bO$, $\bI_i$ (recurrences) & Zero angular-momentum vector (an $s$
  function) and Cartesian unit vector. Here $i$ labels $x,y,z$.
  In block matrices $\bO$ denotes zero;
  $\bI$ denotes an identity matrix.\\
  $Q_{ab}$, $\tau$, $K_\text{sig}$ & Shell-pair screening bound, integral
  threshold, and number of retained significant integrals;
  Section~\ref{sec:integral-screening}.\\
  $Q_A$, $N_A^\text{core}$ & ECP ionic-core charge and removed core-electron
  count, with $Q_A=Z_A-N_A^\text{core}$; not a population-analysis charge $q_A$.\\

  \rowcolor{hfGreenLight}
  \multicolumn{2}{@{}l@{}}{\strut\textcolor{hfGreenDark}{\textbf{Auxiliary functions, fitting, and other local reuses}}}\\*
  $\varphi_A^\text{aux}$, $K_\text{aux}$ & Auxiliary fitting function and
  auxiliary-basis size in RI/DF. Here $A,B$ are auxiliary-function indices,
  not nuclear labels.\\
  $\rho_{\mu\nu}$, $C_{\mu\nu}^A$ & AO pair density and its auxiliary
  expansion coefficients in the real-function RI derivation; the latter
  are not MO coefficients $C_{\mu p}$.\\
  $V_{AB}$, $\bV$ & Auxiliary Coulomb metric $V_{AB}=\rbraket{A}{B}$ in
  RI. In the Cholesky discussion $\bV$ instead denotes the full ERI matrix
  in compound AO-pair indices.\\
  $B_{\mu\nu}^P$, $L_{\mu\nu}^P$ & Three-index RI and Cholesky factors.
  Here $P$ labels an auxiliary direction, not a density matrix or Gaussian center.\\
  $N_\text{CD}$, $\tau_\text{CD}$ & Number of Cholesky factors and
  decomposition threshold; Section~\ref{sec:ri}.\\
  $S$, $\bS$ & Scalar atomic overlap in the H$_2$ example versus AO
  overlap matrix; the scalar $S$ elsewhere denotes total spin.\\
  $K$, $\bK$, $\chK$ & Basis size, exchange matrix, and complex-conjugation
  operator in the HF classification table. These are distinct local uses.\\
  $\Omega_\text{ER}$, $\Omega_\text{BF}$, $\Omega_\text{PM}$ & Localization
  criteria, distinct from the plane-wave volume $\Omega$; their minimization
  or maximization convention is given with each functional.\\

  \rowcolor{hfGreenLight}
  \multicolumn{2}{@{}l@{}}{\strut\textcolor{hfGreenDark}{\textbf{Acronyms and abbreviations}}}\\*
  AO & Atomic orbital.\\
  ASCW & Axial spin current wave.\\
  ASDW & Axial spin density wave.\\
  ASW & Axial spin wave.\\
  BF & Foster--Boys localization, labeled BF in $\Omega_\text{BF}$.\\
  BSSE & Basis-set superposition error.\\
  CBS & Complete basis set.\\
  CCW & Charge current wave.\\
  CD & Cholesky decomposition.\\
  CF & Coulson--Fischer.\\
  CP & Counterpoise correction.\\
  DF & Density fitting; also called RI.\\
  DIIS & Direct inversion in the iterative subspace.\\
  EA & Electron affinity.\\
  ECP & Effective core potential.\\
  ER & Edmiston--Ruedenberg localization.\\
  ERI & Electron-repulsion integral.\\
  ESQC & European Summer School of Quantum Chemistry.\\
  GHF & Generalized Hartree--Fock.\\
  GTO & Gaussian-type orbital.\\
  HF & Hartree--Fock.\\
  HFB & Hartree--Fock--Bogoliubov.\\
  HOMO & Highest occupied molecular orbital.\\
  HRR & Horizontal recurrence relation.\\
  IMOM & Initial maximum-overlap method, with a fixed initial reference.\\
  IP & Ionization potential.\\
  LCAO & Linear combination of atomic orbitals.\\
  LUMO & Lowest unoccupied molecular orbital.\\
  MO & Molecular orbital.\\
  MOM & Maximum-overlap method, with the previous iteration as reference.\\
  MP & M{\o}ller--Plesset perturbation theory.\\
  MP2 & Second-order M{\o}ller--Plesset perturbation theory.\\
  NAO & Numerical atomic orbital.\\
  PES & Potential-energy surface.\\
  PM & Pipek--Mezey localization.\\
  PW & Plane wave.\\
  RHF & Restricted Hartree--Fock.\\
  RI & Resolution of the identity; also called DF.\\
  ROHF & Restricted open-shell Hartree--Fock.\\
  SAD & Superposition of atomic densities.\\
  SCF & Self-consistent field.\\
  STO & Slater-type orbital.\\
  TICS & Time-reversal-invariant closed-shell.\\
  TSCW & Torsional spin current wave.\\
  TSDW & Torsional spin density wave.\\
  TSW & Torsional spin wave.\\
  UHF & Unrestricted Hartree--Fock.\\
  VRR & Vertical recurrence relation.\\
\end{longtable}
\endgroup

\clearpage
\pagenumbering{arabic}


\section{Electronic structure and the Hartree--Fock framework}
\label{sec:hf-framework}

Predicting electronic structure from first principles is one of the central
problems of quantum chemistry. Once the electronic structure of a molecular
system is known, one can in principle determine its energy and rationalize a
broad range of chemical phenomena, from bonding and molecular structure to
reactivity and spectroscopy. At the non-relativistic level, the fundamental
equations governing this problem are known. The difficulty lies not in writing
down the Hamiltonian, but in solving the resulting many-electron
Schr\"odinger equation.

The origin of this difficulty is the electron-electron interaction. Because
the Coulomb repulsion couples the motion of all electrons, the electronic
wavefunction depends simultaneously on all their coordinates and cannot, in
general, be decomposed into independent one-electron contributions. The
problem is therefore intrinsically many-body. Exact solutions are restricted
to exceptionally small systems, and practical electronic structure theory
must rely on approximations that retain the essential physics while reducing
the complexity of the many-electron problem.

\begin{figure}[htbp]
  \centering
  \includegraphics[width=\linewidth]{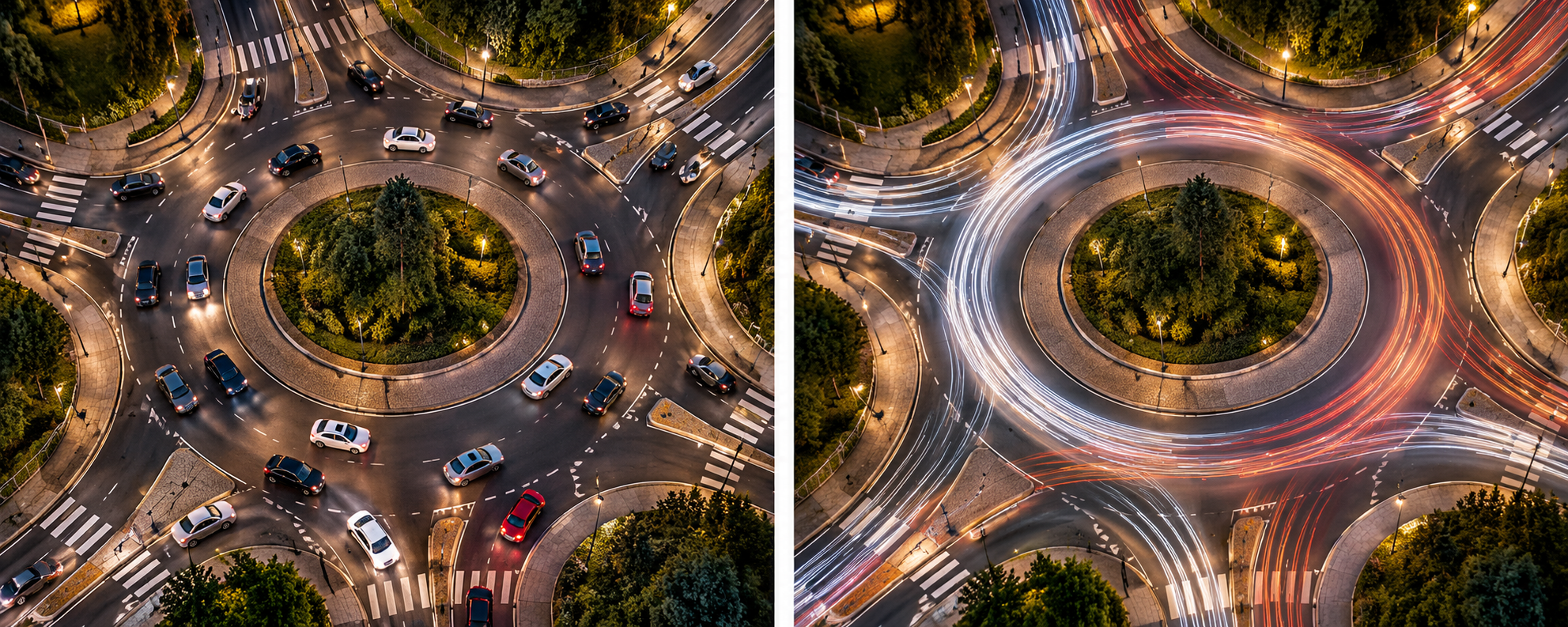}
  \caption{Illustration of the mean-field approximation using traffic on a roundabout. In the left panel, a short-exposure snapshot resolves the instantaneous positions of the individual cars, representing the fully interacting many-particle picture in which the configuration of all particles is explicitly retained. In the right panel, a long-exposure image of the same roundabout replaces the individual cars by smooth light trails, representing the mean-field description: the detailed instantaneous motion of the other particles is no longer followed explicitly, and each particle instead experiences an effective field generated by their average distribution. The comparison emphasizes that the mean field is an averaged environment rather than a blurred or uncertain version of the exact configuration.}
  \label{fig:meanfield}
\end{figure}

Hartree--Fock (HF) theory provides the most fundamental such reduction. Its
central idea is to replace the exact many-electron wavefunction by the best
wavefunction that can be written as a single Slater determinant. This
restriction transforms the interacting many-electron problem into an
effective independent-particle picture: each electron moves in a field
generated by the nuclei and by the average distribution of the other
electrons (see Fig.~\ref{fig:meanfield}).
Importantly, the antisymmetry of the electronic wavefunction is
retained exactly within this approximation. The Pauli principle and the
associated exchange interaction therefore emerge naturally, rather than being
introduced as additional corrections
\citep{Hartree1928,Slater1929,Fock1930}.

Historically, Hartree's self-consistent-field construction dates from 1928.
The incorporation of exchange through antisymmetric wavefunctions was
developed independently by Fock and Slater in 1930
\citep{Fock1930,Slater1930}.

\hfBioHartree
\hfBioWilliamHartree
\hfBioFock

The price of this simplification is equally fundamental. A single determinant
cannot describe the full correlated motion of interacting electrons.
Hartree--Fock theory accounts for exchange exactly within its
single-determinant framework, but neglects electron correlation beyond
exchange. Nevertheless, Hartree--Fock occupies a unique position in electronic structure
theory. It provides a remarkably useful orbital picture of atoms and
molecules, establishes the language of occupied and virtual molecular
orbitals, and serves as the reference state for a large fraction of more
advanced quantum-chemical methods. Understanding both its strengths and its
limitations is therefore essential before moving to correlated theories.

There is, however, a second layer to the problem. The Hartree--Fock equations
are formulated in terms of continuous one-electron functions, whereas an
actual calculation requires a finite numerical representation of these
orbitals. In molecular quantum chemistry, this is usually achieved by
expanding the molecular orbitals in a finite set of basis functions. The
choice of basis determines the flexibility with which the orbitals can be
represented and has a direct impact on both accuracy and computational cost.
It also reduces the continuous Hartree--Fock equations to a matrix problem
whose elements are one- and two-electron molecular integrals. Basis sets and
molecular integrals are therefore not peripheral technical details: they are
an integral part of the practical formulation of electronic structure theory.

This chapter develops these ideas from the ground up. We first formulate the
electronic many-body problem and introduce the approximations that separate
electronic and nuclear motion. We then impose the fermionic antisymmetry of
the wavefunction, introduce Slater determinants, and derive the
Hartree--Fock energy and equations in Section~\ref{sec:hf-framework}.
Section~\ref{sec:finite-basis-hf} develops their finite-basis formulation,
the self-consistent-field procedure, and the role of orbital rotations.
Section~\ref{sec:open-shell-hf} then examines unrestricted and generalized
Hartree--Fock theory alongside the restricted description. We subsequently
turn to the basis functions themselves and examine the principal
representations used in practice (Section~\ref{sec:basis-sets}). Finally,
Sections~\ref{sec:gaussian-integrals} and~\ref{sec:ri} develop the molecular
integrals and low-rank representations required to construct the
Hartree--Fock matrices.

\hfBioSwirles

\subsection{The electronic many-body problem}

We begin by defining the framework in which Hartree--Fock theory will be
developed. We consider the time-independent Schr\"odinger equation and
seek stationary states of molecular systems. Unless stated otherwise, the
treatment is non-relativistic and atomic units are employed. The separation
between nuclear and electronic motion will be introduced through the
Born--Oppenheimer approximation in the next section.

Within this framework, Hartree--Fock is an ab initio theory. Once the
nuclear charges, nuclear positions, and number of electrons have been
specified, the electronic Hamiltonian contains no parameters adjusted to
experimental data. In practical calculations, the molecular orbitals are
represented in a finite basis, and the quality of this representation becomes
an additional approximation that will be examined in detail later in this
chapter.

Within this electronic problem, the Hartree--Fock approximation restricts the
wavefunction of $N$ electrons to a single Slater determinant and optimizes
its orbitals variationally. It does not turn off the electron-electron
interaction: Coulomb repulsion remains present, together with the exchange
contribution required by antisymmetry. Each electron experiences the nuclear
attraction and the mean field generated by the occupied orbitals. Because
that field depends on the orbitals being sought, the resulting one-electron
equations are nonlinear and must be solved self-consistently. The optimal
determinant is therefore a variational independent-particle description,
not an exact correlated solution.

To make this variational construction precise, we first need a molecular
Hamiltonian with the nuclei treated as fixed parameters. Antisymmetry then
determines the allowed determinant form, from which the energy functional
and orbital equations follow. The finite-basis and numerical formulations
will retain these assumptions while changing how the orbitals are represented.

\subsubsection{The Born--Oppenheimer approximation}

A molecule is a quantum system composed of both electrons and nuclei. Before
introducing any electronic structure approximation, the electronic and nuclear
degrees of freedom are therefore treated on equal footing
\citep{BornOppenheimer1927}. For a molecule
containing $N$ electrons and $M$ nuclei, the molecular wavefunction depends on
the electronic coordinates $\{\br_i\}$ and nuclear coordinates $\{\bR_A\}$,
and satisfies the time-independent Schr\"odinger equation
\begin{equation}
  \chH
  \Phi(\{\br_i\},\{\bR_A\})
  =
  \cE
  \Phi(\{\br_i\},\{\bR_A\})
\end{equation}
The full non-relativistic molecular Hamiltonian contains five contributions,
\begin{equation}
  \chH
  =
  \green{\chT_\text{n}}
  +
  \blue{\chT_\text{e}}
  +
  \orange{\chV_\text{ne}}
  +
  \red{\chV_\text{ee}}
  +
  \violet{\chV_\text{nn}}
  \label{eq:total-hamiltonian}
\end{equation}
Here, $\green{\chT_\text{n}}$ and $\blue{\chT_\text{e}}$ are the nuclear
and electronic kinetic-energy operators, respectively.
The operator $\orange{\chV_\text{ne}}$ describes the Coulomb attraction
between nuclei and electrons, while $\red{\chV_\text{ee}}$ and
$\violet{\chV_\text{nn}}$ describe the electron-electron and
nucleus-nucleus Coulomb repulsions.

\begin{figure}[htbp]
  \centering
  \resizebox{\linewidth}{!}{
\begingroup

\tikzset{
 nucleus/.style={circle,draw=hfGreenDark,fill=hfGreen!12,minimum size=7mm},
 electron/.style={circle,draw=hfBlue,fill=hfBlue!10,minimum size=7mm},
 old/.style={hfBlue,densely dashed,thick,-{Latex[length=2mm]}},
 new/.style={hfGreenDark,very thick,-{Latex[length=2mm]}},
 card/.style={draw=hfGreenDark,fill=hfGreen!5,rounded corners=2mm,
   align=center,minimum height=12mm},
 occupied/.style={circle,fill=hfGreenDark,inner sep=2.1pt},
 vacant/.style={circle,draw=hfInk!50,inner sep=2.1pt},
 hole/.style={circle,draw=hfRed,thick,inner sep=2.1pt},
 particle/.style={circle,fill=hfViolet,inner sep=2.1pt}}


\begin{tikzpicture}[x=8mm,y=8mm,font=\small,>=Latex]
 \foreach \shift/\title in {0/Position vectors,10/Relative vectors}{
  \begin{scope}[xshift=\shift cm]
   \node[font=\bfseries,text=hfGreenDark] at (1,6.1) {\title};
   \coordinate (O) at (0,0);
   \node[nucleus] (A) at (-2.3,2.0) {$A$};
   \node[nucleus] (B) at (2.7,1.0) {$B$};
   \node[electron] (i) at (-1.0,4.5) {$i$};
   \node[electron] (j) at (3.5,4.0) {$j$};
   \ifnum\shift=0
    \fill[hfInk] (O) circle (2pt) node[below] {$O$};
    \draw[->,hfInk!40] (O)--(4.4,0) node[right] {$y$};
    \draw[->,hfInk!40] (O)--(0,5.2) node[above] {$z$};
    \draw[->,hfInk!40] (O)--(-1.8,-1.2) node[below] {$x$};
    \draw[->,hfGreenDark,thick] (O)--(A)
      node[midway,left] {$\bR_A$};
    \draw[->,hfGreenDark,thick] (O)--(B)
      node[midway,below] {$\bR_B$};
    \draw[->,hfBlue,thick] (O)--(i)
      node[midway,left] {$\br_i$};
    \draw[->,hfBlue,thick] (O)--(j)
      node[midway,right] {$\br_j$};
   \else
    \draw[->,hfViolet,very thick] (A)--(i)
      node[midway,left] {$\br_{iA}$};
    \draw[->,hfBlue,very thick] (j)--(i)
      node[midway,above] {$\br_{ij}$};
    \draw[->,hfGreenDark,very thick] (B)--(A)
      node[midway,below] {$\bR_{AB}$};
   \fi
  \end{scope}
 }
\end{tikzpicture}
\endgroup

}
  \caption{Electronic and nuclear coordinates entering the molecular
  Hamiltonian. Green circles denote nuclei $A$ and $B$, and blue circles
  denote electrons $i$ and $j$. In the left panel, $O$ is the common origin:
  $\bR_A$ and $\bR_B$ are the nuclear position vectors, while $\br_i$ and
  $\br_j$ are the electronic position vectors. In the right panel,
  $\br_{iA}=\br_i-\bR_A$ points from nucleus $A$ to electron $i$,
  $\br_{ij}=\br_i-\br_j$ points from electron $j$ to electron $i$, and
  $\bR_{AB}=\bR_A-\bR_B$ points from nucleus $B$ to nucleus $A$.
  Their lengths $r_{iA}=\norm*{\br_{iA}}$, $r_{ij}=\norm*{\br_{ij}}$, and
  $R_{AB}=\norm*{\bR_{AB}}$ are the corresponding electron-nucleus,
  electron-electron, and internuclear distances. Both panels use the same
  molecular geometry.}
  \label{fig:molecular-coordinates}
\end{figure}
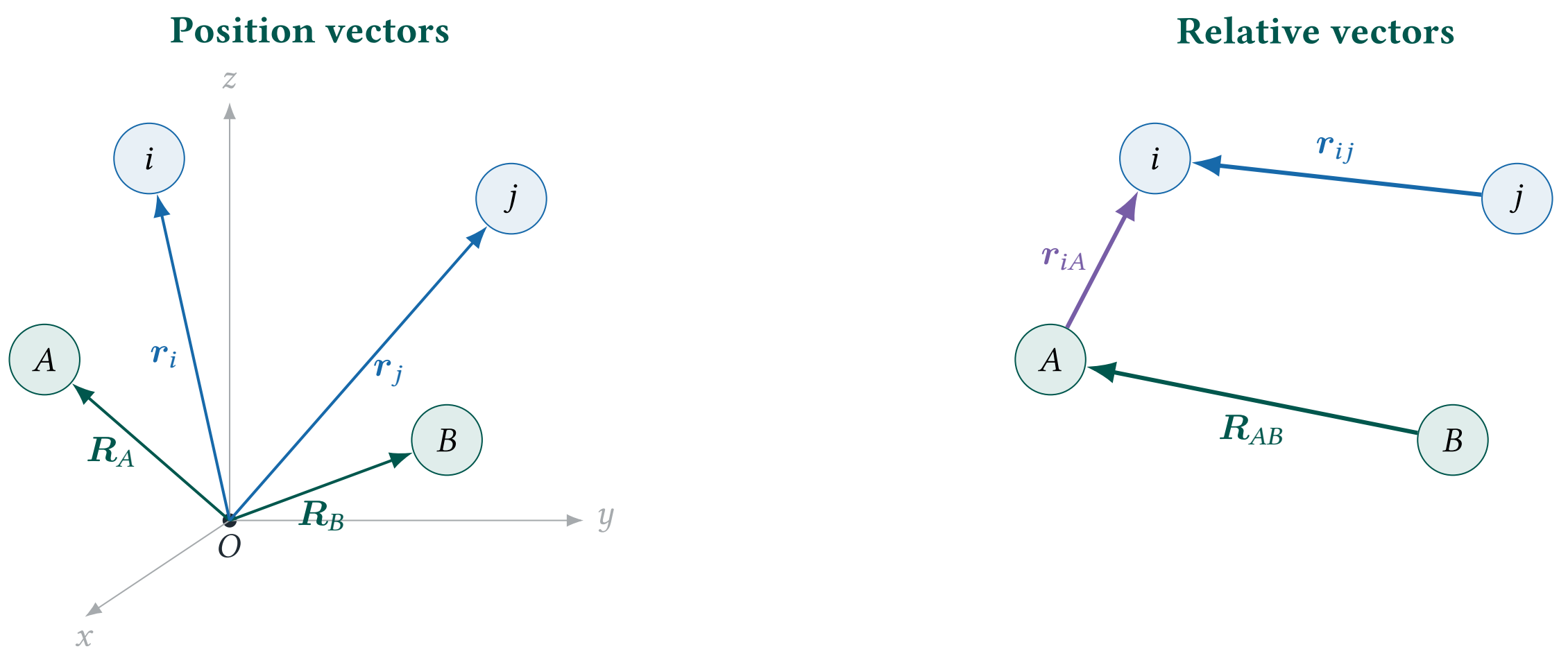

In atomic units, for which $m=e=\hbar=1$ with $m$ the electron mass,
these operators read
\begin{subequations}
\begin{align}
  \green{\chT_\text{n}}
  &=
  -\sum_{A=1}^{M}
  \frac{\laplacian_A}{2M_A}
  &
  \blue{\chT_\text{e}}
  &=
  -\sum_{i=1}^{N}
  \frac{\laplacian_i}{2}
  \\
  \orange{\chV_\text{ne}}
  &=
  -\sum_{A=1}^{M}
  \sum_{i=1}^{N}
  \frac{Z_A}{r_{iA}}
  &
  \red{\chV_\text{ee}}
  &=
  \sum_{i<j}^{N}
  \frac{1}{r_{ij}}
  &
  \violet{\chV_\text{nn}}
  &=
  \sum_{A<B}^{M}
  \frac{Z_A Z_B}{R_{AB}}
\end{align}
\end{subequations}
The quantities $M_A$ and $Z_A$ are the mass and charge of nucleus $A$,
$r_{iA}$ is the distance between electron $i$ and nucleus $A$, $r_{ij}$ is
the distance between electrons $i$ and $j$, and $R_{AB}$ is the distance
between nuclei $A$ and $B$ (see Fig.~\ref{fig:molecular-coordinates}).

Equation~\eqref{eq:total-hamiltonian} defines a formidable quantum
many-body problem. Even if the electron-electron interaction could somehow
be treated exactly, the electronic and nuclear motions would remain coupled.
The key observation behind the Born--Oppenheimer approximation is the large
separation of mass scales between electrons and nuclei. In atomic units, the
electron mass is unity whereas even the lightest nucleus is more than three
orders of magnitude heavier. Correspondingly, the nuclear kinetic-energy
operator contains the small factors $1/M_A$. Nuclear motion is therefore
typically much slower than electronic motion.

This separation of time scales suggests a natural hierarchy. On the
electronic time scale, the nuclei can be regarded as approximately fixed, and
the electrons rapidly adjust to a given nuclear configuration. The first step
is consequently to solve a clamped-nuclei electronic problem. For fixed
nuclear positions $\{\bR_A\}$, one defines the electronic Hamiltonian
\begin{equation}
  \chH_\text{elec}
  =
  \blue{\chT_\text{e}}
  +
  \orange{\chV_\text{ne}}
  +
  \red{\chV_\text{ee}}
  \label{eq:electronic-hamiltonian}
\end{equation}
and solves
\begin{equation}
  \chH_\text{elec}
  \Phi_\text{elec}^{(I)}
  (\{\br_i\};\{\bR_A\})
  =
  \cE_\text{elec}^{(I)}(\{\bR_A\})
  \Phi_\text{elec}^{(I)}
  (\{\br_i\};\{\bR_A\})
  \label{eq:clamped-nuclei}
\end{equation}
Here $I$ labels an electronic state. The semicolon emphasizes the different
roles played by electronic and nuclear
coordinates: the electronic coordinates are genuine variables of
$\Phi_\text{elec}^{(I)}$, whereas the nuclear positions enter
parametrically.

The nuclear repulsion $\chV_\text{nn}$ does not appear in
Eq.~\eqref{eq:electronic-hamiltonian} because, for fixed nuclei, it is simply
a constant with respect to the electronic coordinates. Once the electronic
problem has been solved, it can be added to the electronic energy to define
the energy surface
\begin{equation}
  U_I(\{\bR_A\})
  =
  \cE_\text{elec}^{(I)}(\{\bR_A\})
  +
  \sum_{A<B}^{M}
  \frac{Z_A Z_B}{R_{AB}}
  \label{eq:potential-energy-surface}
\end{equation}
Repeating the electronic calculation for different nuclear geometries traces
out a potential-energy surface (PES). Stable equilibrium structures correspond
to local minima of this surface, while chemical reactions, molecular
vibrations, and other nuclear motions can be interpreted in terms of motion
over it.

It is important to distinguish this \emph{clamped-nuclei approximation} from
the Born--Oppenheimer approximation itself. The connection becomes clear by
expanding the exact molecular wavefunction in the complete set of
clamped-nuclei electronic states,
\begin{equation}
  \Phi(\{\br_i\},\{\bR_A\})
  =
  \sum_I
  \Phi_\text{nucl}^{(I)}(\{\bR_A\})
  \Phi_\text{elec}^{(I)}
  (\{\br_i\};\{\bR_A\})
  \label{eq:born-huang-expansion}
\end{equation}
This Born--Huang expansion is, in principle, exact if the electronic basis is
complete. The electronic and nuclear motions remain coupled because
$\Phi_\text{elec}^{(I)}$ itself changes when the nuclei move. In particular,
the nuclear kinetic-energy operator acts not only on
$\Phi_\text{nucl}^{(I)}$ but also on the parametric dependence of the
electronic wavefunction on $\{\bR_A\}$.

These additional terms generate couplings between different electronic
states. Their leading contributions involve the derivative couplings
\begin{equation}
  \bd_{IJ}^{A}
  =
  \mel*{
    \Phi_\text{elec}^{(I)}
  }{
    \nabla_A
  }{
    \Phi_\text{elec}^{(J)}
  }
  \label{eq:nonadiabatic-coupling}
\end{equation}
which measure how rapidly the electronic wavefunctions change as nucleus $A$
moves. The Born--Oppenheimer approximation consists in neglecting these
nonadiabatic couplings and assuming that nuclear motion takes place on a
single electronic potential-energy surface. In the nuclear equation below,
we also neglect the diagonal correction arising when nuclear derivatives
act on that electronic state; the potential is therefore the clamped-nuclei
energy plus nuclear repulsion, without a derivative correction.

For electronic state $I$, the resulting nuclear Schr\"odinger equation is
\begin{equation}
  \qty[
    \green{\chT_\text{n}}
    +
    U_I(\{\bR_A\})
  ]
  \Phi_\text{nucl}^{(I)}(\{\bR_A\})
  =
  \cE_\text{tot}
  \Phi_\text{nucl}^{(I)}(\{\bR_A\})
  \label{eq:nuclear-BO}
\end{equation}
Equivalently,
\begin{equation}
  \chH_\text{nucl}
  =
  \green{\chT_\text{n}}
  +
  \violet{\chV_\text{nn}}
  +
  \cE_\text{elec}^{(I)}(\{\bR_A\})
\end{equation}
Thus, within the Born--Oppenheimer picture, the electrons determine the
potential on which the nuclei move.

For a single isolated electronic state, the molecular wavefunction is then
approximated by the product
\begin{equation}
  \Phi(\{\br_i\},\{\bR_A\})
  \approx
  \Phi_\text{nucl}(\{\bR_A\})
  \Phi_\text{elec}(\{\br_i\};\{\bR_A\})
  \label{eq:BO-product}
\end{equation}
This familiar factorized form should therefore be viewed as the
\emph{consequence} of neglecting couplings between electronic states rather
than as the starting point of the derivation.

The Born--Oppenheimer approximation is remarkably successful for most
ground-state molecular problems because the electronic wavefunction usually
changes smoothly with the nuclear coordinates and neighboring electronic
states are sufficiently well separated in energy. It becomes less reliable
when two or more electronic states approach one another. Near avoided
crossings, conical intersections, or other regions of strong electronic
near-degeneracy, the derivative couplings in
Eq.~\eqref{eq:nonadiabatic-coupling} can become large and the electronic and
nuclear motions can no longer be treated independently. Such nonadiabatic
effects play a central role, for example, in photochemical processes and
radiationless transitions.

\hfBioBornOppenheimer

The remainder of this chapter is concerned with the electronic problem.
Unless stated otherwise, the nuclear positions will therefore be regarded as
fixed, and we will seek approximate solutions of the clamped-nuclei
electronic Schr\"odinger equation
\begin{equation}
  \chH_\text{elec}
  \Phi_\text{elec}
  =
  \cE_\text{elec}
  \Phi_\text{elec}
  \label{eq:electronic-SE}
\end{equation}
The next question is then not how the nuclei move, but what restrictions must
be imposed on $\Phi_\text{elec}$ because the particles it describes are
indistinguishable fermions. This brings us directly to the Pauli principle
and the antisymmetry of the electronic wavefunction.
In the electronic-only discussions that follow, $\chH$ without a subscript
also denotes the clamped-nuclei electronic Hamiltonian
$\chH_\text{elec}$, rather than the full molecular Hamiltonian.

\subsubsection{The Pauli principle and antisymmetry}

Having separated the electronic and nuclear motions, we now turn to a second
fundamental ingredient of the electronic problem: electrons are
indistinguishable fermions. This statement imposes a strict symmetry
requirement on the electronic wavefunction, independently of any
approximation that will later be made to solve the electronic
Schr\"odinger equation.

The non-relativistic electronic Hamiltonian introduced above is spin
independent: it contains the electronic kinetic energy and Coulomb
interactions, but no explicit spin operator. Nevertheless, a complete
description of an electron must contain both its spatial coordinate and its
intrinsic spin degree of freedom. Electrons have spin $s=1/2$, and their spin
functions satisfy
\begin{align}
  \ket{\omega}&=\ket{s,m_s}
  &
  s^2\ket{s,m_s}&=s(s+1)\ket{s,m_s}
  &
  s_z\ket{s,m_s}&=m_s\ket{s,m_s}
\end{align}
The two one-electron spin functions are
\begin{align}
  \ket{\blue{\alpha}}
  &= \ket{\tfrac{1}{2},+\tfrac{1}{2}}
  &
  \ket{\red{\beta}}
  &= \ket{\tfrac{1}{2},-\tfrac{1}{2}}
\end{align}
corresponding respectively to spin-up and spin-down electrons. They form an
orthonormal basis for the one-electron spin space,
\begin{align}
  \int\blue{\alpha}^*(\omega)\red{\beta}(\omega)\dd{\omega}
  &=\int\red{\beta}^*(\omega)\blue{\alpha}(\omega)\dd{\omega}=0
  &
  \int\blue{\alpha}^*(\omega)\blue{\alpha}(\omega)\dd{\omega}
  &=\int\red{\beta}^*(\omega)\red{\beta}(\omega)\dd{\omega}=1
  \\
  \braket{\blue{\alpha}}{\red{\beta}}
  &= \braket{\red{\beta}}{\blue{\alpha}}=0
  &
  \braket{\blue{\alpha}}{\blue{\alpha}}
  &= \braket{\red{\beta}}{\red{\beta}}=1
\end{align}

It is therefore convenient to combine the spin coordinate $\omega$ and the
spatial coordinate $\br$ into a single composite coordinate,
\begin{equation}
  \violet{\bx}=(\orange{\omega},\red{\br})
\end{equation}
An $N$-electron wavefunction is consequently a function of $N$ such
coordinates,
\begin{equation}
  \Phi(\bx_1,\bx_2,\ldots,\bx_N)
\end{equation}
Although the Hamiltonian is spin independent in the present treatment, the
spin coordinates are indispensable because the symmetry of the full
spin-spatial wavefunction determines whether the state is compatible with
fermionic statistics.

\paragraph{Indistinguishability and permutation symmetry.}

The labels assigned to identical particles carry no physical information.
Interchanging two electron labels must therefore leave every observable
quantity unchanged. For two particles, this implies in particular that the
probability density satisfies
\begin{equation}
  \abs{\Psi(\red{\bx_1},\blue{\bx_2})}^2
  =
  \abs{\Psi(\blue{\bx_2},\red{\bx_1})}^2
\end{equation}
To obtain an exchange rule for the wavefunction, we additionally assume that
exchange leaves the pure-state ray unchanged: the exchanged state differs
from the original by a single, coordinate-independent phase. Equality of the
coordinate probability densities alone would allow a coordinate-dependent
phase and would not establish this stronger statement. With this state-ray
assumption, the phase relation is
\begin{equation}
  \abs{\Psi(\red{\bx_1},\blue{\bx_2})}^2
  =\abs{\Psi(\blue{\bx_2},\red{\bx_1})}^2
  \quad\Rightarrow\quad
  \Psi(\blue{\bx_2},\red{\bx_1})
  =\green{e^{\ii\theta}}\Psi(\red{\bx_1},\blue{\bx_2})
\end{equation}

Introducing the permutation operator $\chP_{12}$, which exchanges particles
$1$ and $2$, gives
\begin{equation}
\begin{split}
  \chP_{12}\Psi(\red{\bx_1},\blue{\bx_2})
  &=
  \Psi(\blue{\bx_2},\red{\bx_1})
  =
  \green{e^{\ii\theta}}
  \Psi(\red{\bx_1},\blue{\bx_2})
  \\
  \chP_{12}^2\Psi(\red{\bx_1},\blue{\bx_2})
  &=
  \green{e^{2\ii\theta}}
  \Psi(\red{\bx_1},\blue{\bx_2})
  =
  \Psi(\red{\bx_1},\blue{\bx_2})
\end{split}
\end{equation}
Because exchanging the same pair twice restores the original labeling,
$\chP_{12}^2=1$, and consequently
\begin{equation}
  \green{e^{2\ii\theta}}=1
  \quad\Rightarrow\quad
  \green{\theta}=n\pi
  \quad\Rightarrow\quad
  \green{e^{\ii\theta}}=\pm1
\end{equation}

The wavefunction can therefore belong to either a symmetric or an
antisymmetric permutation sector. The plus sign defines \violet{bosons},
for which
\begin{equation}
  \violet{
  \Psi(\bx_1,\bx_2)
  =
  \Psi(\bx_2,\bx_1)
  }
\end{equation}
whereas the minus sign defines \orange{fermions},
\begin{equation}
  \orange{
  \Psi(\bx_1,\bx_2)
  =
  -
  \Psi(\bx_2,\bx_1)
  }
\end{equation}
Electrons are fermions and must therefore be described by an antisymmetric
wavefunction \citep{Dirac1926}. The connection between half-integer spin and fermionic
statistics is the content of the spin-statistics theorem; within
non-relativistic quantum mechanics, fermionic antisymmetry is imposed as a
fundamental property of the electronic wavefunction \citep{Pauli1940}.

A notable exception occurs for identical particles confined to two spatial
dimensions. In this case, particle exchanges are described by the braid
group rather than solely by the permutation group, and two successive
exchanges need not be equivalent to the identity. The exchange phase is
therefore not restricted to $\pm1$, allowing \emph{anyonic} statistics.
For Abelian anyons, an exchange produces an arbitrary phase
$e^{\ii\theta}$, with bosons and fermions corresponding to
$\theta=0$ and $\theta=\pi$, respectively. In three spatial dimensions,
only the bosonic and fermionic possibilities occur
\citep{LeinaasMyrheim1977,Wilczek1982}.

For an arbitrary pair of electrons $i$ and $j$, the antisymmetry requirement
can be written compactly as
\begin{equation}
  \chP_{ij}\Phi=-\Phi
\end{equation}
The electronic Schr\"odinger equation and Pauli antisymmetry condition are
therefore
\begin{equation}
  \chH_\text{elec}\Phi(\bx_1,\bx_2,\ldots,\bx_N)
  =
  \cE_\text{elec}\Phi(\bx_1,\bx_2,\ldots,\bx_N)
\end{equation}
and
\begin{equation}
  \Phi(\bx_1,\ldots,\bx_i,\ldots,\bx_j,\ldots,\bx_N)
  =
  -
  \Phi(\bx_1,\ldots,\bx_j,\ldots,\bx_i,\ldots,\bx_N)
  \label{eq:pauli-antisymmetry}
\end{equation}
Since the electronic Hamiltonian is invariant under the exchange of two
electron labels,
\begin{equation}
  \comm{\chH_\text{elec}}{\chP_{ij}}=0
\end{equation}
this antisymmetry is preserved by the electronic dynamics.

\paragraph{The Pauli principle and the exchange hole.}

A particularly important consequence follows immediately from
Eq.~\eqref{eq:pauli-antisymmetry}. If two fermions are assigned the same
composite coordinate, $\bx=\bx_1=\bx_2$, then
\begin{equation}
  \Psi(\bx,\bx)
  =
  -\Psi(\bx,\bx)
\end{equation}
and hence
\begin{equation}
  \Psi(\bx,\bx)=0
\end{equation}
The probability of finding two electrons with the same spin at exactly the
same spatial position therefore vanishes. This is one manifestation of the
Pauli exclusion principle \citep{Pauli1925}.

\begin{hftheorem}{Pauli exclusion principle}
\hftheoremportrait{figures/sketch-pauli.png}{Wolfgang Pauli}
{An electronic wavefunction is antisymmetric under exchange of two complete
space--spin coordinates. Consequently, no two electrons can occupy the same
spin orbital. In particular, the wavefunction vanishes when two electrons
have identical space--spin coordinates \citep{Pauli1925}.}
{Source portrait: Bettina Katzenstein / ETH-Bibliothek Z\"urich, Bildarchiv,
\href{https://commons.wikimedia.org/wiki/File:Wolfgang_Pauli_ETH-Bib_Portr_01042_(cropped).jpg}{Portr 01042 (cropped by Xabier)},
\href{https://creativecommons.org/licenses/by-sa/3.0/}{CC BY-SA 3.0}.
\hfsketchcredit}
\end{hftheorem}

\hfBioPauli

It is worth emphasizing that the fundamental statement is the
\emph{antisymmetry of the many-electron wavefunction}, not merely the familiar
rule that ``two electrons cannot occupy the same state.'' The latter follows
from the former. If two columns or rows of an antisymmetric fermionic
wavefunction become identical, the wavefunction must vanish.

Antisymmetry also has consequences beyond the special case in which the two
electrons occupy exactly the same point. Conditioning on the presence of one
electron reduces the probability of finding another electron of the same
spin in its vicinity. This deficit in the same-spin pair probability is
called the Fermi hole, or exchange hole. It arises entirely from
fermionic antisymmetry and is present even in the absence of dynamical
electron correlation.

We have therefore arrived at a central requirement: any acceptable
many-electron electronic wavefunction must be antisymmetric under the
exchange of two electrons. The next question is how to construct such a
wavefunction from simpler one-electron objects.

\begin{hfexercise}{Two-electron antisymmetry}
\violet{\textit{Given two orthonormal one-electron functions
$\chi_1(\bx)$ and $\chi_2(\bx)$, construct a two-electron fermionic
wavefunction $\Psi(\bx_1,\bx_2)$ and normalize it.}}

\end{hfexercise}
\begin{hfsolution}{Two-electron antisymmetry}
Starting from the two possible assignments of the one-electron functions to
the two electrons, consider a linear combination with coefficients $A$ and
$B$,
\begin{equation}
  \Psi(\bx_1,\bx_2)
  =A\chi_1(\bx_1)\chi_2(\bx_2)
   +B\chi_2(\bx_1)\chi_1(\bx_2)
\end{equation}
Exchanging the coordinates interchanges the two products. Requiring the
result to be $-\Psi(\bx_1,\bx_2)$ therefore gives $B=-A$. The two
products are
\begin{equation}
  \chi_1(\bx_1)\chi_2(\bx_2)
  \qquad\text{and}\qquad
  \chi_2(\bx_1)\chi_1(\bx_2)
\end{equation}
the required antisymmetric combination is obtained by taking their
difference,
\begin{equation}
  \Psi(\red{\bx_1},\blue{\bx_2})
  =
  \frac{1}{\sqrt{2}}\qty[
    \chi_1(\red{\bx_1})\chi_2(\blue{\bx_2})
    -
    \chi_2(\red{\bx_1})\chi_1(\blue{\bx_2})
  ]
  \label{eq:two-electron-antisym}
\end{equation}
Indeed, exchanging the two electron coordinates immediately yields
\begin{equation}
  \Psi(\bx_2,\bx_1)
  =
  -\Psi(\bx_1,\bx_2)
\end{equation}
For orthonormal $\chi_1$ and $\chi_2$, the prefactor $1/\sqrt{2}$ ensures
normalization. Indeed, each product has unit norm and their overlap is
$\braket{\chi_1}{\chi_2}\braket{\chi_2}{\chi_1}=0$, so their difference
has squared norm two. Choosing $A=1/\sqrt{2}$ fixes the normalization;
any additional global phase describes the same physical state. The minus
sign enforces antisymmetry, while the prefactor fixes the norm: these are
separate requirements.

\end{hfsolution}

Slater popularized the determinant representation of this combination
\citep{Slater1929},
\begin{equation}
  \Psi(\red{\bx_1},\blue{\bx_2})
  =
  \frac{1}{\sqrt{2}}
    \mdet{
      \chi_1(\red{\bx_1})&\chi_2(\red{\bx_1})\\
      \chi_1(\blue{\bx_2})&\chi_2(\blue{\bx_2})
    }
  =
  \frac{1}{\sqrt{2}}\qty[
    \chi_1(\red{\bx_1})\chi_2(\blue{\bx_2})
    -
    \chi_2(\red{\bx_1})\chi_1(\blue{\bx_2})
  ]
\end{equation}
The determinant is ideally suited to fermions because interchanging two
electron coordinates corresponds to exchanging two rows of the determinant,
which changes its sign automatically. Likewise, if two one-electron
functions are identical, two columns become identical and the determinant
vanishes. Antisymmetry and the Pauli exclusion principle are therefore built
directly into the determinant structure.

\hfBioSlater

This construction is a Slater determinant. By contrast, the simple
product
\begin{equation}
  \Psi(\bx_1,\bx_2)
  =
  \chi_1(\bx_1)\chi_2(\bx_2)
\end{equation}
is a Hartree product. A Hartree product assigns a particular
one-electron function to each labeled electron and is not, by itself,
antisymmetric under particle exchange. It therefore cannot serve as a
fermionic many-electron wavefunction.

The two-electron determinant provides the prototype for the general
$N$-electron construction. Hartree--Fock theory adopts precisely this idea:
the electronic wavefunction is approximated by a single Slater determinant
built from $N$ one-electron spin orbitals. We now turn to this
single-determinant ansatz.

\subsubsection{The single-determinant ansatz}

The previous section established that an acceptable electronic wavefunction
must be antisymmetric with respect to the exchange of any pair of electrons.
A Slater determinant provides the simplest construction that satisfies this
requirement automatically. Hartree--Fock theory takes the decisive additional
step of assuming that the entire $N$-electron wavefunction can be represented
by a single such determinant. We denote this trial electronic wavefunction by
$\Psi_\text{HF}$ to distinguish it from the exact electronic wavefunction
$\Phi_\text{elec}$.

Thus,
\begin{equation}
\begin{split}
  \Psi_\text{HF}(\bx_1,\ldots,\bx_N)
  &=\boxed{
  \frac{1}{\sqrt{N!}}
  \mdet{
    \chi_1(\bx_1)&\chi_2(\bx_1)&\cdots&\chi_N(\bx_1)\\
    \chi_1(\bx_2)&\chi_2(\bx_2)&\cdots&\chi_N(\bx_2)\\
    \vdots&\vdots&\ddots&\vdots\\
    \chi_1(\bx_N)&\chi_2(\bx_N)&\cdots&\chi_N(\bx_N)
  }
  }
  \\
  &\equiv
  \ket{\chi_1(\bx_1)\chi_2(\bx_2)\cdots\chi_N(\bx_N)}
  \\
  &=
  \blue{\chA}\qty[
    \chi_1(\bx_1)\chi_2(\bx_2)\cdots\chi_N(\bx_N)
  ]
  =
  \blue{\chA}\green{\Pi(\bx_1,\bx_2,\ldots,\bx_N)}
  \label{eq:hf-determinant}
\end{split}
\end{equation}
Here,
\begin{equation}
  \green{
  \Pi(\bx_1,\bx_2,\ldots,\bx_N)
  =
  \chi_1(\bx_1)\chi_2(\bx_2)\cdots\chi_N(\bx_N)
  }
\end{equation}
is the Hartree product introduced above, while $\blue{\chA}$ is the
antisymmetrizer.

The role of $\chA$ can be made explicit by writing
\begin{equation}
  \chA
  =
  \frac{1}{\sqrt{N!}}
  \sum_{\pi\in S_N}
  (-1)^\pi \chP_\pi
\end{equation}
where $S_N$ is the group of all $N!$ permutations of the electron labels,
$\chP_\pi$ is the operator associated with the permutation $\pi$, and
$(-1)^\pi$ is $+1$ for an even permutation and $-1$ for an odd one.
Applying $\chA$ to a Hartree product generates all possible assignments of
the one-electron functions to the electrons, with the signs required by
fermionic antisymmetry.
The factor $1/\sqrt{N!}$ normalizes the resulting determinant when the
spin orbitals are orthonormal. With this convention, $\chA$ is a
normalization-weighted antisymmetrizer, rather than an idempotent projector;
the latter would instead contain the factor $1/N!$.

The determinant notation is simply a compact representation of this
antisymmetrized sum. Its usefulness follows directly from elementary
properties of determinants. Exchanging two electron coordinates exchanges
two rows and changes the sign of the determinant. Exchanging two spin
orbitals exchanges two columns and likewise changes its sign. Finally, if
two spin orbitals become identical, two columns coincide and the determinant
vanishes. The Pauli principle is therefore built into the algebraic
structure of the wavefunction.

The normalization factor $1/\sqrt{N!}$ in
Eq.~\eqref{eq:hf-determinant} assumes that the spin orbitals are
orthonormal. Under this condition,
\begin{equation}
  \braket{\Psi_\text{HF}}{\Psi_\text{HF}}=1
\end{equation}
A normalized Slater determinant is thus completely specified by its $N$
occupied one-electron spin orbitals.

This is a dramatic reduction of the original many-electron problem. Instead
of optimizing an arbitrary function of $N$ electronic coordinates, Hartree--Fock theory
optimizes a set of one-electron functions. The price of this simplification
is that the allowed wavefunctions occupy only a very small subset of the
full antisymmetric $N$-electron Hilbert space. The variational problem is
therefore not to find the exact electronic wavefunction, but to find the
\emph{best single Slater determinant} for the Hamiltonian under
consideration.

\paragraph{Spin orbitals and spatial orbitals.}

Each one-electron function $\chi_p(\bx)$ appearing in the determinant is a
spin orbital: it depends on both the spatial coordinate $\br$ and the
spin coordinate $\omega$. In the collinear spin formulations considered
initially, a spin orbital separates into spin and spatial factors,
\begin{equation}
  \chi_p(\bx)
  =
  \red{\sigma(\omega)}
  \blue{\psi_p^\sigma(\br)}
  \qq{with}
  \sigma(\omega)\in\{\alpha(\omega),\beta(\omega)\}
  \label{eq:spin-orbital-factorization}
\end{equation}
Thus, a spin orbital is obtained by associating a spatial orbital with either
an $\alpha$ or a $\beta$ spin function.

The distinction between \emph{spin orbital} and \emph{spatial orbital} will
be important throughout the chapter. The Slater determinant itself is built
from spin orbitals because antisymmetry applies to the complete
spin-spatial coordinates of the electrons. In practical molecular
calculations, however, it is usually the spatial functions
$\psi_p^\sigma(\br)$ that are expanded and optimized.

\paragraph{Finite one-electron basis.}

The spatial orbitals are not represented directly as arbitrary continuous
functions. Instead, each is expanded in a finite set of $K$ prescribed basis
functions,
\begin{equation}
  \boxed{
  \blue{\psi_p^\sigma(\br)}
  =
  \sum_{\mu=1}^{K}
  C_{\mu p}^{\sigma}
  \green{\phi_\mu(\br)}
  }
  \label{eq:lcao-introduction}
\end{equation}
The coefficients $C_{\mu p}^{\sigma}$ then become the quantities to be
determined variationally.

This expansion is conventionally called the linear combination of
atomic orbitals (LCAO) representation. The terminology is historical:
although the basis functions $\phi_\mu$ are most commonly centered on atoms
in molecular calculations, the essential mathematical point is simply that
the unknown orbitals are represented in a finite one-electron basis.

The finite expansion in Eq.~\eqref{eq:lcao-introduction} introduces a second,
independent approximation on top of the single-determinant ansatz. Even
within Hartree--Fock theory, the exact  solution is recovered only in the
limit of a complete one-electron basis. The construction, properties, and
convergence of such basis sets will be examined in detail later in this
chapter.

\paragraph{Orthonormality.}

The spin orbitals entering a Slater determinant are chosen to be
orthonormal,
\begin{equation}
  \braket*{\chi_p}{\chi_q}
  =
  \int
  \chi_p^*(\bx)
  \chi_q(\bx)
  \dd{\bx}
  =
  \delta_{pq}
  =
  \begin{cases}
    1,&\text{if $p=q$}\\
    0,&\text{otherwise}
  \end{cases}
\end{equation}
where $\delta_{pq}$ is the Kronecker delta.

This does not imply that the underlying basis functions are themselves
orthogonal. In fact, atom-centered basis functions on different atoms
generally overlap. Their pairwise overlaps define the overlap matrix
\begin{equation}
  \braket*{\phi_\mu}{\phi_\nu}
  =
  \int
  \phi_\mu^*(\br)
  \phi_\nu(\br)
  \dd{\br}
  =S_{\mu\nu}
\end{equation}
Consequently, the orthonormality of the molecular orbitals imposes a
constraint on the expansion coefficients rather than on the basis functions
themselves. For example, for a given spin sector,
\begin{equation}
  \braket*{\psi_p^\sigma}{\psi_q^\sigma}
  =
  \sum_{\mu\nu}
  C_{\mu p}^{\sigma *}
  S_{\mu\nu}
  C_{\nu q}^{\sigma}
  =
  \delta_{pq}
\end{equation}
or, in matrix notation,
\begin{equation}
  \qty(\bC^\sigma)^\dagger\cdot\bS\cdot\bC^\sigma=\bI
\end{equation}
This distinction between an orthonormal set of molecular orbitals and a
generally nonorthogonal atomic-orbital basis will become central when we
derive the Roothaan--Hall equations.

\paragraph{Restricted and unrestricted Hartree--Fock.}

The single-determinant ansatz still leaves freedom in how the $\alpha$ and
$\beta$ spin orbitals are related. Different constraints on their spatial
parts lead to different variants of Hartree--Fock theory.

In restricted Hartree--Fock (RHF) theory, appropriate for the usual
closed-shell description, an $\alpha$ electron and a $\beta$ electron in a
pair share the same spatial orbital,
\begin{equation}
  \chi_p^\text{RHF}(\bx)
  =
  \begin{cases}
    \blue{\alpha(\omega)}\psi_p(\br)\\
    \red{\beta(\omega)}\psi_p(\br)
  \end{cases}
  =
  \begin{cases}
    \blue{\psi_p(\br)}\\
    \red{\Bar{\psi}_p(\br)}
  \end{cases}
\end{equation}
The barred notation will be used when it is convenient to distinguish the
$\beta$ spin orbital associated with a given spatial orbital from its
$\alpha$ partner.

For an even number of electrons in a closed-shell RHF state, each occupied
spatial orbital therefore contains two electrons with opposite spins. Only
$N/2$ distinct occupied spatial orbitals are required to construct the
$N$ occupied spin orbitals.

The restriction that the two spin channels share the same spatial orbitals
can be advantageous: it preserves the closed-shell spin symmetry and reduces
the number of independent orbital coefficients. It is also a genuine
constraint on the variational space. In situations where the optimal
$\alpha$ and $\beta$ electrons would prefer different spatial
distributions, RHF cannot accommodate that relaxation.

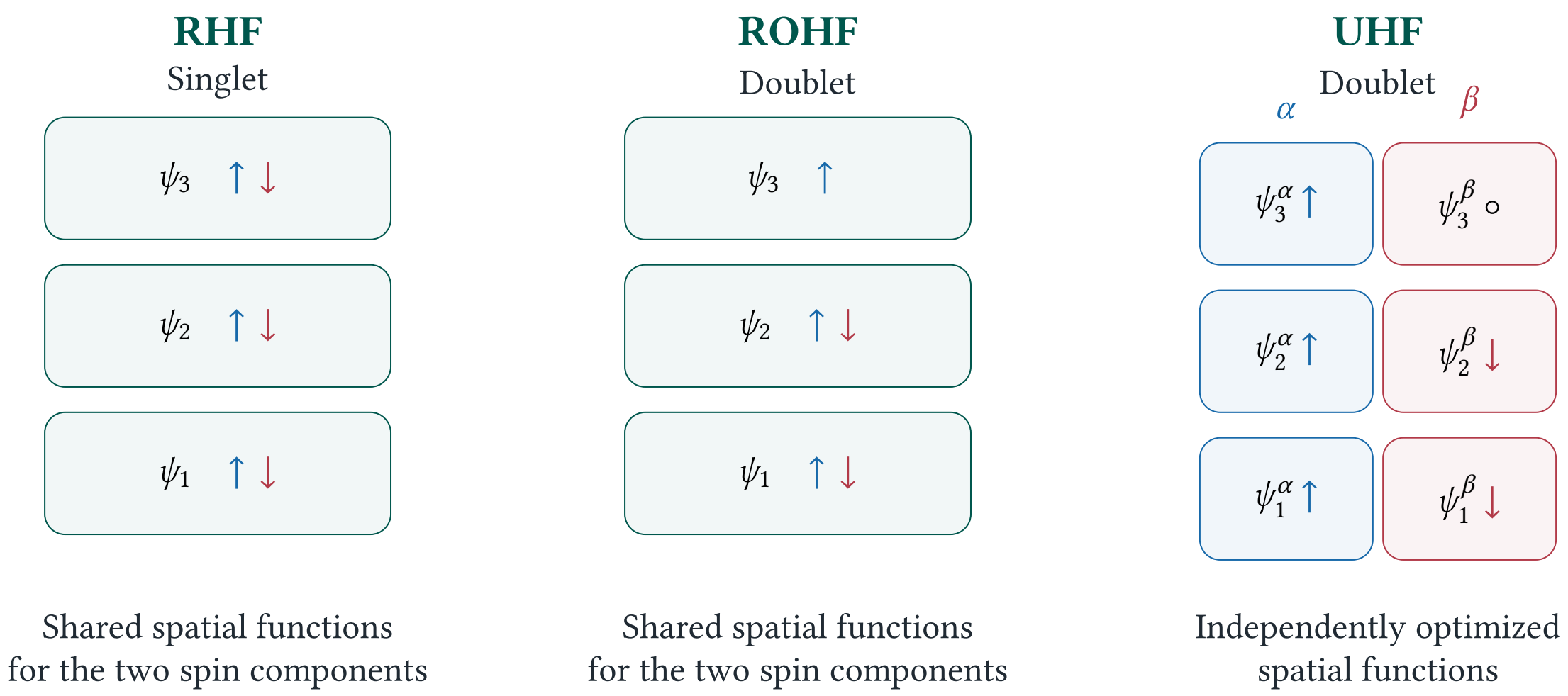
\begin{figure}[htbp]
  \centering
  \resizebox{\linewidth}{!}{
\begingroup

\tikzset{
 nucleus/.style={circle,draw=hfGreenDark,fill=hfGreen!12,minimum size=7mm},
 electron/.style={circle,draw=hfBlue,fill=hfBlue!10,minimum size=7mm},
 old/.style={hfBlue,densely dashed,thick,-{Latex[length=2mm]}},
 new/.style={hfGreenDark,very thick,-{Latex[length=2mm]}},
 card/.style={draw=hfGreenDark,fill=hfGreen!5,rounded corners=2mm,
   align=center,minimum height=12mm},
 occupied/.style={circle,fill=hfGreenDark,inner sep=2.1pt},
 vacant/.style={circle,draw=hfInk!50,inner sep=2.1pt},
 hole/.style={circle,draw=hfRed,thick,inner sep=2.1pt},
 particle/.style={circle,fill=hfViolet,inner sep=2.1pt}}


\begin{tikzpicture}[font=\small,x=1cm,y=1cm]
 \foreach \x/\title/\subtitle in {0/RHF/Singlet,5.7/ROHF/Doublet,11.4/UHF/Doublet}{
  \node[font=\large\bfseries,text=hfGreenDark] at (\x,0) {\title};
  \node[text=hfInk] at (\x,-0.5) {\subtitle};
 }
 \foreach \x in {0,5.7}{
  \foreach \r/\p in {1/3,2/2,3/1}{
   \node[card,minimum width=34mm] at (\x,-1.45*\r)
    {$\psi_{\p}$\quad
      \textcolor{hfBlue}{$\uparrow$}\,
      \ifnum\r=1\ifdim\x pt=0pt
      \textcolor{hfRed}{$\downarrow$}\fi
      \else\textcolor{hfRed}{$\downarrow$}\fi};
  }
  \node[align=center,text=hfInk] at (\x,-6.1)
   {Shared spatial functions\\for the two spin components};
 }
 \node[text=hfBlue,anchor=south] at (10.5,-1) {$\alpha$};
 \node[text=hfRed,anchor=south] at (12.3,-1) {$\beta$};
 \foreach \r/\p in {1/3,2/2,3/1}{
  \node[card,draw=hfBlue,fill=hfBlue!6,minimum width=17mm] at (10.5,-1.45*\r-0.25)
   {$\psi_{\p}^{\alpha}\,\textcolor{hfBlue}{\uparrow}$};
  \node[card,draw=hfRed,fill=hfRed!6,minimum width=17mm] at (12.3,-1.45*\r-0.25)
   {$\psi_{\p}^{\beta}\,\ifnum\r=1\circ\else\textcolor{hfRed}{\downarrow}\fi$};
 }
 \node[align=center,text=hfInk] at (11.4,-6.1)
   {Independently optimized\\spatial functions};
\end{tikzpicture}
\endgroup

}
  \caption{Restricted, restricted open-shell, and unrestricted choices for the spatial parts of the
  spin orbitals. In RHF, paired $\alpha$ and $\beta$ electrons share the same
  spatial orbital, whereas UHF allows the two spin channels to relax
  independently. ROHF retains a common spatial-orbital set with singly
  occupied orbitals. The cards illustrate spatial-orbital sharing, not
  orbital-energy ordering.}
  \label{fig:rhf-uhf-orbitals}
\end{figure}

The closed-shell RHF ansatz does not describe general open-shell systems. In
unrestricted Hartree--Fock (UHF) theory, the $\alpha$ and $\beta$
spin components are allowed to use different spatial orbitals (see Fig.~\ref{fig:rhf-uhf-orbitals}),
\begin{equation}
  \chi_p^\text{UHF}(\bx)
  =
  \begin{cases}
    \alpha(\omega)\blue{\psi_p^\alpha(\br)}\\
    \beta(\omega)\red{\psi_p^\beta(\br)}
  \end{cases}
\end{equation}
This additional variational freedom permits the treatment of open-shell
systems and can also lower the energy of nominally closed-shell systems when
the restricted solution becomes unstable. As we shall see later, however,
this freedom comes at a price: a UHF determinant is generally an eigenfunction
of $\chS_z$, but not necessarily of $\chS^2$. The resulting breaking of
spin symmetry is the origin of spin contamination.

Restricted open-shell Hartree--Fock (ROHF) provides an alternative treatment
of open-shell systems in which the $\alpha$ and $\beta$ electrons retain a
common set of spatial orbitals \citep{Roothaan1960}. We mention this possibility for completeness,
but do not develop the ROHF formalism further here.

For RHF, the spatial molecular orbitals form an orthonormal set,
\begin{equation}
  \braket*{\psi_p}{\psi_q}
  =
  \int
  \psi_p^*(\br)
  \psi_q(\br)
  \dd{\br}
  =
  \delta_{pq}
\end{equation}
For UHF, orthonormality holds separately within each spin sector,
\begin{equation}
\begin{split}
  \braket*{\psi_p^\alpha}{\psi_q^\alpha}
  &=
  \int
  \psi_p^{\alpha*}(\br)
  \psi_q^\alpha(\br)
  \dd{\br}
  =
  \delta_{pq}
  \\
  \braket*{\psi_p^\beta}{\psi_q^\beta}
  &=
  \int
  \psi_p^{\beta*}(\br)
  \psi_q^\beta(\br)
  \dd{\br}
  =
  \delta_{pq}
\end{split}
\end{equation}
whereas opposite-spin spatial orbitals need not be orthogonal,
\begin{equation}
  \braket*{\psi_p^\alpha}{\psi_q^\beta}
  =
  \int
  \psi_p^{\alpha*}(\br)
  \psi_q^\beta(\br)
  \dd{\br}
  =
  S_{pq}^{\alpha\beta}
\end{equation}
There is no contradiction with the orthonormality of the full spin orbitals:
because
\begin{equation}
  \braket{\alpha}{\beta}=0
\end{equation}
an $\alpha$ spin orbital and a $\beta$ spin orbital are orthogonal regardless
of the overlap between their spatial parts.
This distinction is useful to keep in mind. In UHF, one may have
$\braket*{\psi_p^\alpha}{\psi_q^\beta}\neq0$ but
$\braket*{\alpha\psi_p^\alpha}{\beta\psi_q^\beta}=0$.

\paragraph{Orbital spaces and index conventions.}

We conclude by fixing the notation used throughout the remainder of the
chapter. The one-electron spaces are
\begin{equation}
\begin{split}
  \{\phi_\mu\mid\mu=1,\ldots,K\}
  &:\quad
  \green{\text{basis functions or atomic orbitals (AOs)}}
  \\
  \{\psi_p\mid p=1,\ldots,K\}
  &:\quad
  \violet{\text{restricted spatial orbitals or molecular orbitals (MOs)}}
  \\
  \{\chi_p\mid p=1,\ldots,2K\}
  &:\quad
  \orange{\text{spin orbitals}}
\end{split}
\end{equation}
Thus, a basis of $K$ spatial functions generates $K$ spatial molecular
orbitals and, when the two spin functions are included, $2K$ spin orbitals.

Unless otherwise stated, the following index convention will be used:
\begin{equation}
  \begin{array}{ccl}
    \mu,\nu,\lambda,\sigma,\ldots
      &:& \text{basis functions / AOs}
    \\[1mm]
    p,q,r,s,\ldots
      &:& \text{general orbitals}
    \\[1mm]
    i,j,k,l,\ldots
      &:& \text{occupied orbitals}
    \\[1mm]
    a,b,c,d,\ldots
      &:& \text{unoccupied or virtual orbitals}
  \end{array}
\end{equation}
Among the $2K$ spin orbitals, $N$ are occupied in the HF determinant and are
therefore labeled by $i,j,k,l,\ldots$. The remaining $2K-N$ spin orbitals are
unoccupied. They are variously called vacant, unoccupied, or
virtual orbitals and are labeled by $a,b,c,d,\ldots$.

The distinction between occupied and virtual orbitals is relative to the
particular determinant under consideration. Only the occupied orbitals enter
the Slater determinant explicitly, but the virtual orbitals provide the
complementary one-electron space and will become essential when discussing
orbital optimization, excited determinants, response, and correlated
electronic-structure methods.

Finally, a system with two electrons in each occupied spatial orbital
is called closed shell. Otherwise, it is open shell. For the
ground state of a closed-shell system containing an even number $N$ of
electrons, the first $N/2$ spatial orbitals are doubly occupied,
$\psi_1^2\psi_2^2\cdots\psi_{N/2}^2$,
while the remaining $K-N/2$ spatial orbitals are vacant.

The single-determinant ansatz, together with these orthonormality constraints,
defines the variational space of Hartree--Fock theory. The next step is to
determine which determinant within this space is optimal. This requires
evaluating the expectation value of the electronic Hamiltonian as a functional
of the occupied orbitals.

\subsection{The Hartree--Fock energy functional}

The variational content of Hartree--Fock theory is encoded in the expectation
value of the electronic Hamiltonian over a normalized Slater determinant. The
derivation below separates the one- and two-electron contributions and makes
the Coulomb, exchange, and self-interaction terms explicit before introducing
their integral representations.

\subsubsection{Energy expectation value}

We have now specified the class of wavefunctions allowed within
Hartree--Fock theory: normalized single Slater determinants constructed from
orthonormal spin orbitals. The remaining question is variational. Among all
such determinants, which one provides the lowest possible energy?

For fixed nuclear positions, the electronic energy associated with a given
Slater determinant is the expectation value of the electronic Hamiltonian,
\begin{align}
  \EHF
  &=
  \mel{\Psi_\text{HF}}{\chH_\text{elec}}{\Psi_\text{HF}}
  &
  \chH_\text{elec}
  &=
  \blue{\chT_\text{e}}
  +
  \orange{\chV_\text{ne}}
  +
  \red{\chV_\text{ee}}
  \label{eq:hf-expectation-value}
\end{align}
Since $\Psi_\text{HF}$ depends on the occupied spin orbitals, this expression
should be regarded as an energy functional,
\begin{equation}
  \EHF
  =
  \EHF[\chi_1,\chi_2,\ldots,\chi_N]
\end{equation}
Hartree--Fock theory seeks the stationary point---and, for the ground-state
solution, the minimum---of this functional subject to the orthonormality
constraints
\begin{equation}
  \braket*{\chi_i}{\chi_j}=\delta_{ij}
\end{equation}

The variational principle guarantees that the energy of any normalized trial
wavefunction lies above or at the exact ground-state energy,
\begin{equation}
  \mel{\Psi_\text{trial}}{\chH_\text{elec}}{\Psi_\text{trial}}
  \geq
  \cE_\text{elec}^{(0)}
\end{equation}
Hartree--Fock therefore finds the lowest energy accessible within the
restricted manifold of single determinants. The remaining difference between
this optimal single-determinant energy and the exact non-relativistic
electronic energy is associated with electron correlation beyond the HF
description.

\paragraph{One- and two-electron operators.}

It is useful to separate the electronic Hamiltonian into one- and
two-electron parts,
\begin{equation}
  \chH_\text{elec}
  =
  \chO_1+\chO_2
  =
  \sum_{i=1}^{N}\hh(\bx_i)
  +
  \sum_{i<j}^{N}\frac{1}{r_{ij}}
  \label{eq:one-two-electron-hamiltonian}
\end{equation}
The one-electron, or core, Hamiltonian is
\begin{equation}
  \hh(\bx_i)
  =
  -\frac{\laplacian_i}{2}
  -
  \sum_{A=1}^{M}\frac{Z_A}{r_{iA}}
\end{equation}
so that
\begin{equation}
  \chO_1
  =
  \sum_{i=1}^{N}\hh(\bx_i)
\end{equation}
It contains the kinetic energy of electron $i$ and its attraction to all
nuclei. The explicit electron-electron interaction is contained in the
two-electron operator
\begin{equation}
  \chO_2
  =
  \red{\chV_\text{ee}}
  =
  \sum_{i<j}^{N}\frac{1}{r_{ij}}
\end{equation}

The distinction between one- and two-electron operators is important because
their expectation values over a Slater determinant have particularly simple
forms. This simplification is one of the main reasons why the
single-determinant ansatz is computationally tractable.

\paragraph{One-electron contribution.}

For a normalized determinant constructed from orthonormal spin orbitals, the
Slater--Condon rules reduce the expectation value of a one-electron operator
to a sum over the occupied spin orbitals. We use their diagonal expectation
values here; the few-electron examples below derive them explicitly, and
the general off-diagonal rules are developed in
Section~\ref{sec:slater-condon}.
For the one-electron part,
\begin{equation}
  \mel{\Psi_\text{HF}}{\chO_1}{\Psi_\text{HF}}
  =
  \sum_{i=1}^{N}
  \mel{\chi_i}{\hh}{\chi_i}
\end{equation}
Introducing
\begin{equation}
  h_{ij}
  =
  \mel{\chi_i}{\hh}{\chi_j}
\end{equation}
we obtain simply
\begin{equation}
  \mel{\Psi_\text{HF}}{\chO_1}{\Psi_\text{HF}}
  =
  \sum_{i=1}^{N}h_{ii}
  \label{eq:hf-one-electron-energy}
\end{equation}
Thus, the one-electron part of the HF energy is obtained by adding the kinetic
and electron-nuclear energies associated with each occupied spin orbital.

The nucleus-nucleus repulsion does not belong to the electronic Hamiltonian
because, within the clamped-nuclei problem, it contains no electronic
coordinates. Nevertheless, it must be added when the total
Born--Oppenheimer energy is required. For a normalized electronic
wavefunction,
\begin{equation}
  \mel{\Psi_\text{HF}}{\violet{\chV_\text{nn}}}{\Psi_\text{HF}}
  =
  \violet{V_\text{nn}}
  \braket{\Psi_\text{HF}}{\Psi_\text{HF}}
  =
  \violet{V_\text{nn}}
  \qquad\text{if}\qquad
  \braket{\Psi_\text{HF}}{\Psi_\text{HF}}=1
\end{equation}
Hence,
\begin{equation}
  E_\text{tot}^\text{HF}
  =
  \EHF+V_\text{nn}
\end{equation}
In what follows, $\EHF$ will denote the electronic Hartree--Fock energy
unless stated otherwise.

\paragraph{Two-electron contribution: Coulomb and exchange.}

The effect of antisymmetry first appears explicitly in the expectation value
of the electron-electron interaction. For each pair of occupied spin
orbitals, two contributions arise,
\begin{equation}
\begin{split}
  \mel{\Psi_\text{HF}}{\chO_2}{\Psi_\text{HF}}
  &=
  \sum_{i<j}^{N}
  \left[
    \mel{\chi_i(\bx_1)\chi_j(\bx_2)}{r_{12}^{-1}}
        {\chi_i(\bx_1)\chi_j(\bx_2)}
  \right.
  \\
  &\hspace{7em}\left.
    -
    \mel{\chi_i(\bx_1)\chi_j(\bx_2)}{r_{12}^{-1}}
        {\chi_j(\bx_1)\chi_i(\bx_2)}
  \right]
  \\
  &=
  \sum_{i<j}^{N}
  \qty(
    \underbrace{\cJ_{ij}}_{\green{\text{Coulomb}}}
    -
    \underbrace{\cK_{ij}}_{\red{\text{exchange}}}
  )
  \label{eq:hf-two-electron-energy}
\end{split}
\end{equation}

The first term,
\begin{equation}
  \cJ_{ij}
  =
  \mel{\chi_i(\bx_1)\chi_j(\bx_2)}{r_{12}^{-1}}
       {\chi_i(\bx_1)\chi_j(\bx_2)}
\end{equation}
is the \green{Coulomb integral}. Written explicitly,
\begin{equation}
  \cJ_{ij}
  =
  \iint
  \chi_i^*(\bx_1)\chi_j^*(\bx_2)
  \frac{1}{r_{12}}
  \chi_i(\bx_1)\chi_j(\bx_2)
  \dd{\bx_1}\dd{\bx_2}
  \label{eq:coulomb-integral}
\end{equation}
It has a direct classical interpretation: it is the electrostatic repulsion
between the charge distributions associated with orbitals $i$ and $j$.

The second term,
\begin{equation}
  \cK_{ij}
  =
  \mel{\chi_i(\bx_1)\chi_j(\bx_2)}{r_{12}^{-1}}
       {\chi_j(\bx_1)\chi_i(\bx_2)}
\end{equation}
is the \red{exchange integral},
\begin{equation}
  \cK_{ij}
  =
  \iint
  \chi_i^*(\bx_1)\chi_j^*(\bx_2)
  \frac{1}{r_{12}}
  \chi_j(\bx_1)\chi_i(\bx_2)
  \dd{\bx_1}\dd{\bx_2}
  \label{eq:exchange-integral}
\end{equation}
Unlike $\cJ_{ij}$, the exchange contribution has no classical electrostatic
counterpart. It arises entirely from the antisymmetry of the fermionic
wavefunction: it is the energetic signature of the exchange of electron
labels between the two terms of the determinant.

This point is fundamental. The Coulomb interaction itself is the same
$1/r_{12}$ operator as before; there is no separate ``exchange force'' in the
Hamiltonian. Exchange appears because the expectation value of the Coulomb
operator is evaluated with an antisymmetric wavefunction.

Because the spin orbitals factorize into spin and spatial parts, the exchange
integral also carries an important spin selection rule. If $\chi_i$ and
$\chi_j$ have opposite spin,
\begin{equation}
  \braket{\alpha}{\beta}=0
  \qquad\Longrightarrow\qquad
  \cK_{ij}=0
\end{equation}
Exchange therefore acts only between electrons of the same spin in the
collinear RHF and UHF formulations considered here. Coulomb interactions, by
contrast, are present between every pair of electrons regardless of spin.

\paragraph{Self-interaction cancellation.}

The pairwise form of the two-electron energy can equivalently be written as a
double sum,
\begin{equation}
  \mel{\Psi_\text{HF}}{\chO_2}{\Psi_\text{HF}}
  =
  \frac{1}{2}
  \sum_{i=1}^{N}
  \sum_{j=1}^{N}
  \qty(
    \cJ_{ij}-\cK_{ij}
  )
  \label{eq:hf-double-sum}
\end{equation}
The factor $1/2$ avoids counting every distinct electron pair twice.

At first sight, the terms with $i=j$ in this expression might appear to
describe the Coulomb interaction of an electron with itself. However,
  $\red{
  \cJ_{ii}=\cK_{ii}
  }$
because
\begin{equation}
  \cJ_{ii}
  =
  \cK_{ii}
  =
  \iint
  \frac{
  \abs{\chi_i(\bx_1)}^2
  \abs{\chi_i(\bx_2)}^2
  }{r_{12}}
  \dd{\bx_1}\dd{\bx_2}
\end{equation}
Consequently,
\begin{equation}
  \cJ_{ii}-\cK_{ii}=0
\end{equation}
The spurious Coulomb self-interaction is therefore cancelled exactly by
exchange in Hartree--Fock theory.

Combining the one- and two-electron contributions gives the fundamental HF
energy expression
\begin{equation}
  \EHF
  =
  \sum_{i=1}^{N}h_{ii}
  +
  \sum_{i<j}^{N}
  \qty(
    \cJ_{ij}-\cK_{ij}
  )
  \label{eq:hf-energy-spin-orbitals}
\end{equation}
Equivalently,
\begin{equation}
  \boxed{
  \EHF
  =
  \sum_{i=1}^{N}h_{ii}
  +
  \frac{1}{2}
  \sum_{i=1}^{N}
  \sum_{j=1}^{N}
  \qty(
    \cJ_{ij}-\cK_{ij}
  )
  }
  \label{eq:hf-energy-spin-orbitals-double}
\end{equation}

This expression already reveals the central structure of HF theory. The
energy depends on each occupied orbital individually through $h_{ii}$, but
also on every pair of occupied orbitals through the Coulomb and exchange
terms. The orbitals therefore cannot be optimized independently: changing
one orbital changes the effective field experienced by all the others. This
coupling is the origin of the self-consistent character of the Hartree--Fock
equations.

\paragraph{Coulomb operator.}

It is useful to promote the Coulomb and exchange integrals to one-electron
operators. For a fixed occupied spin orbital $\chi_j$, the Coulomb operator
$\chJ_j$ acts on $\chi_i$ as
\begin{equation}
  \chJ_j(\bx_1)\ket{\chi_i(\bx_1)}
  =
  \mel{\chi_j(\bx_2)}{r_{12}^{-1}}{\chi_j(\bx_2)}
  \ket{\chi_i(\bx_1)}
\end{equation}
or explicitly,
\begin{equation}
  \chJ_j(\bx_1)\ket{\chi_i(\bx_1)}
  =
  \qty[
    \int
    \chi_j^*(\bx_2)
    \frac{1}{r_{12}}
    \chi_j(\bx_2)
    \dd{\bx_2}
  ]
  \ket{\chi_i(\bx_1)}
  \label{eq:coulomb-operator}
\end{equation}
The quantity in square brackets is the electrostatic potential generated by
the density of orbital $j$ at the position of electron $1$.

Its expectation value is
\begin{equation}
\begin{split}
  \cJ_{ij}
  &=
  \mel{\chi_i(\bx_1)}
       {\chJ_j(\bx_1)}
       {\chi_i(\bx_1)}
  \\
  &=
  \mel{\chi_i(\bx_1)\chi_j(\bx_2)}
       {r_{12}^{-1}}
       {\chi_i(\bx_1)\chi_j(\bx_2)}
  \\
  &=
  \iint
  \chi_i^*(\bx_1)
  \chi_j^*(\bx_2)
  \frac{1}{r_{12}}
  \chi_i(\bx_1)
  \chi_j(\bx_2)
  \dd{\bx_1}\dd{\bx_2}
\end{split}
\end{equation}

The Coulomb operator is local in the usual one-particle sense. Once the
integration over $\bx_2$ has been carried out, it acts on $\chi_i(\bx_1)$ by
multiplication with a function of $\bx_1$,
\begin{equation}
  \chJ_j(\bx_1)\chi_i(\bx_1)
  =
  J_j(\bx_1)\chi_i(\bx_1)
\end{equation}
Every orbital acted upon by $\chJ_j$ therefore experiences the same
multiplicative Coulomb potential generated by orbital $j$.

\paragraph{Exchange operator.}

The exchange operator has a qualitatively different structure. Its action is
defined by
\begin{equation}
  \chK_j(\bx_1)\ket{\chi_i(\bx_1)}
  =
  \mel{\chi_j(\bx_2)}{r_{12}^{-1}}{\chi_i(\bx_2)}
  \ket{\chi_j(\bx_1)}
\end{equation}
or explicitly,
\begin{equation}
  \chK_j(\bx_1)\ket{\chi_i(\bx_1)}
  =
  \qty[
    \int
    \chi_j^*(\bx_2)
    \frac{1}{r_{12}}
    \chi_i(\bx_2)
    \dd{\bx_2}
  ]
  \ket{\chi_j(\bx_1)}
  \label{eq:exchange-operator}
\end{equation}
Its matrix element is
\begin{equation}
\begin{split}
  \cK_{ij}
  &=
  \mel{\chi_i(\bx_1)}
       {\chK_j(\bx_1)}
       {\chi_i(\bx_1)}
  \\
  &=
  \mel{\chi_i(\bx_1)\chi_j(\bx_2)}
       {r_{12}^{-1}}
       {\chi_j(\bx_1)\chi_i(\bx_2)}
  \\
  &=
  \iint
  \chi_i^*(\bx_1)
  \chi_j^*(\bx_2)
  \frac{1}{r_{12}}
  \chi_j(\bx_1)
  \chi_i(\bx_2)
  \dd{\bx_1}\dd{\bx_2}
\end{split}
\end{equation}

Unlike the Coulomb operator, $\chK_j$ is nonlocal. Its value at
$\bx_1$ depends on the values of the orbital on which it acts,
$\chi_i(\bx_2)$, over all coordinates $\bx_2$. Moreover, the resulting
function is proportional to $\chi_j(\bx_1)$ rather than simply to
$\chi_i(\bx_1)$. Exchange therefore cannot, in general, be represented by a
single multiplicative potential.

This distinction between the local Coulomb operator and the nonlocal exchange
operator is one of the defining features of Hartree--Fock theory. Both arise
from the same electron-electron Coulomb interaction, but the exchange term
is generated solely by the antisymmetry of the Slater determinant.

The HF energy can therefore be regarded as a functional of the occupied
orbitals,
\begin{align}
  \EHF[\{\chi_i\}]
  & =
  \sum_i
  \mel{\chi_i}{\hh}{\chi_i}
  +
  \frac{1}{2}
  \sum_{ij}
  \qty(
    \cJ_{ij}-\cK_{ij}
  )
  &
  \braket{\chi_i}{\chi_j}& =\delta_{ij}
  \label{eq:hf-energy-functional}
\end{align}
The remainder of Hartree--Fock theory consists essentially in minimizing this
functional with respect to the occupied orbitals while preserving their
orthonormality. Before carrying out that minimization, however, it is useful
to introduce a compact notation for the one- and two-electron integrals
appearing throughout the theory.

\subsubsection{One- and two-electron integrals}

The Hartree--Fock energy derived in the previous section is expressed entirely
in terms of matrix elements of one- and two-electron operators. These
quantities occur throughout electronic-structure theory, so it is useful to
introduce their notation carefully at this stage.

There are two widely used conventions for two-electron integrals:
physicists' notation and chemists' notation. They describe
exactly the same mathematical objects, but group the four orbital indices
differently. Confusion between these conventions is a frequent source of
index errors, particularly when moving between quantum-chemistry and
many-body-theory literature.

\paragraph{One-electron integrals.}

For a general one-electron operator $\hh$, the matrix element between spin
orbitals $\chi_i$ and $\chi_j$ is
\begin{equation}
  \mel{i}{\hh}{j}
  =
  h_{ij}
  =
  \mel*{\chi_i}{\hh}{\chi_j}
  =
  \int
  \chi_i^*(\bx_1)
  \hh(\bx_1)
  \chi_j(\bx_1)
  \dd{\bx_1}
  \label{eq:one-electron-integral}
\end{equation}
For the Hartree--Fock problem, $\hh$ is the core Hamiltonian,
\begin{equation}
  \hh(\bx)
  =
  -\frac{\laplacian}{2}
  -
  \sum_A\frac{Z_A}{r_A}
\end{equation}
so that $h_{ij}$ contains both the electronic kinetic energy and the
electron-nuclear attraction.

Since $\hh$ is Hermitian,
\begin{equation}
  h_{ij}=h_{ji}^*
\end{equation}
For real orbitals and a real one-electron operator, this reduces to
\begin{equation}
  h_{ij}=h_{ji}
\end{equation}

In Mulliken notation, the same one-electron integral may equivalently be
written
\begin{equation}
  \rmel{i}{\hh}{j}
  =
  h_{ij}
  =
  \rmel{\chi_i}{\hh}{\chi_j}
  =
  \int
  \chi_i^*(\bx_1)
  \hh(\bx_1)
  \chi_j(\bx_1)
  \dd{\bx_1}
\end{equation}
Thus, for one-electron quantities, the distinction between physicists' and
chemists' notation is essentially cosmetic. The important difference appears
for two-electron integrals.

\paragraph{Electron-repulsion integrals: Dirac or physicists' notation.}

The matrix element of the Coulomb interaction between two-electron product
states defines an electron-repulsion integral (ERI). In Dirac (or physicists')
notation,
\begin{equation}
  \braket{ij}{kl}
  =
  \braket{\chi_i\chi_j}{\chi_k\chi_l}
  =
  \iint
  \chi_i^*(\bx_1)
  \chi_j^*(\bx_2)
  \frac{1}{r_{12}}
  \chi_k(\bx_1)
  \chi_l(\bx_2)
  \dd{\bx_1}\dd{\bx_2}
  \label{eq:eri-physicists}
\end{equation}
The ordering of the indices reflects the underlying braket structure:
$i$ and $j$ label the two orbitals in the bra, whereas $k$ and $l$ label
those in the ket. More importantly, the first and third indices refer to
electron coordinate $\bx_1$, while the second and fourth refer to $\bx_2$:
\begin{equation}
  \braket*{
    \underbrace{i}_{\bx_1}
    \underbrace{j}_{\bx_2}
  }{
    \underbrace{k}_{\bx_1}
    \underbrace{l}_{\bx_2}
  }
\end{equation}

With this notation, the Coulomb and exchange integrals introduced in the
previous section are simply
\begin{align}
  \cJ_{ij}&=\braket{ij}{ij}
  &
  \cK_{ij}&=\braket{ij}{ji}
  \label{eq:JK-physicists}
\end{align}
The only difference between the two is therefore the permutation of the two
indices in the ket.

This observation motivates the introduction of the
antisymmetrized two-electron integral,
\begin{equation}
  \boxed{
  \mel{ij}{}{kl}
  =
  \braket{ij}{kl}
  -
  \braket{ij}{lk}
  }
  \label{eq:antisymmetrized-eri}
\end{equation}
Explicitly,
\begin{equation}
\begin{split}
  \mel{ij}{}{kl}
  &=
  \iint
  \chi_i^*(\bx_1)
  \chi_j^*(\bx_2)
  \frac{1}{r_{12}}
  \qty[
    \chi_k(\bx_1)\chi_l(\bx_2)
    -
    \chi_l(\bx_1)\chi_k(\bx_2)
  ]
  \dd{\bx_1}\dd{\bx_2}
  \\
  &=
  \iint
  \chi_i^*(\bx_1)
  \chi_j^*(\bx_2)
  \frac{1}{r_{12}}
  (1-\chP_{12})
  \chi_k(\bx_1)\chi_l(\bx_2)
  \dd{\bx_1}\dd{\bx_2}
\end{split}
\end{equation}
The antisymmetrized integral combines the direct Coulomb contribution and
its exchange counterpart into a single object. For example,
\begin{equation}
  \mel{ij}{}{ij}
  =
  \cJ_{ij}-\cK_{ij}
\end{equation}
The HF energy can therefore also be written compactly as
\begin{equation}
  \EHF
  =
  \sum_i h_{ii}
  +
  \frac{1}{2}
  \sum_{ij}
  \mel{ij}{}{ij}
\end{equation}

As expected from its definition, the antisymmetrized integral changes sign
upon exchanging either pair of fermionic indices,
\begin{equation}
  \mel{ij}{}{kl}
  =
  -\mel{ji}{}{kl}
  =
  -\mel{ij}{}{lk}
  =
  \mel{ji}{}{lk}
\end{equation}
This notation becomes particularly convenient in post-Hartree--Fock and
many-body perturbation theories, where direct and exchange contributions
repeatedly occur together.

\paragraph{Mulliken or chemists' notation.}

Quantum-chemistry literature very often uses a different grouping of the
same four orbital indices. In Mulliken, or chemists', notation,
\begin{equation}
  \rbraket{ij}{kl}
  =
  \rbraket{\chi_i\chi_j}{\chi_k\chi_l}
  =
  \iint
  \chi_i^*(\bx_1)
  \chi_j(\bx_1)
  \frac{1}{r_{12}}
  \chi_k^*(\bx_2)
  \chi_l(\bx_2)
  \dd{\bx_1}\dd{\bx_2}
  \label{eq:eri-chemists}
\end{equation}
Here, the first pair of indices belongs to electron coordinate $\bx_1$ and
the second pair belongs to electron coordinate $\bx_2$:
\begin{equation}
  \rbraket{
    \underbrace{ij}_{\bx_1}
  }{
    \underbrace{kl}_{\bx_2}
  }
\end{equation}

The two notations therefore differ only by an index permutation. Comparing
Eqs.~\eqref{eq:eri-physicists} and \eqref{eq:eri-chemists} gives
\begin{equation}
  \boxed{
  \braket{ij}{kl}
  =
  \rbraket{ik}{jl}
  }
  \label{eq:phys-chem-map}
\end{equation}
This simple correspondence is worth remembering:
\begin{equation}
  \text{physicists: }\braket{ij}{kl}
  \qquad\longleftrightarrow\qquad
  \text{chemists: }\rbraket{ik}{jl}
\end{equation}

Consequently, the Coulomb and exchange integrals become
\begin{align}
  \cJ_{ij}
  &=
  \rbraket{ii}{jj}
  &
  \cK_{ij}
  &=
  \rbraket{ij}{ji}
  &
  \text{(chemists' notation)}
\end{align}
These expressions also make the physical interpretation transparent:
$\cJ_{ij}$ couples the densities $\chi_i^*\chi_i$ and
$\chi_j^*\chi_j$, whereas $\cK_{ij}$ couples the transition-like products
$\chi_i^*\chi_j$ and $\chi_j^*\chi_i$.

The same notation will be used throughout this chapter for integrals over
spin orbitals and over spatial orbitals; the surrounding context specifies
which type is intended.

\paragraph{Spin selection rules.}

When the $\chi_p$ are collinear spin orbitals,
\begin{equation}
  \chi_p(\bx)
  =
  \sigma_p(\omega)\psi_p(\br)
\end{equation}
integration over the spin coordinates gives
\begin{equation}
  \braket{ij}{kl}
  =
  \delta_{\sigma_i\sigma_k}
  \delta_{\sigma_j\sigma_l}
  \iint
  \psi_i^*(\br_1)
  \psi_j^*(\br_2)
  \frac{1}{r_{12}}
  \psi_k(\br_1)
  \psi_l(\br_2)
  \dd{\br_1}\dd{\br_2}
\end{equation}
A spin-orbital ERI therefore vanishes unless the spin on each electron line
is conserved,
  $\sigma_i=\sigma_k$
  and
  $\sigma_j=\sigma_l$.
For the exchange integral,
\begin{equation}
  \cK_{ij}=\braket{ij}{ji}
\end{equation}
this requires $\sigma_i=\sigma_j$. Hence,
  $\cK_{ij}=0$
  for opposite-spin orbitals.
This is the integral form of the same-spin nature of Hartree--Fock exchange
discussed in the previous section.

\paragraph{Permutation symmetries.}

Electron-repulsion integrals possess extensive permutation symmetries. These
follow from three simple properties: the symmetry of the Coulomb operator
under interchange of the two electrons, its Hermiticity, and, for real
orbitals, the commutativity of the orbital products evaluated at a common
coordinate.

For reference, the physicists' ERI is
\begin{equation}
  \braket{ij}{kl}
  =
  \braket{\chi_i\chi_j}{\chi_k\chi_l}
  =
  \iint
  \chi_i^*(\bx_1)
  \chi_j^*(\bx_2)
  \frac{1}{r_{12}}
  \chi_k(\bx_1)
  \chi_l(\bx_2)
  \dd{\bx_1}\dd{\bx_2}
\end{equation}
For \red{complex-valued orbitals}, physicists' ERIs obey
\begin{equation}
  \braket{ij}{kl}
  =
  \braket{ji}{lk}
  =
  \braket{kl}{ij}^*
  =
  \braket{lk}{ji}^*
  \label{eq:eri-complex-symmetry}
\end{equation}
The first equality corresponds to exchanging the two electron coordinates,
while the complex-conjugated relations follow from Hermiticity.

For \green{real-valued orbitals}, the additional within-coordinate
interchanges give
\begin{equation}
\begin{split}
  \braket{ij}{kl}
  &=
  \braket{ji}{lk}
  =
  \braket{kl}{ij}
  =
  \braket{lk}{ji}
  \\
  &=
  \braket{kj}{il}
  =
  \braket{jk}{li}
  =
  \braket{il}{kj}
  =
  \braket{li}{jk}
  \label{eq:eri-physicists-eightfold}
\end{split}
\end{equation}

In chemists' notation, the defining integral is
\begin{equation}
  \rbraket{ij}{kl}
  =
  \rbraket{\chi_i\chi_j}{\chi_k\chi_l}
  =
  \iint
  \chi_i^*(\bx_1)
  \chi_j(\bx_1)
  \frac{1}{r_{12}}
  \chi_k^*(\bx_2)
  \chi_l(\bx_2)
  \dd{\bx_1}\dd{\bx_2}
\end{equation}
For \green{real-valued orbitals}, its symmetry relations take the familiar
form
\begin{equation}
\begin{split}
  \rbraket{ij}{kl}
  &=
  \rbraket{ji}{kl}
  =
  \rbraket{ij}{lk}
  =
  \rbraket{ji}{lk}
  \\
  &=
  \rbraket{kl}{ij}
  =
  \rbraket{lk}{ij}
  =
  \rbraket{kl}{ji}
  =
  \rbraket{lk}{ji}
  \label{eq:eri-chemists-eightfold}
\end{split}
\end{equation}
Thus, for real-valued orbitals, ERIs possess
eightfold permutation symmetry.
The condition concerns the orbital functions, not merely the numerical
values of their integrals. Complex functions can yield real ERIs without
possessing all eight permutation symmetries, as the plane-wave expressions
in Section~\ref{sec:basis-sets} illustrate.

This symmetry has an important computational consequence. A naive set of
$K$ one-electron basis functions generates $K^4$ four-index ERIs, $\rbraket{\mu\nu}{\lambda\sigma}$.
However, many of these entries are identical by permutation symmetry.
For four distinct indices, up to eight nominally different integrals
represent the same numerical quantity. Modern integral algorithms exploit
these symmetries extensively to reduce both the amount of computation and
the storage required for the two-electron integrals.

The apparent $K^4$ growth nevertheless makes the treatment of ERIs one of
the central computational challenges of conventional electronic-structure
theory. Later in this chapter, when the basis functions themselves are
introduced in detail, we will return to the evaluation of these integrals
and to the algorithms that make their computation practical.

\subsubsection{Worked examples for few-electron determinants}

The general expressions derived above can appear rather compact when first
encountered. It is therefore useful to recover them explicitly for systems
containing only a few electrons. These examples illustrate how the
orthonormality of the spin orbitals and the antisymmetry of the Slater
determinant combine to produce the characteristic Hartree--Fock energy
expression.

We begin with two electrons occupying the orthonormal spin orbitals
$\chi_1$ and $\chi_2$. Throughout the following examples, the shorthand
\begin{equation}
  \chi_p(i)\equiv\chi_p(\bx_i)
\end{equation}
will be used, where $i$ labels the complete spin-spatial coordinate of
electron $i$.


\begin{hfexercise}{Normalization of the HF wavefunction}

\violet{\textit{Show that the HF wavefunction built with two normalized,
orthogonal spin orbitals $\chi_1$ and $\chi_2$ is normalized.}}

\end{hfexercise}
\begin{hfsolution}{Normalization of the HF wavefunction}

Take two orthonormal spin orbitals, allowing them to be complex. The
two-electron Slater determinant with the proposed normalization factor is
\begin{equation}
  \Psi_\text{HF}
  =
  \frac{1}{\sqrt{2}}
  \mdet{
    \chi_1(1) & \chi_2(1)\\
    \chi_1(2) & \chi_2(2)
  }
\end{equation}
or, after expanding the determinant,
\begin{equation}
  \Psi_\text{HF}
  =
  \frac{
    \chi_1(1)\chi_2(2)
    -
    \chi_2(1)\chi_1(2)
  }{\sqrt{2}}
  \label{eq:two-electron-determinant-example}
\end{equation}

Its norm is
\begin{equation}
\begin{split}
  \braket{\Psi_\text{HF}}{\Psi_\text{HF}}
  &=
  \frac{1}{2}
  \braket{
    \chi_1(1)\chi_2(2)-\chi_2(1)\chi_1(2)
  }{
    \chi_1(1)\chi_2(2)-\chi_2(1)\chi_1(2)
  }
  \\
  &=
  \frac{1}{2}
  \Big[
    \braket{\chi_1(1)\chi_2(2)}
            {\chi_1(1)\chi_2(2)}
    -
    \braket{\chi_1(1)\chi_2(2)}
            {\chi_2(1)\chi_1(2)}
  \\
  &
    -
    \braket{\chi_2(1)\chi_1(2)}
            {\chi_1(1)\chi_2(2)}
    +
    \braket{\chi_2(1)\chi_1(2)}
            {\chi_2(1)\chi_1(2)}
  \Big]
  \label{eq:two-electron-norm-expansion}
\end{split}
\end{equation}

Because the coordinates of the two electrons are independent, each
two-electron overlap factorizes into products of one-electron overlaps.
For example,
\begin{equation}
  \braket{\chi_1(1)\chi_2(2)}
          {\chi_1(1)\chi_2(2)}
  =
  \braket{\chi_1(1)}{\chi_1(1)}
  \braket{\chi_2(2)}{\chi_2(2)}
  =
  1
\end{equation}
Likewise,
\begin{equation}
\begin{split}
  \braket{\chi_1(1)\chi_2(2)}
          {\chi_2(1)\chi_1(2)}
  &=
  \braket{\chi_1}{\chi_2}
  \braket{\chi_2}{\chi_1}
  \\
  &=0
\end{split}
\end{equation}
since the spin orbitals are orthogonal. The same argument applies to the
other two terms, and therefore
\begin{equation}
  \braket{\Psi_\text{HF}}{\Psi_\text{HF}}
  =
  \frac{1}{2}[1-0-0+1]
  =
  \boxed{1}
\end{equation}

This example explains the origin of the normalization factor $1/\sqrt{2}$
for a two-electron determinant. More generally, a determinant built from
$N$ orthonormal spin orbitals contains $N!$ mutually orthogonal Hartree
products and is normalized by the factor $1/\sqrt{N!}$.

\end{hfsolution}


\begin{hfexercise}{Core Hamiltonian}

\violet{\textit{Evaluate the one-electron energy of the normalized
two-electron determinant and explain the general occupied-orbital sum
\[
  \mel{\Psi_\text{HF}}{\chO_1}{\Psi_\text{HF}}
  =
  \sum_{i=1}^{N}h_{ii}
\]
}}

\end{hfexercise}
\begin{hfsolution}{Core Hamiltonian}

For two electrons, the one-electron part of the Hamiltonian is
\begin{equation}
  \chO_1
  =
  \hh(1)+\hh(2)
\end{equation}
Using Eq.~\eqref{eq:two-electron-determinant-example},
\begin{equation}
\begin{split}
  &
  \mel{\Psi_\text{HF}}
      {\hh(1)+\hh(2)}
      {\Psi_\text{HF}}
  \\
  &=
  \frac{1}{2}
  \mel{
    \chi_1(1)\chi_2(2)-\chi_2(1)\chi_1(2)
  }
  {\hh(1)+\hh(2)}
  {
    \chi_1(1)\chi_2(2)-\chi_2(1)\chi_1(2)
  }
  \\
  &=
  \frac{1}{2}
  \Big[
  \mel{\chi_1(1)\chi_2(2)}
      {\hh(1)+\hh(2)}
      {\chi_1(1)\chi_2(2)}
  -
  \mel{\chi_1(1)\chi_2(2)}
      {\hh(1)+\hh(2)}
      {\chi_2(1)\chi_1(2)}
  \\
  &-
  \mel{\chi_2(1)\chi_1(2)}
      {\hh(1)+\hh(2)}
      {\chi_1(1)\chi_2(2)}
  +
  \mel{\chi_2(1)\chi_1(2)}
      {\hh(1)+\hh(2)}
      {\chi_2(1)\chi_1(2)}
  \Big]
  \label{eq:core-two-electron-expansion}
\end{split}
\end{equation}

Consider first the diagonal matrix element
\begin{equation}
  \mel{\chi_1(1)\chi_2(2)}
      {\hh(1)+\hh(2)}
      {\chi_1(1)\chi_2(2)}
\end{equation}
Because $\hh(1)$ acts only on the coordinates of electron $1$,
\begin{equation}
\begin{split}
  \mel{\chi_1(1)\chi_2(2)}
      {\hh(1)}
      {\chi_1(1)\chi_2(2)}
  &=
  \mel{\chi_1}{\hh}{\chi_1}
  \braket{\chi_2}{\chi_2}
  \\
  &=h_{11}
\end{split}
\end{equation}
Similarly,
\begin{equation}
  \mel{\chi_1(1)\chi_2(2)}
      {\hh(2)}
      {\chi_1(1)\chi_2(2)}
  =
  h_{22}
\end{equation}
and hence
\begin{equation}
  \mel{\chi_1(1)\chi_2(2)}
      {\hh(1)+\hh(2)}
      {\chi_1(1)\chi_2(2)}
  =
  h_{11}+h_{22}
\end{equation}

The off-diagonal terms vanish. For example,
\begin{equation}
  \mel{\chi_1(1)\chi_2(2)}
      {\hh(1)}
      {\chi_2(1)\chi_1(2)}
  =
  \mel{\chi_1}{\hh}{\chi_2}
  \braket{\chi_2}{\chi_1}
  =
  h_{12}\times 0
  =
  0
\end{equation}
and
\begin{equation}
  \mel{\chi_1(1)\chi_2(2)}
      {\hh(2)}
      {\chi_2(1)\chi_1(2)}
  =
  \braket{\chi_1}{\chi_2}
  \mel{\chi_2}{\hh}{\chi_1}
  =
  0\times h_{21}
  =
  0
\end{equation}
The cancellation therefore does \emph{not} require $h_{12}=0$; it follows
from the orthogonality of the spectator orbital.

Collecting all terms gives
\begin{equation}
\begin{split}
  \mel{\Psi_\text{HF}}{\chO_1}{\Psi_\text{HF}}
  &=
  \frac{1}{2}
  [h_{11}+h_{22}-0-0+h_{22}+h_{11}]
  \\
  &=
  h_{11}+h_{22}
\end{split}
\end{equation}

This is the two-electron instance of the general Slater--Condon rule
\begin{equation}
  \mel{\Psi_\text{HF}}{\chO_1}{\Psi_\text{HF}}
  =
  \sum_{i=1}^{N}h_{ii}
\end{equation}
Only diagonal matrix elements over the occupied spin orbitals survive.
To see why the same argument holds for $N$ electrons, expand the bra and
ket determinants into their $N!$ Hartree products. A one-electron operator
leaves $N-1$ spectator overlaps. These force the corresponding orbital
assignments to agree; because each product contains the same occupied set,
the remaining assignment must then agree as well. Thus only identical
permutations survive. Each gives the sum of the $N$ occupied diagonal
matrix elements, and the $N!$ equal contributions cancel the normalization
factor $1/N!$. No assumption that the occupied orbitals diagonalize
$\hh$ is needed.

\end{hfsolution}


\begin{hfexercise}{Two-electron Hamiltonian}

\violet{\textit{Evaluate the two-electron interaction for the same
normalized determinant, explain the general pair sum}}
\begin{equation}
  \mel{\Psi_\text{HF}}{\chO_2}{\Psi_\text{HF}}
  =
  \sum_{i<j}^{N}
  (\cJ_{ij}-\cK_{ij})
\end{equation}
\violet{\textit{and write down the HF energy.}}

\end{hfexercise}
\begin{hfsolution}{Two-electron Hamiltonian}

For a two-electron system there is only one distinct electron pair, so
\begin{equation}
  \chO_2
  =
  \frac{1}{r_{12}}
\end{equation}
Direct expansion gives
\begin{equation}
\begin{split}
  \mel{\Psi_\text{HF}}{r_{12}^{-1}}{\Psi_\text{HF}}
  &=
  \frac{1}{2}
  \mel{\chi_1\chi_2-\chi_2\chi_1}
      {r_{12}^{-1}}
      {\chi_1\chi_2-\chi_2\chi_1}
  \\
  &=
  \frac{1}{2}
  \Big[
    \mel{\chi_1\chi_2}
        {r_{12}^{-1}}
        {\chi_1\chi_2}
    -
    \mel{\chi_1\chi_2}
        {r_{12}^{-1}}
        {\chi_2\chi_1}
    -
    \mel{\chi_2\chi_1}
        {r_{12}^{-1}}
        {\chi_1\chi_2}
    +
    \mel{\chi_2\chi_1}
        {r_{12}^{-1}}
        {\chi_2\chi_1}
  \Big]
  \label{eq:two-electron-two-body-expansion}
\end{split}
\end{equation}

The first term is the direct, or Coulomb, contribution,
\begin{equation}
  \mel{\chi_1\chi_2}
      {r_{12}^{-1}}
      {\chi_1\chi_2}
  =
  \cJ_{12}
\end{equation}
whereas the second is the exchange contribution,
\begin{equation}
  \mel{\chi_1\chi_2}
      {r_{12}^{-1}}
      {\chi_2\chi_1}
  =
  \cK_{12}
\end{equation}
By exchanging the dummy electron coordinates,
\begin{equation}
  \mel{\chi_2\chi_1}
      {r_{12}^{-1}}
      {\chi_2\chi_1}
  =
  \mel{\chi_1\chi_2}
      {r_{12}^{-1}}
      {\chi_1\chi_2}
  =
  \cJ_{12}
\end{equation}
and similarly,
\begin{equation}
  \mel{\chi_2\chi_1}
      {r_{12}^{-1}}
      {\chi_1\chi_2}
  =
  \cK_{12}
\end{equation}

Consequently,
\begin{equation}
\begin{split}
  \mel{\Psi_\text{HF}}{r_{12}^{-1}}{\Psi_\text{HF}}
  &=
  \frac{1}{2}
  [
    \cJ_{12}
    -
    \cK_{12}
    -
    \cK_{12}
    +
    \cJ_{12}
  ]
  \\
  &=
  \cJ_{12}-\cK_{12}
\end{split}
\end{equation}

This simple calculation contains the essential physics of Hartree--Fock
theory. The Coulomb term would already be present for a product
wavefunction. The exchange term appears solely because the wavefunction has
been antisymmetrized.
For $N$ electrons, the same reasoning applies to each pair of occupied
spin orbitals. The $N-2$ spectator overlaps fix all other assignments.
The two active orbitals can then be matched directly, giving
$\cJ_{ij}$, or exchanged, giving $-\cK_{ij}$. Summing over $i<j$
counts each distinct pair once; a double sum over all $i,j$ instead
requires a factor of one half. The terms with $i=j$ vanish because
$\cJ_{ii}=\cK_{ii}$.

Combining the one- and two-electron contributions gives the electronic
Hartree--Fock energy
\begin{equation}
  E_\text{HF}
  =
  h_{11}
  +
  h_{22}
  +
  \cJ_{12}
  -
  \cK_{12}
  \label{eq:two-electron-hf-energy}
\end{equation}

If the two occupied spin orbitals have opposite spin, as in a closed-shell
two-electron system,
\begin{align}
  \chi_1(\bx)&=\psi(\br)\alpha(\omega)
  &
  \chi_2(\bx)&=\psi(\br)\beta(\omega)
\end{align}
then their exchange integral vanishes because
$\braket{\alpha}{\beta}=0$, i.e., $\cK_{12}=0$.
This does not mean that exchange is absent from Hartree--Fock theory; rather,
exchange couples only same-spin spin orbitals. The corresponding
closed-shell spatial-orbital expression will be derived explicitly in the
next subsection.

\end{hfsolution}


\begin{hfexercise}{Three-electron system}

\violet{\textit{Find the HF energy of a three-electron system composed of
the orthonormal spin orbitals $\chi_1$, $\chi_2$, and $\chi_3$.}}

\end{hfexercise}
\begin{hfsolution}{Three-electron system}

Assume that $\chi_1$, $\chi_2$, and $\chi_3$ are orthonormal and that
the electronic energy is sought at a fixed nuclear geometry. The
normalized determinant is
\begin{equation}
  \Psi_\text{HF}
  =\frac{1}{\sqrt{3!}}
  \mdet{
    \chi_1(1)&\chi_2(1)&\chi_3(1)\\
    \chi_1(2)&\chi_2(2)&\chi_3(2)\\
    \chi_1(3)&\chi_2(3)&\chi_3(3)
  }
\end{equation}
Its six Hartree products are mutually orthogonal, so their squared
coefficients sum to $6/3!=1$. We may therefore use the one- and
two-electron expectation values established in the preceding exercises.
The one-electron operator contains one contribution for each electron,
\begin{equation}
  \chO_1
  =
  \hh(\bx_1)
  +
  \hh(\bx_2)
  +
  \hh(\bx_3)
\end{equation}
while the two-electron interaction contains one term for every distinct
electron pair,
\begin{equation}
  \chO_2
  =
  r_{12}^{-1}
  +
  r_{13}^{-1}
  +
  r_{23}^{-1}
\end{equation}

There are therefore three one-electron contributions,
\begin{equation}
  h_{11}+h_{22}+h_{33}
\end{equation}
and three distinct orbital pairs: (1,2), (1,3), and (2,3).
Each pair contributes one Coulomb term and one exchange term. Applying the
preceding two-electron result to each pair gives
\begin{equation}
  E_\text{HF}
  =
  h_{11}+h_{22}+h_{33}
  +
  \cJ_{12}+\cJ_{13}+\cJ_{23}
  -
  \cK_{12}-\cK_{13}-\cK_{23}
  \label{eq:three-electron-hf-energy}
\end{equation}

Equivalently,
\begin{equation}
  E_\text{HF}
  =
  \sum_{i=1}^{3}h_{ii}
  +
  \sum_{i<j}^{3}
  (\cJ_{ij}-\cK_{ij})
\end{equation}
The pattern is now clear: extending the result to an arbitrary number of
electrons gives
\begin{equation}
  E_\text{HF}
  =
  \sum_{i=1}^{N}h_{ii}
  +
  \sum_{i<j}^{N}
  (\cJ_{ij}-\cK_{ij})
\end{equation}
For three electrons, the interaction contains exactly
$\binom{3}{2}=3$ distinct pairs. This count is a useful check: there is
neither a self-interaction term nor an additional factor of one half in
the explicit pair expression. The spin dependence remains inside the
spin-orbital exchange integrals; no particular spin occupation was assumed.

\end{hfsolution}

These few-electron examples provide an explicit derivation of the
Slater--Condon expressions used above. They also reveal the structure that
will remain throughout Hartree--Fock theory: one-electron contributions are
associated with individual occupied spin orbitals, while the
electron-electron interaction generates pairwise Coulomb and exchange
terms.

So far, all expressions have been written in terms of spin orbitals. For
closed-shell systems, however, the $\alpha$ and $\beta$ electrons occur in
pairs that share the same spatial orbital. Exploiting this structure leads
to the more familiar restricted spatial-orbital formulation developed next.

\subsection{Spin-orbital and spatial-orbital formulations}

The spin-orbital energy expression is general, but closed-shell calculations
are more naturally interpreted in terms of doubly occupied spatial orbitals.
This subsection performs that reduction explicitly and then applies it to
closed- and open-shell atomic examples.

\subsubsection{From spin orbitals to spatial orbitals}

The Hartree--Fock energy has so far been written in its most general
spin-orbital form,
\begin{equation}
  \EHF
  =
  \sum_{i=1}^{N}\mel{i}{\hh}{i}
  +
  \frac{1}{2}
  \sum_{i=1}^{N}\sum_{j=1}^{N}
  \qty[
    \braket{ij}{ij}
    -
    \braket{ij}{ji}
  ]
  \label{eq:hf-energy-spin-general}
\end{equation}
This expression is compact and applies equally well to restricted and
unrestricted determinants. In many molecular applications, however, it is
more natural to work with spatial orbitals. The passage from one
representation to the other is obtained simply by carrying out the spin
integrations explicitly.

The essential ingredients are the orthonormality relations
\begin{align}
  \braket{\alpha}{\alpha}
  =
  \braket{\beta}{\beta}
  &=
  1
  &
  \braket{\alpha}{\beta}
  =
  \braket{\beta}{\alpha}
  &=
  0
  \label{eq:spin-orthogonality-spatial-reduction}
\end{align}
As we shall see, the first two relations preserve the Coulomb interaction
between all electron pairs, whereas the last two eliminate exchange between
opposite-spin electrons.


\subsubsection{Closed-shell \ce{H2} in a minimal basis}

The simplest example is closed-shell \ce{H2} in a minimal basis. There is
one doubly occupied spatial orbital, $\psi_1(\br)$, from which the two
occupied spin orbitals are constructed:
\begin{equation}
\begin{split}
  \chi_1(\bx)
  &\equiv
  \psi_1(\bx)
  =
  \psi_1(\br)\alpha(\omega)
  \\
  \chi_2(\bx)
  &\equiv
  \Bar{\psi}_1(\bx)
  =
  \psi_1(\br)\beta(\omega)
  \label{eq:h2-spin-orbitals}
\end{split}
\end{equation}
The bar therefore denotes the $\beta$ spin partner of the corresponding
$\alpha$ spin orbital; both contain the same spatial function.

In the spin-orbital representation, the Hartree--Fock energy is
\begin{equation}
  \EHF
  =
  \mel{\chi_1}{\hh}{\chi_1}
  +
  \mel{\chi_2}{\hh}{\chi_2}
  +
  \braket{\chi_1\chi_2}{\chi_1\chi_2}
  -
  \braket{\chi_1\chi_2}{\chi_2\chi_1}
  \label{eq:h2-spin-orbital-energy}
\end{equation}
Using the barred notation, the same expression can be written in
spatial-orbital-style notation as
\begin{equation}
  \EHF
  =
  \rmel{\psi_1}{\hh}{\psi_1}
  +
  \rmel{\Bar{\psi}_1}{\hh}{\Bar{\psi}_1}
  +
  \rbraket{\psi_1\psi_1}{\Bar{\psi}_1\Bar{\psi}_1}
  -
  \rbraket{\psi_1\Bar{\psi}_1}{\Bar{\psi}_1\psi_1}
\end{equation}
We now evaluate each contribution by separating the spin and spatial
coordinates.

\paragraph{One-electron contribution.}

The core Hamiltonian contains no spin dependence. Consequently, its matrix
elements factorize into independent spin and spatial integrals. For the
$\alpha$ spin orbital,
\begin{equation}
\begin{split}
  \rmel{\chi_1}{\hh}{\chi_1}
  &=
  \int
  \chi_1^*(\bx)
  \hh(\bx)
  \chi_1(\bx)
  \dd{\bx}
  \\
  &=
  \iint
  \alpha^*(\omega)
  \psi_1^*(\br)
  \hh(\br)
  \alpha(\omega)
  \psi_1(\br)
  \dd{\omega}\dd{\br}
  \\
  &=
  \underbrace{
  \qty[
    \int
    \alpha^*(\omega)\alpha(\omega)
    \dd{\omega}
  ]}_{=1}
  \underbrace{
  \qty[
    \int
    \psi_1^*(\br)
    \hh(\br)
    \psi_1(\br)
    \dd{\br}
  ]}_{\rmel{\psi_1}{\hh}{\psi_1}}
\end{split}
\end{equation}
Thus,
\begin{equation}
  \rmel{\chi_1}{\hh}{\chi_1}
  =
  \rmel{\psi_1}{\hh}{\psi_1}
\end{equation}

Exactly the same argument applies to the $\beta$ spin orbital,
\begin{equation}
\begin{split}
  \rmel{\chi_2}{\hh}{\chi_2}
  &=
  \int
  \chi_2^*(\bx)
  \hh(\bx)
  \chi_2(\bx)
  \dd{\bx}
  \\
  &=
  \iint
  \beta^*(\omega)
  \psi_1^*(\br)
  \hh(\br)
  \beta(\omega)
  \psi_1(\br)
  \dd{\omega}\dd{\br}
  \\
  &=
  \underbrace{
  \qty[
    \int
    \beta^*(\omega)\beta(\omega)
    \dd{\omega}
  ]}_{=1}
  \underbrace{
  \qty[
    \int
    \psi_1^*(\br)
    \hh(\br)
    \psi_1(\br)
    \dd{\br}
  ]}_{\rmel{\psi_1}{\hh}{\psi_1}}
\end{split}
\end{equation}
and therefore
\begin{equation}
  \rmel{\chi_2}{\hh}{\chi_2}
  =
  \rmel{\psi_1}{\hh}{\psi_1}
\end{equation}

The two electrons thus contribute identically to the one-electron energy:
\begin{equation}
  \rmel{\chi_1}{\hh}{\chi_1}
  +
  \rmel{\chi_2}{\hh}{\chi_2}
  =
  2\rmel{\psi_1}{\hh}{\psi_1}
\end{equation}
The factor of two is simply the occupation number of the doubly occupied
spatial orbital.

\paragraph{Coulomb contribution.}

The Coulomb term contains only diagonal spin overlaps and therefore survives
the spin integration. Explicitly,
\begin{equation}
\begin{split}
  \rbraket{\chi_1\chi_1}{\chi_2\chi_2}
  &=
  \iint
  \chi_1^*(\bx_1)
  \chi_1(\bx_1)
  \frac{1}{r_{12}}
  \chi_2^*(\bx_2)
  \chi_2(\bx_2)
  \dd{\bx_1}\dd{\bx_2}
  \\
  &=
  \iiiint
  \alpha^*(\omega_1)
  \psi_1^*(\br_1)
  \alpha(\omega_1)
  \psi_1(\br_1)
  \frac{1}{r_{12}}
  \\
  &\times
  \beta^*(\omega_2)
  \psi_1^*(\br_2)
  \beta(\omega_2)
  \psi_1(\br_2)
  \dd{\omega_1}\dd{\br_1}
  \dd{\omega_2}\dd{\br_2}
  \\
  &=
  \underbrace{
  \qty[
    \int
    \alpha^*(\omega_1)\alpha(\omega_1)
    \dd{\omega_1}
  ]}_{=1}
  \underbrace{
  \qty[
    \int
    \beta^*(\omega_2)\beta(\omega_2)
    \dd{\omega_2}
  ]}_{=1}
  \\
  &\times
  \underbrace{
  \qty[
    \iint
    \psi_1^*(\br_1)\psi_1(\br_1)
    \frac{1}{r_{12}}
    \psi_1^*(\br_2)\psi_1(\br_2)
    \dd{\br_1}\dd{\br_2}
  ]}_{\rbraket{\psi_1\psi_1}{\psi_1\psi_1}}
\end{split}
\end{equation}
Hence,
\begin{equation}
  \rbraket{\chi_1\chi_1}{\chi_2\chi_2}
  =
  \rbraket{\psi_1\psi_1}{\psi_1\psi_1}
\end{equation}

This is the electrostatic repulsion between the two electrons occupying the
same spatial orbital. The Coulomb interaction is insensitive to whether the
electrons have parallel or antiparallel spins.

\paragraph{Exchange contribution.}

The exchange term behaves differently because the spin functions are crossed:
\begin{equation}
\begin{split}
  \rbraket{\chi_1\chi_2}{\chi_2\chi_1}
  &=
  \iint
  \chi_1^*(\bx_1)
  \chi_2(\bx_1)
  \frac{1}{r_{12}}
  \chi_2^*(\bx_2)
  \chi_1(\bx_2)
  \dd{\bx_1}\dd{\bx_2}
  \\
  &=
  \iiiint
  \alpha^*(\omega_1)
  \psi_1^*(\br_1)
  \beta(\omega_1)
  \psi_1(\br_1)
  \frac{1}{r_{12}}
  \\
  &\times
  \beta^*(\omega_2)
  \psi_1^*(\br_2)
  \alpha(\omega_2)
  \psi_1(\br_2)
  \dd{\omega_1}\dd{\br_1}
  \dd{\omega_2}\dd{\br_2}
  \\
  &=
  \underbrace{
  \qty[
    \int
    \alpha^*(\omega_1)\beta(\omega_1)
    \dd{\omega_1}
  ]}_{=0}
  \underbrace{
  \qty[
    \int
    \beta^*(\omega_2)\alpha(\omega_2)
    \dd{\omega_2}
  ]}_{=0}
  \\
  &\times
  \underbrace{
  \qty[
    \iint
    \psi_1^*(\br_1)\psi_1(\br_1)
    \frac{1}{r_{12}}
    \psi_1^*(\br_2)\psi_1(\br_2)
    \dd{\br_1}\dd{\br_2}
  ]}_{\rbraket{\psi_1\psi_1}{\psi_1\psi_1}}
  \\
  &=0
\end{split}
\end{equation}

Thus,
\begin{equation}
  \rbraket{\chi_1\chi_2}{\chi_2\chi_1}=0
\end{equation}
The spatial part of the integral is perfectly finite; it is the
\emph{orthogonality of the spin functions} that makes the complete
spin-orbital exchange integral vanish.

This is an important point. Opposite-spin electrons still repel each other
through the Coulomb interaction, but there is no HF exchange contribution
between them. Exchange survives only for pairs of electrons with the same
spin.

Combining the preceding results gives the spatial-orbital expression for
minimal-basis \ce{H2},
\begin{equation}
  \EHF
  =
  2\rmel{\psi_1}{\hh}{\psi_1}
  +
  \rbraket{\psi_1\psi_1}{\psi_1\psi_1}
  =
  2\rmel{1}{\hh}{1}
  +
  \rbraket{11}{11}
  \label{eq:h2-rhf-energy}
\end{equation}

The simple \ce{H2} example contains the entire logic needed for the general
closed-shell case: every occupied spatial orbital contributes twice to the
one-electron energy, Coulomb interactions occur between all spin
combinations, and exchange occurs only between same-spin electrons.


\subsubsection{General closed-shell determinants}

Consider now a system containing an even number $N$ of electrons described
by a closed-shell RHF determinant. There are $N/2$ occupied spatial orbitals,
and each spatial orbital $\psi_i$ generates an $\alpha$ and a $\beta$ spin
orbital,
  $\psi_i(\br)\alpha(\omega)$
  and
  $\psi_i(\br)\beta(\omega)$.

To avoid ambiguity, in the remainder of this derivation the indices
$i,j=1,\ldots,N/2$ label \emph{occupied spatial orbitals}, while
$\Bar{i}$ denotes the $\beta$ spin partner of spatial orbital $i$.

Starting from the general spin-orbital expression,
\begin{equation}
  \EHF
  =
  \sum_{i=1}^{N}\mel{i}{\hh}{i}
  +
  \frac{1}{2}
  \sum_{i=1}^{N}\sum_{j=1}^{N}
  \qty[
    \braket{ij}{ij}
    -
    \braket{ij}{ji}
  ]
  \label{eq:rhf-start-spin}
\end{equation}
we now sum explicitly over the two spin components associated with each
occupied spatial orbital.

\paragraph{One-electron contribution.}

The spin-orbital sum can be separated into its $\alpha$ and $\beta$
components:
\begin{equation}
\begin{split}
  \sum_{i=1}^{N}\mel{i}{\hh}{i}
  &=
  \sum_{i=1}^{N/2}\mel{i}{\hh}{i}
  +
  \sum_{i=1}^{N/2}\mel{\Bar{i}}{\hh}{\Bar{i}}
  =
  \sum_{i=1}^{N/2}\rmel{i}{\hh}{i}
  +
  \sum_{i=1}^{N/2}\rmel{i}{\hh}{i}
  \\
  &=
  2\sum_{i=1}^{N/2}\rmel{i}{\hh}{i}
  \label{eq:rhf-one-electron-spatial}
\end{split}
\end{equation}
Again, the factor of two simply reflects the double occupation of every
occupied spatial orbital.

\paragraph{Two-electron contribution.}

For every pair of occupied spatial orbitals $i$ and $j$, four spin
combinations are possible:
  $\alpha\alpha$,
  $\alpha\beta$,
  $\beta\alpha$, and
  $\beta\beta$.
Separating these four possibilities gives
\begin{equation}
\begin{split}
  \frac{1}{2}
  \sum_{i=1}^{N}\sum_{j=1}^{N}
  \qty[
    \braket*{ij}{ij}
    -
    \braket*{ij}{ji}
  ]
  & =
  \frac{1}{2}
  \Bigg\{
  \sum_{i=1}^{N/2}\sum_{j=1}^{N/2}
  \qty(
    \braket*{ij}{ij}
    -
    \braket*{ij}{ji}
  )
  +
  \sum_{i=1}^{N/2}\sum_{j=1}^{N/2}
  \qty(
    \braket*{\Bar{i}j}{\Bar{i}j}
    -
    \braket*{\Bar{i}j}{j\Bar{i}}
  )
  \\
  &+
  \sum_{i=1}^{N/2}\sum_{j=1}^{N/2}
  \qty(
    \braket*{i\Bar{j}}{i\Bar{j}}
    -
    \braket*{i\Bar{j}}{\Bar{j}i}
  )
  +
  \sum_{i=1}^{N/2}\sum_{j=1}^{N/2}
  \qty(
    \braket*{\Bar{i}\Bar{j}}{\Bar{i}\Bar{j}}
    -
    \braket*{\Bar{i}\Bar{j}}{\Bar{j}\Bar{i}}
  )
  \Bigg\}
  \label{eq:rhf-four-spin-combinations}
\end{split}
\end{equation}

The four Coulomb contributions all survive because they involve diagonal
spin overlaps. After integration over spin,
\begin{equation}
  \braket*{ij}{ij}
  =
  \braket*{\Bar{i}j}{\Bar{i}j}
  =
  \braket*{i\Bar{j}}{i\Bar{j}}
  =
  \braket*{\Bar{i}\Bar{j}}{\Bar{i}\Bar{j}}
  =
  \rbraket{ii}{jj}
\end{equation}

Exchange behaves differently. The two same-spin combinations survive,
\begin{equation}
  \braket{ij}{ji}
  =
  \braket*{\Bar{i}\Bar{j}}{\Bar{j}\Bar{i}}
  =
  \rbraket{ij}{ji}
\end{equation}
whereas the two opposite-spin exchange integrals vanish,
\begin{equation}
  \braket*{\Bar{i}j}{j\Bar{i}}
  =
  \braket*{i\Bar{j}}{\Bar{j}i}
  =
  0
\end{equation}

For each pair of spatial orbitals, the four spin combinations therefore
produce
\begin{equation}
  \frac{1}{2}
  \qty[
    4\rbraket{ii}{jj}
    -
    2\rbraket{ij}{ji}
  ]
  =
  2\rbraket{ii}{jj}
  -
  \rbraket{ij}{ji}
\end{equation}
Consequently,
\begin{equation}
  \frac{1}{2}
  \sum_{i=1}^{N}\sum_{j=1}^{N}
  \qty[
    \braket{ij}{ij}
    -
    \braket{ij}{ji}
  ]
  =
  \sum_{i=1}^{N/2}\sum_{j=1}^{N/2}
  \qty[
    2\rbraket{ii}{jj}
    -
    \rbraket{ij}{ji}
  ]
  \label{eq:rhf-two-electron-spatial}
\end{equation}

Combining the one- and two-electron contributions gives the familiar
closed-shell RHF energy,
\begin{equation}
  \EHF
  =
  2\sum_{i=1}^{N/2}
  \rmel{i}{\hh}{i}
  +
  \sum_{i=1}^{N/2}
  \sum_{j=1}^{N/2}
  \qty[
    2\rbraket{ii}{jj}
    -
    \rbraket{ij}{ji}
  ]
  \label{eq:rhf-energy-spatial}
\end{equation}

Introducing the spatial-orbital Coulomb and exchange integrals
\begin{align}
  J_{ij}
  &\equiv
  \rbraket{ii}{jj}
  &
  K_{ij}
  &\equiv
  \rbraket{ij}{ji}
\end{align}
the same expression takes the compact form
\begin{equation}
  \boxed{
  \EHF
  =
  2\sum_{i=1}^{N/2}h_{ii}
  +
  \sum_{i=1}^{N/2}
  \sum_{j=1}^{N/2}
  \qty(
    2J_{ij}-K_{ij}
  )
  }
  \label{eq:rhf-energy-jk}
\end{equation}

The origin of the numerical factors is now transparent. Each occupied
spatial orbital contains two electrons, giving the factor of two in the
one-electron contribution. For each pair of spatial orbitals, all four spin
combinations contribute to the Coulomb interaction, whereas only the two
same-spin combinations contribute to exchange. After the overall factor
$1/2$ removing double counting, the result is precisely
\begin{equation}
  2J_{ij}-K_{ij}
\end{equation}

Notice also that the terms with $i=j$ are perfectly well behaved:
\begin{equation}
  2J_{ii}-K_{ii}
  =
  J_{ii}
\end{equation}
since $J_{ii}=K_{ii}$. Thus, a doubly occupied spatial orbital contains one
physical Coulomb repulsion between its $\alpha$ and $\beta$ electrons; the
spurious same-spin self-interactions cancel through exchange.


\begin{hfexercise}{HF energy of the \ce{Li} atom}

\violet{\textit{Find the HF energy of the \ce{Li} atom with occupation
$\psi_1^2\psi_2^1$ in terms of spatial orbitals.}}

\end{hfexercise}
\begin{hfsolution}{HF energy of the \ce{Li} atom}

For the determinant considered here, the three occupied spin orbitals are
\begin{align}
  \chi_1&=\alpha\psi_1
  &
  \chi_2&=\beta\psi_1
  &
  \chi_3&=\alpha\psi_2
\end{align}
The first spatial orbital is therefore doubly occupied, whereas the second
contains a single $\alpha$ electron. Thus $N^\alpha=2$ and
$N^\beta=1$, with orthonormal spatial orbitals $\psi_1$ and $\psi_2$.
The closed-shell core uses the same spatial function for its two spins.

The one-electron contribution is immediately
\begin{equation}
  E_1
  =
  2h_{11}+h_{22}
\end{equation}

There are three electron pairs. The two electrons in $\psi_1$ have opposite
spin and contribute
  $J_{11}$.
The $\alpha$ electron in $\psi_2$ interacts Coulombically with both
electrons in $\psi_1$, giving
  $2J_{12}$.
Exchange occurs only with the $\alpha$ electron in $\psi_1$, and therefore
contributes
  $-K_{12}$.
Explicitly, the three spin-orbital pair contributions reduce to
\begin{align}
  \cJ_{12}-\cK_{12}&=J_{11}\\
  \cJ_{13}-\cK_{13}&=J_{12}-K_{12}\\
  \cJ_{23}-\cK_{23}&=J_{12}
\end{align}
The indices on $\cJ$ and $\cK$ label the three spin orbitals, whereas
those on $J$ and $K$ label spatial orbitals. In particular, the surviving
$J_{11}$ is the repulsion between two different, opposite-spin electrons,
not a one-electron self-interaction.
The total energy is consequently
\begin{equation}
  E_\text{HF}
  =
  2h_{11}
  +
  h_{22}
  +
  J_{11}
  +
  2J_{12}
  -
  K_{12}
\end{equation}

This example provides a useful way of reading spatial-orbital HF energies:
Coulomb terms count interacting electron pairs, whereas exchange terms
count same-spin pairs.

\end{hfsolution}
\begin{hfexercise}{HF energy of the \ce{B} atom}

\violet{\textit{Find the HF energy of the \ce{B} atom with occupation
$\psi_1^2\psi_2^2\psi_3^1$ in terms of spatial orbitals.}}

\end{hfexercise}
\begin{hfsolution}{HF energy of the \ce{B} atom}

For the occupation considered here, the first two spatial orbitals are
doubly occupied and the third contains a single $\alpha$ electron,
corresponding schematically to
\begin{equation}
  \psi_1^2\psi_2^2\psi_3^1
\end{equation}
The five occupied spin orbitals are $\alpha\psi_1$, $\beta\psi_1$,
$\alpha\psi_2$, $\beta\psi_2$, and $\alpha\psi_3$, so
$N^\alpha=3$ and $N^\beta=2$. As in the lithium exercise, the doubly
occupied orbitals share their spatial parts between the two spins.
The one-electron contribution is therefore
\begin{equation}
  E_1
  =
  2h_{11}
  +
  2h_{22}
  +
  h_{33}
\end{equation}

The two electrons occupying $\psi_1$ contribute $J_{11}$, while those in
$\psi_2$ contribute $J_{22}$. Between the two doubly occupied orbitals there
are four Coulomb pairs but only two same-spin exchange pairs, giving
\begin{equation}
  4J_{12}-2K_{12}
\end{equation}

The singly occupied $\psi_3$ orbital interacts with both electrons in each
doubly occupied orbital. Hence,
  $2J_{13}-K_{13}$
arises from the interaction between $\psi_1$ and $\psi_3$, and
  $2J_{23}-K_{23}$
from the interaction between $\psi_2$ and $\psi_3$.

Collecting all terms gives
\begin{equation}
    E_\text{HF}
    =
    2h_{11}
    +
    2h_{22}
    +
    h_{33}
    +
    J_{11}
    +
    4J_{12}
    +
    J_{22}
    -
    2K_{12}
    +
    2J_{13}
    +
    2J_{23}
    -
    K_{13}
    -
    K_{23}
\end{equation}
There are $\binom{5}{2}=10$ Coulomb pairs: one within each doubly
occupied orbital, four between these two orbitals, and two between each
of them and $\psi_3$. Exchange survives for
$\binom{N^\alpha}{2}+\binom{N^\beta}{2}=3+1=4$ pairs, explaining
the coefficients $2$, $1$, and $1$ of $K_{12}$, $K_{13}$, and $K_{23}$.
The pair counts provide an independent check on the spatial-orbital result.

\end{hfsolution}

The \ce{Li} and \ce{B} examples illustrate that the reduction from spin
orbitals to spatial orbitals is not restricted to closed shells. The
closed-shell formula in Eq.~\eqref{eq:rhf-energy-spatial} benefits from the
particularly simple double-occupation pattern of RHF, but for a general
determinant the same procedure can always be carried out by keeping track of
which spatial orbitals are occupied by $\alpha$ and $\beta$ electrons.

We now return to spin orbitals so that individual substitutions can be
specified without ambiguity. The spin selection rules just derived continue
to determine which exchange matrix elements survive when occupied orbitals
are replaced by virtual ones.

\subsection{Excited determinants and Hamiltonian matrix elements}

The Hartree--Fock determinant also provides a reference from which alternative
occupation patterns can be generated. The excitation rank of these
determinants determines which Hamiltonian matrix elements can be nonzero,
leading directly to the Slater--Condon rules and Brillouin's theorem.

\subsubsection{Excited determinants}

The Hartree--Fock solution provides not only an energy but also a complete
orthonormal set of occupied and virtual spin orbitals. Once this orbital basis
has been obtained, many other Slater determinants can be generated by
replacing one or more occupied spin orbitals of the Hartree--Fock determinant by virtual
ones. These determinants play a central role in configuration-interaction
\citep{Shavitt1998}, coupled-cluster \citep{Cizek1966,BartlettMusial2007},
perturbation \citep{MollerPlesset1934,Cremer2011}, and excited-state theories.

We therefore introduce the language of orbital substitutions, or
excitations, relative to the Hartree--Fock reference determinant.

\paragraph{Reference determinant.}

For a conventional ground-state HF solution, the electrons occupy the $N$
lowest spin orbitals according to the Aufbau principle. The reference
determinant is
\begin{equation}
  \ket{\Psi_0}
  =
  \ket{
    \chi_1
    \ldots
    \chi_{\green{i}}
    \chi_{\green{j}}
    \ldots
    \chi_N
  }
  \label{eq:hf-reference-determinant}
\end{equation}
Here and in the following,
  $\green{i,j,k,l,\ldots}$
  and
  $\red{a,b,c,d,\ldots}$
denote occupied and virtual spin orbitals, respectively.

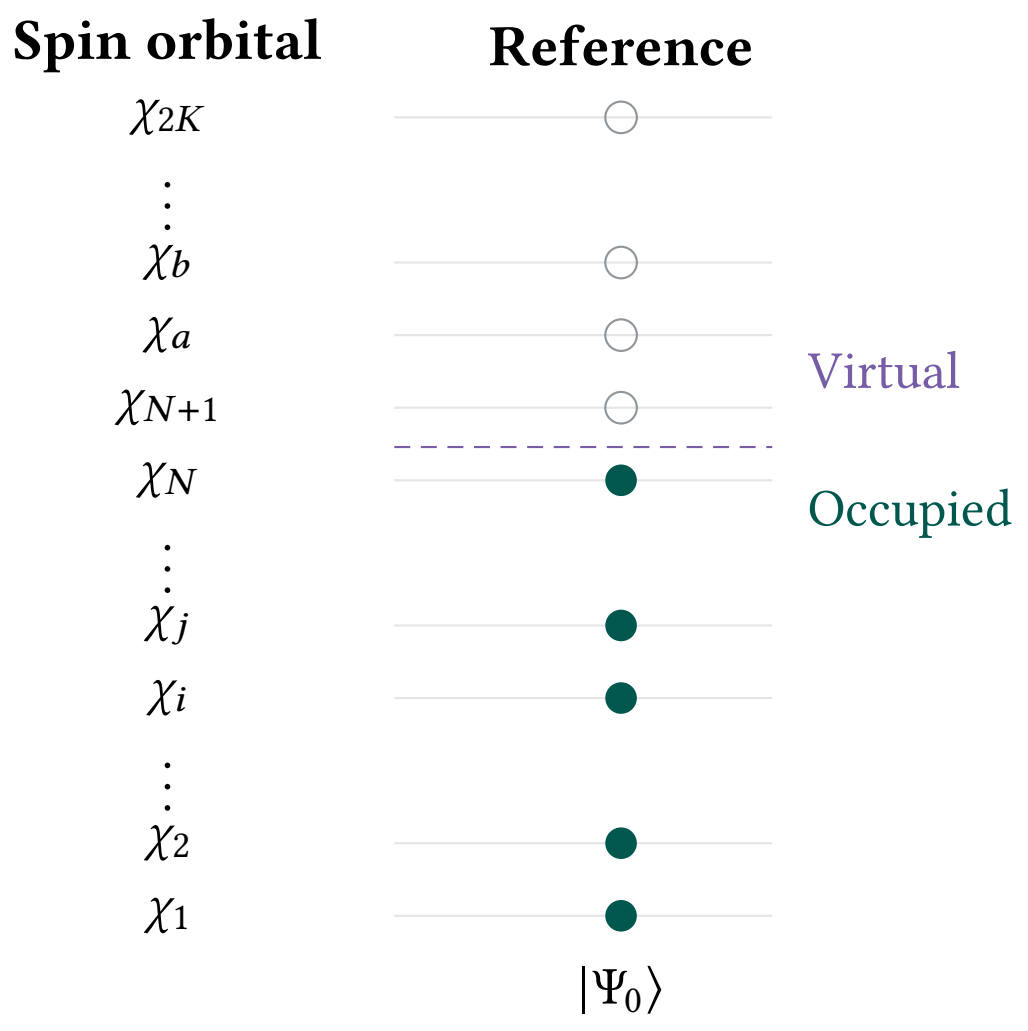
\begin{figure}[htbp]
  \centering

\begingroup

\tikzset{
 nucleus/.style={circle,draw=hfGreenDark,fill=hfGreen!12,minimum size=7mm},
 electron/.style={circle,draw=hfBlue,fill=hfBlue!10,minimum size=7mm},
 old/.style={hfBlue,densely dashed,thick,-{Latex[length=2mm]}},
 new/.style={hfGreenDark,very thick,-{Latex[length=2mm]}},
 card/.style={draw=hfGreenDark,fill=hfGreen!5,rounded corners=2mm,
   align=center,minimum height=12mm},
 occupied/.style={circle,fill=hfGreenDark,inner sep=2.1pt},
 vacant/.style={circle,draw=hfInk!50,inner sep=2.1pt},
 hole/.style={circle,draw=hfRed,thick,inner sep=2.1pt},
 particle/.style={circle,fill=hfViolet,inner sep=2.1pt}}


\begin{tikzpicture}[font=\small,x=1cm,y=1cm]
 \node[font=\bfseries] at (0,0) {Spin orbital};
 \node[font=\bfseries] at (3,0) {Reference};
 \foreach \r/\label in {1/{\chi_{2K}},2/{\vdots},3/{\chi_b},4/{\chi_a},
   5/{\chi_{N+1}},6/{\chi_N},7/{\vdots},8/{\chi_j},9/{\chi_i},
   10/{\vdots},11/{\chi_2},12/{\chi_1}}{
  \node at (0,-0.48*\r) {$\label$};
  \ifnum\r=2\else\ifnum\r=7\else\ifnum\r=10\else
   \draw[hfInk!12] (1.5,-0.48*\r)--(4,-0.48*\r);
   \foreach \x in {3}{
    \ifnum\r<6 \node[vacant] at (\x,-0.48*\r) {};
    \else \node[occupied] at (\x,-0.48*\r) {};\fi
   }
  \fi\fi\fi
 }
 \draw[densely dashed,hfViolet] (1.5,-2.66)--(4,-2.66);
 \node[font=\small,text=hfViolet,anchor=west] at (4.1,-2.15) {Virtual};
 \node[font=\small,text=hfGreenDark,anchor=west] at (4.1,-3.1) {Occupied};
 \node at (3,-6.25) {$\ket{\Psi_0}$};
\end{tikzpicture}
\endgroup


  \caption{Reference Hartree--Fock determinant constructed from the occupied
  spin orbitals. Filled green circles indicate occupied spin orbitals; empty
  circles indicate virtual ones. Adapted from Szabo and Ostlund \citep{SzaboOstlund1996}.}
  \label{fig:hf-reference-determinant}
\end{figure}

The distinction is always made \emph{relative to the reference determinant}:
an occupied orbital contains an electron in $\ket{\Psi_0}$, whereas a virtual
orbital does not.
Figure~\ref{fig:hf-reference-determinant} illustrates this partition of the
spin-orbital space into occupied and virtual levels.

Because the spin orbitals form an orthonormal set, the occupied and virtual
subspaces are mutually orthogonal,
\begin{equation}
  \braket{\chi_i}{\chi_a}=0
\end{equation}
This simple property will have important consequences for matrix elements
between different determinants.


\paragraph{Singly excited determinants.}

A singly excited determinant is obtained by removing an electron from
an occupied spin orbital $i$ and placing it in a virtual spin orbital $a$.
Symbolically,
\begin{equation}
  i\longrightarrow a
\end{equation}
The resulting determinant is denoted
\begin{equation}
  \ket{\Psi_{\green{i}}^{\red{a}}}
\end{equation}
In determinant notation,
\begin{equation}
  \ket{\Psi_{\green{i}}^{\red{a}}}
  =
  \ket*{
    \chi_1
    \ldots
    \chi_{\red{a}}
    \chi_{\green{j}}
    \ldots
    \chi_N
  }
\end{equation}
Thus, $\ket{\Psi_i^a}$ differs from the reference determinant by exactly one
spin orbital.

The same substitution can be expressed particularly compactly using
creation and annihilation operators,
\begin{equation}
  \ket{\Psi_{\green{i}}^{\red{a}}}
  =
  \cre{\red{a}}
  \ani{\green{i}}
  \ket{\Psi_0}
  \label{eq:single-excited-determinant}
\end{equation}
The annihilation operator $\ani{i}$ first removes the electron occupying
spin orbital $i$, while the creation operator $\cre{a}$ places an electron
in the previously unoccupied spin orbital $a$.

It is useful to interpret this substitution as the simultaneous creation of
a \green{hole} in the occupied orbital $i$ and a \red{particle} in the
virtual orbital $a$:
\begin{equation}
  i\rightarrow a
  \qquad\Longleftrightarrow\qquad
  \text{one hole }i
  +
  \text{one particle }a
\end{equation}
This particle-hole language will reappear naturally in excited-state and
many-body formulations.

Because one occupied orbital has been replaced by an orthogonal virtual
orbital,
\begin{equation}
  \braket{\Psi_0}{\Psi_i^a}=0
\end{equation}
Thus, every singly excited determinant is orthogonal to the Hartree--Fock reference.


\paragraph{Doubly excited determinants.}

A doubly excited determinant is obtained by replacing two occupied
spin orbitals, $i$ and $j$, by two virtual spin orbitals, $a$ and $b$:
\begin{equation}
  ij\longrightarrow ab
\end{equation}
In determinant notation,
\begin{equation}
  \ket*{\Psi_{\green{ij}}^{\red{ab}}}
  =
  \ket{
    \chi_1
    \ldots
    \chi_{\red{a}}
    \chi_{\red{b}}
    \ldots
    \chi_N
  }
\end{equation}
In second-quantized notation,
\begin{equation}
  \ket*{\Psi_{\green{ij}}^{\red{ab}}}
  =
  \cre{\red{a}}
  \cre{\red{b}}
  \ani{\green{j}}
  \ani{\green{i}}
  \ket{\Psi_0}
  \label{eq:double-excited-determinant}
\end{equation}
The determinant now differs from the reference by two spin orbitals and
contains two holes, $i$ and $j$, together with two particles, $a$ and $b$.

The ordering of the fermionic creation and annihilation operators matters:
interchanging two such operators changes the sign because
\begin{align}
  \cre{p}\cre{q}
  &=
  -\cre{q}\cre{p}
  &
  \ani{p}\ani{q}
  &=
  -\ani{q}\ani{p}
\end{align}
A consistent ordering convention therefore fixes the overall phase of each
excited determinant. This phase has no effect on diagonal expectation
values, but it is essential when evaluating off-diagonal matrix elements
between different determinants.

\begin{figure}[htbp]
  \centering

\begingroup

\tikzset{
 nucleus/.style={circle,draw=hfGreenDark,fill=hfGreen!12,minimum size=7mm},
 electron/.style={circle,draw=hfBlue,fill=hfBlue!10,minimum size=7mm},
 old/.style={hfBlue,densely dashed,thick,-{Latex[length=2mm]}},
 new/.style={hfGreenDark,very thick,-{Latex[length=2mm]}},
 card/.style={draw=hfGreenDark,fill=hfGreen!5,rounded corners=2mm,
   align=center,minimum height=12mm},
 occupied/.style={circle,fill=hfGreenDark,inner sep=2.1pt},
 vacant/.style={circle,draw=hfInk!50,inner sep=2.1pt},
 hole/.style={circle,draw=hfRed,thick,inner sep=2.1pt},
 particle/.style={circle,fill=hfViolet,inner sep=2.1pt}}


\begin{tikzpicture}[font=\small,x=1cm,y=1cm]
 \node[font=\bfseries] at (0,0) {Spin orbital};
 \node[font=\bfseries] at (3,0) {Single};
 \node[font=\bfseries] at (6,0) {Double};
 \foreach \r/\label in {1/{\chi_{2K}},2/{\vdots},3/{\chi_b},4/{\chi_a},
   5/{\chi_{N+1}},6/{\chi_N},7/{\vdots},8/{\chi_j},9/{\chi_i},
   10/{\vdots},11/{\chi_2},12/{\chi_1}}{
  \node at (0,-0.48*\r) {$\label$};
  \ifnum\r=2\else\ifnum\r=7\else\ifnum\r=10\else
   \draw[hfInk!12] (1.5,-0.48*\r)--(7,-0.48*\r);
   \foreach \x in {3,6}{
    \ifnum\r<6 \node[vacant] at (\x,-0.48*\r) {};
    \else \node[occupied] at (\x,-0.48*\r) {};\fi
   }
  \fi\fi\fi
 }
 \path[fill=white] (2.82,-4.48) rectangle (3.18,-4.17);
 \node[hole] at (3,-4.32) {};
 \node[particle] at (3,-1.92) {};
 \foreach \y in {-4.32,-3.84}{
  \path[fill=white] (5.82,\y-0.16) rectangle (6.18,\y+0.16);
  \node[hole] at (6,\y) {};
 }
 \node[particle] at (6,-1.92) {};
 \node[particle] at (6,-1.44) {};
 \draw[densely dashed,hfViolet] (1.5,-2.66)--(7,-2.66);
 \node[font=\small,text=hfViolet,anchor=west] at (7.1,-2.15) {Virtual};
 \node[font=\small,text=hfGreenDark,anchor=west] at (7.1,-3.1) {Occupied};
 \node at (3,-6.25) {$\ket{\Psi_i^a}$};
 \node at (6,-6.25) {$\ket{\Psi_{ij}^{ab}}$};
\end{tikzpicture}
\endgroup


  \caption{Singly excited (left) and doubly excited (right) determinants
  obtained by replacing occupied spin orbitals of the Hartree--Fock
  reference by virtual ones. Filled green circles indicate unchanged
  occupations, violet circles promoted electrons, and red outlines holes.
  Empty circles are unoccupied; unshown occupations remain unchanged.
  Adapted from Szabo and Ostlund \citep{SzaboOstlund1996}.}
  \label{fig:excited-determinants}
\end{figure}
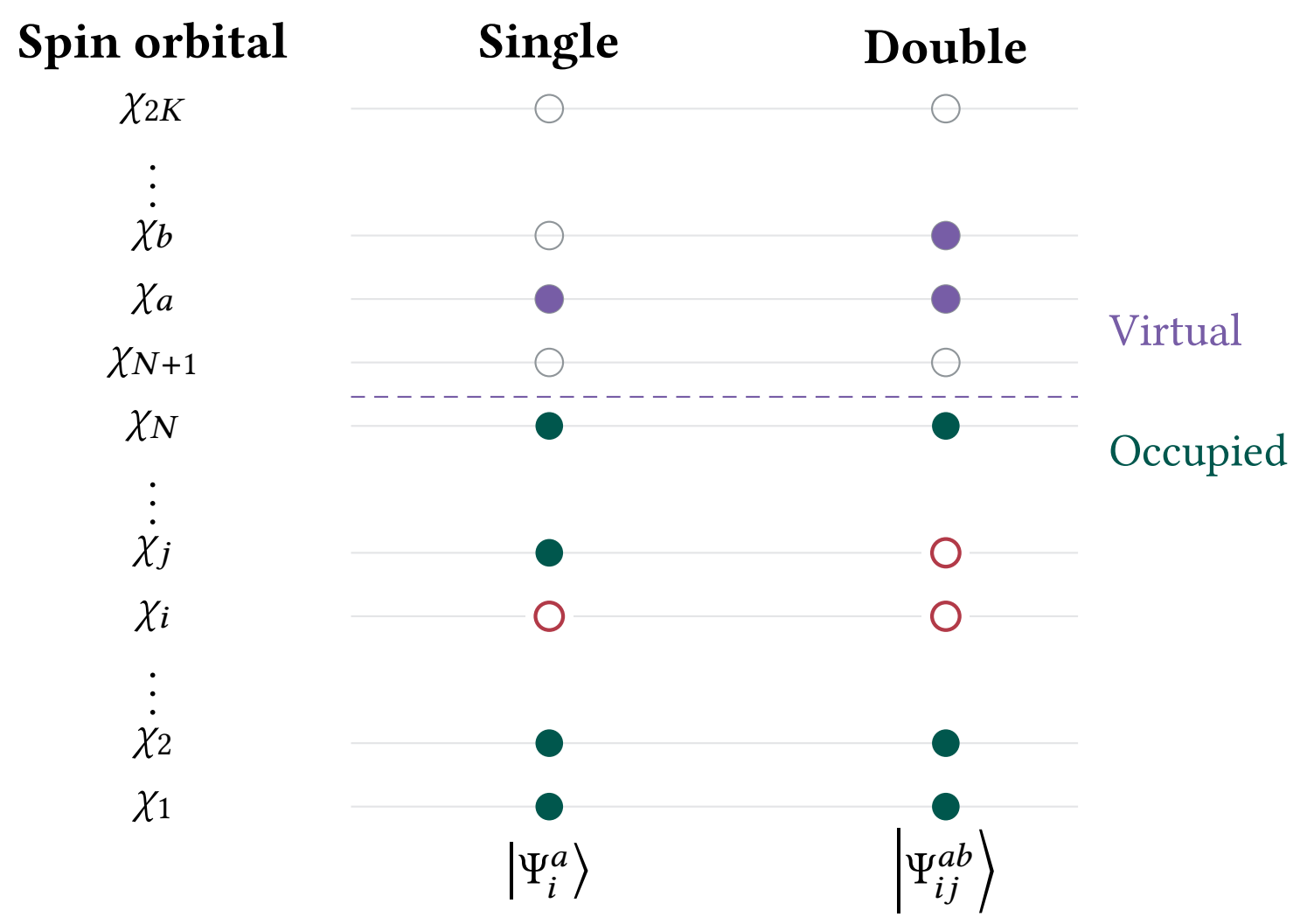

Figure~\ref{fig:excited-determinants} compares the holes and particles created
by single and double substitutions relative to the reference occupation.


\paragraph{Excitation rank.}

The construction extends immediately to higher substitutions. A triple
excitation replaces three occupied orbitals,
\begin{equation}
  ijk\longrightarrow abc
\end{equation}
and may be written
\begin{equation}
  \ket*{\Psi_{ijk}^{abc}}
  =
  \cre{a}\cre{b}\cre{c}
  \ani{k}\ani{j}\ani{i}
  \ket{\Psi_0}
\end{equation}
More generally, an $m$-fold excited determinant differs from the reference
by $m$ occupied-to-virtual substitutions. The integer $m$ is called the
excitation rank:
\begin{equation}
  \begin{array}{ccl}
    m=0 &:& \text{reference determinant}\\
    m=1 &:& \text{single excitation}\\
    m=2 &:& \text{double excitation}\\
    m=3 &:& \text{triple excitation}\\
    &\vdots&
  \end{array}
\end{equation}

With $N$ occupied spin orbitals and $2K-N$ virtual spin orbitals, the number
of singly excited determinants is
\begin{equation}
  N_\text{S}
  =
  N(2K-N)
\end{equation}
whereas the number of doubly excited determinants is
\begin{equation}
  N_\text{D}
  =
  \binom{N}{2}
  \binom{2K-N}{2}
\end{equation}
The number of possible determinants therefore grows extremely rapidly with
the size of the orbital space. This combinatorial growth is one of the
fundamental computational difficulties encountered by correlated
wavefunction methods.


\paragraph{Excited determinants are not necessarily excited states.}

An important distinction must be made at this point. The terminology
``singly excited determinant'' describes how a determinant is constructed
\emph{relative to the HF reference}; it does not imply that this determinant
is itself an eigenfunction of the electronic Hamiltonian.

In general,
\begin{equation}
  \chH_\text{elec}\ket*{\Psi_i^a}
  \neq
  E_i^a\ket*{\Psi_i^a}
\end{equation}
and similarly for doubly and higher excited determinants. The exact excited
states are, in general, linear combinations of many determinants,
\begin{equation}
  \ket*{\Psi_I}
  =
  C_0^{(I)}\ket{\Psi_0}
  +
  \sum_{ia}
  C_i^{a,(I)}
  \ket*{\Psi_i^a}
  +
  \frac{1}{4}
  \sum_{ijab}
  C_{ij}^{ab,(I)}
  \ket*{\Psi_{ij}^{ab}}
  +
  \cdots
\end{equation}
The determinants introduced here should therefore be viewed as a convenient
many-electron basis built from the Hartree--Fock orbitals.

This distinction is central to post-Hartree--Fock theory: the reference and
its excited determinants provide the building blocks, while the electronic
Hamiltonian determines how these configurations interact and combine.


\paragraph{Why excitation rank matters.}

Two Slater determinants may differ by zero, one, two, or more spin orbitals.
For Hamiltonians containing only one- and two-electron operators, the matrix
element between two determinants depends very strongly on this number.

Schematically,
\begin{equation*}
  \boxed{
  \text{number of orbital substitutions}
  \quad\Longrightarrow\quad
  \text{structure of the Hamiltonian matrix element}
  }
\end{equation*}
A one-electron operator can connect determinants differing by at most one
spin orbital, whereas a two-electron operator can connect determinants
differing by at most two. Consequently, determinants separated by three or
more substitutions have a vanishing Hamiltonian matrix element.

These selection rules are precisely the content of the
Slater--Condon rules, to which we now turn.

\subsection{Slater--Condon rules and Brillouin's theorem}
\label{sec:slater-condon}

The determinants introduced in the previous subsection form an orthonormal
many-electron basis constructed from a common set of orthonormal spin
orbitals. The next question is therefore how the electronic Hamiltonian acts
in this determinant basis.

At first sight, evaluating a matrix element between two $N$-electron Slater
determinants might appear to require an explicit expansion of both
determinants into their $N!$ Hartree products. The
Slater--Condon rules \citep{Slater1929,Condon1930} avoid this combinatorial expansion. They express
matrix elements of one- and two-electron operators directly in terms of
one- and two-electron orbital integrals.

\hfBioCondon

The central principle is simple: the value of a matrix element depends only
on the number of spin orbitals by which the two determinants differ. This
number is often called their degree of substitution. Because a
one-electron operator acts on only one electron at a time, it can connect
determinants differing by at most one spin orbital. Similarly, a
two-electron operator can connect determinants differing by at most two spin
orbitals.

Throughout this subsection, we assume that the determinants are written
with a consistent orbital ordering so that their relative fermionic phase
has already been fixed.


\subsubsection{Matrix elements of one-electron operators}

Consider a general one-electron operator,
\begin{equation}
  \chO_1
  =
  \sum_{i=1}^{N}\hh(\bx_i)
  \label{eq:slater-condon-one-body}
\end{equation}
The Slater--Condon rule depends on how many spin orbitals differ between the
two determinants.

\paragraph{Case 1: determinants differ by zero spin orbitals.}

If the two determinants are identical,
\begin{equation}
  \ket{\red{K}}
  =
  \ket{\cdots i j\cdots}
\end{equation}
then the matrix element is simply the sum of the one-electron expectation
values over all occupied spin orbitals:
\begin{equation}
  \mel{\red{K}}{\chO_1}{\red{K}}
  =
  \sum_{i\in K}\mel{i}{\hh}{i}
  \label{eq:sc-one-zero}
\end{equation}
For the Hartree--Fock reference,
\begin{equation}
  \mel{\Psi_0}{\chO_1}{\Psi_0}
  =
  \sum_{i=1}^{N}h_{ii}
\end{equation}

This is the result already encountered in the derivation of the HF energy:
each occupied spin orbital contributes one diagonal matrix element of the
one-electron operator.

\paragraph{Case 2: determinants differ by one spin orbital.}

Consider
\begin{align}
  \ket{\red{K}}
  &=
  \ket{\cdots i j\cdots}
  &
  \ket{\blue{L}}
  &=
  \ket{\cdots k j\cdots}
\end{align}
All occupied spin orbitals are common to the two determinants except for the
replacement
\begin{equation}
  i\longleftrightarrow k
\end{equation}
The spectator orbitals integrate to unity, while orthogonality eliminates
all terms in which the one-electron operator acts on a common orbital.
Consequently,
\begin{equation}
  \mel{\red{K}}{\chO_1}{\blue{L}}
  =
  \mel{i}{\hh}{k}
  \label{eq:sc-one-one}
\end{equation}
For example, between the HF reference and a singly excited determinant,
\begin{equation}
  \mel*{\Psi_0}{\chO_1}{\Psi_{\green{i}}^{\red{a}}}
  =
  h_{\green{i}\red{a}}
\end{equation}

Thus a one-electron operator can directly remove an electron from one spin
orbital and place it in another.

\paragraph{Case 3: determinants differ by two or more spin orbitals.}

Let
\begin{align}
  \ket{\red{K}}
  &=
  \ket{\cdots i j\cdots}
  &
  \ket{\blue{L}}
  &=
  \ket{\cdots k l\cdots}
\end{align}
so that at least two occupied spin orbitals differ. Since $\chO_1$ can act
on only one electron, it cannot simultaneously repair both substitutions.
At least one overlap between different orthogonal spin orbitals therefore
remains, and the matrix element vanishes:
\begin{equation}
  \mel{\red{K}}{\chO_1}{\blue{L}}
  =
  0
  \label{eq:sc-one-two}
\end{equation}
In particular,
\begin{equation}
  \mel*{\Psi_0}{\chO_1}
  {\Psi_{\green{ij}}^{\red{ab}}}
  =
  0
\end{equation}

More generally,
\begin{equation}
  \boxed{
  \chO_1
  \quad\text{connects determinants differing by at most one spin orbital}
  }
\end{equation}


\subsubsection{Matrix elements of two-electron operators}

Consider now the electron-electron interaction,
\begin{equation}
  \chO_2
  =
  \sum_{i<j}^{N}\frac{1}{r_{ij}}
  \label{eq:slater-condon-two-body}
\end{equation}
Because this operator acts simultaneously on two electrons, its connectivity
extends one excitation rank further than that of a one-electron operator.

\paragraph{Case 1: determinants differ by zero spin orbitals.}

For identical determinants,
\begin{equation}
  \mel{\red{K}}{\chO_2}{\red{K}}
  =
  \frac{1}{2}
  \sum_{i\in K}\sum_{j\in K}
  \mel{ij}{}{ij}
  \label{eq:sc-two-zero}
\end{equation}
The antisymmetrized integral
\begin{equation}
  \mel{ij}{}{ij}
  =
  \braket{ij}{ij}
  -
  \braket{ij}{ji}
\end{equation}
contains the Coulomb and exchange contributions associated with the occupied
pair $(i,j)$. The factor $1/2$ removes the double counting of electron
pairs.

For the HF reference,
\begin{equation}
  \mel*{\Psi_0}{\chO_2}{\Psi_0}
  =
  \frac{1}{2}
  \sum_{ij}^{N}
  \mel{ij}{}{ij}
\end{equation}

\paragraph{Case 2: determinants differ by one spin orbital.}

Consider again
\begin{align}
  \ket{\red{K}}
  &=
  \ket{\cdots i j\cdots}
  &
  \ket{\blue{L}}
  &=
  \ket{\cdots k j\cdots}
\end{align}
One electron undergoes the substitution $i\rightarrow k$, while the second
electron acted upon by $\chO_2$ may occupy any spin orbital common to the two
determinants. Summing over these spectator occupied orbitals gives
\begin{equation}
  \mel{\red{K}}{\chO_2}{\blue{L}}
  =
  \sum_{j\in K\cap L}
  \mel{ij}{}{kj}
  \label{eq:sc-two-one}
\end{equation}
For the reference and a singly excited determinant,
\begin{equation}
  \mel*{\Psi_0}{\chO_2}
  {\Psi_{\green{i}}^{\red{a}}}
  =
  \sum_{k}^{N}
  \mel{\green{i}k}{}{\red{a}k}
  \label{eq:sc-reference-single-two-body}
\end{equation}

Physically, the promoted electron is coupled through the two-electron
interaction to each of the remaining occupied electrons.

\paragraph{Case 3: determinants differ by two spin orbitals.}

Let
\begin{align}
  \ket{\red{K}}
  &=
  \ket{\cdots i j\cdots}
  &
  \ket{\blue{L}}
  &=
  \ket{\cdots k l\cdots}
\end{align}
The two-electron operator can now act precisely on the two orbitals that
differ, giving
\begin{equation}
  \mel{\red{K}}{\chO_2}{\blue{L}}
  =
  \mel{ij}{}{kl}
  \label{eq:sc-two-two}
\end{equation}
Thus,
\begin{equation}
  \mel*{\Psi_0}{\chO_2}
  {\Psi_{\green{ij}}^{\red{ab}}}
  =
  \mel{\green{ij}}{}{\red{ab}}
\end{equation}

This result is particularly important in correlated wavefunction theory:
the two-electron interaction couples the HF reference directly to doubly
excited determinants.

\paragraph{Case 4: determinants differ by three or more spin orbitals.}

A two-electron operator cannot change three independent orbital occupations
simultaneously. Therefore, $\mel{\red{K}}{\chO_2}{\blue{L}} = 0$
if $K$ and $L$ differ by three or more spin orbitals.

Hence,
\begin{equation}
  \boxed{
  \chO_2
  \quad\text{connects determinants differing by at most two spin orbitals}
  }
\end{equation}


\subsubsection{Matrix elements of the electronic Hamiltonian}

Because
\begin{equation}
  \chH_\text{elec}
  =
  \chO_1+\chO_2
\end{equation}
the preceding results may be summarized directly for the electronic
Hamiltonian.

For identical determinants,
\begin{equation}
  \mel{K}{\chH_\text{elec}}{K}
  =
  \sum_{i\in K}h_{ii}
  +
  \frac{1}{2}
  \sum_{i,j\in K}
  \mel{ij}{}{ij}
  \label{eq:sc-h-zero}
\end{equation}

For determinants differing by a single substitution $i\rightarrow a$,
\begin{equation}
  \mel*{\Psi_0}{\chH_\text{elec}}{\Psi_i^a}
  =
  h_{ia}
  +
  \sum_{k}^{N}
  \mel{ik}{}{ak}
  \label{eq:sc-h-single}
\end{equation}

For determinants differing by a double substitution $ij\rightarrow ab$,
\begin{equation}
  \mel*{\Psi_0}{\chH_\text{elec}}{\Psi_{ij}^{ab}}
  =
  \mel{ij}{}{ab}
  \label{eq:sc-h-double}
\end{equation}

Finally,
\begin{equation}
  \mel*{\Psi_0}{\chH_\text{elec}}{\Psi_{ijk}^{abc}}
  =
  0
\end{equation}
and similarly for all higher excitation ranks.

The Hamiltonian matrix in a determinant basis is therefore highly
structured rather than dense: two determinants can interact directly only
if they differ by at most two spin orbitals. This sparsity is a direct
consequence of the fact that the nonrelativistic electronic Hamiltonian
contains at most two-electron operators.


\paragraph{Excited determinant versus excited HF solution.}

The terminology ``excited determinant'' must be distinguished carefully from
a separately optimized excited-state Hartree--Fock solution.

An excited determinant is an orbital substitution within a fixed set of
reference orbitals. It is a valid many-electron basis function and serves as a
building block for configuration-interaction and coupled-cluster expansions
\citep{Shavitt1998,BartlettMusial2007}.
In general, however, an excited determinant is not necessarily a separately
optimized HF state, a stationary point of the HF energy, or an eigenfunction
of $\chS^2$.
Thus,
\begin{equation}
  \boxed{
  \text{orbital excitation}
  \neq
  \text{state-specific orbital optimization}
  }
\end{equation}

In the first procedure, the reference orbitals remain frozen and only their
occupation pattern changes. In the second, the orbitals themselves are
reoptimized for the targeted electronic state. The distinction is important
because orbital relaxation can be substantial, particularly for states with
a qualitatively different electronic distribution from the ground state.


\subsubsection{Brillouin's theorem}

For a reference and a singly excited determinant, the Slater--Condon rules
give
\begin{equation}
\begin{split}
  \mel{\Psi_0}{\chH_\text{elec}}{\Psi_i^a}
  &=
  h_{ia}
  +
  \sum_{k}^{N}
  \mel{ik}{}{ak}
\end{split}
\end{equation}
The right-hand side is precisely an occupied-virtual matrix element of the
Fock operator, introduced here in anticipation of its variational derivation
in Section~\ref{sec:fock-operator}:
\begin{equation}
  f_{ia}
  =
  h_{ia}
  +
  \sum_{k}^{N}
  \mel{ik}{}{ak}
\end{equation}

\begin{hftheorem}{Brillouin's theorem}
\hftheoremportrait{figures/sketch-brillouin.png}{L\'eon Brillouin}
{At a stationary Hartree--Fock determinant, the Hamiltonian matrix elements
to singly substituted determinants vanish for the occupied-virtual
rotations admitted by the chosen variational space. Stationarity, not the
requirement of an energy minimum, is the essential assumption.}
{Source portrait: Benjamin Couprie, Institut International de Physique Solvay,
\href{https://commons.wikimedia.org/wiki/File:L\%C3\%A9on_Brillouin_Solvay_conference_1927.jpg}{1927 Solvay photograph (crop: Setreset)}, public domain.
\hfsketchcredit}

At a stationary Hartree--Fock solution, occupied-virtual orbital rotations
cannot lower the energy to first order. The corresponding Fock matrix
elements therefore vanish:
\begin{equation}
  f_{ia}=0
\end{equation}
Hence,
\begin{equation}
  \mel{\Psi_0}{\chH_\text{elec}}
  {\Psi_{\green{i}}^{\red{a}}}
  =
  h_{\green{i}\red{a}}
  +
  \sum_{k}^{N}
  \mel{\green{i}k}{}{\red{a}k}
  =
  f_{\green{i}\red{a}}
  =
  0
  \label{eq:brillouin-theorem}
\end{equation}
This result is Brillouin's theorem \citep{Brillouin1932}. The derivation of the HF equations below
will also establish the stationarity condition used here.
\end{hftheorem}

\hfBioBrillouin

Brillouin's theorem does not state that singly excited determinants are
unimportant. Rather, it states that a \emph{stationary HF determinant does
not couple directly to singles through the Hamiltonian}. This has several
important consequences. In a nondegenerate perturbative expansion about a
canonical HF reference, the direct first-order mixing of singles with the
reference vanishes. Singles can nevertheless enter indirectly through
coupling to other configurations; the theorem does not remove them from
correlated wavefunctions.

The theorem also provides a useful physical interpretation of HF
stationarity: the orbitals have already been optimized so that infinitesimal
occupied-virtual mixing produces no first-order lowering of the energy.


\subsection{Spin symmetry and spin-adapted determinants}
\label{sec:spin-symmetry}

\subsubsection{Spin quantum numbers and collinear determinants}

Excited determinants also provide a useful setting in which to distinguish
the spin projection $M_S$ from the total spin $S$.

Within the nonrelativistic spin-free framework, the electronic Hamiltonian
commutes with both the total-spin operator and its projection along the
chosen quantization axis,
\begin{align}
  \comm{\chH}{\purple{\chS^2}} & =0
  &
  \comm{\chH}{\violet{\chS_z}}& =0
\end{align}
The exact eigenfunctions of $\chH$ may therefore be chosen as simultaneous
eigenfunctions of $\chS^2$ and $\chS_z$:
\begin{align}
  \purple{\chS^2}\ket{S,M_S}
  &=
  S(S+1)\ket{S,M_S}
  &
  \violet{\chS_z}\ket{S,M_S}
  &=
  M_S\ket{S,M_S}
\end{align}

The quantum number $S$ determines the spin multiplicity $2S+1$.
Thus,
\begin{equation}
  S=0,\quad\frac{1}{2},\quad1,\quad\frac{3}{2},\ldots
\end{equation}
correspond to singlets, doublets, triplets, quartets, and so forth. For a
given value of $S$, the spin projection takes the values
\begin{equation}
  M_S=-S,-S+1,\ldots,S-1,S
\end{equation}

For a collinear determinant containing $N^\alpha$ spin-up and $N^\beta$
spin-down electrons,
\begin{align}
  \violet{\chS_z}\ket{\Psi}
  & =
  M_S\ket{\Psi}
  &
  M_S
  & =
  \frac{N^\alpha-N^\beta}{2}
  \label{eq:ms-determinant}
\end{align}
Here, \emph{collinear} means that all occupied spin orbitals are expressed
with respect to the same spin quantization axis.

Every RHF or collinear UHF determinant therefore has a definite value of
$M_S$. Having a definite $M_S$, however, does \emph{not} imply having a
definite total spin $S$.

\subsubsection{Spin purity and spin contamination}

\paragraph{Closed-shell RHF.}

For a closed-shell RHF determinant,
\begin{align}
  N^\alpha=N^\beta&=\frac{N}{2}
  &
  M_S&=0
\end{align}
Because every spatial orbital is occupied by one $\alpha$ and one $\beta$
electron, the determinant is a pure singlet:
\begin{equation}
  \purple{\chS^2}\ket{\Psi_\text{RHF}}
  =
  0
\end{equation}
Thus, a closed-shell RHF determinant is simultaneously an eigenfunction of
$\chS_z$ and $\chS^2$.

\paragraph{Open-shell UHF.}

For UHF, the $\alpha$ and $\beta$ electrons are allowed to occupy different
spatial orbitals. The determinant remains an eigenfunction of $\chS_z$, but,
in general,
\begin{equation}
  \purple{\chS^2}\ket{\Psi_\text{UHF}}
  \neq
  S(S+1)\ket{\Psi_\text{UHF}}
\end{equation}
Hence, every collinear RHF or UHF determinant has definite $M_S$, but
only special determinants have definite $S$.

For $N^\alpha\geq N^\beta$, the expectation value of $\chS^2$ for a UHF
determinant is
\begin{align}
  \expval{\purple{\chS^2}}_\text{UHF}
  &=
  M_S(M_S+1)
  +
  N^\beta
  -
  \sum_{i=1}^{N^\alpha}
  \sum_{j=1}^{N^\beta}
  \abs{S_{ij}^{\alpha\beta}}^2
  &
  \underbrace{
  S_{ij}^{\alpha\beta}
  }_{\text{overlap}}
  &=
  \braket{\psi_i^\alpha}{\psi_j^\beta}
  \label{eq:uhf-spin-contamination}
\end{align}

If the $\alpha$ and $\beta$ spatial orbitals coincide pairwise, the overlap
terms take their maximal values and the spin-pure restricted result is
recovered. As the two spin manifolds relax independently, the value of
$\expval*{\chS^2}$ may depart from the desired eigenvalue.

For a target high-spin state with $S=M_S$,
the spin contamination may be quantified as
\begin{equation}
  \Delta S^2
  =
  \expval{\purple{\chS^2}}_\text{UHF}
  -
  \expval{\purple{\chS^2}}_\text{exact}
  \geq0
\end{equation}
More explicitly, a singlet may be contaminated by triplets, quintets, and
higher-spin states; a doublet may be contaminated by quartets, sextets, and
higher-spin states; and a triplet may be contaminated by quintets, septets,
and higher-spin states. Restoring spin symmetry generally requires
spin projection or a linear combination of determinants.
The last point is particularly important: a linear combination of
determinants can possess a well-defined total spin even when the individual
determinants do not.


\subsubsection{Spin adaptation of singly excited determinants}

Consider a closed-shell singlet reference and promote one electron from the
doubly occupied spatial orbital $i$ to the virtual spatial orbital $a$.
There are two distinct $M_S=0$ determinants,
\begin{align}
  \ket{D_\alpha}
  & =
  \cre{a\alpha}\ani{i\alpha}\ket{\Psi_0}
  &
  \ket{D_\beta}
  & =
  \cre{a\beta}\ani{i\beta}\ket{\Psi_0}
  \label{eq:spin-adapted-determinants}
\end{align}
The first promotes the $\alpha$ electron and leaves the $\beta$ electron in
orbital $i$, whereas the second promotes the $\beta$ electron.
Both determinants contain one $\alpha$ and one $\beta$ electron in the two
open-shell orbitals. Consequently,
\begin{equation}
  \violet{\chS_z}\ket{D_\sigma}
  =
  0
\end{equation}
and each determinant has
\begin{equation}
  \mel{D_\sigma}{\purple{\chS^2}}{D_\sigma}
  =
  1
\end{equation}
The value $1$ is neither the singlet eigenvalue $0$ nor the triplet
eigenvalue $2$. Therefore, neither determinant is individually an
eigenfunction of $\chS^2$.

Figure~\ref{fig:spin-substitutions} compares the occupations of the two
open-shell determinants.
\begin{center}
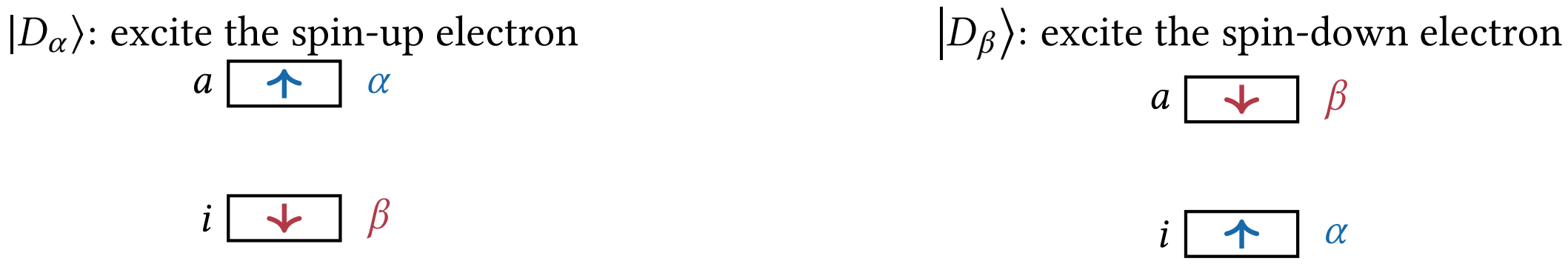

\begin{minipage}[t]{0.43\textwidth}
  \centering
  $\ket{D_\alpha}$: excite the spin-up electron

  \begin{tikzpicture}[x=1cm,y=1cm]
    \node[left] at (0,0.35) {$i$};
    \node[left] at (0,1.55) {$a$};
    \draw[thick] (0,0.15) rectangle (1.0,0.55);
    \draw[thick] (0,1.35) rectangle (1.0,1.75);
    \draw[hfRed,very thick,->] (0.50,0.48)--(0.50,0.22);
    \draw[hfBlue,very thick,->] (0.50,1.42)--(0.50,1.68);
    \node[hfRed,right] at (1.10,0.35) {$\beta$};
    \node[hfBlue,right] at (1.10,1.55) {$\alpha$};
  \end{tikzpicture}
\end{minipage}
\hfill
\begin{minipage}[t]{0.43\textwidth}
  \centering
  $\ket{D_\beta}$: excite the spin-down electron

  \begin{tikzpicture}[x=1cm,y=1cm]
    \node[left] at (0,0.35) {$i$};
    \node[left] at (0,1.55) {$a$};
    \draw[thick] (0,0.15) rectangle (1.0,0.55);
    \draw[thick] (0,1.35) rectangle (1.0,1.75);
    \draw[hfBlue,very thick,->] (0.50,0.22)--(0.50,0.48);
    \draw[hfRed,very thick,->] (0.50,1.68)--(0.50,1.42);
    \node[hfBlue,right] at (1.10,0.35) {$\alpha$};
    \node[hfRed,right] at (1.10,1.55) {$\beta$};
  \end{tikzpicture}
\end{minipage}
\captionof{figure}{Two spin-conserving single substitutions from a closed-shell reference.
The determinants $\ket{D_\alpha}$ and $\ket{D_\beta}$ each have $M_S=0$
but are not individually eigenfunctions of $\chS^2$.}
\label{fig:spin-substitutions}
\end{center}

Each determinant has $M_S=0$, but contains equal singlet and triplet
character. This provides a concrete demonstration that being an eigenfunction
of $\violet{\chS_z}$ does not imply being an eigenfunction of
$\purple{\chS^2}$. In particular, an open-shell determinant need not be a pure spin function.

In the determinant basis
$\qty{
  \ket{D_\alpha},
  \ket{D_\beta}
  }$,
the matrix representation of $\chS^2$ is
\begin{equation}
  [\purple{\chS^2}]_{\{D_\alpha,D_\beta\}}
  =
  \pmqty{
    1 & -1\\
    -1 & 1
  }
\end{equation}
The determinant basis is therefore not spin adapted. Diagonalizing this
$2\times2$ matrix produces eigenvectors with definite values of $S$.
The symmetric combination has eigenvalue zero,
\begin{align}
  \ket{{}^1\Psi_i^a}
  &=
  \frac{
    \ket{D_\alpha}
    \red{+}
    \ket{D_\beta}
  }{\sqrt{2}}
  &
  \purple{\chS^2}
  \ket{{}^1\Psi_i^a}
  &=
  0
\end{align}
and is therefore a pure singlet $S=0$.
The antisymmetric combination has eigenvalue two,
\begin{equation}
  \ket{{}^{3,0}\Psi_i^a}
  =
  \frac{
    \ket{D_\alpha}
    \red{-}
    \ket{D_\beta}
  }{\sqrt{2}}
\end{equation}
with
\begin{equation}
  \purple{\chS^2}
  \ket{{}^{3,0}\Psi_i^a}
  =
  2
  \ket{{}^{3,0}\Psi_i^a}
\end{equation}
and therefore $S=1$.
It is the $M_S=0$ component of a triplet.
Thus, spin adaptation is the diagonalization of $\chS^2$ in a determinant space.

\paragraph{Spatial and spin symmetry.}

The same result can be understood directly from the permutation symmetry of
the two open-shell electrons. The following factors display only these two
electrons; a common closed-shell core and antisymmetrization with that core
are implicit for a larger system. Here $\psi_i$ and $\psi_a$ denote the
two spatial molecular orbitals, not members of the AO basis.

For two open-shell spatial orbitals $i$ and $a$, the singlet may be written
as
\begin{equation}
  \ket{{}^1\Psi_i^a}
  =
  \frac{1}{2}
  \underbrace{
  \qty[
    \ket{\psi_i(1)\psi_a(2)}
    +
    \ket{\psi_a(1)\psi_i(2)}
  ]}_{\text{spatially symmetric}}
  \underbrace{
  \qty[
    \alpha(1)\beta(2)
    -
    \beta(1)\alpha(2)
  ]}_{\text{spin antisymmetric}}
\end{equation}
The spatial factor is \green{symmetric} under exchange of the two electrons,
whereas the spin factor is \red{antisymmetric}. Their product is therefore
antisymmetric, as required for fermions.
For the $M_S=0$ triplet,
\begin{equation}
  \ket{{}^{3,0}\Psi_i^a}
  =
  \frac{1}{2}
  \underbrace{
  \qty[
    \ket{\psi_i(1)\psi_a(2)}
    -
    \ket{\psi_a(1)\psi_i(2)}
  ]}_{\text{spatially antisymmetric}}
  \underbrace{
  \qty[
    \alpha(1)\beta(2)
    +
    \beta(1)\alpha(2)
  ]}_{\text{spin symmetric}}
\end{equation}
Here the spatial factor is \red{antisymmetric} and the spin factor is
\green{symmetric}. Again,
\begin{equation}
  \text{the total electronic wavefunction is antisymmetric}
\end{equation}
This complementary symmetry between spatial and spin parts is a direct
consequence of the Pauli principle introduced earlier in the chapter.

\paragraph{The three triplet components.}

A state with $S=1$ contains three components corresponding to
$M_S=+1,0,-1$.
For the two open-shell orbitals $i$ and $a$, these are
\begin{align}
  \ket{{}^{3,+1}\Psi_i^a}
  &=
  \ket{\cdots i^\alpha a^\alpha\cdots}
  &
  \ket{{}^{3,0}\Psi_i^a}
  &=
  \frac{\ket{D_\alpha}-\ket{D_\beta}}{\sqrt{2}}
  &
  \ket{{}^{3,-1}\Psi_i^a}
  &=
  \ket{\cdots i^\beta a^\beta\cdots}
\end{align}
All three belong to the same spin multiplet and satisfy
\begin{align}
  \purple{\chS^2}\ket{1,M_S}
  &=
  2\ket{1,M_S}
  &
  \violet{\chS_z}\ket{1,M_S}
  &=
  M_S\ket{1,M_S}
\end{align}
The components are connected by the spin ladder operators,
\begin{equation}
  \underbrace{
  \chS_\pm
  }_{\text{ladder operator}}
  \ket{S,M_S}
  =
  \sqrt{
    S(S+1)-M_S(M_S\pm1)
  }
  \ket{S,M_S\pm1}
\end{equation}
The $M_S=\pm1$ triplet components are each represented by a single
spin-pure determinant. By contrast, the $M_S=0$ component requires a linear
combination of two determinants.

This example illustrates a general lesson that will recur throughout
electronic-structure theory: determinants are a convenient antisymmetric
basis, but states with the correct physical symmetries often require
specific linear combinations of determinants.

\subsection{The Fock operator and orbital energies}

The preceding discussion used the occupied-virtual Fock matrix to express
stationarity. We now derive that operator by varying the Hartree--Fock energy
under the orbital orthonormality constraints. Its eigenfunctions
define the canonical orbitals and their orbital energies, whose physical
interpretation is developed here together with Koopmans' theorem.

\subsubsection{The Fock operator}
\label{sec:fock-operator}

The Hartree--Fock approximation has so far been formulated as an energy
functional of a set of orthonormal occupied spin orbitals,
\begin{equation}
  E_\text{HF}[\{\chi_i\}]
  =
  \sum_{i=1}^{N} h_{ii}
  +
  \frac{1}{2}
  \sum_{i=1}^{N}\sum_{j=1}^{N}
  \mel{ij}{}{ij}
  \label{eq:hf-energy-orbital-functional}
\end{equation}
The Hartree--Fock problem consists in finding the set of spin orbitals that
makes this energy stationary while preserving their orthonormality,
\begin{equation}
  \braket{\chi_i}{\chi_j}
  =
  \delta_{ij}
\end{equation}

This is a constrained variational problem. Since the orthonormality
conditions couple every pair of occupied spin orbitals, a single Lagrange
multiplier is not sufficient. Instead, one introduces a complete matrix of
multipliers $\beps$ and defines the Hartree--Fock Lagrangian
\begin{equation}
  \mathcal{L}[\{\chi_i\}]
  =
  E_\text{HF}[\{\chi_i\}]
  -
  \sum_{ij}
  \underbrace{\epsilon_{ji}}_{
    \substack{\text{\red{Lagrange}}\\
              \text{\red{multipliers}}}
  }
  \underbrace{
    \qty(
      \braket{\chi_i}{\chi_j}
      -
      \delta_{ij}
    )
  }_{
    \substack{\text{\green{orthonormality}}\\
              \text{\green{constraint}}}
  }
  \label{eq:hf-lagrangian}
\end{equation}

The orbitals and their complex conjugates may be treated as independent
variables in the variation. Stationarity with respect to
$\chi_i^*(\bx)$ requires
\begin{equation}
  \fdv{\mathcal{L}}{\chi_i^*(\bx)}
  =
  0
  \label{eq:hf-stationarity}
\end{equation}

\paragraph{Functional derivative of the HF energy.}

The one-electron part is straightforward:
\begin{equation}
  \fdv{}{\chi_i^*(\bx)}
  \sum_j
  \mel{\chi_j}{\hh}{\chi_j}
  =
  \hh(\bx)\chi_i(\bx)
\end{equation}

The derivative of the two-electron contribution generates the Coulomb and
exchange operators. Using the symmetry of the double sum, the factor
$1/2$ disappears because the orbital $\chi_i$ may occur in either member
of an interacting pair:
\begin{equation}
  \fdv{}{\chi_i^*(\bx)}
  \qty[
    \frac{1}{2}
    \sum_{jk}
    \mel{jk}{}{jk}
  ]
  =
  \sum_{j=1}^{N}
  \qty[
    \chJ_j(\bx)
    -
    \chK_j(\bx)
  ]
  \chi_i(\bx)
\end{equation}
Consequently,
\begin{equation}
  \fdv{E_\text{HF}}{\chi_i^*(\bx)}
  =
  \qty{
    \hh(\bx)
    +
    \sum_{j=1}^{N}
    \qty[
      \chJ_j(\bx)-\chK_j(\bx)
    ]
  }
  \chi_i(\bx)
\end{equation}
This motivates the definition of the Fock operator,
\begin{equation}
  \boxed{
  \hf(\bx)
  =
  \hh(\bx)
  +
  \underbrace{
    \sum_{j=1}^{N}
    \qty[
      \chJ_j(\bx)-\chK_j(\bx)
    ]
  }_{
    \Hat{v}^{\text{HF}}(\bx)
    \text{ = Hartree--Fock potential}
  }
  }
  \label{eq:fock-operator}
\end{equation}

The Fock operator is therefore an effective one-electron operator composed
of the core Hamiltonian and the average interaction with all occupied spin
orbitals. The Coulomb term represents the classical electrostatic
interaction with the electronic distribution, whereas the exchange term is
a purely fermionic contribution arising from antisymmetry.
As discussed previously, the Coulomb operator is local while the exchange
operator is nonlocal. The Fock operator is therefore, in general, a
nonlocal one-electron operator.

Notice also that an electron does not interact with itself. When the Fock
operator acts on $\chi_i$,
\begin{equation}
  \qty(
    \chJ_i-\chK_i
  )
  \ket{\chi_i}
  =
  0
\end{equation}
because the Coulomb and exchange self-interaction contributions are
identical and cancel exactly.

\paragraph{The Hartree--Fock equations.}

Taking the functional derivative of the constraint term in
Eq.~\eqref{eq:hf-lagrangian} gives
\begin{equation}
  \fdv{}{\chi_i^*(\bx)}
  \sum_{jk}
  \epsilon_{kj}
  \braket{\chi_j}{\chi_k}
  =
  \sum_j
  \epsilon_{ji}
  \chi_j(\bx)
\end{equation}
The stationarity condition therefore becomes
\begin{equation}
  \hf(\bx)\chi_i(\bx)
  =
  \sum_j
  \epsilon_{ji}\chi_j(\bx)
  \label{eq:general-hf-equations}
\end{equation}
These are the general Hartree--Fock equations. An important feature is that
the Fock operator acting on a given occupied spin orbital need not return
that same orbital. It is only required to remain within the space spanned
by the occupied spin orbitals.

Multiplying Eq.~\eqref{eq:general-hf-equations} from the left by
$\bra{\chi_k}$ and using orthonormality gives
\begin{equation}
  \epsilon_{ki}
  =
  \mel{\chi_k}{\hf}{\chi_i}
  \label{eq:lagrange-fock}
\end{equation}
Since $\hf$ is Hermitian,
\begin{equation}
  \mel{\chi_k}{\hf}{\chi_i}
  =
  \mel{\chi_i}{\hf}{\chi_k}^*
\end{equation}
and hence
\begin{equation}
  \epsilon_{ki}
  =
  \epsilon_{ik}^*
\end{equation}
The matrix of Lagrange multipliers is therefore Hermitian,
\begin{equation}
  \beps^\dag=\beps
\end{equation}
This property is crucial because every Hermitian matrix can be diagonalized
by a unitary transformation.

\paragraph{Why the HF equations are self-consistent.}

Equation~\eqref{eq:general-hf-equations} resembles an ordinary one-electron
eigenvalue problem, but there is a fundamental difference. The Fock
operator itself depends on the occupied orbitals through the Coulomb and
exchange operators:
\begin{equation}
  \hf
  =
  \hf[\chi_1,\chi_2,\ldots,\chi_N]
\end{equation}
The orbitals are therefore required to satisfy an equation containing an
operator constructed from those same orbitals.
Schematically,
\begin{equation}
  \{\chi_i\}
  \quad\longrightarrow\quad
  \hf[\{\chi_i\}]
  \quad\longrightarrow\quad
  \{\chi_i\}
\end{equation}
The Hartree--Fock equations are consequently nonlinear and must be
solved self-consistently. This is the origin of the term
self-consistent field (SCF).

In practice, one starts from a trial set of orbitals, constructs the
corresponding Fock operator, obtains a new set of orbitals, and repeats the
procedure until the orbitals, density, and energy are self-consistent.
The variational equations establish a stationary point of the Hartree--Fock energy.
They do not, by themselves, guarantee that a particular SCF solution is the
global minimum: multiple HF stationary solutions may exist.

Exercise~\ref{ex:hf-orbital-stationarity} supplies the functional-derivative steps behind this
stationarity condition, including the separate variations of the two
electron coordinates in the Coulomb and exchange terms.

\begin{hfexercise}[phantom={\label{ex:hf-orbital-stationarity}}]{Stationarity of the Hartree--Fock energy}

\violet{\textit{Derive the Hartree--Fock equations by making the spin-orbital
energy stationary under the orbital orthonormality constraints.}}

\end{hfexercise}
\begin{hfsolution}{Stationarity of the Hartree--Fock energy}

We vary the occupied spin orbitals without imposing restricted spatial
pairing. Thus the derivation establishes the general spin-orbital
stationarity condition; restricted variations retain only the rotations
allowed by their additional constraints. The nuclei and one-electron
Hamiltonian are fixed throughout the variation.

Write the energy as the sum of one-electron, Coulomb, and exchange
contributions,
\begin{equation}
  E_\text{HF}=E_1+E_\text{H}+E_\text{X}
\end{equation}
with
\begin{align}
  E_1&=\sum_j\int\chi_j^*(\bx_1)\hh(\bx_1)
    \chi_j(\bx_1)\dd{\bx_1}
  \\
  E_\text{H}&=\frac12\sum_{jk}\iint
    \chi_j^*(\bx_1)\chi_j(\bx_1)\frac{1}{r_{12}}
    \chi_k^*(\bx_2)\chi_k(\bx_2)\dd{\bx_1}\dd{\bx_2}
  \\
  E_\text{X}&=-\frac12\sum_{jk}\iint
    \chi_j^*(\bx_1)\chi_k(\bx_1)\frac{1}{r_{12}}
    \chi_k^*(\bx_2)\chi_j(\bx_2)\dd{\bx_1}\dd{\bx_2}
\end{align}
The sums run over occupied spin orbitals. For complex functions we take
$\chi_i$ and $\chi_i^*$ as independent variables, with the conjugate
variation supplying the conjugate equation. The elementary differentiation
rule is
\begin{equation}
  \fdv{\chi_j^*(\bx_1)}{\chi_i^*(\bx)}
  =\delta_{ij}\delta(\bx_1-\bx)
\end{equation}
Here the delta function in the combined coordinate denotes a spatial Dirac
delta and the identity on the spin coordinate. It selects both the orbital
label and the electron coordinate at which the variation is taken.

For the one-electron term only $j=i$ contributes, so
\begin{equation}
\begin{split}
  \fdv{E_1}{\chi_i^*(\bx)}
  &=\sum_j\int\delta_{ij}\delta(\bx_1-\bx)
    \hh(\bx_1)\chi_j(\bx_1)\dd{\bx_1}
  \\
  &=\hh(\bx)\chi_i(\bx)
\end{split}
\end{equation}

In the Coulomb term the varied conjugate can occur at either coordinate.
Differentiating each occurrence gives
\begin{equation}
\begin{split}
  \fdv{E_\text{H}}{\chi_i^*(\bx)}
  &=\frac12\sum_k\int
    \frac{\chi_i(\bx)\chi_k^*(\bx')\chi_k(\bx')}
      {\abs{\br-\br'}}\dd{\bx'}
  \\
  &+\frac12\sum_j\int
    \frac{\chi_j^*(\bx')\chi_j(\bx')\chi_i(\bx)}
      {\abs{\br'-\br}}\dd{\bx'}
  \\
  &=\sum_j\qty[\int
    \frac{\abs{\chi_j(\bx')}^2}{\abs{\br-\br'}}\dd{\bx'}]
    \chi_i(\bx)
  \\
  &=\sum_j\chJ_j(\bx)\chi_i(\bx)
\end{split}
\end{equation}
The first contribution comes from $j=i$ at electron 1 and the second from
$k=i$ at electron 2. Relabeling the dummy orbital index makes them identical.
Their sum cancels the original factor of one half. The denominator involves
the spatial coordinates $\br$ and $\br'$, not the spin coordinate.

The exchange term has the same two possible conjugate occurrences, but
the unstarred orbitals are crossed between the coordinates:
\begin{equation}
\begin{split}
  \fdv{E_\text{X}}{\chi_i^*(\bx)}
  &=-\frac12\sum_k\int
    \frac{\chi_k(\bx)\chi_k^*(\bx')\chi_i(\bx')}
      {\abs{\br-\br'}}\dd{\bx'}
  \\
  &-\frac12\sum_j\int
    \frac{\chi_j^*(\bx')\chi_i(\bx')\chi_j(\bx)}
      {\abs{\br'-\br}}\dd{\bx'}
  \\
  &=-\sum_j\chi_j(\bx)\int
    \frac{\chi_j^*(\bx')\chi_i(\bx')}
      {\abs{\br-\br'}}\dd{\bx'}
  \\
  &=-\sum_j\chK_j(\bx)\chi_i(\bx)
\end{split}
\end{equation}
Again, the two halves are equal, but the overall exchange minus sign remains.
Unlike the Coulomb potential, the resulting operator samples the varied
orbital at $\bx'$ and therefore acts nonlocally. Combining the three pieces
recovers
\begin{equation}
  \fdv{E_\text{HF}}{\chi_i^*(\bx)}
  =\qty[\hh(\bx)+\sum_j\qty(\chJ_j(\bx)-\chK_j(\bx))]
    \chi_i(\bx)
  =\hf(\bx)\chi_i(\bx)
\end{equation}

The energy derivative alone cannot be set to zero because arbitrary orbital
variations do not preserve orthonormality. Using the Lagrangian of
Eq.~\eqref{eq:hf-lagrangian}, differentiation of the constraint term yields
\begin{equation}
\begin{split}
  \fdv{}{\chi_i^*(\bx)}
    \sum_{jk}\epsilon_{kj}\qty(\braket{\chi_j}{\chi_k}-\delta_{jk})
  &=\sum_{jk}\epsilon_{kj}\delta_{ij}\chi_k(\bx)
  \\
  &=\sum_j\epsilon_{ji}\chi_j(\bx)
\end{split}
\end{equation}
Variation with respect to the multipliers independently restores the overlap
constraints. Variation with respect to the conjugate orbitals gives
\begin{equation}
  \fdv{\mathcal L}{\chi_i^*(\bx)}
  =\hf(\bx)\chi_i(\bx)-\sum_j\epsilon_{ji}\chi_j(\bx)=0
\end{equation}
Thus the Fock operator maps every occupied orbital back into the occupied
space. Projecting onto an occupied $\chi_k$ identifies the multipliers,
\begin{equation}
  \epsilon_{ki}=\mel{\chi_k}{\hf}{\chi_i}
  =\epsilon_{ik}^*
\end{equation}
Their Hermiticity allows the occupied-space unitary canonicalization
developed next; it is not an extra assumption in the variational problem.

To express the same condition directly as an orbital gradient, consider an
infinitesimal occupied-virtual rotation,
\begin{equation}
  \delta\chi_i=\sum_a\chi_a\kappa_{ai}
\end{equation}
where the $\kappa_{ai}$ are independent complex rotation parameters and the
complementary virtual-orbital rotation preserves orthonormality to first
order. The first-order energy change is
\begin{equation}
\begin{split}
  \delta E_\text{HF}
  &=2\Re\sum_i\int\delta\chi_i^*(\bx)
    \hf(\bx)\chi_i(\bx)\dd{\bx}
  \\
  &=2\Re\sum_{ia}\kappa_{ai}^*\mel{\chi_a}{\hf}{\chi_i}
  =2\Re\sum_{ia}\kappa_{ai}^*f_{ai}
\end{split}
\end{equation}
Both the real and imaginary parts of each allowed rotation may be varied.
Stationarity therefore requires
\begin{equation}
  \boxed{f_{ai}=0}
\end{equation}
Equivalently, the projection of the general HF equations onto the virtual
space vanishes. Hermiticity gives $f_{ia}=f_{ai}^*=0$, and the
Slater--Condon rule then recovers
\begin{equation}
  \mel{\Psi_0}{\chH_\text{elec}}{\Psi_i^a}=f_{ia}=0
\end{equation}
This is Brillouin's theorem. Occupied-occupied rotations merely change the
representation of the determinant, as shown in the next subsection.
Finally, a vanishing first variation establishes only a stationary point.
Its stability depends on the second variation, so an SCF solution may be a
minimum or an unstable stationary solution rather than the global HF minimum.

\end{hfsolution}

\subsubsection{Unitary invariance and canonical orbitals}

\hfBioDirac

\paragraph{Unitary invariance of the occupied orbitals.}

The HF determinant is not associated with a unique set of occupied
orbitals. Consider a unitary transformation among the occupied spin
orbitals,
\begin{align}
  \widetilde{\chi}_i(\bx)
  &=
  \sum_j
  \chi_j(\bx)
  \purple{U}_{ji}
  &
  \purple{\bU}^\dag\cdot\purple{\bU}
  &=
  \bI
  \label{eq:occupied-unitary-rotation}
\end{align}
The transformed orbitals remain orthonormal:
\begin{equation}
  \braket{\widetilde{\chi}_i}{\widetilde{\chi}_j}
  =
  \sum_{kl}
  U_{ki}^*
  U_{lj}
  \braket{\chi_k}{\chi_l}
  =
  \sum_k
  U_{ki}^*U_{kj}
  =
  \delta_{ij}
\end{equation}
More importantly, the occupied one-particle projector is unchanged:
\begin{equation}
\begin{split}
  \sum_i
  \widetilde{\chi}_i(\bx)
  \widetilde{\chi}_i^*(\bx')
  &=
  \sum_{ijk}
  \chi_j(\bx)
  U_{ji}
  U_{ki}^*
  \chi_k^*(\bx')
  =
  \sum_{jk}
  \chi_j(\bx)
  \delta_{jk}
  \chi_k^*(\bx')
  \\
  &=
  \sum_j
  \chi_j(\bx)
  \chi_j^*(\bx')
\end{split}
\end{equation}
Since the Fock operator depends only on the occupied one-particle density,
it is invariant under this transformation. Exercise~\ref{ex:operator-unitary-invariance}
establishes this explicitly for its Coulomb and exchange contributions.

The determinant itself transforms as
\begin{equation}
  \boxed{
  \ket*{\widetilde{\Psi}_\text{HF}}
  =
  \det(\purple{\bU})
  \ket*{\Psi_\text{HF}}
  }
  \label{eq:det-unitary-transform}
\end{equation}
Because $\purple{\bU}$ is unitary,
\begin{equation}
  \purple{\bU}^{\dag}\cdot\purple{\bU}
  =
  \bI
  \quad\Rightarrow\quad
  \det(
    \purple{\bU}^{\dag}\cdot\purple{\bU}
  )
  =
  1
\end{equation}
and therefore
\begin{equation}
  \abs{\det(\purple{\bU})}^2
  =
  1
  \quad\Rightarrow\quad
  \det(\purple{\bU})
  =
  e^{\ii\theta}
\end{equation}
Hence,
\begin{equation}
  \ket*{\widetilde{\Psi}_\text{HF}}
  =
  e^{\ii\theta}
  \ket*{\Psi_\text{HF}}
\end{equation}
The two determinants differ only by an irrelevant global phase and describe
the same physical state.
Thus, unitary rotations within the occupied space leave
the energy $E_\text{HF}$ and the physical state unchanged, with the
wavefunction itself defined only up to that phase.

\begin{hfexercise}[phantom={\label{ex:operator-unitary-invariance}}]{Unitary invariance of the HF operators}
\violet{\textit{Show that the total Coulomb, exchange, and Fock operators
are unchanged by a unitary transformation of the occupied spin orbitals.}}
\end{hfexercise}
\begin{hfsolution}{Unitary invariance of the HF operators}
Use the occupied-space transformation in
Eq.~\eqref{eq:occupied-unitary-rotation}, allowing complex coefficients.
All orbital sums below run over the $N$ occupied spin orbitals. Since
$\bU$ is a square unitary matrix, the contraction needed in both proofs is
\begin{equation}
  \sum_i U_{ji}U_{ki}^*
  =(\bU\cdot\bU^\dag)_{jk}
  =\delta_{jk}
\end{equation}
Define the total operators by $\chJ=\sum_i\chJ_i$ and
$\chK=\sum_i\chK_i$; tildes denote operators constructed from the
transformed occupied orbitals. To prove an operator identity, let these
operators act on an arbitrary one-electron test function $g(\bx)$, not
necessarily an occupied orbital.
For the Coulomb operator, substitution into
Eq.~\eqref{eq:coulomb-operator} gives
\begin{equation}
\begin{split}
  (\widetilde{\chJ}g)(\bx_1)
  &=\sum_i\qty[\int
    \widetilde{\chi}_i^*(\bx_2)\frac{1}{r_{12}}
    \widetilde{\chi}_i(\bx_2)\dd{\bx_2}]g(\bx_1)\\
  &=\sum_{jk}\underbrace{\qty(\sum_iU_{ji}U_{ki}^*)}_{\delta_{jk}}
    \qty[\int\chi_k^*(\bx_2)\frac{1}{r_{12}}
      \chi_j(\bx_2)\dd{\bx_2}]g(\bx_1)\\
  &=\sum_j\qty[\int\chi_j^*(\bx_2)\frac{1}{r_{12}}
      \chi_j(\bx_2)\dd{\bx_2}]g(\bx_1)\\
  &=(\chJ g)(\bx_1)
\end{split}
\end{equation}
Unitarity therefore leaves the total density, and hence its Coulomb
potential, unchanged. The exchange operator depends on the nonlocal
one-particle density rather than only its diagonal. Using
Eq.~\eqref{eq:exchange-operator}, its action becomes
\begin{equation}
\begin{split}
  (\widetilde{\chK}g)(\bx_1)
  &=\sum_i\widetilde{\chi}_i(\bx_1)
    \int\widetilde{\chi}_i^*(\bx_2)\frac{1}{r_{12}}
      g(\bx_2)\dd{\bx_2}\\
  &=\sum_{jk}\underbrace{\qty(\sum_iU_{ji}U_{ki}^*)}_{\delta_{jk}}
    \chi_j(\bx_1)\int\chi_k^*(\bx_2)\frac{1}{r_{12}}
      g(\bx_2)\dd{\bx_2}\\
  &=\sum_j\chi_j(\bx_1)\int\chi_j^*(\bx_2)\frac{1}{r_{12}}
      g(\bx_2)\dd{\bx_2}\\
  &=(\chK g)(\bx_1)
\end{split}
\end{equation}
Since $g$ is arbitrary, these equalities establish the operator identities
\begin{align}
  \widetilde{\chJ}&=\chJ
  &\widetilde{\chK}&=\chK
\end{align}
The core Hamiltonian is independent of the occupied orbitals, so
\begin{equation}
  \boxed{
  \widetilde{\hf}
  =\hh+\widetilde{\chJ}-\widetilde{\chK}
  =\hh+\chJ-\chK
  =\hf
  }
\end{equation}
The invariance applies to the occupied sums: individual $\chJ_i$ and
$\chK_i$ generally change when their generating orbitals are rotated.
It does not require self-consistency. Rotations among virtual orbitals also
leave these operators unchanged because the virtuals do not enter their
construction, whereas occupied-virtual mixing generally changes the density
and the operators.

Operator invariance must not be confused with invariance of its matrix
elements in a changing orbital basis. For occupied indices $i,j$,
\begin{equation}
\begin{split}
  \widetilde f_{ij}
  &=\mel{\widetilde\chi_i}{\hf}{\widetilde\chi_j}\\
  &=\sum_{kl}U_{ki}^*\mel{\chi_k}{\hf}{\chi_l}U_{lj}\\
  &=\sum_{kl}U_{ki}^*f_{kl}U_{lj}
\end{split}
\end{equation}
Thus the occupied block obeys
\begin{equation}
  \widetilde{\bff}_\text{oo}
  =\bU^\dag\cdot\bff_\text{oo}\cdot\bU
\end{equation}
Its entries need not remain unchanged, even though the underlying Fock
operator does. This is precisely the freedom used in canonicalization.
\end{hfsolution}

\paragraph{Canonical Hartree--Fock orbitals.}

This unitary freedom can be exploited to choose a particularly convenient
set of occupied orbitals. Since the Lagrange-multiplier matrix is Hermitian,
there exists a unitary matrix $\purple{\bU}$ such that
\begin{equation}
  \widetilde{\beps}
  =
  \purple{\bU}^{\dag}
  \cdot
  \beps
  \cdot
  \purple{\bU}
\end{equation}
is diagonal, i.e.,
$\widetilde{\epsilon}_{ij} = \epsilon_i\delta_{ij}$.

Applying the same unitary transformation to the occupied orbitals converts
the general HF equations,
\begin{equation}
  \hf\ket{\chi_i}
  =
  \sum_j
  \epsilon_{ji}\ket{\chi_j}
\end{equation}
into
\begin{equation}
  \hf\ket{\widetilde{\chi}_i}
  =
  \epsilon_i
  \ket{\widetilde{\chi}_i}
\end{equation}
Dropping the tildes, one obtains the
canonical Hartree--Fock equations,
\begin{equation}
  \boxed{
  \hf(\bx)\chi_i(\bx)
  =
  \epsilon_i\chi_i(\bx)
  }
  \label{eq:canonical-hf}
\end{equation}
The eigenfunctions $\{\chi_i\}$ are called
canonical spin orbitals, and the corresponding eigenvalues
$\{\epsilon_i\}$ are called orbital energies.
Canonicalization does not produce a new Hartree--Fock wavefunction. It is
only a convenient choice of basis within the same occupied orbital space.
At self-consistency, the Fock operator may also be diagonalized in the
orthogonal complement of the occupied space. This defines the
canonical virtual orbitals. The complete canonical orbital set then
satisfies
\begin{equation}
  \mel{\chi_p}{\hf}{\chi_q}
  =
  \epsilon_p\delta_{pq}
  \label{eq:canonical-fock-diagonal}
\end{equation}
In particular, the occupied-virtual block vanishes,
\begin{equation}
  f_{ia}
  =
  \mel{i}{\hf}{a}
  =
  0
  \label{eq:occupied-virtual-fock-zero}
\end{equation}
This is the orbital form of the stationarity condition encountered in
Brillouin's theorem:
\begin{equation}
  \mel{\Psi_0}{\chH_\text{elec}}{\Psi_i^a}
  =
  f_{ia}
  =
  0
\end{equation}


\begin{hfexercise}{Fock matrix elements in the spin-orbital basis}

\violet{\textit{Express the Fock matrix elements in terms of one- and
two-electron integrals,
\[
  f_{ij}
  =
  \mel{\chi_i}{\hf}{\chi_j}
\]
}}

\end{hfexercise}
\begin{hfsolution}{Fock matrix elements in the spin-orbital basis}

Use orthonormal spin orbitals and keep the occupied-orbital sum explicit;
the indices $i,j$ specify the matrix element, while $k$ runs over occupied
spin orbitals. Canonical orbitals are not required for this expansion.
Expanding the Fock operator gives
\begin{equation}
\begin{split}
  f_{ij}
  &=
  \mel{\chi_i}
      {
        \hh
        +
        \sum_k
        \qty(
          \chJ_k-\chK_k
        )
      }
      {\chi_j}
  \\
  &=
  \mel{\chi_i}{\hh}{\chi_j}
  +
  \sum_k
  \qty[
    \mel{\chi_i}{\chJ_k}{\chi_j}
    -
    \mel{\chi_i}{\chK_k}{\chi_j}
  ]
\end{split}
\end{equation}

The Coulomb contribution is
\begin{equation}
  \mel{\chi_i}{\chJ_k}{\chi_j}
  =
  \braket{ik}{jk}
\end{equation}
whereas the exchange contribution is
\begin{equation}
  \mel{\chi_i}{\chK_k}{\chi_j}
  =
  \braket{ik}{kj}
\end{equation}
These index arrangements follow directly from the actions of the two
operators. Writing $\chi_p(n)=\chi_p(\bx_n)$, their matrix elements are
\begin{align}
  \mel{\chi_i}{\chJ_k}{\chi_j}
  &=\iint\chi_i^*(1)\chi_k^*(2)\frac{1}{r_{12}}
    \chi_j(1)\chi_k(2)\dd{\bx_1}\dd{\bx_2}\\
  \mel{\chi_i}{\chK_k}{\chi_j}
  &=\iint\chi_i^*(1)\chi_k^*(2)\frac{1}{r_{12}}
    \chi_k(1)\chi_j(2)\dd{\bx_1}\dd{\bx_2}
\end{align}
The Coulomb operator leaves $\chi_j$ at coordinate 1, whereas exchange
places it at coordinate 2 and returns $\chi_k$ at coordinate 1. Reading
the bra and ket indices in physicists' notation gives the two integrals
above; their difference is the antisymmetrized integral.
Therefore,
\begin{equation}
\begin{split}
  f_{ij}
  &=
  \mel{i}{\hh}{j}
  +
  \sum_k
  \qty[
    \braket{ik}{jk}
    -
    \braket{ik}{kj}
  ]
  \\
  &=
  \mel{i}{\hh}{j}
  +
  \sum_k
  \mel{ik}{}{jk}
  \label{eq:fock-matrix-spin-orbital}
\end{split}
\end{equation}

Thus, each Fock matrix element contains the bare one-electron contribution
plus the Coulomb and exchange interactions with all occupied spin orbitals.

For an occupied-virtual matrix element,
\begin{equation}
  f_{ia}
  =
  h_{ia}
  +
  \sum_k
  \mel{ik}{}{ak}
\end{equation}
which is precisely the quantity appearing in Brillouin's theorem.
For a self-consistent stationary determinant this occupied-virtual element
vanishes, but the formula itself holds for any chosen orthonormal
spin-orbital basis with the stated occupied set.

\end{hfsolution}


\begin{hfexercise}{Orbital energies in the spin-orbital basis}

\violet{\textit{Deduce the energy $\epsilon_i$ of a canonical spin
orbital from the diagonal Fock matrix element.}}

\end{hfexercise}
\begin{hfsolution}{Orbital energies in the spin-orbital basis}

For a canonical orbital,
\begin{equation}
  \hf\ket{\chi_i}
  =
  \epsilon_i\ket{\chi_i}
\end{equation}
Multiplying by $\bra{\chi_i}$ gives
\begin{equation}
\begin{split}
  \mel{\chi_i}{\hf}{\chi_i}
  &=
  \epsilon_i
  \braket{\chi_i}{\chi_i}
  \\
  &=
  \epsilon_i
\end{split}
\end{equation}
because the orbital is normalized. Therefore,
\begin{equation}
\begin{split}
  \epsilon_i
  &=
  f_{ii}
  \\
  &=
  \mel{i}{\hh}{i}
  +
  \sum_k
  \qty[
    \braket{ik}{ik}
    -
    \braket{ik}{ki}
  ]
  \\
  &=
  \mel{i}{\hh}{i}
  +
  \sum_k
  \mel{ik}{}{ik}
  \label{eq:hf-orbital-energy}
\end{split}
\end{equation}

The term with $k=i$ vanishes,
\begin{equation}
  \mel{ii}{}{ii}
  =
  \braket{ii}{ii}
  -
  \braket{ii}{ii}
  =
  0
\end{equation}
which again reflects the exact cancellation of one-electron
self-interaction in Hartree--Fock theory.
Here $i$ labels an occupied canonical spin orbital, so its own term in the
occupied sum over $k$ cancels as shown. The analogous expression for a
virtual canonical orbital $a$ follows by replacing $i$ with $a$, while
$k$ still runs only over occupied orbitals; there is then no $k=a$ term.
Canonicalization is
essential to interpreting $f_{ii}$ as an orbital eigenvalue: a diagonal
expectation value in a noncanonical orbital is not, in general, one of the
canonical orbital energies.

\end{hfsolution}
\subsubsection{Orbital energies and the total Hartree--Fock energy}

The orbital energies are Lagrange multipliers associated with the
orthonormality constraints. They should therefore not be interpreted simply
as independent one-electron contributions to the total energy.

Indeed, summing the occupied orbital energies gives
\begin{equation}
\begin{split}
  \sum_i\epsilon_i
  &=
  \sum_i h_{ii}
  +
  \sum_{ij}
  \mel{ij}{}{ij}
\end{split}
\end{equation}
By comparison, the Hartree--Fock energy is
\begin{equation}
  E_\text{HF}
  =
  \sum_i h_{ii}
  +
  \frac{1}{2}
  \sum_{ij}
  \mel{ij}{}{ij}
\end{equation}
Thus,
\begin{equation}
  \boxed{
  E_\text{HF}
  =
  \sum_i\epsilon_i
  -
  \frac{1}{2}
  \sum_{ij}
  \mel{ij}{}{ij}
  }
  \label{eq:hf-energy-orbital-energies}
\end{equation}
Equivalently,
\begin{equation}
  E_\text{HF}
  =
  \frac{1}{2}
  \sum_i
  \qty(
    h_{ii}+\epsilon_i
  )
  \label{eq:hf-energy-half-sum}
\end{equation}

The reason the total HF energy is not simply the sum of the occupied
orbital energies is that each $\epsilon_i$ contains the interaction of
electron $i$ with every other occupied electron. Summing the orbital
energies therefore counts each electron pair twice, whereas the total
energy must count each pair only once.
Although orbital energies are therefore not independent electron energies,
they nevertheless acquire an important physical interpretation within the
frozen-orbital approximation. This observation leads directly to
Koopmans' theorem, discussed next.

\subsubsection{Koopmans' theorem}

The canonical Hartree--Fock orbital energies are Lagrange multipliers
associated with the orthonormality constraints, but they also acquire an
important physical interpretation when electrons are removed from or added
to the system while the orbitals are kept fixed.
This is the content of Koopmans' theorem \citep{Koopmans1934}. Its central approximation is
the frozen-orbital approximation: when the number of electrons is
changed, the spin orbitals of the original $N$-electron HF calculation are
not allowed to relax.
For fixed nuclear coordinates, Koopmans' theorem therefore provides
vertical removal and attachment energies directly from the canonical
HF orbital energies.

\begin{hftheorem}{Koopmans' theorem: frozen-orbital removal and attachment}
\hftheoremportrait{figures/sketch-koopmans.png}{Tjalling Koopmans}
{For fixed nuclei and canonical orbitals of the $N$-electron HF reference,
removing an electron from occupied spin orbital $i$, or adding one to virtual
spin orbital $a$, gives the frozen-orbital energy differences
\citep{Koopmans1934}. These are not relaxed or correlated energy differences.}
{Source portrait: Eric Koch / Anefo, Nationaal Archief,
\href{https://commons.wikimedia.org/wiki/File:TjallingKoopmans1967_(cropped).jpg}{Leuven, 1967 (crop: Wabbuh)},
\href{https://creativecommons.org/publicdomain/zero/1.0/}{CC0 1.0}.
\hfsketchcredit}
\begin{equation}
\begin{split}
  \mathrm{IP}_i^\text{frozen}
  &=E_{N-1}^{\text{frozen},i}-E_N^\text{HF}=-\epsilon_i
  \\
  \mathrm{EA}_a^\text{frozen}
  &=E_N^\text{HF}-E_{N+1}^{\text{frozen},a}=-\epsilon_a
\end{split}
\label{eq:koopmans-summary}
\end{equation}
Orbital relaxation and correlation modify these estimates, as the complete
decompositions below make explicit.
\end{hftheorem}

\hfBioKoopmans


\paragraph{Electron removal.}

Consider the $N$-electron HF ground state. Its electronic energy is
\begin{equation}
  {}^N E_0
  =
  \sum_i h_{ii}
  +
  \frac{1}{2}
  \sum_{ij}
  \mel{ij}{}{ij}
  \label{eq:koopmans-neutral-energy}
\end{equation}
All sums run over the occupied spin orbitals of the $N$-electron reference.
Suppose now that the electron occupying spin orbital $k$ is removed while
all remaining orbitals are kept frozen. The energy of this frozen
$(N-1)$-electron determinant is
\begin{equation}
  {}^{N-1}\widetilde{E}_k
  =
  \sum_{i\ne k}h_{ii}
  +
  \frac{1}{2}
  \sum_{i\ne k}
  \sum_{j\ne k}
  \mel{ij}{}{ij}
  \label{eq:koopmans-cation-frozen}
\end{equation}
where the tilde emphasizes that the orbitals have \emph{not} been
reoptimized for the cation.
The corresponding frozen-orbital ionization potential is
\begin{equation}
  \text{IP}_k^\text{K}
  =
  {}^{N-1}\widetilde{E}_k
  -
  {}^N E_0
\end{equation}
Subtracting Eqs.~\eqref{eq:koopmans-neutral-energy} and
\eqref{eq:koopmans-cation-frozen} gives
\begin{equation}
\begin{split}
  \text{IP}_k^\text{K}
  &=
  -h_{kk}
  -
  \frac{1}{2}
  \sum_i
  \mel{ik}{}{ik}
  -
  \frac{1}{2}
  \sum_j
  \mel{kj}{}{kj}
  \\
  &=
  -h_{kk}
  -
  \sum_i
  \mel{ki}{}{ki}
\end{split}
\end{equation}
where the symmetry of the two-electron integrals has been used.
From the expression of the canonical orbital energy,
\begin{equation}
  \epsilon_k
  =
  h_{kk}
  +
  \sum_i
  \mel{ki}{}{ki}
\end{equation}
we immediately obtain
\begin{equation}
  \text{IP}_k^\text{K}
  =
  -\epsilon_k
  \label{eq:koopmans-ip}
\end{equation}
Thus, within the frozen-orbital approximation, the energy required to remove
an electron from an occupied canonical HF spin orbital is the negative of
its orbital energy.

For the first ionization potential, electron removal usually occurs from the
highest occupied molecular orbital (HOMO), giving the familiar estimate
\begin{equation}
  \text{IP}
  \approx
  -\epsilon_\text{HOMO}
\end{equation}

This result also provides a useful interpretation of the occupied HF
orbital energies: $-\epsilon_i$ is the energy required to create a frozen
hole in occupied spin orbital $i$.


\begin{hfexercise}{Koopmans' theorem for electron affinity}

\violet{\textit{Derive the frozen-orbital electron affinity for attachment
to a vacant canonical spin orbital and relate it to that orbital's energy.}}

\end{hfexercise}
\begin{hfsolution}{Koopmans' theorem for electron affinity}

Consider the attachment of an electron to a vacant canonical spin orbital
$c$. All energies refer to the same nuclear geometry and the occupied
orbitals of the $N$-electron reference remain frozen. The original
one-electron and occupied-pair contributions are therefore unchanged.
The new orbital contributes $h_{cc}$ and interacts with every occupied
spin orbital. In the double-sum energy expression this gives
\begin{equation}
\begin{split}
  {}^{N+1}\widetilde E^c
  &={}^N E_0+h_{cc}
    +\frac12\sum_i\qty[\mel{ci}{}{ci}+\mel{ic}{}{ic}]
    +\frac12\mel{cc}{}{cc}
\end{split}
\end{equation}
The two terms for each old--new pair are equal by electron interchange,
and the self-pair term is zero by antisymmetry. Thus the factor of one
half disappears when each added pair is counted once. The energy of the resulting
$(N+1)$-electron determinant is
\begin{equation}
  {}^{N+1}\widetilde{E}^{c}
  =
  {}^N E_0
  +
  h_{cc}
  +
  \sum_i
  \mel{ci}{}{ci}
\end{equation}
The frozen-orbital electron affinity is defined as
\begin{equation}
  \text{EA}_c^\text{K}
  =
  {}^N E_0
  -
  {}^{N+1}\widetilde{E}^{c}
\end{equation}
Consequently,
\begin{equation}
\begin{split}
  \text{EA}_c^\text{K}
  &=
  -h_{cc}
  -
  \sum_i
  \mel{ci}{}{ci}
  \\
  &=
  \boxed{-\epsilon_c}
  \label{eq:koopmans-ea}
\end{split}
\end{equation}

For attachment into the lowest unoccupied molecular orbital (LUMO),
\begin{equation}
  \text{EA}
  \approx
  -\epsilon_\text{LUMO}
\end{equation}
With this sign convention, a positive electron affinity means that the
attached determinant lies below the reference in energy. The orbital
energy $\epsilon_c$ is evaluated using the reference Fock operator, not a
Fock operator reoptimized for the anion. Nuclear repulsion cancels from
the energy difference because the geometry is fixed.

The algebra is therefore completely analogous to that for electron
removal. There is, however, an important practical asymmetry between the
two cases. Occupied HF orbitals have been variationally optimized to
describe the $N$-electron ground state, whereas the virtual orbitals have
not been optimized to describe an additional electron. Consequently,
Koopmans-type electron affinities are usually considerably less reliable
than the corresponding ionization potentials.

In particular, a neutral molecule may have a positive HF LUMO energy,
leading to a negative Koopmans electron affinity even when the exact anion
is bound.

\end{hfsolution}


\paragraph{The frozen-orbital approximation.}

The crucial assumption behind Eqs.~\eqref{eq:koopmans-ip} and
\eqref{eq:koopmans-ea} is not Hartree--Fock theory itself, but the decision
to retain the orbitals of the neutral $N$-electron system after changing
the electron number.
In reality, removing or adding an electron changes the effective potential
felt by all the remaining electrons. The orbitals therefore relax.

For ionization from orbital $k$, let
\begin{equation}
  {}^{N-1}\widetilde{E}_k
\end{equation}
denote the frozen-orbital cation energy and
\begin{equation}
  {}^{N-1}E_k^\text{HF}
\end{equation}
the energy obtained after a separate HF optimization of the cation.
All comparisons in this discussion use the same nuclear geometry and
one-electron space. Provided that the variational space for the ion contains
the frozen determinant and that its energy is minimized in that space,
optimization can only lower the energy:
\begin{equation}
  {}^{N-1}E_k^\text{HF}
  \leq
  {}^{N-1}\widetilde{E}_k
\end{equation}
It is therefore useful to define the ionization relaxation correction as
\begin{equation}
  \Delta_\text{relax}^\text{IP}
  =
  {}^{N-1}E_k^\text{HF}
  -
  {}^{N-1}\widetilde{E}_k
  \leq0
\end{equation}
Orbital relaxation thus tends to \emph{lower} the ionization potential
relative to the frozen-orbital Koopmans value.

For electron attachment, the relaxed anion satisfies the corresponding bound
under the same variational assumption:
\begin{equation}
  {}^{N+1}E_c^\text{HF}
  \leq
  {}^{N+1}\widetilde{E}^{c}
\end{equation}
However, because the electron affinity is defined with the opposite energy
difference,
\begin{equation}
  \text{EA}
  =
  {}^N E_0
  -
  {}^{N+1}E_c
\end{equation}
the corresponding relaxation correction is
\begin{equation}
  \Delta_\text{relax}^\text{EA}
  =
  {}^{N+1}\widetilde{E}^{c}
  -
  {}^{N+1}E_c^\text{HF}
  \geq0
\end{equation}
Relaxation therefore tends to \emph{increase} the electron affinity
relative to the frozen-orbital estimate.


\paragraph{Relaxation, correlation, and error compensation.}

Koopmans' theorem neglects not only orbital relaxation but also electron
correlation beyond Hartree--Fock theory.

In the following decomposition, the superscript $M$ denotes the electron
number, and $E_\text{corr}^{M}$ is the correlation energy of that system,
defined as the difference between its exact and optimized HF electronic
energies at the same nuclear geometry \citep{Lowdin1959}. Writing
\begin{equation}
  E_\text{exact}^{M}
  =
  E_\text{HF}^{M}
  +
  E_\text{corr}^{M}
\end{equation}
the exact ionization potential may be decomposed as
\begin{equation}
\begin{split}
  \text{IP}_k
  &=
  E_\text{exact}^{N-1}
  -
  E_\text{exact}^{N}
  \\
  &=
  E_\text{HF}^{N-1}
  -
  E_\text{HF}^{N}
  +
  E_\text{corr}^{N-1}
  -
  E_\text{corr}^{N}
  \\
  &=\boxed{
  -\epsilon_k
  +
  \underbrace{
    \Delta_\text{relax}^\text{IP}
  }_{\leq0}
  +
  \underbrace{
    \Delta_\text{corr}^\text{IP}
  }_{
    E_\text{corr}^{N-1}-E_\text{corr}^{N}
  }
  }
  \label{eq:koopmans-ip-corrections}
\end{split}
\end{equation}

For many ordinary molecular ionizations, the neutral system has a somewhat
larger magnitude of correlation energy than the corresponding cation, so
that
\begin{equation}
  \Delta_\text{corr}^\text{IP}>0
\end{equation}
In such cases, relaxation and correlation corrections have opposite signs
and partially compensate:
\begin{equation}
  \underbrace{\Delta_\text{relax}^\text{IP}}_{<0}
  \quad+\quad
  \underbrace{\Delta_\text{corr}^\text{IP}}_{>0}
  \quad\longrightarrow\quad
  \text{partial error compensation}
\end{equation}
This error compensation is one reason why Koopmans' theorem can yield
surprisingly reasonable ionization potentials despite neglecting both
effects.
The sign of the correlation correction is, however, \emph{not} guaranteed
by any general theorem. The statement above is a common chemical trend, not
a mathematical requirement.

For electron attachment,
\begin{equation}
\begin{split}
  \text{EA}_c
  &=
  E_\text{exact}^{N}
  -
  E_\text{exact}^{N+1}
  \\
  &=
  E_\text{HF}^{N}
  -
  E_\text{HF}^{N+1}
  +
  E_\text{corr}^{N}
  -
  E_\text{corr}^{N+1}
  \\
  &=\boxed{
  -\epsilon_c
  +
  \underbrace{
    \Delta_\text{relax}^\text{EA}
  }_{\geq0}
  +
  \underbrace{
    \Delta_\text{corr}^\text{EA}
  }_{
    E_\text{corr}^{N}-E_\text{corr}^{N+1}
  }
  }
  \label{eq:koopmans-ea-corrections}
\end{split}
\end{equation}
For many bound molecular anions, the $(N+1)$-electron system gains more
correlation energy than the neutral system, so that
\begin{equation}
  \Delta_\text{corr}^\text{EA}>0
\end{equation}
In this common situation, relaxation and correlation act in the same
direction rather than compensating one another:
\begin{equation}
  \underbrace{\Delta_\text{relax}^\text{EA}}_{>0}
  \quad+\quad
  \underbrace{\Delta_\text{corr}^\text{EA}}_{>0}
  \quad\longrightarrow\quad
  \text{no analogous error compensation}
\end{equation}

Again, the sign of the correlation contribution is system-dependent.
Nevertheless, the combination of stronger orbital-relaxation effects,
unoptimized virtual orbitals, and the frequent absence of favorable error
compensation explains why Koopmans' theorem is usually more reliable
for IPs than for EAs.


\subsubsection{The auxiliary Hartree--Fock Hamiltonian}

The interpretation of orbital energies becomes clearer if we distinguish
carefully between the physical electronic Hamiltonian and the auxiliary
independent-particle Hamiltonian associated with the converged Hartree--Fock
potential.

Define the $N$-electron Hartree--Fock Hamiltonian as
\begin{equation}
  \boxed{
  \underbrace{
    \chH^\text{HF}
  }_{\text{\red{$N$-electron auxiliary operator}}}
  =
  \sum_{i=1}^{N}
  \underbrace{
    \hf(\bx_i)
  }_{\text{one-electron Fock operator}}
  =
  \sum_{i=1}^{N}
  \qty[
    \hh(\bx_i)
    +
    \Hat{v}^\text{HF}(\bx_i)
  ]
  }
  \label{eq:hf-auxiliary-hamiltonian}
\end{equation}
Once the Hartree--Fock potential has been converged, $\chH^\text{HF}$ is a sum of
one-electron operators. The Slater determinant constructed from the
canonical occupied orbitals is therefore an eigenfunction of this
\emph{auxiliary} Hamiltonian:
\begin{equation}
  \chH^\text{HF}
  \ket{\Psi_\text{HF}}
  =
  \qty(
    \sum_{i=1}^{N}\epsilon_i
  )
  \ket{\Psi_\text{HF}}
\end{equation}

Three quantities must not be confused.
First,
\begin{align}
  \chH^\text{HF}
  \ket{\Psi_\text{HF}}
  &\green{=}
  \qty(
    \sum_{i=1}^{N}\epsilon_i
  )
  \ket{\Psi_\text{HF}}
  &
  \sum_i\epsilon_i
  &\red{\neq}
  E_\text{HF}
\end{align}
The Hartree--Fock determinant is an eigenfunction of the auxiliary Hartree--Fock Hamiltonian, but
its eigenvalue is the sum of the occupied orbital energies, not the Hartree--Fock energy.
Second,
\begin{equation}
  \chH_\text{elec}
  \ket{\Psi_\text{HF}}
  \red{\neq}
  E_\text{HF}
  \ket{\Psi_\text{HF}}
\end{equation}
The Hartree--Fock determinant is, in general, \red{not} an eigenfunction of the true
interacting electronic Hamiltonian.
Third,
\begin{equation}
  \mel{\Psi_\text{HF}}
      {\chH_\text{elec}}
      {\Psi_\text{HF}}
  \green{=}
  E_\text{HF}
\end{equation}
The Hartree--Fock energy is the \emph{expectation value} of the physical
electronic Hamiltonian over the optimized HF determinant.
Thus,
\begin{align}
    \chH^\text{HF}\Psi_\text{HF}
    &=
    \qty(\sum_i\epsilon_i)\Psi_\text{HF}
    &
    \mel{\Psi_\text{HF}}
        {\chH_\text{elec}}
        {\Psi_\text{HF}}
    &=
    E_\text{HF}
    &
    \chH_\text{elec}\Psi_\text{HF}
    &\neq
    E_\text{HF}\Psi_\text{HF}
\end{align}


\paragraph{Why the sum of orbital energies is not the HF energy.}

For occupied canonical HF orbitals,
\begin{equation}
  \epsilon_i
  =
  h_{ii}
  +
  \sum_j
  \qty(
    \cJ_{ij}-\cK_{ij}
  )
\end{equation}
Summing over all occupied orbitals gives
\begin{equation}
  \sum_{i=1}^{N}\epsilon_i
  =
  \sum_{i=1}^{N}h_{ii}
  +
  \sum_{ij}^{N}
  \qty(
    \cJ_{ij}-\cK_{ij}
  )
\end{equation}
By contrast,
\begin{equation}
  E_\text{HF}
  =
  \sum_{i=1}^{N}h_{ii}
  +
  \frac{1}{2}
  \sum_{ij}^{N}
  \qty(
    \cJ_{ij}-\cK_{ij}
  )
\end{equation}
Hence,
\begin{equation}
  \sum_{i=1}^{N}\epsilon_i
  \red{\neq}
  E_\text{HF}
\end{equation}

This recovers the double-counting correction in
Eq.~\eqref{eq:hf-energy-orbital-energies}: the orbital energy $\epsilon_i$
contains the interaction of electron $i$ with every other occupied electron,
so their sum counts each pair twice. Here the same correction distinguishes
the eigenvalue of the auxiliary Hamiltonian from the expectation value of the
physical Hamiltonian.
Equivalently,
\begin{equation}
  E_\text{HF}
  =
  \sum_i\epsilon_i
  -
  \frac{1}{2}
  \sum_{ij}
  \qty(
    \cJ_{ij}-\cK_{ij}
  )
  \label{eq:hf-energy-vs-orbital-sum}
\end{equation}


\subsubsection{Connection with many-body perturbation theory}
\label{sec:mp2-connection}

This distinction between the sum of the orbital energies and the HF energy
also provides the natural starting point for
\red{M{\o}ller--Plesset perturbation theory} \citep{MollerPlesset1934,Cremer2011}.

\hfBioMollerPlesset

The electronic Hamiltonian can be partitioned as
\begin{equation}
  \boxed{
  \chH_\text{elec}
  =
  \underbrace{
    \sum_{i=1}^{N}\hf(\bx_i)
  }_{\chH^{(0)}}
  +
  \underbrace{
    \qty[
      \sum_{i<j}\frac{1}{r_{ij}}
      -
      \sum_{i=1}^{N}
      \Hat{v}^\text{HF}(\bx_i)
    ]
  }_{\chV}
  }
  \label{eq:mp-partition}
\end{equation}
The zeroth-order Hamiltonian is therefore the sum of the Fock operators,
\begin{equation}
  \chH^{(0)}
  =
  \chH^\text{HF}
\end{equation}
and its eigenvalue for the HF reference determinant is
\begin{equation}
  E^{(0)}
  =
  \sum_{i=1}^{N}\epsilon_i
\end{equation}
For an Aufbau reference occupying the lowest spin orbitals, this is also the
ground-state eigenvalue of the auxiliary operator. For a non-Aufbau
reference, such as those targeted by the maximum-overlap method below, it
is the eigenvalue of the selected determinant instead.
The first-order correction is the expectation value of the perturbation over
the Hartree--Fock determinant:
\begin{equation}
\begin{split}
  E^{(1)}
  &=
  \mel{\Psi_\text{HF}}
      {\chV}
      {\Psi_\text{HF}}
  \\
  &=
  \mel{\Psi_\text{HF}}
  {
    \sum_{i<j}\frac{1}{r_{ij}}
    -
    \sum_{i=1}^{N}
    \Hat{v}^\text{HF}(\bx_i)
  }
  {\Psi_\text{HF}}
  \\
  &=
  -\frac{1}{2}
  \sum_{ij}^{N}
  \qty(
    \cJ_{ij}-\cK_{ij}
  )
\end{split}
\end{equation}
Consequently,
\begin{equation}
  E^{(0)}+E^{(1)}
  =
  E_\text{HF}
\end{equation}

Thus, in the conventional M{\o}ller--Plesset partitioning, the sum of the
occupied HF orbital energies is the zeroth-order energy, while the
first-order correction removes the double counting of the electron-electron
interaction. Correlation beyond Hartree--Fock first appears at second order,
leading to the MP2 correlation energy \citep{Cremer2011}.

\paragraph{The MP2 correlation energy.}

The HF determinant is stationary against single substitutions, but it still
couples to doubly substituted determinants through the two-electron
interaction. MP2 turns these couplings into the first correction beyond the
mean-field energy. For a nondegenerate canonical Hartree--Fock reference,
second-order perturbation theory gives
\begin{equation}
  E^{(2)}
  =\sum_{n\ne0}
    \frac{\abs*{\mel{\Psi_n}{\chV}{\Psi_0}}^2}
      {E^{(0)}-E_n^{(0)}}
\end{equation}
Here $\ket{\Psi_n}$ are excited determinants constructed from the canonical orbital
basis, eigenfunctions of $\chH^{(0)}$, and the reference is
$\ket{\Psi_0}=\ket{\Psi_\text{HF}}$. Brillouin's theorem eliminates
singles, and the two-electron character of $\chV$ eliminates couplings to
triples and higher substitutions. Therefore, only doubly-excited determinants
remain. For a double substitution, the one-electron Hartree--Fock potential cannot
connect the two determinants, so the remaining matrix element is the antisymmetrized
ERI. Consequently,
\begin{equation}
  \green{\Ec^{(2)}}
  =\sum_{i<j}^{\text{occ}}\sum_{a<b}^{\text{virt}}
    \frac{\abs{\mel{ij}{}{ab}}^2}
      {\epsilon_i+\epsilon_j-\epsilon_a-\epsilon_b}
  =\frac14\sum_{ij}^{\text{occ}}\sum_{ab}^{\text{virt}}
    \frac{\abs{\mel{ij}{}{ab}}^2}
      {\epsilon_i+\epsilon_j-\epsilon_a-\epsilon_b}
  \label{eq:mp2-spin-orbital}
\end{equation}
These indices label spin orbitals. The unrestricted sums count each pair of
occupied labels and each pair of virtual labels twice, explaining the factor
$1/4$; coincident labels contribute zero by antisymmetry. The absolute square
also makes this expression valid for complex canonical orbitals. The MP2
electronic energy is
\begin{equation}
  E_\text{MP2}=E_\text{HF}+\Ec^{(2)}
\end{equation}
For an Aufbau reference with a positive occupied-virtual gap, every
denominator in Eq.~\eqref{eq:mp2-spin-orbital} is negative, so the
second-order correction is nonpositive. Small excitation gaps, however,
make the correction sensitive to the denominators; the nondegenerate
perturbative construction should not be interpreted as a remedy for a
nearly degenerate HF reference.

For implementation, it is useful to specialize to a closed-shell RHF
reference with \emph{real} spatial orbitals.
Let $O=N/2$ and $V=K-O$ be the numbers of occupied and virtual spatial
orbitals. As usual, $i,j$ label the $O$ occupied
spatial orbitals, and $a,b$ the $V$ virtual spatial orbitals. Carrying out the spin
sum gives
\begin{equation}
\begin{split}
  \green{\Ec^{(2)}}
  &=\sum_{\green{ij}}^{O}\sum_{\red{ab}}^{V}
    \frac{\braket{\green{ij}}{\red{ab}}
      \qty(2\braket{\green{ij}}{\red{ab}}
        -\braket{\green{ij}}{\red{ba}})}
      {\epsilon_i+\epsilon_j-\epsilon_a-\epsilon_b}
  \\
  &=\underbrace{2\sum_{ij}^{O}\sum_{ab}^{V}
    \frac{\braket{ij}{ab}^2}
      {\epsilon_i+\epsilon_j-\epsilon_a-\epsilon_b}}
      _{\text{direct contribution}}
    -\underbrace{\sum_{ij}^{O}\sum_{ab}^{V}
    \frac{\braket{ij}{ab}\braket{ij}{ba}}
      {\epsilon_i+\epsilon_j-\epsilon_a-\epsilon_b}}
      _{\text{exchange contribution}}
  \label{eq:mp2-rhf}
\end{split}
\end{equation}
The factor of two in the direct term is the result of spin summation, not
an additional empirical weight. More explicitly, if the spatial direct
and exchanged integrals are denoted temporarily by $A=\braket{ij}{ab}$
and $B=\braket{ij}{ba}$, summing the squared spin-orbital integral over
its four spin labels gives $4A^2+4B^2-4AB$. Multiplication by $1/4$ and
interchange of the dummy virtual labels in the $B^2$ sum yield
$2A^2-AB$ after summation.

\paragraph{Why a four-loop energy expression leads to a fifth-order method.}

If the MO integrals are already available, evaluating
Eq.~\eqref{eq:mp2-rhf} requires only a loop over each occupied pair and
each virtual pair. One initializes the energy accumulator to zero, forms
the denominator for each quartet, and adds its direct-minus-exchange
contribution. With
\begin{equation}
  \purple{\Delta_{ij}^{ab}}
  =\epsilon_i+\epsilon_j-\epsilon_a-\epsilon_b
\end{equation}
the accumulation step is
\begin{equation}
  \Ec^{(2)}\longleftarrow\Ec^{(2)}
  +\frac{2\braket{ij}{ab}^2-\braket{ij}{ab}\braket{ij}{ba}}
    {\purple{\Delta_{ij}^{ab}}}
\end{equation}
There are $O^2V^2$ such quartets. Thus the final energy summation costs
$\order{O^2V^2}$, or $\order{K^4}$ when both orbital populations grow
proportionally to $K$. The accumulator itself requires only scalar storage.
This quartic summation does not explain the familiar fifth-order cost of
conventional MP2 because it assumes that the transformed integrals have
already been obtained. That preparation is the missing computational step.

The distinction between integral conventions matters here. The spatial
integrals required by the MP2 expression are
\begin{align}
  \braket{ij}{ab}&=\rbraket{ia}{jb}
  &
  \braket{ij}{ba}&=\rbraket{ib}{ja}
\end{align}
Thus MP2 needs the occupied-virtual pair block of the chemists' ERI tensor,
not the block $\rbraket{ij}{ab}$. This follows from the physicists'-to-chemists'
index conversion established earlier.

\paragraph{The naive AO-to-MO transformation.}

The integrals generated by a molecular program are initially expressed in
the AO basis. Although the finite-basis equations are developed in the next
section, the required change of representation can already be understood
from $\psi_p=\sum_\mu C_{\mu p}\phi_\mu$. Retaining the real-orbital
convention and using chemists' notation, substitution of the four orbital
expansions gives
\begin{equation}
  \underbrace{\rbraket{pq}{rs}}_{\text{MO integral}}
  =\sum_{\mu\nu\lambda\sigma}^{K}
    C_{\mu p}C_{\nu q}C_{\lambda r}C_{\sigma s}
    \underbrace{\rbraket{\mu\nu}{\lambda\sigma}}_{\text{AO integral}}
  \label{eq:ao-mo-four-index}
\end{equation}
Here $p,q,r,s$ are free MO labels, whereas the four Greek AO labels are
summed. For a full transformation there are $K^4$ output integrals, and
computing each independently requires $K^4$ AO terms. A literal algorithm
therefore loops over $p,q,r,s$, initializes that output entry to zero, and
accumulates it through four further loops over $\mu,\nu,\lambda,\sigma$.
The resulting eight nested loops cost $\order{K^8}$. This is the cost of
a poor contraction order, not an intrinsic cost of changing orbital basis.

\paragraph{Four successive one-index transformations.}

A semi-direct construction avoids repeating the same partial
sums by transforming one AO index at a time
\citep{FrischHeadGordonPople1990}. The four stages are
\begin{align}
  \violet{I_{\mu\nu\lambda s}}
    &=\sum_\sigma C_{\sigma s}\rbraket{\mu\nu}{\lambda\sigma}
    \label{eq:ao-mo-quarter-one}
  \\
  \orange{J_{\mu\nu rs}}
    &=\sum_\lambda C_{\lambda r}\violet{I_{\mu\nu\lambda s}}
    \label{eq:ao-mo-quarter-two}
  \\
  \blue{I_{\mu qrs}}
    &=\sum_\nu C_{\nu q}\orange{J_{\mu\nu rs}}
    \label{eq:ao-mo-quarter-three}
  \\
  \red{\rbraket{pq}{rs}}
    &=\sum_\mu C_{\mu p}\blue{I_{\mu qrs}}
    \label{eq:ao-mo-quarter-four}
\end{align}
The labels $I$ and $J$ denote temporary four-index arrays, not Coulomb
operators or identity matrices. An accumulation implementation allocates
and zeros $I$ before the first stage and $J$ before the second.
The first stage contracts $\sigma$ for
each fixed $(\mu,\nu,\lambda,s)$ and stores the result in $I$. It has
four output loops and one contraction loop, so costs $\order{K^5}$ and
requires $\order{K^4}$ storage. The second stage contracts $\lambda$ and
writes $J$; its cost and storage have the same orders.

Once $J$ is complete, the original contents of $I$ are no longer needed.
The third stage therefore reuses that buffer, now with index pattern
$(\mu,q,r,s)$, to contract $\nu$. Its entries must be reset before
accumulation; no new intermediate buffer is required. Finally, contraction
of $\mu$ writes the MO integral array. Both final stages again have five
loops and cost $\order{K^5}$. The output is also quartic in size, but it
does not require another kind of intermediate. The AO data, output tensor,
and temporary buffers together retain a quartic storage requirement.

Substitution of each intermediate into the following stage recovers
Eq.~\eqref{eq:ao-mo-four-index} exactly. The improvement is therefore
purely algebraic: four fifth-order contractions replace an eighth-order
one without approximating any integral. In this full-tensor construction,
\begin{align}
  \text{transformation work}&=\order{K^5}
  &
  \text{tensor storage}&=\order{K^4}
\end{align}
The transformed integrals can then be used in the quartic MP2 energy
summation. This explains the conventional fifth-order scaling. In an
actual MP2 calculation, only the $\rbraket{ia}{jb}$ block is required,
so the output can be restricted to $O^2V^2$ entries rather than retaining
every MO integral.

\section{Finite-basis Hartree--Fock and self-consistent-field methods}
\label{sec:finite-basis-hf}

The orbital equations derived above define the Hartree--Fock stationary
condition, but they are not yet in a form suited to routine molecular
calculations. This section first projects the continuous equations into a
finite nonorthogonal basis, then develops the iterative procedure required by
the density dependence of the Fock operator. The same matrix formulation also
provides the natural setting for convergence acceleration, state-specific SCF
solutions, and the evaluation and interpretation of molecular properties.

\subsection{The Roothaan--Hall equations}

Projection onto a finite nonorthogonal basis converts each orbital equation
into a generalized matrix eigenvalue problem. The construction proceeds from
the basis expansion and overlap metric to orthogonalization, Fock-matrix
assembly, and the finite-basis energy expression
\citep{Roothaan1951,Hall1951}.

\hfBioRoothaan
\hfBioHall

\subsubsection{The basis-set approximation}

The Hartree--Fock equations derived in the previous section are formally
equations for unknown functions,
\begin{equation}
  \hf(\br)\psi_i(\br)
  =
  \epsilon_i\psi_i(\br)
  \label{eq:hf-continuous}
\end{equation}
Solving these equations directly in continuous three-dimensional space is
possible in special cases, most notably for atoms, but is inconvenient for
general molecules. In practical molecular calculations, the spatial
orbitals are therefore expanded in a finite set of known one-electron basis
functions.
This introduces a second approximation, distinct from the Hartree--Fock
single-determinant approximation itself. Hartree--Fock restricts the
many-electron wavefunction to a single determinant, whereas the
basis-set approximation restricts the one-electron functions entering
that determinant to a finite-dimensional space.

\paragraph{Expansion of the molecular orbitals.}

Let
\begin{equation}
  \qty{
    \phi_\mu(\br)
  }_{\mu=1}^{K}
\end{equation}
be a set of $K$ basis functions. Each molecular orbital is expanded as
\begin{equation}
  \psi_i(\br)
  =
  \sum_{\mu=1}^{K}
  C_{\mu i}\phi_\mu(\br)
  \qquad
  \equiv
  \qquad
  \ket{i}
  =
  \sum_{\mu=1}^{K}
  C_{\mu i}\ket{\mu}
  \label{eq:lcao-expansion}
\end{equation}
The unknown functions $\psi_i(\br)$ have therefore been replaced by the
unknown expansion coefficients $C_{\mu i}$.
This is the linear combination of atomic orbitals (LCAO)
representation when the basis functions are centered on the nuclei, as is
usually the case in molecular electronic-structure calculations. More
generally, however, the derivation requires only that the functions span a
suitable finite one-electron space.
For a fixed basis, the Hartree--Fock problem is therefore no longer a
variational problem over arbitrary functions, but over the coefficients
contained in the matrix $\bC$.

\paragraph{Nonorthogonality of the basis functions.}

The molecular orbitals are required to be orthonormal,
\begin{equation}
  \braket{\psi_i}{\psi_j}
  =
  \delta_{ij}
\end{equation}
but the underlying basis functions generally are not:
\begin{equation}
  S_{\mu\nu}
  =
  \braket{\mu}{\nu}
  =
  \int
  \phi_\mu^*(\br)
  \phi_\nu(\br)
  \dd{\br}
  \label{eq:ao-overlap}
\end{equation}
The matrix $\bS$ is called the overlap matrix.
Substituting the basis expansion into the MO orthonormality condition gives
\begin{equation}
  \braket{\psi_i}{\psi_j}
  =
  \sum_{\mu\nu}
  C_{\mu i}^*
  C_{\nu j}
  \braket{\mu}{\nu}
  =
  \sum_{\mu\nu}
  C_{\mu i}^*
  S_{\mu\nu}
  C_{\nu j}
  =
  \delta_{ij}
\end{equation}
In matrix notation,
\begin{equation}
  \boxed{
  \bC^\dag\cdot\bS\cdot\bC
  =
  \bI
  }
  \label{eq:mo-orthonormality-matrix}
\end{equation}
Thus, the coefficient matrix is generally \emph{not} unitary:
\begin{equation}
  \bC^\dag\cdot\bC
  \neq
  \bI
\end{equation}
Instead, the molecular orbitals are orthonormal with respect to the
metric defined by $\bS$.

If the basis functions are linearly independent, $\bS$ is Hermitian and
positive definite. If two or more basis functions become nearly linearly
dependent, one or more eigenvalues of $\bS$ become very small, which can
lead to numerical difficulties. This issue becomes particularly relevant
for large and diffuse basis sets.


\paragraph{From the HF equation to the Roothaan--Hall equations.}

We now insert the finite-basis expansion into the canonical HF equation,
\begin{equation}
  \hf\ket{i}
  =
  \epsilon_i\ket{i}
\end{equation}
Using
\begin{equation}
  \ket{i}
  =
  \sum_\nu
  C_{\nu i}\ket{\nu}
\end{equation}
gives
\begin{equation}
  \hf
  \sum_\nu
  C_{\nu i}\ket{\nu}
  =
  \epsilon_i
  \sum_\nu
  C_{\nu i}\ket{\nu}
\end{equation}
Projecting from the left with $\bra{\mu}$ yields
\begin{equation}
\begin{split}
  \sum_\nu
  C_{\nu i}
  \mel{\mu}{\hf}{\nu}
  &=
  \epsilon_i
  \sum_\nu
  C_{\nu i}
  \braket{\mu}{\nu}
\end{split}
\end{equation}
Defining the Fock matrix
\begin{equation}
  F_{\mu\nu}
  =
  \mel{\mu}{\hf}{\nu}
  \label{eq:fock-matrix-definition}
\end{equation}
and recalling
\begin{equation}
  S_{\mu\nu}
  =
  \braket{\mu}{\nu}
\end{equation}
we obtain
\begin{equation}
  \sum_\nu
  F_{\mu\nu}
  C_{\nu i}
  =
  \sum_\nu
  S_{\mu\nu}
  C_{\nu i}
  \epsilon_i
  \label{eq:roothaan-hall-index}
\end{equation}

The complete derivation can therefore be summarized as
\begin{equation}
\begin{split}
  \hf\ket{i}
  =
  \epsilon_i\ket{i}
  &\Rightarrow\quad
  \hf
  \sum_\nu
  C_{\nu i}\ket{\nu}
  =
  \epsilon_i
  \sum_\nu
  C_{\nu i}\ket{\nu}
  \\
  &\xRightarrow{\ \bra{\mu}\ }\quad
  \sum_\nu
  C_{\nu i}
  \underbrace{
    \mel{\mu}{\hf}{\nu}
  }_{F_{\mu\nu}}
  =
  \epsilon_i
  \sum_\nu
  C_{\nu i}
  \underbrace{
    \braket{\mu}{\nu}
  }_{S_{\mu\nu}}
  \\
  &\Rightarrow\quad
  \sum_\nu
  F_{\mu\nu}C_{\nu i}
  =
  \sum_\nu
  S_{\mu\nu}C_{\nu i}\epsilon_i
\end{split}
\end{equation}

Collecting all molecular orbitals simultaneously gives
\begin{equation}
  \boxed{
  \bF\cdot\bC
  =
  \bS\cdot\bC\cdot\beps
  }
  \label{eq:roothaan-hall}
\end{equation}
These are the Roothaan--Hall equations.
Here,
\begin{equation}
  \bC
  =
  \pmqty{
    C_{11}&C_{12}&\cdots&C_{1K}\\
    C_{21}&C_{22}&\cdots&C_{2K}\\
    \vdots&\vdots&\ddots&\vdots\\
    C_{K1}&C_{K2}&\cdots&C_{KK}
  }
\end{equation}
contains the MO coefficients, while
\begin{equation}
  \beps
  =
  \pmqty{
    \epsilon_1&0&\cdots&0\\
    0&\epsilon_2&\cdots&0\\
    \vdots&\vdots&\ddots&\vdots\\
    0&0&\cdots&\epsilon_K
  }
\end{equation}
contains the canonical orbital energies.

Equation~\eqref{eq:roothaan-hall} is the finite-dimensional counterpart of
the canonical HF equation. The basis-set expansion has converted the
integro-differential orbital problem into a matrix problem.


\paragraph{A generalized eigenvalue problem.}

If the basis were orthonormal, i.e., $\bS=\bI$,
the Roothaan--Hall equations would reduce to the ordinary eigenvalue problem
\begin{equation}
  \bF\cdot\bC
  =
  \bC\cdot\beps
\end{equation}
For an atomic-orbital basis, however, $\bS\neq\bI$,
and the equations instead constitute a
generalized eigenvalue problem,
\begin{equation}
  \bF\cdot\bC
  =
  \bS\cdot\bC\cdot\beps
\end{equation}

For an individual orbital,
\begin{equation}
  \bF\cdot\bC_i
  =
  \epsilon_i\bS\cdot\bC_i
\end{equation}
where $\bC_i$ is the $i$th column of $\bC$. Nontrivial solutions satisfy
the secular equation
\begin{equation}
  \det\qty(
    \bF-\epsilon_i\bS
  )
  =
  0
  \label{eq:secular-equation}
\end{equation}

The presence of $\bS$ is not a peculiarity of Hartree--Fock theory. It
simply reflects the fact that the coordinates used to represent the
one-electron functions form a nonorthogonal basis.


\subsubsection{Orthogonalization and the generalized eigenvalue problem}
\label{sec:ao-orthogonalization}

A generalized Hermitian eigenvalue problem can be converted into an
ordinary Hermitian eigenvalue problem by transforming to an orthonormal
basis.

We seek a nonsingular matrix $\bX$ satisfying
\begin{equation}
  \boxed{
  \bX^\dag\cdot\bS\cdot\bX
  =
  \bI
  }
  \label{eq:orthogonalizer-condition}
\end{equation}
The transformed basis functions may be viewed as
\begin{equation}
  \ket{\mu'}
  =
  \sum_\nu
  \ket{\nu}X_{\nu\mu}
\end{equation}
Their overlaps are
\begin{equation}
  \braket{\mu'}{\nu'}
  =
  \qty(
    \bX^\dag\cdot\bS\cdot\bX
  )_{\mu\nu}
  =
  \delta_{\mu\nu}
\end{equation}
so the transformed basis is orthonormal by construction.

Writing
\begin{equation}
  \bC
  =
  \bX\cdot\bC'
  \label{eq:coefficient-transformation}
\end{equation}
and substituting into the Roothaan--Hall equations gives
\begin{equation}
\begin{split}
  \bF\cdot\bC
  &=
  \bS\cdot\bC\cdot\beps
  \\
  \bF\cdot\bX\cdot\bC'
  &=
  \bS\cdot\bX\cdot\bC'\cdot\beps
\end{split}
\end{equation}
Multiplying from the left by $\bX^\dag$ gives
\begin{equation}
  \bX^\dag\cdot\bF\cdot\bX\cdot\bC'
  =
  \bX^\dag\cdot\bS\cdot\bX\cdot\bC'\cdot\beps
\end{equation}
Using Eq.~\eqref{eq:orthogonalizer-condition},
\begin{equation}
\begin{split}
  \bF'\cdot\bC'
  &=
  \bC'\cdot\beps
  \\
  \bF'
  &=
  \bX^\dag\cdot\bF\cdot\bX
\end{split}
  \label{eq:orthogonalized-fock}
\end{equation}

Thus,
\begin{align}
  \bF\cdot\bC
  &=
  \underbrace{\bS}_{\text{metric}}
  \cdot\bC\cdot\beps
  &
  \bF'\cdot\bC'
  &=
  \bC'\cdot\beps
  \\
  \bF'
  &=
  \bX^\dag\cdot\bF\cdot\bX
  &
  \bC
  &=
  \bX\cdot\bC'
  &
  \bX^\dag\cdot\bS\cdot\bX
  &=
  \bI
\end{align}

Every step from the generalized problem to the standard problem is
\begin{gather}
  \bF\cdot\bC
  =
  \underbrace{\bS}_{\text{metric}}
  \cdot\bC\cdot\beps
  \qquad\Longleftrightarrow\qquad
  \text{generalized eigenvalue problem}
  \\
  \Downarrow
  \\
  \bX^\dag\cdot\bF\cdot
  \underbrace{\bC}_{\bX\cdot\bC'}
  =
  \bX^\dag\cdot\bS\cdot
  \underbrace{\bC}_{\bX\cdot\bC'}\cdot\beps
  \\
  \Downarrow
  \\
  \underbrace{
    \bX^\dag\cdot\bF\cdot\bX
  }_{\bF'}
  \cdot\bC'
  =
  \underbrace{
    \bX^\dag\cdot\bS\cdot\bX
  }_{\bI}
  \cdot\bC'\cdot\beps
  \\
  \Downarrow
  \\
  \bF'\cdot\bC'
  =
  \bC'\cdot\beps
  \qquad\Longleftrightarrow\qquad
  \text{standard eigenvalue problem}
\end{gather}
Thus, orthogonalization converts the generalized problem into a standard
eigenvalue problem.

Because $\bF$ is Hermitian and $\bX$ is nonsingular,
\begin{equation}
  \bF'^\dag
  =
  \qty(
    \bX^\dag\cdot\bF\cdot\bX
  )^\dag
  =
  \bX^\dag\cdot\bF\cdot\bX
  =
  \bF'
\end{equation}
so $\bF'$ is also Hermitian. It may therefore be diagonalized by a unitary
transformation, and its eigenvalues are real.


\paragraph{Choice of the orthogonalization matrix.}

The matrix $\bX$ is not unique. Since the overlap matrix is Hermitian and
positive definite, it may be diagonalized as
\begin{equation}
  \bS
  =
  \bU\cdot\bs\cdot\bU^\dag
  \label{eq:overlap-diagonalization}
\end{equation}
where
\begin{align}
  \bs
  &=
  \operatorname{diag}(s_1,s_2,\ldots,s_K)
  &s_p&>0
\end{align}

A particularly common choice is the
\red{symmetric, or L\"owdin, orthogonalizer} \citep{Lowdin1950},
\begin{equation}
  \boxed{
  \bX
  =
  \bS^{-1/2}
  =
  \bU\cdot\bs^{-1/2}\cdot\bU^\dag
  }
  \label{eq:lowdin-orthogonalizer}
\end{equation}
Indeed,
\begin{equation}
  \bX^\dag\cdot\bS\cdot\bX
  =
  \bS^{-1/2}\cdot
  \bS\cdot
  \bS^{-1/2}
  =
  \bI
\end{equation}

Another frequently used choice is the canonical orthogonalizer \citep{Lowdin1970},
\begin{equation}
  \bX
  =
  \bU\cdot\bs^{-1/2}
\end{equation}
Both choices span the same one-electron space and produce the same
generalized eigenvalues. They simply correspond to different orthonormal
representations of that space.

\hfBioLowdin

If some eigenvalues $s_p$ of the overlap matrix are extremely small, the
factor $s_p^{-1/2}$ becomes very large. In practical implementations,
directions associated with sufficiently small overlap eigenvalues may
therefore be removed to avoid numerical instabilities caused by near-linear
dependencies.


\paragraph{Recovering the molecular-orbital coefficients.}

After diagonalizing the transformed Fock matrix,
\begin{equation}
  \bF'\cdot\bC'
  =
  \bC'\cdot\beps
\end{equation}
the coefficients in the original nonorthogonal basis are recovered through
\begin{equation}
  \bC
  =
  \bX\cdot\bC'
\end{equation}

Since the eigenvectors of the Hermitian matrix $\bF'$ may be chosen
orthonormal,
\begin{equation}
  \bC'^\dag\cdot\bC'
  =
  \bI
\end{equation}
we recover the correct generalized normalization,
\begin{equation}
\begin{split}
  \bC^\dag\cdot\bS\cdot\bC
  &=
  \bC'^\dag\cdot
  \bX^\dag\cdot\bS\cdot\bX\cdot
  \bC'
  \\
  &=
  \bC'^\dag\cdot\bC'
  \\
  &=
  \bI
\end{split}
\end{equation}

Thus the orthogonalization procedure does not alter the physical molecular
orbitals; it merely provides a convenient orthonormal coordinate system in
which the generalized eigenvalue problem can be solved using standard
Hermitian eigensolvers.


\paragraph{The matrix problem is still self-consistent.}

At first sight,
\begin{equation}
  \bF\cdot\bC
  =
  \bS\cdot\bC\cdot\beps
\end{equation}
looks like an ordinary generalized eigenvalue problem. There is, however,
one essential complication: the Fock matrix depends on the occupied
molecular orbitals and therefore on the coefficient matrix itself,
\begin{equation}
  \bF
  =
  \bF(\bC)
\end{equation}
The actual Roothaan--Hall problem is therefore
\begin{equation}
  \bF(\bC)\cdot\bC
  =
  \bS\cdot\bC\cdot\beps
  \label{eq:nonlinear-roothaan-hall}
\end{equation}
It is a nonlinear generalized eigenvalue problem, not a single matrix
diagonalization.

Schematically,
\begin{equation}
  \bC
  \longrightarrow
  \bP
  \longrightarrow
  \bF[\bP]
  \longrightarrow
  \bC
\end{equation}
where $\bP$ is the one-particle density matrix constructed from the occupied
molecular orbitals \citep{McWeeny1960}.

\hfBioMcWeeny

This circular dependence is the finite-basis manifestation of the
self-consistency already encountered in the continuous HF equations:
How do we solve these HF equations?
The answer is the self-consistent-field procedure developed below.


\subsubsection{The variational meaning of the finite basis}

Restricting the orbitals to a finite basis reduces the variational freedom
available to the Hartree--Fock determinant. Consequently, for a given
Hamiltonian,
\begin{hftheorem}{Variational bounds and the Hartree--Fock limit}
\begin{equation}
  E_\text{HF}^{(K)}
  \geq
  E_\text{HF}^{\text{CBS}}
  \geq
  E_\text{exact}
  \label{eq:hf-basis-variational-hierarchy}
\end{equation}
Here, $E_\text{HF}^{(K)}$ is the minimum HF energy in a finite
$K$-dimensional one-electron space, whereas
$E_\text{HF}^{\text{CBS}}$ denotes the complete-basis-set (CBS) HF limit.
The bounds concern the global minima within the same chosen determinant
class, not arbitrary stationary SCF solutions.
\end{hftheorem}

The first inequality reflects basis-set incompleteness: a finite basis
cannot describe all possible orbital shapes. The second reflects the
single-determinant approximation: even with a complete one-electron
basis, Hartree--Fock theory explores only the manifold of Slater
determinants, whereas the exact wavefunction belongs to the much larger
space of antisymmetric many-electron functions.

For a nested sequence of basis spaces,
\begin{equation}
  \mathcal{B}_1
  \subset
  \mathcal{B}_2
  \subset
  \cdots
\end{equation}
the variational principle implies
\begin{equation}
  E_\text{HF}^{(1)}
  \geq
  E_\text{HF}^{(2)}
  \geq
  \cdots
  \geq
  E_\text{HF}^{\text{CBS}}
\end{equation}
Enlarging the variational space can lower, or leave unchanged, the optimized
HF energy, but cannot raise it.

It is useful to keep these two approximations conceptually separate:
\begin{equation}
  \boxed{\text{finite basis}
  \quad+\quad
  \text{single determinant}
  \quad\Longrightarrow\quad
  \text{practical finite-basis Hartree--Fock}}
\end{equation}
The following sections examine first how the Fock matrix is constructed in
this finite basis and then how the resulting nonlinear Roothaan--Hall
equations are solved self-consistently.

\subsubsection{The Fock matrix}

The Roothaan--Hall equations become operational only after the matrix elements
of the Fock operator have been expressed in the atomic-orbital basis. This is
the point at which the one-electron Hamiltonian, the electron--electron
interaction, and the occupied molecular orbitals are brought together in a
single matrix quantity.

\paragraph{From the orbital operator to AO matrix elements.}

The spin-orbital Fock operator is a one-electron operator even though it
contains the mean effect of the two-electron interaction. Its matrix element
between two AO basis functions therefore separates naturally into a fixed
one-electron contribution and a density-dependent contribution,
\begin{equation}
  F_{\mu\nu}
  =
  H_{\mu\nu}
  +
  G_{\mu\nu}[\bP]
\end{equation}
The following worked example makes the contraction hidden in
$G_{\mu\nu}[\bP]$ explicit.
It first uses the general spin-orbital formulation. In this example and the
general AO-energy exercise below, $\chi_\mu^\text{AO}(\bx)$ denotes a
basis function formed from a spatial AO and a spin function. The Greek
basis label temporarily collects both labels and runs over $2K$ spin-AO
functions; the coefficients $C_{\mu i}$ refer to this spin-AO expansion.
The closed-shell specialization subsequently returns to the $K$ spatial
AO functions $\phi_\mu(\br)$ and a spin-summed density. Thus $\phi$
continues to denote a spatial basis function and $\chi$ a spin orbital.

\begin{hfexercise}{AO Fock matrix}

\violet{\textit{Express the Fock matrix in the spin-AO basis in terms of
the one- and two-electron integrals and the occupied-orbital density matrix.}}

\end{hfexercise}
\begin{hfsolution}{AO Fock matrix}

Use the spin-AO expansion defined above, with one occupied spin orbital per
electron and no closed-shell factor of two. Allow complex coefficients,
so the coefficient from the bra orbital must be conjugated.
Expanding the occupied spin orbitals in the basis exposes how their
coefficients enter the AO Fock matrix:
\begin{equation}
\begin{split}
  F_{\mu\nu}
  &=\mel{\mu}{\hh+\sum_{i=1}^{N}(\chJ_i-\chK_i)}{\nu}
  \\
  &=\underbrace{\mel{\mu}{\hh}{\nu}}_{\red{H_{\mu\nu}}}
  +\sum_{i=1}^{N}\mel{\mu}{\chJ_i-\chK_i}{\nu}
  \\
  &=H_{\mu\nu}+\sum_{i=1}^{N}\qty(
    \mel{\mu\chi_i}{r_{12}^{-1}}{\nu\chi_i}
    -\mel{\mu\chi_i}{r_{12}^{-1}}{\chi_i\nu}
  )
  \\
  &=H_{\mu\nu}
  +\sum_{i=1}^{N}\sum_{\lambda\sigma}
    C_{\lambda i}C_{\sigma i}^*
    \qty(
      \mel{\mu\sigma}{r_{12}^{-1}}{\nu\lambda}
      -\mel{\mu\sigma}{r_{12}^{-1}}{\lambda\nu}
    )
  \\
  &=H_{\mu\nu}
  +\sum_{\lambda\sigma}\red{P_{\lambda\sigma}}
    \qty(\braket{\mu\sigma}{\nu\lambda}
    -\braket{\mu\sigma}{\lambda\nu})
  \\
  &=H_{\mu\nu}
  +\underbrace{\sum_{\lambda\sigma}\red{P_{\lambda\sigma}}
    \mel{\mu\sigma}{}{\nu\lambda}}_{\red{G_{\mu\nu}}}
  =H_{\mu\nu}+G_{\mu\nu}
\end{split}
\end{equation}

The fourth line follows by expanding both occurrences of the occupied orbital
$\chi_i$,
\begin{equation}
\begin{split}
  \chi_i
  &=
  \sum_\lambda C_{\lambda i}\chi_\lambda^\text{AO}
  \\
  \chi_i^*
  &=
  \sum_\sigma C_{\sigma i}^*\chi_\sigma^{\text{AO}*}
\end{split}
\end{equation}
The sum over occupied orbitals can then be performed before the integral
contraction:
\begin{equation}
  \sum_{i=1}^{N}C_{\lambda i}C_{\sigma i}^*
  =
  P_{\lambda\sigma}
\end{equation}
Thus the occupied orbitals enter the Fock matrix only through their
one-particle density matrix.
In particular, $P_{\lambda\sigma}$ contains the ket coefficient
$C_{\lambda i}$ and the conjugated bra coefficient $C_{\sigma i}^*$.
This ordering fixes the integral indices even when the coefficients are
complex. No overlap matrix is inserted in this contraction: the AO
integrals already contain the basis functions, while their overlap metric
enters orbital normalization and the generalized eigenvalue problem.

The three quantities introduced in this derivation are the core Hamiltonian,
the density matrix, and the two-electron part:
\begin{equation}
\begin{split}
  H_{\mu\nu}&=\mel{\mu}{\hh}{\nu}
  \\
  P_{\lambda\sigma}&=\sum_{i=1}^{N}C_{\lambda i}C_{\sigma i}^*
  \\
  G_{\mu\nu}&=\sum_{\lambda\sigma}P_{\lambda\sigma}
  \mel{\mu\sigma}{}{\nu\lambda}
\end{split}
\end{equation}
The density matrix $\bP$ thus collects all occupied-orbital coefficients
needed for the two-electron contraction. Its explicit form also makes the
nonlinearity of the Roothaan--Hall equations apparent.

The separation
\begin{equation}
  \bF=\bH+\bG[\bP]
\end{equation}
is worth emphasizing. The core Hamiltonian $\bH$ is fixed once the molecular
geometry and basis set have been chosen. By contrast, $\bG$ must be rebuilt
whenever the density changes. Thus, although each diagonalization is an
ordinary linear-algebra operation, the complete Hartree--Fock problem remains
self-consistent.

\end{hfsolution}
\paragraph{Interpretation of the AO contraction.}

For every pair $(\mu,\nu)$, the two-electron part is obtained by contracting
the four-index integral tensor with the two-index density matrix. The density
selects the occupied part of the one-electron space, while the integral tensor
supplies the Coulomb interaction and its exchange counterpart. Schematically,
\begin{equation}
  \boxed{\text{density}
  \quad+\quad
  \text{two-electron integrals}
  \quad\longrightarrow\quad
  \text{mean-field potential}}
\end{equation}
This contraction is the central repeated operation in a conventional SCF
calculation.

\subsubsection{The Hartree--Fock energy}

The matrix form of the energy is obtained from the same ingredients. It is
particularly useful in an implementation because it allows the energy to be
evaluated directly from the density, core Hamiltonian, and the Fock matrix
constructed from that density, without transforming all integrals to the
molecular-orbital basis.

\begin{hfexercise}{AO Hartree--Fock energy}

\violet{\textit{Express the electronic HF energy in the spin-AO basis
using the density matrix, and derive its trace form in terms of $\bH$ and
$\bF$.}}

\end{hfexercise}
\begin{hfsolution}{AO Hartree--Fock energy}

Retain the spin-AO convention of the preceding exercise: $\bP$ counts
each occupied spin orbital once, and the coefficients need not be real.
The energy below excludes nuclear repulsion.
Starting from the spin-orbital energy and explicitly substituting each basis
expansion gives
\begin{equation}
\begin{split}
  E_\text{HF}
  &=\sum_{i=1}^{N}h_{ii}
    +\frac{1}{2}\sum_{ij}^{N}\qty(\cJ_{ij}-\cK_{ij})
  \\
  &=\sum_{i=1}^{N}
    \mel{\sum_\mu C_{\mu i}\chi_\mu^\text{AO}}{\hh}
        {\sum_\nu C_{\nu i}\chi_\nu^\text{AO}}
    \\
  &+\frac{1}{2}\sum_{ij}^{N}
    \mel{
      \qty(\sum_\mu C_{\mu i}\chi_\mu^\text{AO})
      \qty(\sum_\lambda C_{\lambda j}\chi_\lambda^\text{AO})
    }{}{
      \qty(\sum_\nu C_{\nu i}\chi_\nu^\text{AO})
      \qty(\sum_\sigma C_{\sigma j}\chi_\sigma^\text{AO})
    }
  \\
  &=\sum_{\mu\nu}P_{\nu\mu}
    \qty(H_{\mu\nu}
    +\frac{1}{2}\sum_{\lambda\sigma}P_{\sigma\lambda}
      \mel{\mu\lambda}{}{\nu\sigma})
  \\
  &=\frac{1}{2}\sum_{\mu\nu}P_{\nu\mu}
    \qty(H_{\mu\nu}+F_{\mu\nu})
  \\
  &=\boxed{\frac{1}{2}\Tr\qty[\bP\cdot(\bH+\bF)]
  }
  \label{eq:hf-energy-matrix}
\end{split}
\end{equation}

The reduction from orbital coefficients to density matrices can be seen
separately for the one- and two-electron terms. For the first,
\begin{equation}
\begin{split}
  \sum_i h_{ii}
  &=\sum_{\mu\nu}\qty(\sum_i C_{\mu i}^*C_{\nu i})H_{\mu\nu}\\
  &=\sum_{\mu\nu}P_{\nu\mu}H_{\mu\nu}
\end{split}
\end{equation}
For the second, the two occupied-orbital sums are independent:
\begin{equation}
\begin{split}
  \sum_{ij}C_{\mu i}^*C_{\lambda j}^*C_{\nu i}C_{\sigma j}
  &=\qty(\sum_i C_{\nu i}C_{\mu i}^*)
    \qty(\sum_j C_{\sigma j}C_{\lambda j}^*)\\
  &=P_{\nu\mu}P_{\sigma\lambda}
\end{split}
\end{equation}
Consequently, the two-electron energy is
\begin{equation}
\begin{split}
  \frac12\sum_{ij}\qty(\cJ_{ij}-\cK_{ij})
  &=\frac12\sum_{\mu\nu\lambda\sigma}
    P_{\nu\mu}P_{\sigma\lambda}\mel{\mu\lambda}{}{\nu\sigma}\\
  &=\frac12\sum_{\mu\nu}P_{\nu\mu}G_{\mu\nu}
\end{split}
\end{equation}
The prefactor removes the double counting of electron pairs; it does not
halve the one-electron energy. These contractions also explain why the
index order must be retained rather than silently replacing a Hermitian
density by a real symmetric one.
To see the origin of the trace expression, define the two-electron matrix
$\bG=\bF-\bH$. The preceding result may then be written as
\begin{equation}
\begin{split}
  E_\text{HF}
  &=
  \sum_{\mu\nu}P_{\nu\mu}H_{\mu\nu}
  +
  \frac12\sum_{\mu\nu}P_{\nu\mu}G_{\mu\nu}
  \\
  &=
  \frac12\sum_{\mu\nu}P_{\nu\mu}
  \qty(H_{\mu\nu}+H_{\mu\nu}+G_{\mu\nu})
  \\
  &=
  \frac12\sum_{\mu\nu}P_{\nu\mu}
  \qty(H_{\mu\nu}+F_{\mu\nu})
\end{split}
\end{equation}
which is precisely Eq.~\eqref{eq:hf-energy-matrix}. The ordering of the
indices $P_{\nu\mu}$ ensures that the component sum is the matrix trace.
This identity applies in a nonorthogonal basis as well: the overlap metric
is already accounted for in the occupied-orbital normalization, and the
energy is a contraction of the AO density coefficients with the AO
Hamiltonian matrix elements. In an SCF implementation, $\bF$ must be built
from the same $\bP$ used in the energy trace. The spin-summed closed-shell
specialization follows below.

\end{hfsolution}
\paragraph{Closed-shell specialization.}

We now carry out the spin sum and return to spatial AO indices. For a
closed-shell system, the density matrix is
\begin{equation}
\begin{split}
  P_{\red{\mu\nu}}
  &=2\sum_{i=1}^{N/2}C_{\red{\mu}i}C_{\red{\nu}i}^*
  \\
  \bP&=\boxed{2\bC_\text{o}\cdot\bC_\text{o}^\dag}
\end{split}
\end{equation}
where $\bC_\text{o}$ contains the occupied MO coefficients. In chemists'
notation, the closed-shell AO Fock matrix becomes
\begin{equation}
  \boxed{
  F_{\red{\mu\nu}}=H_{\red{\mu\nu}}
  +\underbrace{\sum_{\blue{\lambda\sigma}}
    P_{\blue{\sigma\lambda}}
    \rbraket{\red{\mu\nu}}{\blue{\lambda\sigma}}}
    _{J_{\red{\mu\nu}}=\text{ Coulomb}}
  -\underbrace{\frac{1}{2}\sum_{\blue{\lambda\sigma}}
    P_{\blue{\sigma\lambda}}
    \rbraket{\red{\mu}\blue{\sigma}}{\blue{\lambda}\red{\nu}}}
    _{K_{\red{\mu\nu}}=\text{ exchange}}
  }
  \label{eq:rhf-fock-ao}
\end{equation}
The factor of two in $\bP$ accounts for the two electrons occupying each
spatial orbital: $\bP$ here is the spin-summed density, whereas a density for
one spin component counts each occupied orbital once. Correspondingly,
the exchange term in
Eq.~\eqref{eq:rhf-fock-ao} carries a factor of one half when the
spin-summed closed-shell density is used.
The corresponding energy is
\begin{equation}
  E_\text{HF}=\sum_{\red{\mu\nu}}
    P_{\red{\nu\mu}}H_{\red{\mu\nu}}
  +\frac{1}{2}\sum_{\red{\mu\nu}\blue{\lambda\sigma}}
    P_{\red{\nu\mu}}
    \qty[
      \rbraket{\red{\mu\nu}}{\blue{\lambda\sigma}}
      -\frac{1}{2}
       \rbraket{\red{\mu}\blue{\sigma}}{\red{\lambda}\blue{\nu}}
    ]
    P_{\blue{\sigma\lambda}}
\end{equation}

The factor of one half multiplying the two-electron contribution removes the
double counting that would otherwise arise when the interaction of each
electron pair is summed from the one-electron Fock operators. Equivalently,
the compact trace formula contains one full copy of the one-electron energy
and one half of the mean-field electron-electron energy.

Exercise~\ref{ex:ao-density-metric} makes explicit how the overlap metric enters the
normalization and projector properties of the density matrix. These
identities also provide useful checks on an implementation.

\begin{hfexercise}[phantom={\label{ex:ao-density-metric}}]{Density matrices in a nonorthogonal basis}
\violet{\textit{Derive the electron-counting and idempotency relations for
RHF and UHF density matrices in a nonorthogonal AO basis.}}
\end{hfexercise}
\begin{hfsolution}{Density matrices in a nonorthogonal basis}
Consider $K$ linearly independent spatial AOs with overlap matrix $\bS$.
For a collinear determinant, let $\bC_\text{o}^\sigma$ contain the
$N^\sigma$ occupied spatial-orbital coefficients of spin
$\sigma=\alpha,\beta$, with
\begin{equation}
  \qty(\bC_\text{o}^\sigma)^\dag\cdot\bS\cdot\bC_\text{o}^\sigma=\bI
\end{equation}
Define the single-spin density by
\begin{equation}
  \bP^\sigma=\bC_\text{o}^\sigma\cdot
  \qty(\bC_\text{o}^\sigma)^\dag
\end{equation}
Taking the Hermitian adjoint reverses the order of the two coefficient
matrices and conjugates each factor. Thus
\begin{equation}
  \qty(\bP^\sigma)^\dag
  =\qty[\bC_\text{o}^\sigma\cdot
    \qty(\bC_\text{o}^\sigma)^\dag]^\dag
  =\bC_\text{o}^\sigma\cdot
    \qty(\bC_\text{o}^\sigma)^\dag
  =\bP^\sigma
\end{equation}
Hermiticity therefore does not require the AOs to be orthogonal or the
coefficients to be real.

The spin density in real space is
\begin{equation}
  \rho^\sigma(\br)
  =\sum_{i=1}^{N^\sigma}\abs{\psi_i^\sigma(\br)}^2
  =\sum_{\mu\nu}P_{\mu\nu}^\sigma
    \phi_\mu(\br)\phi_\nu^*(\br)
\end{equation}
Integrating the AO products gives $S_{\nu\mu}$, rather than a Kronecker
delta. Consequently,
\begin{equation}
\begin{split}
  \int\rho^\sigma(\br)\dd\br
  &=\sum_{\mu\nu}P_{\mu\nu}^\sigma S_{\nu\mu}\\
  &=\Tr(\bP^\sigma\cdot\bS)\\
  &=\Tr\qty[\qty(\bC_\text{o}^\sigma)^\dag
    \cdot\bS\cdot\bC_\text{o}^\sigma]\\
  &=\boxed{N^\sigma}
\end{split}
\end{equation}
The cyclic trace identity moves the rectangular coefficient matrices into
the occupied space, where the identity matrix has dimension $N^\sigma$.
The unweighted trace $\Tr\bP^\sigma$ is not generally an electron count.

The same occupied-orbital orthonormality gives
\begin{equation}
\begin{split}
  \bP^\sigma\cdot\bS\cdot\bP^\sigma
  &=\bC_\text{o}^\sigma\cdot
    \underbrace{\qty(\bC_\text{o}^\sigma)^\dag\cdot\bS
      \cdot\bC_\text{o}^\sigma}_{\bI}
    \cdot\qty(\bC_\text{o}^\sigma)^\dag\\
  &=\boxed{\bP^\sigma}
\end{split}
\end{equation}
This is metric idempotency: the overlap matrix is required between the two
density factors.

For RHF, the spin-summed matrix is Hermitian because it is twice the
Hermitian single-spin density. Its electron count is
\begin{equation}
  \Tr(\bP\cdot\bS)
  =2\Tr(\bC_\text{o}^\dag\cdot\bS\cdot\bC_\text{o})
  =2\frac{N}{2}=N
\end{equation}
Its metric product instead contains two factors of two:
\begin{equation}
\begin{split}
  \bP\cdot\bS\cdot\bP
  &=4\bC_\text{o}\cdot
    \underbrace{\bC_\text{o}^\dag\cdot\bS\cdot\bC_\text{o}}_{\bI}
    \cdot\bC_\text{o}^\dag\\
  &=4\bC_\text{o}\cdot\bC_\text{o}^\dag\\
  &=\boxed{2\bP}
\end{split}
\end{equation}
Even in an orthonormal AO basis, the spin-summed RHF matrix therefore
satisfies $\bP^2=2\bP$, not $\bP^2=\bP$. It is the density divided by two
that represents a projector in that orthonormal basis.

To see the role of the metric directly, apply the one-spin occupied-space
projector to an arbitrary spatial orbital
$\psi_v=\sum_\nu v_\nu\phi_\nu$. Its projection is
\begin{equation}
\begin{split}
  \sum_{i=1}^{N^\sigma}\psi_i^\sigma(\br)
    \braket{\psi_i^\sigma}{\psi_v}
  &=\sum_{\mu\nu\lambda}\phi_\mu(\br)
    P_{\mu\lambda}^\sigma S_{\lambda\nu}v_\nu
\end{split}
\end{equation}
Thus its action on AO coefficient vectors is represented by
$\bP^\sigma\cdot\bS$, and
\begin{equation}
  \qty(\bP^\sigma\cdot\bS)^2
  =\bP^\sigma\cdot\bS\cdot\bP^\sigma\cdot\bS
  =\bP^\sigma\cdot\bS
\end{equation}
The coefficient-space projector is ordinarily idempotent even though
$\bP^\sigma$ alone generally is not. For RHF, the corresponding projector
is $\bP\cdot\bS/2$. For a general collinear open-shell determinant the
two spin projectors need not coincide, so the RHF relation
$\bP\cdot\bS\cdot\bP=2\bP$ must not be imposed on their sum.
\end{hfsolution}

At this stage the finite-basis Hartree--Fock problem is completely specified:
$\bP$ determines $\bF$, the generalized eigenvectors of $\bF$ determine a new
$\bP$, and the energy provides a scalar measure of the resulting state. The
remaining task is to solve this circular dependence.

\subsection{The SCF algorithm and convergence}

The self-consistent-field procedure solves the nonlinear
Roothaan--Hall equations by fixed-point iteration. One begins from an
approximate density, constructs the Fock matrix generated by that density,
solves the associated generalized eigenvalue problem, and uses the occupied
eigenvectors to construct a new density. The process is repeated until the
input and output densities describe the same mean field.
Algorithm~\ref{alg:rhf-scf-summary} collects the working steps in one place,
separating the one-time setup from the iterative loop and the final output.

\subsubsection{Fixed quantities and the SCF iteration}

The geometry, basis functions, overlap matrix, core Hamiltonian, and
electron-repulsion integrals remain fixed throughout an SCF calculation. The
density, Fock matrix, orbital coefficients, orbital energies, and total energy
change from one iteration to the next. Separating these two groups clarifies
why, in the conventional stored-integral algorithm described here, the
expensive integral generation is performed before the iterative cycle,
whereas the Fock build and diagonalization are repeated. Direct SCF, discussed
in Section~\ref{sec:integral-screening}, instead regenerates the required
integrals during each Fock build.

\paragraph{The SCF algorithm.}

In practice, a restricted Hartree--Fock calculation begins by
specifying the molecule through $\{\bR_A\}$ and $\{Z_A\}$ and choosing
the basis set $\{\phi_\mu\}$. The integrals $S_{\mu\nu}$,
$H_{\mu\nu}$, and $\braket{\mu\nu}{\lambda\sigma}$ are then calculated,
after which $\bS$ is diagonalized to obtain $\bX$.

Once an initial density matrix $\bP$ has been chosen, the iterative
cycle begins. The matrix $\bG$ is calculated and used to form
$\bF=\bH+\bG$. The Fock matrix is transformed according to
$\bF'=\bX^\dag\cdot\bF\cdot\bX$ and diagonalized to obtain $\bC'$ and
$\beps$. The AO coefficients are recovered as
$\bC=\bX\cdot\bC'$, and a new density matrix is formed from
$\bP=2\bC_\text{o}\cdot\bC_\text{o}^\dag$. The calculation is then tested
for convergence. If it has not converged, the cycle returns to the
construction of $\bG$ and $\bF$; otherwise, the desired observables,
including $\EHF$, are evaluated.

\begin{hfscfsummary}[phantom={\label{alg:rhf-scf-summary}}]
\textcolor{hfInk}{Closed-shell RHF in a fixed spatial AO basis;
$\bP$ is the spin-summed density. The electronic energy excludes nuclear repulsion.}
\medskip
\colorlet{hfSCFStep}{hfGreenDark}
\algrenewcommand\alglinenumber[1]{%
  \tikz[baseline=(scfnumber.base)]
    \node[circle,fill=hfSCFStep,text=white,inner sep=0pt,
      minimum size=4.4mm,font=\scriptsize\bfseries] (scfnumber) {#1};}
\begin{algorithmic}[1]
  \Statex \colorbox{hfGreenLight}{\strut\textcolor{hfGreenDark}{\bfseries
    SETUP\enspace\textbar\ Prepare once at a fixed geometry}}
  \State \textbf{Specify the calculation.}\phantomsection\label{algstep:scf-input}
    Choose $\{\bR_A,Z_A\}$, an even electron number $N$, the basis
    $\{\phi_\mu\}$, and convergence thresholds.
  \State \textbf{Prepare integrals and the AO metric.}\phantomsection\label{algstep:scf-metric}
    Calculate $\bS$, $\bH$, and the ERIs; form an orthogonalizer
    $\bX$ satisfying $\bX^\dag\cdot\bS\cdot\bX=\bI$.
  \State \textbf{Choose an initial density.}\phantomsection\label{algstep:scf-guess}
    Supply $\bP^{(0)}$ and set $n=0$.
  \Statex \smallskip
    \colorbox{hfBlue!7}{\strut\textcolor{hfBlue}{\bfseries
      ITERATION\enspace\textbar\ Repeat Steps 4--7 until self-consistency}}
  \colorlet{hfSCFStep}{hfBlue}
  \State \textbf{Build the physical Fock matrix.}\phantomsection\label{algstep:scf-build}
    Use the current density in Eq.~\eqref{eq:rhf-fock-ao}:
  \Statex $\displaystyle
    \bF^{(n)}=\bF[\bP^{(n)}]=\bH+\bG[\bP^{(n)}]$
  \State \textbf{Evaluate the energy and test convergence.}\phantomsection\label{algstep:scf-test}
    Use this same density--Fock pair:
  \Statex $\displaystyle
    \EHF^{(n)}=\frac12\Tr\qty[\bP^{(n)}\cdot(\bH+\bF^{(n)})]$
  \Statex $\displaystyle
    \be^{(n)}=\bF^{(n)}\cdot\bP^{(n)}\cdot\bS
      -\bS\cdot\bP^{(n)}\cdot\bF^{(n)}$
  \Statex After at least one orbital update, require
    $\max\abs*{\be^{(n)}}<\tau_\text{SCF}$ and the chosen energy/density
    change criteria. If satisfied, go to Step~\ref{algstep:scf-output}.
  \State \textbf{Solve for new orbitals.}\phantomsection\label{algstep:scf-solve}
    Transform, diagonalize, and return to AO coefficients:
  \Statex $\displaystyle
    (\bF^{(n)})'
      =\bX^\dag\cdot\bF^{(n)}\cdot\bX$
  \Statex $\displaystyle
    (\bF^{(n)})'\cdot(\bC^{(n+1)})'
      =(\bC^{(n+1)})'\cdot\beps^{(n+1)}$
  \Statex $\displaystyle
    \bC^{(n+1)}=\bX\cdot(\bC^{(n+1)})'$
  \State \textbf{Occupy and update the density.}\phantomsection\label{algstep:scf-update}
    Occupy the $N/2$ lowest spatial orbitals and form
  \Statex $\displaystyle
    \bP^{(n+1)}=2\bC_\text{o}^{(n+1)}\cdot
      (\bC_\text{o}^{(n+1)})^\dag$
  \Statex Set $n\leftarrow n+1$ and return to
    Step~\ref{algstep:scf-build}, rebuilding the physical Fock matrix.
  \Statex \smallskip
    \colorbox{hfGreenLight}{\strut\textcolor{hfGreenDark}{\bfseries
      OUTPUT\enspace\textbar\ Only after convergence}}
  \colorlet{hfSCFStep}{hfGreenDark}
  \State \textbf{Evaluate the converged observables.}\phantomsection\label{algstep:scf-output}
    Report $\EHF$, the density, orbitals, and required properties.
    Add the fixed nuclear repulsion for the total molecular energy.
\end{algorithmic}
\end{hfscfsummary}

At a given iteration the diagonalization solves exactly the linear
eigenvalue problem defined by the current Fock matrix. It does not solve the
nonlinear Hartree--Fock problem until the occupied eigenvectors reproduce the
density from which that Fock matrix was constructed.

\paragraph{The orthogonalization matrix.}

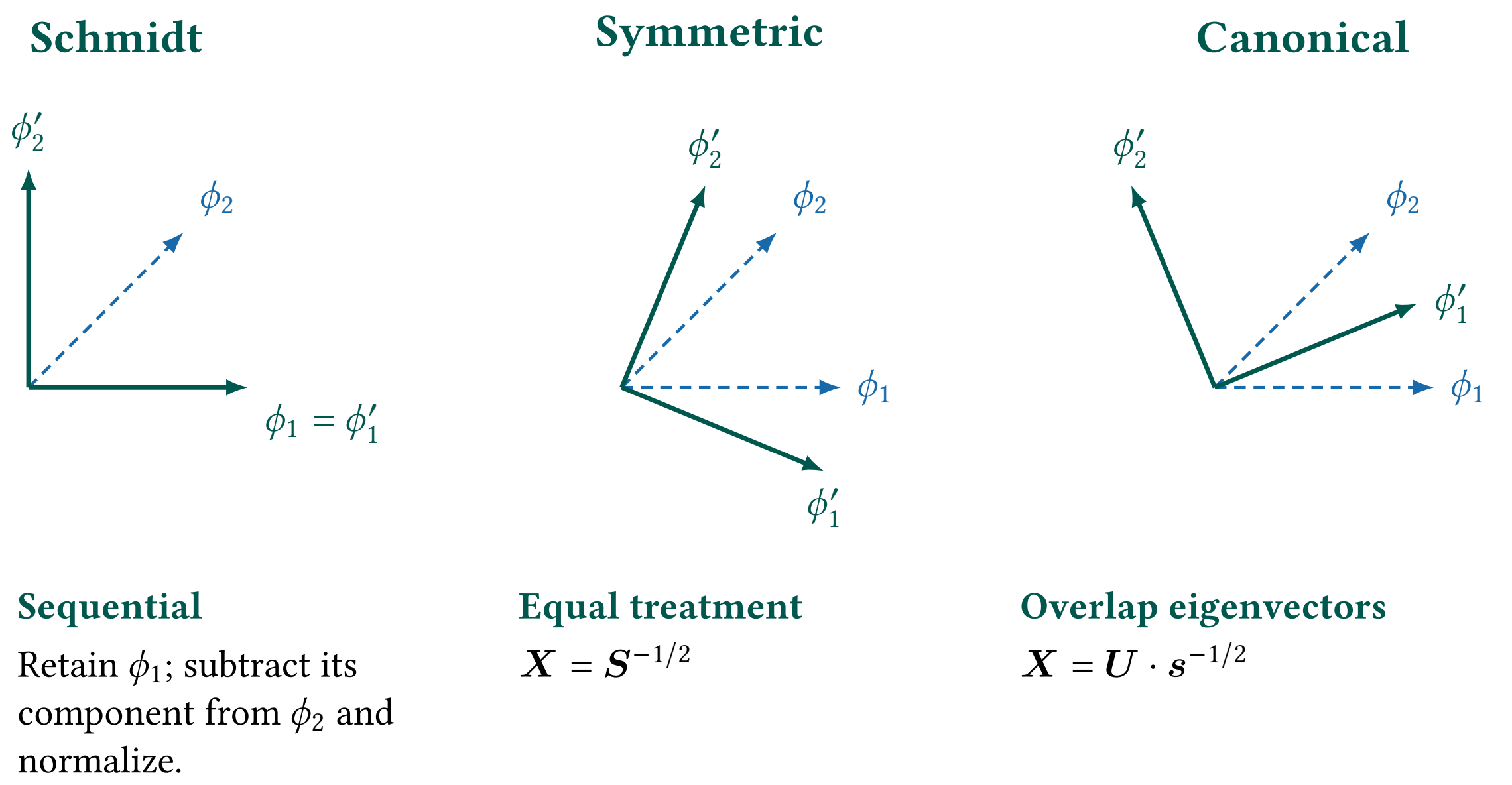
\begin{figure}[htbp]
  \centering

\begingroup

\tikzset{
 nucleus/.style={circle,draw=hfGreenDark,fill=hfGreen!12,minimum size=7mm},
 electron/.style={circle,draw=hfBlue,fill=hfBlue!10,minimum size=7mm},
 old/.style={hfBlue,densely dashed,thick,-{Latex[length=2mm]}},
 new/.style={hfGreenDark,very thick,-{Latex[length=2mm]}},
 card/.style={draw=hfGreenDark,fill=hfGreen!5,rounded corners=2mm,
   align=center,minimum height=12mm},
 occupied/.style={circle,fill=hfGreenDark,inner sep=2.1pt},
 vacant/.style={circle,draw=hfInk!50,inner sep=2.1pt},
 hole/.style={circle,draw=hfRed,thick,inner sep=2.1pt},
 particle/.style={circle,fill=hfViolet,inner sep=2.1pt}}


\resizebox{\linewidth}{!}{%
\begin{tikzpicture}[x=1cm,y=1cm,font=\small]
 \foreach \x/\method in {0/Schmidt,5.4/Symmetric,10.8/Canonical}{
  \begin{scope}[xshift=\x cm]
   \node[font=\large\bfseries,text=hfGreenDark] at (0.8,3.2) {\method};
   \ifdim\x pt=0pt
    \draw[old] (0,0)--(2,0);
   \else
    \draw[old] (0,0)--(2,0) node[right] {$\phi_1$};
   \fi
   \draw[old] (0,0)--(1.414,1.414) node[above right] {$\phi_2$};
   \ifdim\x pt=0pt
    \draw[new] (0,0)--(2,0) node[below right] {$\phi_1=\phi'_1$};
    \draw[new] (0,0)--(0,2) node[above] {$\phi_2'$};
   \else\ifdim\x pt=5.4pt
    \draw[new] (0,0)--(1.848,-0.765) node[below] {$\phi_1'$};
    \draw[new] (0,0)--(0.765,1.848) node[above] {$\phi_2'$};
   \else
    \draw[new] (0,0)--(1.848,0.765) node[right] {$\phi_1'$};
    \draw[new] (0,0)--(-0.765,1.848) node[above] {$\phi_2'$};
   \fi\fi
  \end{scope}
 }
\end{tikzpicture}}
\par\vspace{4mm}
\begin{tabularx}{\linewidth}{>{\raggedright\arraybackslash}X
 >{\raggedright\arraybackslash}X >{\raggedright\arraybackslash}X}
\textcolor{hfGreenDark}{\textbf{Sequential}}&
\textcolor{hfGreenDark}{\textbf{Equal treatment}}&
\textcolor{hfGreenDark}{\textbf{Overlap eigenvectors}}\\[1mm]
Retain $\phi_1$; subtract its component from $\phi_2$ and normalize.&
$\bX=\bS^{-1/2}$&
$\bX=\bU\cdot\bs^{-1/2}$\\
\end{tabularx}
\endgroup


  \caption{Orthogonalization of an illustrative pair of real normalized
  atomic-orbital functions. Dashed blue vectors show the original functions;
  solid green vectors show orthonormal functions. The three methods select
  different orthonormal coordinates in the same space. Adapted from Szabo
  and Ostlund \citep{SzaboOstlund1996}.}
  \label{fig:ao-orthogonalization}
\end{figure}

Step~\ref{algstep:scf-metric} of Algorithm~\ref{alg:rhf-scf-summary} uses
the orthogonalizer derived in Section~\ref{sec:ao-orthogonalization}.
For reference, the two constructions and their checks are collected here.
The required condition is
\begin{equation}
  \red{\bX^\dag\cdot\bS\cdot\bX=\bI}
\end{equation}
For the symmetric, or \red{L\"owdin}, orthogonalization,
\begin{equation}
  \bX=\bS^{-1/2}
  =\bU\cdot\bs^{-1/2}\cdot\bU^\dag
\end{equation}
is one solution. Direct substitution verifies it:
\begin{equation}
  \bX^\dag\cdot\bS\cdot\bX
  =\bS^{-1/2}\cdot\bS\cdot\bS^{-1/2}
  =\bI
  \qquad\green{\checkmark}
\end{equation}
The canonical orthogonalization,
\begin{equation}
  \bX=\bU\cdot\bs^{-1/2}
\end{equation}
is another solution and is useful when linear dependencies are present. In
this case,
\begin{equation}
  \bX^\dag\cdot\bS\cdot\bX
  =\bs^{-1/2}\cdot
   \underbrace{\bU^\dag\cdot\bS\cdot\bU}_{\bs}
   \cdot\bs^{-1/2}
  =\bI
  \qquad\green{\checkmark}
\end{equation}
When all overlap eigenvectors are retained, the two constructions differ
only by a unitary transformation within the orthonormalized space. Discarding
near-dependent directions instead reduces the represented orbital space.
The symmetric choice changes the original basis as
little as possible in the L\"owdin sense, whereas the canonical construction
makes it particularly transparent how directions associated with very small
overlap eigenvalues can be removed.
Figure~\ref{fig:ao-orthogonalization} compares the resulting orthonormal bases
with a Gram--Schmidt construction in a two-function example.

\subsubsection{Initial guesses}

Because the Fock matrix cannot be built before a density is available, every
SCF calculation requires an initial approximation
(Step~\ref{algstep:scf-guess} of Algorithm~\ref{alg:rhf-scf-summary}).
The quality of this guess
does not change the equations being solved, but it can strongly influence the
number of iterations and, when several stationary solutions exist, the
solution to which the procedure converges.
Several initial density matrices are possible. Setting $\bP=\bO$, so
that $\bF=\bH$, gives the core guess; it is usually poor, but it is
easy to implement. An inexpensive, though now less common, alternative is to
use extended H\"uckel theory or a semi-empirical method. A
superposition-of-atomic-densities (SAD) guess is obtained from
tabulated atomic densities. Finally, it is very common and useful to
read the molecular orbitals from a previous calculation \citep{Lehtola2019}.

\subsubsection{Convergence criteria}

Small changes in the energy alone do not necessarily imply that the orbitals
are self-consistent. It is therefore common to monitor both energetic changes
and a matrix residual that tests whether the occupied space is stationary
with respect to the current Fock operator.
This is the convergence test in Step~\ref{algstep:scf-test} of
Algorithm~\ref{alg:rhf-scf-summary}; the density update in
Step~\ref{algstep:scf-update} must be followed by a new physical Fock build
before that test is repeated.
The SCF procedure defines the iterative map
\begin{equation}
    \bP^{(n)}
    \longrightarrow\bF[\bP^{(n)}]
    \longrightarrow\bP^{(n+1)}
\end{equation}
Here $n$ labels the SCF iteration and $\bP^\star$ below denotes a
self-consistent density. In an orthonormal basis, the corresponding
stationarity condition is simply that the Fock
and density matrices commute. The factors of $\bS$ in the AO expression are
the metric factors required by the nonorthogonality of the original basis.
At self-consistency, the occupied and virtual subspaces therefore do not mix
under the Fock operator.
At convergence,
\begin{equation}
  \bP^{(n+1)}=\bP^{(n)}=\bP^\star
\end{equation}
One may check the energy and/or the density matrix. A more direct test of
stationarity uses the commutator residual $\be^{(n)}$, which vanishes at
self-consistency:
\begin{equation}
  \boxed{
    \be^{(n)}
    =\bF^{(n)}\cdot\bP^{(n)}\cdot\bS
    -\bS\cdot\bP^{(n)}\cdot\bF^{(n)}
  }
\end{equation}
Accordingly,
\begin{equation}
  \bP^{(n)}=\bP^\star
  \qquad\Longrightarrow\qquad
  \be^{(n)}=\bO
  \qquad\Longrightarrow\qquad
  \max\abs*{\be^{(n)}}<\tau_\text{SCF}
\end{equation}
In practice, the largest absolute residual element is compared with the
positive convergence threshold $\tau_\text{SCF}$.
Finally, DIIS (direct inversion in the iterative subspace) is usually
used to accelerate convergence. It extrapolates the Fock matrix from the
$m$ previous iterations according to
\begin{equation}
    \bF^{(n+1)}=\sum_{i=n-m+1}^{n}c_i\bF^{(i)}
\end{equation}
The ordinary SCF iteration should therefore be understood as the underlying
fixed-point map. Damping, level shifting \citep{SaundersHillier1973}, and DIIS modify how successive
approximations are generated, but convergence is always judged against the
same self-consistency conditions.

\begin{hfexercise}[phantom={\label{ex:scf-convergence}}]{What does SCF convergence mean?}
\violet{\textit{Derive the SCF stationarity condition in a nonorthogonal AO
basis. Is a small energy change sufficient to establish convergence?}}
\end{hfexercise}
\begin{hfsolution}{What does SCF convergence mean?}
Assume linearly independent AOs and Hermitian overlap and Fock matrices.
For the RHF density $\bP=2\bC_\text{o}\cdot\bC_\text{o}^\dag$, define
the residual
\begin{equation}
  \be=\bF[\bP]\cdot\bP\cdot\bS
  -\bS\cdot\bP\cdot\bF[\bP]
\end{equation}
At self-consistency the Fock matrix constructed from $\bP$ has occupied
eigenvectors that generate this same density. Writing $\beps_\text{o}$ for
the real diagonal matrix of occupied orbital energies, the occupied
Roothaan--Hall equations and their adjoint are
\begin{align}
  \bF\cdot\bC_\text{o}
  &=\bS\cdot\bC_\text{o}\cdot\beps_\text{o}\\
  \bC_\text{o}^\dag\cdot\bF
  &=\beps_\text{o}\cdot\bC_\text{o}^\dag\cdot\bS
\end{align}
Substitution of the first equation gives
\begin{equation}
\begin{split}
  \bF\cdot\bP\cdot\bS
  &=2\bF\cdot\bC_\text{o}\cdot\bC_\text{o}^\dag\cdot\bS\\
  &=2\bS\cdot\bC_\text{o}\cdot\beps_\text{o}
    \cdot\bC_\text{o}^\dag\cdot\bS
\end{split}
\end{equation}
Using the adjoint equation for the other term gives precisely the same
matrix:
\begin{equation}
\begin{split}
  \bS\cdot\bP\cdot\bF
  &=2\bS\cdot\bC_\text{o}\cdot\bC_\text{o}^\dag\cdot\bF\\
  &=2\bS\cdot\bC_\text{o}\cdot\beps_\text{o}
    \cdot\bC_\text{o}^\dag\cdot\bS
\end{split}
\end{equation}
Their difference therefore yields $\be=\bO$. Hermiticity also gives
$\be^\dag=-\be$, as expected for a commutator represented with the AO
metric.

For an arbitrary trial determinant, complete the occupied coefficient
matrix with virtual columns to form
$\bC=(\bC_\text{o},\bC_\text{v})$, with
$\bC^\dag\cdot\bS\cdot\bC=\bI$. These orbitals need not diagonalize
the current $\bF[\bP]$. In this orthonormal MO basis, define
\begin{align}
  \bff&=\bC^\dag\cdot\bF[\bP]\cdot\bC\\
  \bP_\text{MO}
  &=\bC^\dag\cdot\bS\cdot\bP\cdot\bS\cdot\bC
  =\pmqty{2\bI_\text{o}&\bO\\\bO&\bO}
\end{align}
The occupied identity has dimension $N/2$ and the virtual block has
dimension $K-N/2$. Because the AO basis is linearly independent and
$\bC$ is complete, $\bC\cdot\bC^\dag=\bS^{-1}$. Inserting this identity
between the factors gives
\begin{equation}
\begin{split}
  \bC^\dag\cdot\be\cdot\bC
  &=\qty(\bC^\dag\cdot\bF\cdot\bC)
    \cdot\qty(\bC^\dag\cdot\bS\cdot\bP\cdot\bS\cdot\bC)\\
  &-\qty(\bC^\dag\cdot\bS\cdot\bP\cdot\bS\cdot\bC)
    \cdot\qty(\bC^\dag\cdot\bF\cdot\bC)\\
  &=\bff\cdot\bP_\text{MO}-\bP_\text{MO}\cdot\bff\\
  &=\pmqty{\bO&-2\bff_\text{ov}\\2\bff_\text{vo}&\bO}
\end{split}
\end{equation}
In particular, for occupied $i$ and virtual $a$,
\begin{align}
  \qty(\bC^\dag\cdot\be\cdot\bC)_{ai}&=+2f_{ai}
  &
  \qty(\bC^\dag\cdot\be\cdot\bC)_{ia}&=-2f_{ia}
\end{align}
The residual vanishes exactly when the occupied-virtual Fock blocks
vanish. These are the matrix elements that drive first-order changes of
the energy under occupied-virtual orbital rotations. The condition is
therefore one of orbital stationarity, not merely a statement about a
small difference between two scalar energies. The occupied-occupied and
virtual-virtual blocks can then be diagonalized independently to obtain
canonical orbitals without changing the determinant's occupied space.

A small energy change does not impose this condition. For a real orbital
rotation parameter $q$ and a small step $\delta q$, a Taylor expansion gives
\begin{equation}
  E(q+\delta q)-E(q)
  =\dv{E}{q}\delta q
  +\frac12\dv[2]{E}{q}(\delta q)^2+\cdots
\end{equation}
The energy change can be arbitrarily small because the step is small,
even when the derivative is not. Near a stationary point the leading
energy error is generally quadratic in the orbital error, whereas the
orbital gradient is generally linear. An apparent energy plateau can
therefore hide an appreciable residual. Energy changes, density changes,
and a consistently constructed residual provide complementary information.

There is a related distinction within a single iteration. Diagonalizing
$\bF^{(n)}=\bF[\bP^{(n)}]$ constructs a new density that satisfies
\begin{equation}
  \bF^{(n)}\cdot\bP^{(n+1)}\cdot\bS
  -\bS\cdot\bP^{(n+1)}\cdot\bF^{(n)}=\bO
\end{equation}
This mixed-iteration expression vanishes for an exact diagonalization,
whether or not the SCF problem has converged. The new density must instead
be tested against its own rebuilt Fock matrix, or the old density against
its own Fock matrix as in the stated residual. The same caution applies
when the matrix diagonalized is a DIIS extrapolation rather than the Fock
matrix rebuilt from the density being tested.

Finally, vanishing occupied-virtual blocks constrain the first variation
of the energy, not the second. A stationary determinant may be a local
minimum or an unstable stationary point; stability requires examining the
curvature under the allowed orbital rotations. Even a stable local minimum
need not be the global minimum, and stability within RHF does not establish
stability against the additional variations allowed in UHF or GHF.
\end{hfsolution}

\subsection{Direct inversion in the iterative subspace (DIIS)}

Near a self-consistent solution, several recent Fock matrices and their error
vectors contain information about the direction in which the residual is
decreasing. DIIS uses this information to form an extrapolated Fock matrix
whose estimated residual is as small as possible within the affine space
spanned by the stored iterations \citep{Pulay1980,Pulay1982}.

\hfBioPulay

\subsubsection{Residual minimization and extrapolation}

Using the $m$ stored iterations, DIIS constructs an extrapolated Fock matrix
and approximates its error vector as follows. Here $i$ labels a stored
iteration, not an occupied orbital; $\be^{(i)}$ is its commutator residual,
and $c_i$ is its extrapolation weight.
\begin{align}
  \bF_\text{DIIS}^{(m+1)}
  &=\sum_{i=1}^{m}c_i\bF^{(i)}
  &
  \be_\text{DIIS}^{(m+1)}
  &\red{\approx}\sum_{i=1}^{m}c_i\be^{(i)}
\end{align}
The coefficients minimize the squared Frobenius norm of the residual, that
is, the sum of the squared magnitudes of its matrix elements
\begin{equation}
    \min_{\bc}
    \norm{\sum_{i=1}^{m}c_i\be^{(i)}}_F^2
\end{equation}
subject to the affine constraint
\begin{equation}
    \sum_{i=1}^{m}c_i=1
\end{equation}
Define the DIIS overlap matrix by
\begin{equation}
  B_{ij}
  =\braket*{\be^{(i)}}{\be^{(j)}}
  =\Tr\qty[\be^{(i)\dagger}\cdot\be^{(j)}]
\end{equation}
The elements of $\bB$ are therefore pairwise inner products between residual
matrices; the extrapolation weights $c_i$ are taken to be real. For the
Hermitian Fock, density, and overlap matrices used here, the residual is
anti-Hermitian and these inner products are real even when the orbitals are
complex. If two iterations have very similar errors, the corresponding rows
of $\bB$ become nearly linearly dependent; this observation explains the
possible numerical singularity discussed below.
The squared residual norm is then
\begin{equation}
  \norm{\be_\text{DIIS}^{(m+1)}}_F^2
  =\sum_{ij}c_iB_{ij}c_j
  =\bc^\mathsf{T}\cdot\bB\cdot\bc
\end{equation}

\paragraph{DIIS weights and extrapolated Fock matrix.}

The constrained minimization is described by the DIIS Lagrangian, in which
$\lambda$ is the multiplier enforcing the normalization of the weights:
\begin{equation}
  \mathcal{L}(\bc,\lambda)
  =\bc^\mathsf{T}\cdot\bB\cdot\bc
  -2\lambda\qty(\sum_{i=1}^{m}c_i-1)
\end{equation}
Its stationarity conditions are
\begin{align}
  \pdv{\mathcal{L}(\bc,\lambda)}{\bc}&=\bO
  &
  \pdv{\mathcal{L}(\bc,\lambda)}{\lambda}&=0
\end{align}
Writing the first condition component by component gives
\begin{align}
  \pdv{\mathcal L}{c_k}
  &=
  2\sum_{j=1}^{m}B_{kj}c_j-2\lambda
  =
  0
  &
  \pdv{\mathcal L}{\lambda}
  &=
  -2\qty(\sum_{i=1}^{m}c_i-1)
  =
  0
\end{align}
Hence every row obeys
\begin{equation}
  \sum_{j=1}^{m}B_{kj}c_j-\lambda=0
\end{equation}
together with the normalization constraint. These relations imply
\begin{align}
  \bB\cdot\bc-\lambda\boldsymbol{1}&=\bO
  &
  \sum_{i=1}^{m}c_i&=1
\end{align}
These equations form the augmented DIIS linear system
\begin{equation}
  \boxed{
  \pmqty{
    B_{11}&\cdots&B_{1m}&-1\\
    \vdots&\ddots&\vdots&\vdots\\
    B_{m1}&\cdots&B_{mm}&-1\\
    -1&\cdots&-1&0
  }
  \cdot
  \pmqty{
    c_1\\ \vdots\\ c_m\\ \lambda
  }
  =
  \pmqty{
    0\\ \vdots\\ 0\\ -1
  }
  }
\end{equation}
After solving for the coefficients, the extrapolated Fock matrix is
\begin{equation}
    \bF_\text{DIIS}^{(m+1)}
    =\sum_{i=1}^{m}c_i\bF^{(i)}
\end{equation}
The linear system has dimension only \red{$(m+1)\times(m+1)$}.
The normalization constraint $\sum_i c_i=1$ makes the extrapolation affine
rather than arbitrary. In particular, if all stored Fock matrices were equal,
their extrapolated combination would reproduce that same matrix. The
Lagrange multiplier is the algebraic device that enforces this constraint.

\subsubsection{The DIIS algorithm and practical considerations}

The complete algorithm starts from the current density $\bP^{(n)}$, from
which the corresponding Fock matrix
$\bF^{(n)}=\bF[\bP^{(n)}]$ is constructed. The commutator residual is then
evaluated as
\begin{equation}
  \be^{(n)}
  =\bF^{(n)}\cdot\bP^{(n)}\cdot\bS
  -\bS\cdot\bP^{(n)}\cdot\bF^{(n)}
\end{equation}
The pair $\{\bF^{(n)},\be^{(n)}\}$ is stored and the DIIS subspace is
updated. The matrix $\bB$ is constructed, the augmented linear system is
solved, and the coefficients $\{c_i\}$ are obtained. These coefficients
define the extrapolated Fock matrix
\begin{equation}
  \bF^\text{DIIS}=\sum_i c_i\bF^{(i)}
\end{equation}
This is the working matrix used for orbital generation in
Step~\ref{algstep:scf-solve} of Algorithm~\ref{alg:rhf-scf-summary};
it does not replace $\bF[\bP]$ in the convergence test.
Finally, the generalized eigenvalue problem
\begin{equation}
  \bF^\text{DIIS}\cdot\bC
  =\bS\cdot\bC\cdot\beps
\end{equation}
is solved and the next density $\bP^{(n+1)}$ is formed.
The subspace is normally updated as a rolling history: after its maximum size
has been reached, the oldest Fock matrix and residual are discarded when a new
pair is added. If the residuals become nearly linearly dependent, restarting
with a smaller history restores a well-conditioned extrapolation problem.
Figure~\ref{fig:scf-cycle} summarizes the SCF cycle and indicates where
convergence acceleration can enter.
\begin{center}
\begin{tikzpicture}[
  >=stealth,
  node distance=0.55cm and 0.65cm,
  scfstep/.style={
    draw=hfGreen,
    rounded corners=2pt,
    fill=hfGreenLight,
    minimum width=2.45cm,
    minimum height=0.72cm,
    align=center
  },
  decision/.style={
    draw=hfRed,
    rounded corners=8pt,
    fill=hfPale,
    minimum width=2.45cm,
    minimum height=0.72cm,
    align=center
  }
]
  \node[scfstep] (mol) {Molecule\\and basis set};
  \node[scfstep,right=of mol] (ints) {Compute\\integrals};
  \node[scfstep,right=of ints] (guess) {Initial density\\$\bP^{(0)}$};
  \node[scfstep,right=of guess] (fock)
    {Build Fock matrix\\$\bF[\bP^{(n)}]$};

  \node[scfstep,below=1.15cm of fock] (solve)
    {Solve\\$\bF\cdot\bC=\bS\cdot\bC\cdot\beps$};
  \node[scfstep,left=of solve] (density)
    {New density\\$\bP^{(n+1)}$};
  \node[decision,left=of density] (conv) {Converged?};
  \node[scfstep,left=of conv] (properties) {Energy and\\properties};

  \draw[->,thick] (mol)--(ints);
  \draw[->,thick] (ints)--(guess);
  \draw[->,thick] (guess)--(fock);
  \draw[->,thick] (fock)--(solve);
  \draw[->,thick] (solve)--(density);
  \draw[->,thick] (density)--(conv);
  \draw[->,thick,draw=hfGreen]
    (conv)--(properties)
    node[midway,above,text=hfGreen] {yes};
  \draw[->,thick,draw=hfRed]
    (conv.north)--++(0,0.45)-|(fock.south)
    node[pos=0.18,above,text=hfRed] {no};
  \node[below=0.18cm of density,text=hfViolet]
    {\small optional damping or DIIS};
\end{tikzpicture}
\captionof{figure}{The self-consistent-field cycle: fixed molecular data and integrals,
iterative Fock construction, orbital solution, density update, and
convergence testing. Damping or DIIS may accelerate the iteration.}
\label{fig:scf-cycle}
\end{center}

\paragraph{Practical considerations.}

The key idea is that DIIS accelerates SCF convergence by finding a linear
combination of Fock matrices. The practical consequences are straightforward.
A small rolling subspace, commonly
$m\approx6$--$10$, is normally sufficient, and DIIS can also be used for
many other methods. The acceleration nevertheless requires additional
storage of approximately $2mK^2$. Moreover, the extrapolated energy is
not guaranteed to decrease monotonically, and $\bB$ can become
singular when the residuals are nearly linearly dependent, in which
case a restart may be required.

DIIS changes neither the Hartree--Fock energy functional nor the final
self-consistency condition. It is an acceleration strategy for reaching a
stationary solution of the same equations; the convergence and stability
distinctions in Exercise~\ref{ex:scf-convergence} still apply.

\subsection{Excited-state SCF and the maximum-overlap method}

The nonlinear character of the Hartree--Fock equations permits multiple
stationary determinants. Some correspond to non-Aufbau occupations and can be
used as state-specific approximations to electronically excited states. The
challenge is not writing such an occupation initially, but preventing the SCF
iterations from replacing it by a lower-energy Aufbau occupation.
The maximum-overlap method (MOM) provides a practical occupation rule for following
such a state-specific solution \citep{Gilbert2008}.

\subsubsection{Variational collapse}

An excited determinant can be generated by promoting one or more electrons,
but the usual Aufbau occupation does not preserve that configuration.
Figure~\ref{fig:variational-collapse} illustrates the loss of the promoted
occupation under an Aufbau SCF iteration.
\begin{center}
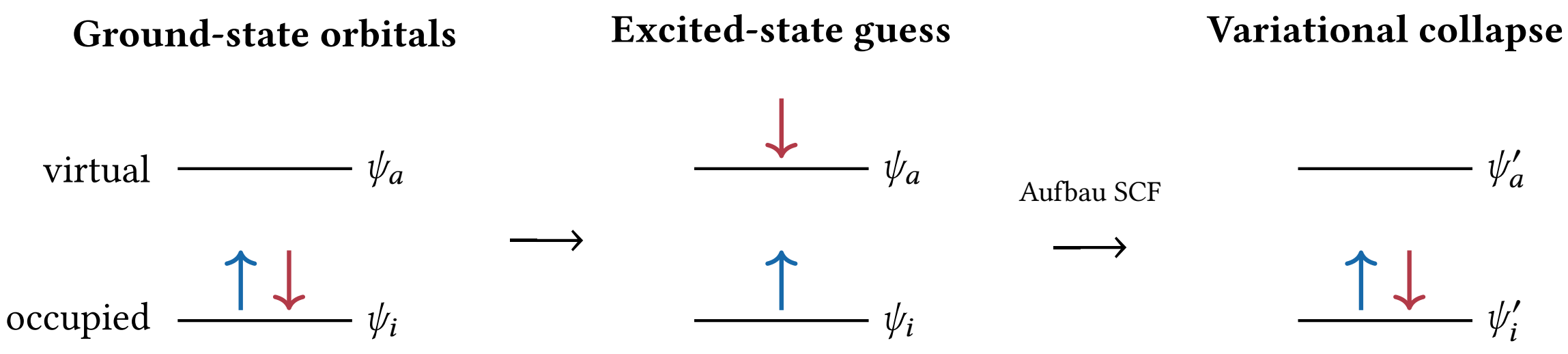

\begin{tikzpicture}[x=1cm,y=0.78cm,scale=1.3]
  \node[font=\bfseries] at (-4.6,2.75) {Ground-state orbitals};
  \draw[thick] (-5.25,0)--(-3.95,0) node[right] {$\psi_i$};
  \draw[thick] (-5.25,1.45)--(-3.95,1.45) node[right] {$\psi_a$};
  \draw[hfBlue,very thick,->] (-4.78,0.10)--(-4.78,0.67);
  \draw[hfRed,very thick,->] (-4.42,0.67)--(-4.42,0.10);
  \node[left] at (-5.35,0) {occupied};
  \node[left] at (-5.35,1.45) {virtual};

  \node at (-2.5,0.75) {\Large$\longrightarrow$};

  \node[font=\bfseries] at (-0.75,2.75) {Excited-state guess};
  \draw[thick] (-1.40,0)--(-0.10,0) node[right] {$\psi_i$};
  \draw[thick] (-1.40,1.45)--(-0.10,1.45) node[right] {$\psi_a$};
  \draw[hfBlue,very thick,->] (-0.75,0.10)--(-0.75,0.67);
  \draw[hfRed,very thick,->] (-0.75,2.12)--(-0.75,1.55);

  \node[align=center] at (1.55,0.95) {\scriptsize Aufbau SCF\\[1mm]
    \Large$\longrightarrow$};

  \node[font=\bfseries] at (3.75,2.75) {Variational collapse};
  \draw[thick] (3.10,0)--(4.40,0) node[right] {$\psi_i'$};
  \draw[thick] (3.10,1.45)--(4.40,1.45) node[right] {$\psi_a'$};
  \draw[hfBlue,very thick,->] (3.57,0.10)--(3.57,0.67);
  \draw[hfRed,very thick,->] (3.93,0.67)--(3.93,0.10);
\end{tikzpicture}
\captionof{figure}{Variational collapse of an excited-state guess. After orbital
reoptimization, an Aufbau occupation can refill the lowest orbital and
destroy the initially promoted configuration.}
\label{fig:variational-collapse}
\end{center}

After each diagonalization, Aufbau occupies the lowest-energy orbitals.
Consequently, the SCF procedure tends to return to a lower-energy
solution.
This behavior is called variational collapse. The variational principle
is not being violated: an unconstrained minimization is doing precisely what
it is designed to do. To follow a chosen excited configuration, the occupation
rule must retain the character of the targeted orbitals rather than simply
selecting the lowest eigenvalues.

\subsubsection{The maximum-overlap criterion}

Let $\bC_\text{o}^\text{ref}$ contain the coefficients of the occupied
reference spin orbitals defining the target configuration. For every new
orbital $\chi_p^\text{new}$, compute
\begin{equation}
\begin{split}
  O_{ip}
  &=\braket{\chi_i^\text{ref}}{\chi_p^\text{new}}
  =\sum_{\mu\nu}C_{\mu i}^{\text{ref}*}
    S_{\mu\nu}C_{\nu p}^\text{new}
  \\
  w_p&=\sum_{i\in\mathcal{O}_\text{ref}}\abs{O_{ip}}^2
\end{split}
\end{equation}
In matrix form, the overlap matrix between the reference occupied orbitals and
the new orbitals is
\begin{equation}
  \boldsymbol{O}
  =
  \bC_\text{o}^{\text{ref}\dagger}
  \cdot\bS\cdot
  \bC^\text{new}
\end{equation}
The weight $w_p$ is the squared norm of column $p$ of this matrix. It therefore
measures how much of the new orbital lies in the occupied reference subspace.
The MOM procedure computes $w_p$ for every new orbital and occupies the $N$
spin orbitals with the largest $w_p$. In MOM, the reference occupied
space is taken from the previous iteration, whereas in the initial
maximum-overlap method (IMOM) the
reference orbitals remain fixed to the initial guess \citep{Barca2018}.
Figure~\ref{fig:mom-weights} illustrates selection by overlap weight
rather than by orbital energy.
\begin{center}
\begin{tikzpicture}[x=0.78cm,y=0.72cm,font=\scriptsize]
  \node[
    draw=hfGreen,
    very thick,
    rounded corners=2pt,
    fill=hfGreenLight,
    align=center,
    text width=4.1cm,
    minimum height=0.75cm
  ] at (2.0,5.15) {
    Occupied reference space\\
    $\mathcal{O}_\text{ref}
    =\operatorname{span}\{\chi_i^\text{ref}\}$
  };
  \node[hfInk] at (2.0,3.80) {overlap with $\mathcal{O}_\text{ref}$};

  \node[anchor=east] at (0.20,3.10) {$\chi_1^\text{new}$};
  \path[fill=hfPale,draw=hfNavy!25,rounded corners=1pt]
    (0.35,2.90) rectangle (3.55,3.30);
  \path[fill=hfTeal,rounded corners=1pt]
    (0.35,2.90) rectangle (0.61,3.30);
  \node[anchor=west] at (3.70,3.10) {$0.08$};

  \node[anchor=east] at (0.20,2.25) {$\chi_2^\text{new}$};
  \path[fill=hfPale,draw=hfNavy!25,rounded corners=1pt]
    (0.35,2.05) rectangle (3.55,2.45);
  \path[fill=hfTeal,rounded corners=1pt]
    (0.35,2.05) rectangle (3.33,2.45);
  \node[anchor=west] at (3.70,2.25) {$0.93$};
  \node[anchor=west,hfTeal,font=\scriptsize\bfseries]
    at (4.35,2.25) {select};

  \node[anchor=east] at (0.20,1.40) {$\chi_3^\text{new}$};
  \path[fill=hfPale,draw=hfNavy!25,rounded corners=1pt]
    (0.35,1.20) rectangle (3.55,1.60);
  \path[fill=hfTeal,rounded corners=1pt]
    (0.35,1.20) rectangle (0.73,1.60);
  \node[anchor=west] at (3.70,1.40) {$0.12$};

  \node[anchor=east] at (0.20,0.55) {$\chi_4^\text{new}$};
  \path[fill=hfPale,draw=hfNavy!25,rounded corners=1pt]
    (0.35,0.35) rectangle (3.55,0.75);
  \path[fill=hfTeal,rounded corners=1pt]
    (0.35,0.35) rectangle (3.10,0.75);
  \node[anchor=west] at (3.70,0.55) {$0.86$};
  \node[anchor=west,hfTeal,font=\scriptsize\bfseries]
    at (4.35,0.55) {select};
\end{tikzpicture}

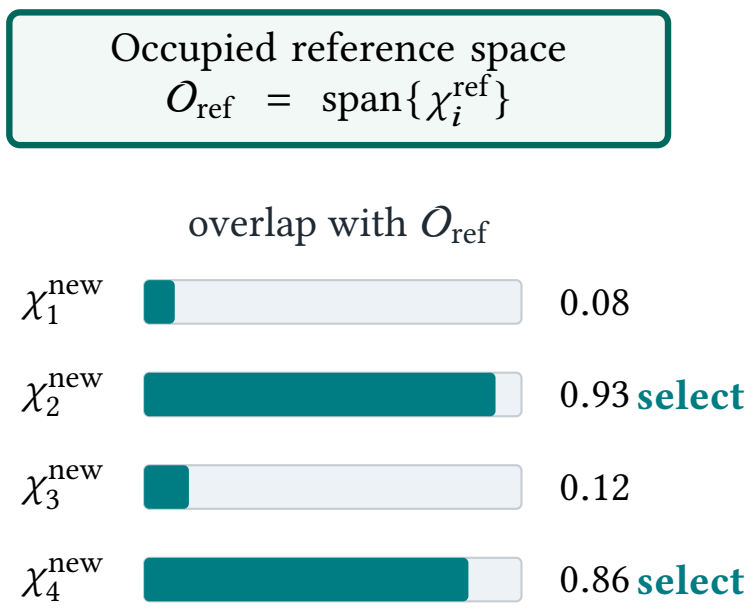
\captionof{figure}{Illustrative maximum-overlap weights $w_p$ for new orbitals projected
onto the occupied reference space $\mathcal{O}_\text{ref}$. The orbitals
with the largest weights are selected for occupation.}
\label{fig:mom-weights}
\end{center}

Thus, MOM selects orbitals by orbital character, not by energy.
The weight $w_p$ measures the total squared projection of a new orbital onto
the reference occupied subspace. It is insensitive to unitary rotations within
that reference space, so the selection is based on the subspace representing
the target configuration rather than on an arbitrary labeling of individual
reference orbitals.

\subsubsection{MOM-based SCF and interpretation}

The MOM-SCF cycle is shown in Fig.~\ref{fig:mom-cycle}.
\begin{center}
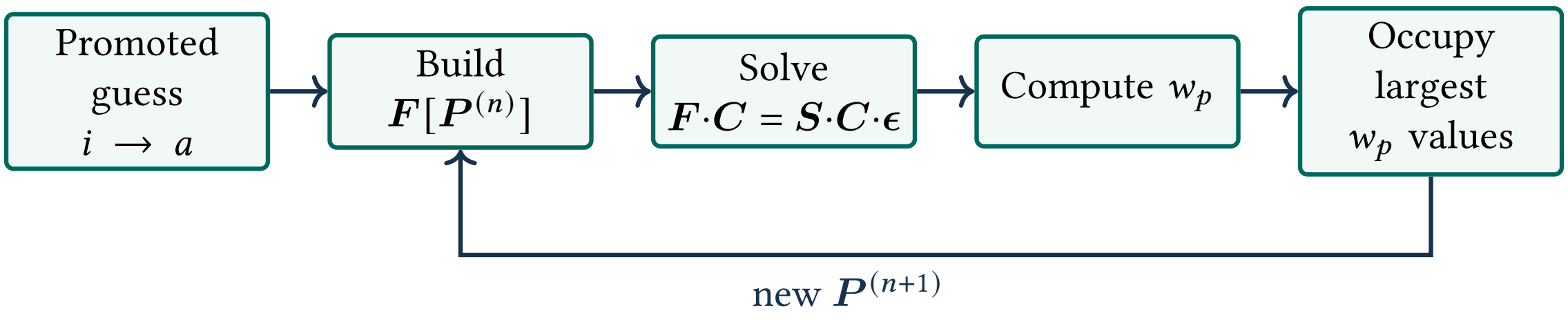

\begin{tikzpicture}[
  node distance=0.55cm and 0.55cm,
  box/.style={
    draw=hfGreen,
    very thick,
    rounded corners=2pt,
    fill=hfGreenLight,
    minimum height=1.05cm,
    text width=2.25cm,
    align=center
  },
  arrow/.style={->,very thick,hfNavy}
]
  \node[box] (guess) {Promoted guess\\$i\rightarrow a$};
  \node[box,right=of guess] (fock) {Build\\$\bF[\bP^{(n)}]$};
  \node[box,right=of fock] (solve)
    {Solve\\$\bF\cdot\bC=\bS\cdot\bC\cdot\beps$};
  \node[box,right=of solve] (overlap) {Compute $w_p$};
  \node[box,right=of overlap] (occupy) {Occupy largest\\$w_p$ values};

  \draw[arrow] (guess)--(fock);
  \draw[arrow] (fock)--(solve);
  \draw[arrow] (solve)--(overlap);
  \draw[arrow] (overlap)--(occupy);
  \draw[arrow] (occupy.south)--++(0,-0.75)-|(fock.south)
    node[pos=0.30,below] {new $\bP^{(n+1)}$};
\end{tikzpicture}
\captionof{figure}{The maximum-overlap SCF cycle. After solving the orbital equations,
the new occupied space is selected by its overlap weights with the
reference space, rather than by the Aufbau rule.}
\label{fig:mom-cycle}
\end{center}

Conventional SCF occupies the lowest orbital energies,
\begin{equation}
  \epsilon_1\leq\epsilon_2\leq\cdots\leq\epsilon_N
\end{equation}
and thereby applies the Aufbau occupation rule at each iteration. This rule
alone does not guarantee convergence or selection of the global minimum.
MOM-SCF instead occupies the
largest overlap weights,
\begin{equation}
  w_1\geq w_2\geq\cdots\geq w_N
\end{equation}
and thereby attempts to follow the chosen occupied space. For collinear
calculations with fixed $N^\alpha$ and $N^\beta$, the selection is made
separately in each spin sector. Thus, MOM changes the
occupation to target non-Aufbau stationary solutions of the Hartree--Fock equations.

\paragraph{What does a MOM calculation give?}

When successful, MOM provides a targeted stationary solution of the HF
equations, often higher in energy than the conventional Aufbau solution,
with full orbital relaxation and a corresponding
state-specific density and set of molecular properties. It can yield useful
excitation energies for states dominated by a single electronic
configuration. Its formal cost per SCF iteration is the same as that of a
conventional calculation, although convergence may be more difficult.

Several cautions remain. The result is still a single determinant, spin
symmetry may be broken, and different SCF states are not necessarily
orthogonal. Convergence depends on the initially promoted orbitals, and the
MOM iteration may drift or converge to an unintended solution.
MOM should consequently be viewed as an occupation-following prescription,
not as a new electronic-structure approximation. It selects a particular
stationary solution of the Hartree--Fock equations and allows the orbitals of
that determinant to relax self-consistently.

\subsection{Molecular properties and orbital analysis}

Once a self-consistent density matrix has been obtained, one-electron
properties can be evaluated by contracting that density with the appropriate
property integrals. The same density also provides the starting point for
population analyses. The HF determinant does not select a unique orbital
representation: unitary rotations within the occupied or virtual subspaces
preserve the state, although the rotated orbitals need not remain canonical. This
freedom motivates localized-orbital representations.

\subsubsection{Dipole moments}

The dipole moment illustrates the general structure of a one-electron
property: a quantum-mechanical electronic expectation value is combined with
the corresponding classical nuclear contribution.
For point charges $q_i$ at positions $\br_i$, the classical dipole vector is
\begin{equation}
  \boldsymbol{\mu}=(\mu_x,\mu_y,\mu_z)
  =\underbrace{\green{\sum_i q_i\br_i}}
    _{\text{classical definition}}
\end{equation}
The electronic term follows from the one-particle density. In the spatial
AO contraction, $\bP$ is the spin-summed density; integration over the spin
coordinate is implicit in the reduction from the spin-orbital sum:
\begin{equation}
\begin{split}
  -\sum_i\mel{\chi_i}{\br}{\chi_i}
  &=
  -\sum_i\sum_{\mu\nu}
  C_{\mu i}^*C_{\nu i}
  \rmel{\mu}{\br}{\nu}
  \\
  &=
  -\sum_{\mu\nu}P_{\nu\mu}
  \rmel{\mu}{\br}{\nu}
\end{split}
\end{equation}
Thus the many-electron expectation value reduces to a contraction of the AO
density with one-electron dipole integrals.
For a molecule, the corresponding quantum-mechanical expression separates
the electronic and nuclear contributions:
\begin{equation}
\begin{split}
  \boldsymbol{\mu}
  &=(\mu_x,\mu_y,\mu_z)
  \\
  &=\underbrace{\red{\mel{\Psi_0}{-\sum_{i=1}^{N}\br_i}{\Psi_0}}}
      _{\text{\red{electrons}}}
    +\underbrace{\blue{\sum_{A=1}^{M}Z_A\bR_A}}
      _{\text{\blue{nuclei}}}
  \\
  &=\red{-\sum_{\mu\nu}P_{\mu\nu}\rmel{\nu}{\br}{\mu}}
    +\blue{\sum_{A=1}^{M}Z_A\bR_A}
\end{split}
\end{equation}
For example, its $x$ component is
\begin{equation}
\begin{split}
  \mu_x
  &=\red{-\sum_{\mu\nu}P_{\mu\nu}\rmel{\nu}{x}{\mu}}
  +\blue{\sum_{A=1}^{M}Z_AX_A}
  \\
  \underbrace{\rmel{\nu}{x}{\mu}}_{\text{one-electron integrals}}
  &=\int\phi_\nu^*(\br)\,x\,\phi_\mu(\br)\dd{\br}
\end{split}
\end{equation}

\subsubsection{Charge analysis}

An atomic charge is not itself a direct observable of the molecular
wavefunction. Population analyses define such charges by adopting a rule for
partitioning the density among atom-centered basis functions. Mulliken and
L\"owdin analyses differ precisely in this partitioning convention. For the
population formulas below, the AOs and orbital coefficients are taken to be
real.

\hfBioMulliken

The electron density is
\begin{align}
  \rho(\br)&=\sum_{\mu\nu}\phi_\mu(\br)P_{\mu\nu}\phi_\nu^*(\br)
  &
  \int\rho(\br)\dd{\br}&=N
\end{align}
The identity $N=\Tr(\bP\cdot\bS)$ expresses electron-number conservation in a
nonorthogonal basis. A population analysis partitions the diagonal
contributions to this trace among the atoms.
Consequently,
\begin{equation}
  N=\sum_{\mu\nu}P_{\mu\nu}S_{\nu\mu}
  =\sum_\mu(\bP\cdot\bS)_{\mu\mu}
  =\Tr(\bP\cdot\bS)
\end{equation}
Assuming that the basis functions are atom-centered, the Mulliken population
analysis \citep{Mulliken1955} assigns the net charge
\begin{equation}
  \underbrace{\blue{q_A^\text{Mulliken}}}_{\text{net charge on $A$}}
  =Z_A-\sum_{\mu\in A}(\bP\cdot\bS)_{\mu\mu}
\end{equation}
For L\"owdin population analysis \citep{Lowdin1950}, cyclic invariance of the trace,
$\Tr(\bA\cdot\bB)=\Tr(\bB\cdot\bA)$, gives, for any real partition
parameter $\alpha$,
\begin{equation}
  N=\sum_\mu
  (\bS^\alpha\cdot\bP\cdot\bS^{1-\alpha})_{\mu\mu}
\end{equation}
For \red{$\alpha=1/2$}, this becomes
\begin{equation}
  N=\sum_\mu
  (\bS^{1/2}\cdot\bP\cdot\bS^{1/2})_{\mu\mu}
  \qquad\Rightarrow\qquad
  \red{q_A^\text{L\"owdin}}
  =Z_A-\sum_{\mu\in A}
  (\bS^{1/2}\cdot\bP\cdot\bS^{1/2})_{\mu\mu}
\end{equation}

\subsubsection{Localized orbitals}

We retain the real-orbital convention for the population-based localization
criteria below. Canonical orbitals diagonalize the occupied-occupied and virtual-virtual
blocks of the Fock matrix, but they are often delocalized over an entire
molecule. Because unitary rotations within either subspace leave the Slater
determinant invariant, the same Hartree--Fock state can be represented by
orbitals chosen to satisfy a localization criterion.
At HF convergence and after canonicalization and localization, the Fock
matrix passes through the sequence
\begin{equation}
  \bff=
  \mqty(\bff_\text{oo}&\bff_\text{ov}\\
        \bff_\text{vo}&\bff_\text{vv})
  \xrightarrow{\text{HF convergence}}
  \mqty(\bff_\text{oo}&\bO\\
        \bO&\bff_\text{vv})
  \xrightarrow{\text{canonicalization}}
  \mqty(\beps_\text{o}&\bO\\
        \bO&\beps_\text{v})
  \xrightarrow{\text{localization}}
  \mqty(\widetilde{\bff}_\text{oo}&\bO\\
        \bO&\widetilde{\bff}_\text{vv})
\end{equation}
The localized occupied orbitals are obtained through a unitary
transformation,
\begin{equation}
  \begin{split}
    \widetilde{\bC}_\text{o}
    &=\bC_\text{o}\cdot\purple{\bU}
    \\
    \purple{\bU}^\dagger\cdot\purple{\bU}&=\bI
    \quad\Longleftrightarrow\quad
    \text{\purple{unitary transformation}}
  \end{split}
\end{equation}
For a closed-shell determinant, the density matrix is unchanged:
\begin{equation}
  \widetilde{\bP}
  =2\widetilde{\bC}_\text{o}
   \cdot\widetilde{\bC}_\text{o}^\dagger
  =2\bC_\text{o}\cdot\bC_\text{o}^\dagger
  =\bP
\end{equation}
The determinant changes only by a phase,
\begin{equation}
  \ket*{\widetilde{\Psi}_0}
  =\det(\purple{\bU})\ket{\Psi_0}
\end{equation}
so the density, energy, and every expectation value are
unchanged.
Figure~\ref{fig:orbital-localization} contrasts schematic canonical and
localized representations of the same occupied space.
\begin{center}
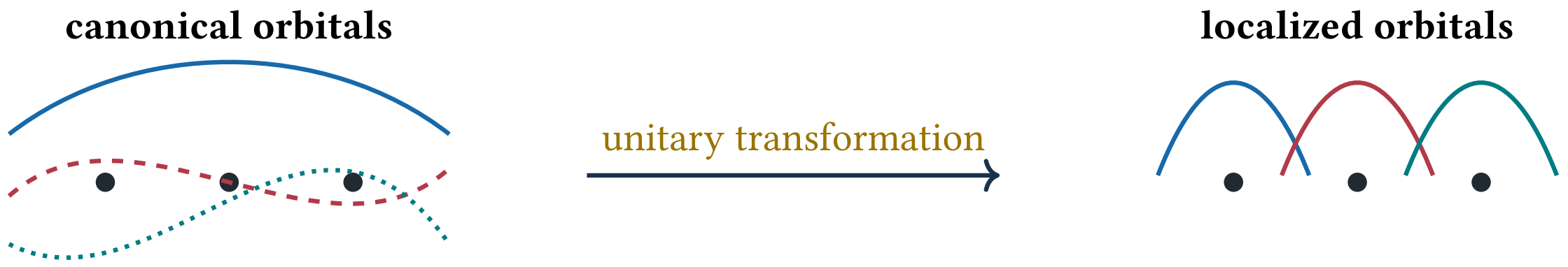

\begin{tikzpicture}[x=1cm,y=1cm,scale=1.3]
  \node[font=\bfseries] at (-4.1,1.15) {canonical orbitals};
  \node[font=\bfseries] at (4.1,1.15) {localized orbitals};
  \foreach \x in {-5.0,-4.1,-3.2}{\fill[hfInk] (\x,0) circle (0.07);}
  \draw[hfBlue,very thick]
    (-5.7,0.35)..controls(-4.8,1.05)and(-3.4,1.05)..(-2.5,0.35);
  \draw[hfRed,very thick,dashed]
    (-5.7,-0.10)..controls(-4.8,0.70)and(-3.4,-0.70)..(-2.5,0.10);
  \draw[hfTeal,very thick,dotted]
    (-5.7,-0.45)..controls(-4.6,-1.00)and(-3.5,1.00)..(-2.5,-0.45);
  \draw[hfNavy,very thick,->] (-1.5,0.05)--(1.5,0.05)
    node[midway,above] {\purple{unitary transformation}};
  \foreach \x in {3.2,4.1,5.0}{\fill[hfInk] (\x,0) circle (0.07);}
  \draw[hfBlue,very thick]
    (2.65,0.05)..controls(3.0,0.95)and(3.4,0.95)..(3.75,0.05);
  \draw[hfRed,very thick]
    (3.55,0.05)..controls(3.9,0.95)and(4.3,0.95)..(4.65,0.05);
  \draw[hfTeal,very thick]
    (4.45,0.05)..controls(4.8,0.95)and(5.2,0.95)..(5.55,0.05);
\end{tikzpicture}
\captionof{figure}{Schematic localization of occupied orbitals by a unitary
transformation. The individual orbital shapes change, but the occupied
space, determinant up to a phase, density, and energy do not.}
\label{fig:orbital-localization}
\end{center}

Only occupied-occupied or virtual-virtual rotations are
invariant. Mixing occupied and virtual orbitals changes the wavefunction
and the energy. Localization functionals are generally nonconvex, so
different guesses may converge to different local extrema.
This invariance separates two logically distinct operations. Orbital
optimization mixes occupied and virtual spaces and changes the physical
determinant; localization rotates only within an already optimized subspace
and changes its representation without changing the state.

\paragraph{Localization procedures.}

Different localization procedures choose different scalar measures of orbital
compactness. Since all admissible transformations are unitary within the
selected subspace, optimizing any of these measures leaves the Hartree--Fock
state itself unchanged.
The Edmiston--Ruedenberg functional \citep{EdmistonRuedenberg1963} is
\begin{equation}
    \Omega_\text{ER}
    =\sum_i\rbraket{\psi_i\psi_i}{\psi_i\psi_i}
    \quad\longrightarrow\quad
    \max\ \text{orbital self-repulsion}
\end{equation}
The Foster--Boys functional \citep{FosterBoys1960} may be written in two equivalent forms:
\begin{equation}
\begin{split}
  \Omega_\text{BF}
  &=\frac{1}{2}\sum_i
    \qty[\mel{\psi_i\psi_i}{r_{12}^2}{\psi_i\psi_i}]
    \quad\longrightarrow\quad
    \min\ \text{spatial extent}
  \\
  &\red{\equiv}\sum_i\qty[
    \mel{\psi_i}{r^2}{\psi_i}
    -\abs{\mel{\psi_i}{\br}{\psi_i}}^2]
    \quad\longrightarrow\quad
    \min\ \text{sum of orbital variances}
\end{split}
\end{equation}
Within a fixed orbital subspace, $\sum_i\mel{\psi_i}{r^2}{\psi_i}$ is
invariant under unitary rotations. Minimizing $\Omega_\text{BF}$ is therefore
equivalent to maximizing $\sum_i\abs{\mel{\psi_i}{\br}{\psi_i}}^2$.
Since the sum of orbital centroids is also invariant, this maximizes their
total pairwise squared separation. Both displayed expressions for
$\Omega_\text{BF}$ are minimized; the centroid criterion is maximized.
Its strengths are
that it requires only one-electron integrals and produces compact bonds and
lone pairs. Its limitations are that it does not explicitly preserve
$\sigma/\pi$ separation and that diffuse virtual orbitals may be difficult
to localize.

For an atomic population $q_i^A$ of orbital $i$ on atom $A$, the original
Pipek--Mezey criterion \citep{PipekMezey1989} maximizes
\begin{equation}
    \Omega_\text{PM}=\sum_A\sum_i\abs*{q_i^A}^2
\end{equation}
using the Mulliken populations
\begin{equation}
  q_i^A
  =\sum_{\mu\in A}\sum_\nu
    C_{\mu i}^*S_{\mu\nu}C_{\nu i}
\end{equation}
Its strengths are
that it often preserves the separation between $\sigma$ and $\pi$ orbitals
and produces intuitive bonds, lone pairs, and core orbitals. Its principal
limitation is that the original Mulliken form can be basis-set dependent;
L\"owdin populations or intrinsic atomic orbitals provide more robust
alternatives.
There is no universally best localized-orbital set; the criterion should
match the intended interpretation or computational use.

\paragraph{Summary of orbital localization.}

The occupied subspace and physical determinant state remain unchanged, as do the
density matrix, electron density, HF energy, and molecular properties. The
occupied-virtual Fock block also remains zero:
\begin{equation}
  \widetilde{\bff}_\text{ov}=\bff_\text{ov}=\bO
\end{equation}
What changes is that localized orbitals are generally not eigenfunctions of
the Fock operator:
\begin{equation}
  \widetilde{\bff}_\text{oo}
  =\purple{\bU}^\dagger\cdot\beps_\text{o}\cdot\purple{\bU}
\end{equation}
Canonical orbital energies therefore cannot be assigned to individual
localized orbitals.
Localization is useful for identifying bonds, lone pairs, and fragments; for
enabling local correlation and embedding methods; for selecting active spaces
and domains; and for constructing Wannier functions in periodic systems. In
practice, the occupied and virtual spaces should be localized separately, and
it is often also advantageous to separate the core and valence orbitals. One
must check for multiple extrema and discontinuous solutions and remember that
spatial symmetry may be lost.

\section{Open-shell and generalized Hartree--Fock methods}
\label{sec:open-shell-hf}

The restricted closed-shell ansatz is efficient and preserves spin symmetry,
but it is too restrictive whenever the two spin populations require different
spatial descriptions. Open-shell species and bond-breaking problems therefore
motivate a hierarchy of increasingly flexible single-determinant
wavefunctions. Unrestricted Hartree--Fock relaxes the requirement that
$\alpha$ and $\beta$ electrons share spatial orbitals, while generalized
Hartree--Fock also removes the requirement that each orbital possess a fixed
spin projection.

\subsection{Unrestricted Hartree--Fock theory}

Unrestricted Hartree--Fock introduces separate variational spaces for the
$\alpha$ and $\beta$ electrons. The resulting energy, density matrices, and
coupled Pople--Nesbet--Berthier equations are developed before the unrestricted SCF
procedure is applied to bond dissociation.

\subsubsection{The unrestricted ansatz}

In restricted Hartree--Fock, a doubly occupied spatial orbital is shared by
one $\alpha$ and one $\beta$ electron. Unrestricted Hartree--Fock removes this
pairing constraint. For open-shell systems, one uses unrestricted spin
orbitals,
\begin{equation}
  \chi_p^\text{UHF}(\bx)
  =\begin{cases}
    \alpha(\omega)\blue{\psi_p^\alpha(\br)}\\
    \beta(\omega)\red{\psi_p^\beta(\br)}
  \end{cases}
\end{equation}
The $\alpha$ and $\beta$ spatial orbitals are therefore allowed to differ.
As a consequence, UHF may break spin symmetry: the determinant need not be an
eigenfunction of $\purple{\chS^2}$, and the HF wavefunction is no longer a
pure spin function. Singlets may be contaminated by triplets, quintets, and
higher-spin states, while doublets may be contaminated by quartets, sextets,
and higher-spin states.
Generalized Hartree--Fock (GHF) also permits spin mixing and need not preserve
$\violet{\chS_z}$. Its orbitals have the form
\begin{equation}
  \chi_p^\text{GHF}(\bx)
  =\alpha(\omega)\blue{\psi_p^\alpha(\br)}
  +\beta(\omega)\red{\psi_p^\beta(\br)}
  \qquad\Rightarrow\qquad
  \text{two-component spinor}
\end{equation}
The additional variational freedom can lower the energy, but it changes the
symmetry content of the determinant. In particular, the distinction between a
useful broken-symmetry energy and a spin-pure wavefunction must be kept in mind
throughout the following discussion.

\subsubsection{The unrestricted Hartree--Fock equations and energy}

The UHF determinant contains two independently optimized orthonormal orbital
sets. Variation with respect to these two sets gives two coupled orbital
equations. To minimize the UHF energy, the unrestricted spatial orbitals must be
eigenfunctions of the separate \red{$\alpha$ and $\beta$ Fock operators}:
\begin{align}
  &
    \blue{\hf^\alpha(\br_1)\psi_i^\alpha(\br_1)
    =\epsilon_i^\alpha\psi_i^\alpha(\br_1)}
  &
  &
    \red{\hf^\beta(\br_1)\psi_i^\beta(\br_1)
    =\epsilon_i^\beta\psi_i^\beta(\br_1)}
\end{align}
The two Fock operators are
\begin{equation}
\begin{split}
  \blue{\hf^\alpha(\br_1)}
  &=\hh(\br_1)
    +\sum_i^{N^\alpha}
      \qty[\blue{\chJ_i^\alpha(\br_1)-\chK_i^\alpha(\br_1)}]
    +\sum_i^{N^\beta}\red{\chJ_i^\beta(\br_1)}
  \\
  \red{\hf^\beta(\br_1)}
  &=\hh(\br_1)
    +\sum_i^{N^\beta}
      \qty[\red{\chJ_i^\beta(\br_1)-\chK_i^\beta(\br_1)}]
    +\sum_i^{N^\alpha}\blue{\chJ_i^\alpha(\br_1)}
\end{split}
\end{equation}
The \green{Coulomb} and \red{exchange} operators are defined by
\begin{equation}
\begin{split}
  \blue{\chJ_i^\sigma(\br_1)}
  &=\int\psi_i^{\sigma *}(\br_2)r_{12}^{-1}
    \psi_i^\sigma(\br_2)\dd{\br_2}
  \\
  \violet{\chK_i^\sigma(\br_1)}\psi_j^\sigma(\br_1)
  &=\qty[\int\psi_i^{\sigma *}(\br_2)r_{12}^{-1}
    \psi_j^\sigma(\br_2)\dd{\br_2}]
    \psi_i^\sigma(\br_1)
\end{split}
\end{equation}
Each spin population contributes a Coulomb field felt by electrons of either
spin. Exchange, however, acts only between electrons carrying the same spin
function because opposite spin functions are orthogonal. Consequently, the
two Fock operators differ whenever the $\alpha$ and $\beta$ densities differ,
yet they remain coupled through their mutual Coulomb contributions.

\paragraph{The UHF energy.}

The UHF energy comprises three contributions:
\begin{equation}
  \boxed{
  E_\text{UHF}
  =\blue{E_\text{UHF}^{\alpha\alpha}}
  +\red{E_\text{UHF}^{\beta\beta}}
  +\violet{E_\text{UHF}^{\alpha\beta}}
  }
\end{equation}
Explicitly,
\begin{equation}
  E_\text{UHF}
  =\blue{\sum_i^{N^\alpha}h_{ii}^\alpha
    +\frac{1}{2}\sum_{ij}^{N^\alpha}
      \qty(J_{ij}^{\alpha\alpha}-K_{ij}^{\alpha\alpha})}
  \\
  +\red{\sum_i^{N^\beta}h_{ii}^\beta
    +\frac{1}{2}\sum_{ij}^{N^\beta}
      \qty(J_{ij}^{\beta\beta}-K_{ij}^{\beta\beta})}
  \\
  +\violet{\sum_i^{N^\alpha}\sum_j^{N^\beta}
    J_{ij}^{\alpha\beta}}
\end{equation}
The matrix elements are
\begin{align}
  h_{ii}^\sigma
  &=\mel{\psi_i^\sigma}{\hh}{\psi_i^\sigma}
  &
  J_{ij}^{\sigma\sigma'}
  &=\braket*{\psi_i^\sigma\psi_j^{\sigma'}}
    {\psi_i^\sigma\psi_j^{\sigma'}}
  &
  K_{ij}^{\sigma\sigma}
  &=\braket*{\psi_i^\sigma\psi_j^\sigma}
    {\psi_j^\sigma\psi_i^\sigma}
\end{align}
There is no exchange between opposite-spin electrons.
The factors of one half in the same-spin contributions remove double counting
of electron pairs. The opposite-spin Coulomb term is written only once as a
double sum over the two populations and therefore requires no such factor.

\begin{hfexercise}{UHF energy of the \ce{Li} atom}

\violet{\textit{Write down the UHF energy of the doublet state of the
lithium atom.}}
Figure~\ref{fig:li-uhf-occupation} specifies the occupation for this
exercise.
\begin{center}
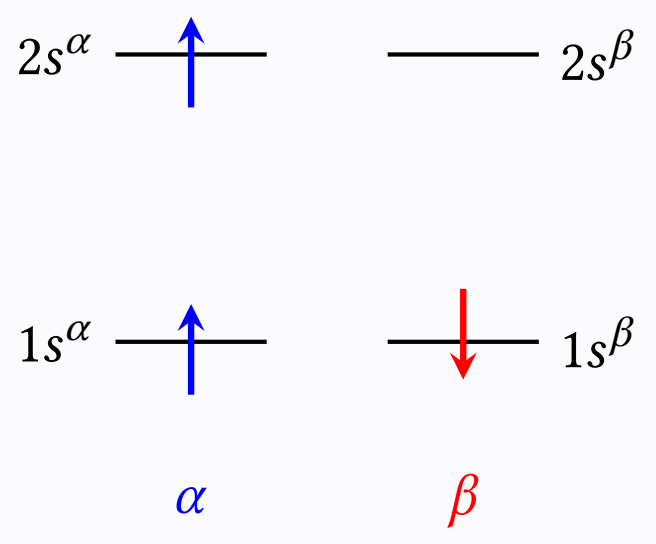

\begin{tikzpicture}[>=stealth]
  \draw[thick] (-1.4,2.2)--(-0.4,2.2);
  \draw[thick] (0.4,2.2)--(1.4,2.2);
  \node[left] at (-1.4,2.2) {$2s^\alpha$};
  \node[right] at (1.4,2.2) {$2s^\beta$};

  \draw[thick] (-1.4,0.3)--(-0.4,0.3);
  \draw[thick] (0.4,0.3)--(1.4,0.3);
  \node[left] at (-1.4,0.3) {$1s^\alpha$};
  \node[right] at (1.4,0.3) {$1s^\beta$};

  \draw[->,very thick,blue] (-0.9,1.85)--(-0.9,2.45);
  \draw[->,very thick,blue] (-0.9,-0.05)--(-0.9,0.55);
  \draw[->,very thick,red] (0.9,0.65)--(0.9,0.05);
  \node[blue] at (-0.9,-0.75) {$\alpha$};
  \node[red] at (0.9,-0.75) {$\beta$};
\end{tikzpicture}
\captionof{figure}{Occupation of the doublet lithium determinant used in the UHF
exercise: $1s^\alpha$, $1s^\beta$, and $2s^\alpha$ are occupied,
whereas $2s^\beta$ is empty.}
\label{fig:li-uhf-occupation}
\end{center}

\end{hfexercise}
\begin{hfsolution}{UHF energy of the \ce{Li} atom}

The occupied spatial indices are $i=1,2$ for $\alpha$ spin and $i=1$
for $\beta$ spin, so $N^\alpha=2$ and $N^\beta=1$. The functions
$\psi_1^\alpha$ and $\psi_1^\beta$ need not coincide. Each spatial
function is normalized, and the two occupied $\alpha$ functions are
mutually orthogonal. No spatial orthogonality between opposite-spin
orbitals is required because their spin functions are orthogonal. The
one-electron contribution is
\begin{equation}
  E_1=h_{11}^\alpha+h_{11}^\beta+h_{22}^\alpha
\end{equation}
For the $\alpha$--$\alpha$ interaction, the diagonal terms cancel through
$J_{ii}^{\alpha\alpha}=K_{ii}^{\alpha\alpha}$, while the two ordered
off-diagonal pairs give the same contribution:
\begin{equation}
\begin{split}
  E_2^{\alpha\alpha}
  &=\frac12\sum_{i,j=1}^{2}
    \qty(J_{ij}^{\alpha\alpha}-K_{ij}^{\alpha\alpha})\\
  &=\frac12\qty[
    J_{12}^{\alpha\alpha}-K_{12}^{\alpha\alpha}
    +J_{21}^{\alpha\alpha}-K_{21}^{\alpha\alpha}]\\
  &=J_{12}^{\alpha\alpha}-K_{12}^{\alpha\alpha}
\end{split}
\end{equation}
There is only one $\beta$ electron, so its same-spin two-electron
contribution is zero. The opposite-spin interaction contains two distinct
pairs and no exchange:
\begin{equation}
  E_2^{\alpha\beta}
  =\sum_{i=1}^{2}J_{i1}^{\alpha\beta}
  =J_{11}^{\alpha\beta}+J_{21}^{\alpha\beta}
\end{equation}
Adding these contributions gives, for the occupation shown above,
\begin{equation}
  E_\text{UHF}
  =h_{11}^\alpha+h_{11}^\beta+h_{22}^\alpha
  +J_{12}^{\alpha\alpha}-K_{12}^{\alpha\alpha}
  +J_{11}^{\alpha\beta}+J_{21}^{\alpha\beta}
\end{equation}
The same-spin interaction between the $1s^\alpha$ and $2s^\alpha$ electrons
contains both Coulomb and exchange contributions. The two interactions with
the $1s^\beta$ electron are purely Coulombic, exactly as required by the spin
selection rule above.
As a check, impose the common-spatial-orbital limit
$\psi_1^\alpha=\psi_1^\beta=\psi_1$ and $\psi_2^\alpha=\psi_2$.
Then $J_{11}^{\alpha\beta}=J_{11}$ and
$J_{21}^{\alpha\beta}=J_{12}$, and the energy reduces to
\begin{equation}
  E_\text{UHF}
  =2h_{11}+h_{22}+J_{11}+2J_{12}-K_{12}
\end{equation}
This recovers the earlier lithium exercise. The unrestricted expression
differs by allowing the two spin populations to optimize their spatial
functions independently, not by changing which electron pairs interact.

\end{hfsolution}
\subsubsection{The Pople--Nesbet--Berthier matrix equations}

\hfBioPopleNesbet
\hfBioBerthier

The continuous UHF equations are reduced to matrix form in the same way as
the restricted equations, except that a separate coefficient matrix and Fock
matrix are required for each spin. The unrestricted spatial orbitals are
expanded in the same AO basis with
separate spin-dependent coefficients:
\begin{align}
  \psi_i^\alpha(\br)
  &=\sum_{\mu=1}^{K}\blue{C_{\mu i}^\alpha}\phi_\mu(\br)
  &
  \psi_i^\beta(\br)
  &=\sum_{\mu=1}^{K}\red{C_{\mu i}^\beta}\phi_\mu(\br)
\end{align}
The resulting Pople--Nesbet--Berthier equations, independently formulated
in 1954 \citep{PopleNesbet1954,Berthier1954}, are
\begin{align}
  &\boxed{\blue{\bF^\alpha}\cdot\blue{\bC^\alpha}
  =\bS\cdot\blue{\bC^\alpha}\cdot\beps^\alpha}
  \\
  &\boxed{
  \red{\bF^\beta}\cdot\red{\bC^\beta}
  =\bS\cdot\red{\bC^\beta}\cdot\beps^\beta}
\end{align}
Their spin-dependent Fock matrices are
\begin{align}
  \blue{F_{\mu\nu}^\alpha}
  &=H_{\mu\nu}
  +\sum_{\lambda\sigma}\blue{P_{\lambda\sigma}^\alpha}
    \qty[
      \rbraket{\mu\nu}{\sigma\lambda}
      -\rbraket{\mu\lambda}{\sigma\nu}
    ]
  +\sum_{\lambda\sigma}\red{P_{\lambda\sigma}^\beta}
    \rbraket{\mu\nu}{\sigma\lambda}
  \\
  \red{F_{\mu\nu}^\beta}
  &=H_{\mu\nu}
  +\sum_{\lambda\sigma}\red{P_{\lambda\sigma}^\beta}
    \qty[
      \rbraket{\mu\nu}{\sigma\lambda}
      -\rbraket{\mu\lambda}{\sigma\nu}
    ]
  +\sum_{\lambda\sigma}\blue{P_{\lambda\sigma}^\alpha}
    \rbraket{\mu\nu}{\sigma\lambda}
\end{align}
Both $\blue{\bF^\alpha}$ and $\red{\bF^\beta}$ depend on
$\blue{\bC^\alpha}$ and $\red{\bC^\beta}$. Therefore, the $\alpha$
and $\beta$ orbitals are coupled.
Thus the two generalized eigenvalue problems cannot be solved independently.
Updating either spin density changes the Coulomb potential in both Fock
matrices and changes the exchange potential in the Fock matrix of the same
spin.

\paragraph{Unrestricted density matrices.}

The spin-up and spin-down density matrices are
\begin{align}
  &
    \blue{P_{\mu\nu}^\alpha}
    =\sum_{i=1}^{N^\alpha}C_{\mu i}^\alpha C_{\nu i}^{\alpha *}
    \quad\Longleftrightarrow\quad\blue{\bP^\alpha}
  &
  &
    \red{P_{\mu\nu}^\beta}
    =\sum_{i=1}^{N^\beta}C_{\mu i}^\beta C_{\nu i}^{\beta *}
    \quad\Longleftrightarrow\quad\red{\bP^\beta}
\end{align}
For $\sigma=\alpha$ or $\beta$, the corresponding real-space density and
its normalization are
\begin{align}
  \rho^\sigma(\br)
  &=\sum_{\mu\nu}\phi_\mu(\br)P_{\mu\nu}^\sigma
    \phi_\nu^*(\br)
  &
  \int\rho^\sigma(\br)\dd{\br}
  &=N^\sigma
\end{align}
Finally, the total and spin density matrices are
\begin{align}
  \underbrace{\blue{\bP^\text{T}}}_{\blue{\text{charge density}}}
  &=\blue{\bP^\alpha}+\red{\bP^\beta}
  &
  \underbrace{\violet{\bP^\text{S}}}_{\violet{\text{spin density}}}
  &=\blue{\bP^\alpha}-\red{\bP^\beta}
\end{align}
The sum $\bP^\text{T}$ determines the total charge density. The difference
$\bP^\text{S}$ measures the spatial imbalance between the two spin
populations and is the matrix representation of the collinear spin density.

\subsubsection{The unrestricted SCF procedure}

The practical UHF SCF algorithm mirrors the restricted procedure, but both
spin densities must be propagated until they are mutually self-consistent. It
is summarized in Algorithm~\ref{alg:uhf-scf-summary}, using the same step
organization as Algorithm~\ref{alg:rhf-scf-summary}. The calculation
begins by specifying the molecule through $\{\bR_A\}$ and $\{Z_A\}$
and choosing the basis set $\{\phi_\mu\}$, as in RHF. The integrals
$S_{\mu\nu}$, $H_{\mu\nu}$, and
$\braket{\mu\nu}{\lambda\sigma}$ are calculated, and $\bS$ is diagonalized
to obtain $\bX$.

After initial density matrices $\bP^\alpha$ and $\bP^\beta$ have been
chosen (Step~\ref{algstep:uhf-guess}), the two spin-dependent matrix builds
in Step~\ref{algstep:uhf-build} are performed:
$\bG^\alpha$ gives $\bF^\alpha=\bH+\bG^\alpha$, while $\bG^\beta$ gives
$\bF^\beta=\bH+\bG^\beta$. The transformed matrices are
$(\bF^\alpha)'=\bX^\dag\cdot\bF^\alpha\cdot\bX$ and
$(\bF^\beta)'=\bX^\dag\cdot\bF^\beta\cdot\bX$. Diagonalizing them yields
$(\bC^\alpha)'$ and $\beps^\alpha$, together with $(\bC^\beta)'$ and
$\beps^\beta$. The AO coefficients are then recovered from
$\bC^\alpha=\bX\cdot(\bC^\alpha)'$ and
$\bC^\beta=\bX\cdot(\bC^\beta)'$. These coefficients define the
new density matrices $\bP^\alpha$ and $\bP^\beta$, from which
$\bP^\text{T}=\bP^\alpha+\bP^\beta$ is formed. If the convergence test is
not satisfied, the cycle returns to the construction of $\bG^\alpha$ and
$\bG^\beta$; otherwise, the desired observables, including
$E_\text{UHF}$, are evaluated.

\begin{hfscfsummary}[title={Algorithm~\thetcbcounter\enspace\textbar\ The UHF self-consistent-field procedure},phantom={\label{alg:uhf-scf-summary}}]
\textcolor{hfInk}{Collinear UHF with fixed $N^\alpha$ and $N^\beta$;
$\sigma=\blue{\alpha},\red{\beta}$. Each $\bP^\sigma$ counts only its own
spin population, with no factor of two.}
\medskip
\colorlet{hfSCFStep}{hfGreenDark}
\algrenewcommand\alglinenumber[1]{%
  \tikz[baseline=(scfnumber.base)]
    \node[circle,fill=hfSCFStep,text=white,inner sep=0pt,
      minimum size=4.4mm,font=\scriptsize\bfseries] (scfnumber) {#1};}
\begin{algorithmic}[1]
  \Statex \colorbox{hfGreenLight}{\strut\textcolor{hfGreenDark}{\bfseries
    SETUP\enspace\textbar\ Prepare once at a fixed geometry}}
  \State \textbf{Specify the calculation.}\phantomsection\label{algstep:uhf-input}
    Choose the geometry, AO basis, spin populations
    $N^\alpha+N^\beta=N$, and convergence thresholds.
  \State \textbf{Prepare integrals and the common AO metric.}\phantomsection\label{algstep:uhf-metric}
    Calculate $\bS$, $\bH$, and the ERIs; use the same $\bX$ for both spins,
    with $\bX^\dag\cdot\bS\cdot\bX=\bI$.
  \State \textbf{Choose two initial densities.}\phantomsection\label{algstep:uhf-guess}
    Supply $\bP^{\alpha,(0)}$ and $\bP^{\beta,(0)}$; set $n=0$.
  \Statex \smallskip
    \colorbox{hfBlue!7}{\strut\textcolor{hfBlue}{\bfseries
      ITERATION\enspace\textbar\ Repeat Steps 4--7 until self-consistency}}
  \colorlet{hfSCFStep}{hfBlue}
  \State \textbf{Build both physical Fock matrices.}\phantomsection\label{algstep:uhf-build}
    Both densities contribute to Coulomb; exchange is same-spin only:
  \Statex $\displaystyle
    \bF^{\sigma,(n)}
      =\bF^\sigma[\bP^{\alpha,(n)},\bP^{\beta,(n)}]$
  \State \textbf{Evaluate the energy and test both spin channels.}\phantomsection\label{algstep:uhf-test}
    Use the physical matrices from Step~\ref{algstep:uhf-build}:
  \Statex $\displaystyle
    E_\text{UHF}^{(n)}=\frac12\sum_{\sigma=\alpha,\beta}
      \Tr\qty[\bP^{\sigma,(n)}\cdot(\bH+\bF^{\sigma,(n)})]$
  \Statex $\displaystyle
    \be^{\sigma,(n)}
      =\bF^{\sigma,(n)}\cdot\bP^{\sigma,(n)}\cdot\bS
      -\bS\cdot\bP^{\sigma,(n)}\cdot\bF^{\sigma,(n)}$
  \Statex After at least one orbital update, require
    $\max_{\sigma=\alpha,\beta}\max\abs*{\be^{\sigma,(n)}}<\tau_\text{SCF}$
    and the chosen energy/density change criteria for both spins.
    If satisfied, go to Step~\ref{algstep:uhf-output}.
  \State \textbf{Solve for new orbitals in each spin channel.}\phantomsection\label{algstep:uhf-solve}
  \Statex $\displaystyle
    (\bF^{\sigma,(n)})'
      =\bX^\dag\cdot\bF^{\sigma,(n)}\cdot\bX$
  \Statex $\displaystyle
    (\bF^{\sigma,(n)})'\cdot(\bC^{\sigma,(n+1)})'
      =(\bC^{\sigma,(n+1)})'\cdot\beps^{\sigma,(n+1)}$
  \Statex $\displaystyle
    \bC^{\sigma,(n+1)}=\bX\cdot(\bC^{\sigma,(n+1)})'$
  \State \textbf{Occupy and update both densities.}\phantomsection\label{algstep:uhf-update}
    Occupy the $N^\sigma$ lowest orbitals of each spin and form
  \Statex $\displaystyle
    \bP^{\sigma,(n+1)}=\bC_\text{o}^{\sigma,(n+1)}\cdot
      (\bC_\text{o}^{\sigma,(n+1)})^\dag$
  \Statex Set $n\leftarrow n+1$ and return to
    Step~\ref{algstep:uhf-build}, rebuilding both physical Fock matrices.
  \Statex \smallskip
    \colorbox{hfGreenLight}{\strut\textcolor{hfGreenDark}{\bfseries
      OUTPUT\enspace\textbar\ Only after convergence}}
  \colorlet{hfSCFStep}{hfGreenDark}
  \State \textbf{Evaluate the converged observables.}\phantomsection\label{algstep:uhf-output}
    Report $E_\text{UHF}$, orbitals, charge/spin densities, and
    $\expval*{\chS^2}$. Add nuclear repulsion for the total molecular energy.
\end{algorithmic}
\end{hfscfsummary}

Convergence therefore requires a fixed point of the coupled map
\begin{equation}
  (\bP^\alpha,\bP^\beta)
  \longrightarrow
  (\bF^\alpha,\bF^\beta)
  \longrightarrow
  (\bP^\alpha,\bP^\beta)
\end{equation}
Different initial spin densities can lead to different stationary solutions,
especially where restricted spin symmetry becomes unstable.
The joint test in Step~\ref{algstep:uhf-test} is essential: convergence of
only one spin channel does not solve the UHF equations. Updating either
density in Step~\ref{algstep:uhf-update} requires rebuilding both physical
Fock matrices before testing the pair again.

\subsection{Symmetry breaking in the dissociation of \ce{H2}}

The dissociation of the hydrogen molecule is the simplest illustration of
both the usefulness and the limitation of unrestricted Hartree--Fock. It
shows explicitly how relaxing the equality of the two spatial orbitals repairs
the restricted dissociation energy while producing a wavefunction that is no
longer a pure eigenfunction of $\chS^2$.
Start from normalized $1s$ functions $\phi_A$ and $\phi_B$, one on each atom.
Their symmetric and antisymmetric combinations define the gerade bonding
orbital $\psi_g$ and ungerade antibonding orbital $\psi_u$:
\begin{align}
  \psi_g&=\frac{\phi_A+\phi_B}{\sqrt{2(1+S)}}
  &
  \psi_u&=\frac{\phi_A-\phi_B}{\sqrt{2(1-S)}}
  &
  S&=\braket{\phi_A}{\phi_B}
\end{align}
Here $S$ is the overlap between the two atomic functions, not the total-spin
quantum number; $R$ below is the internuclear distance. For two electrons in
two spatial orbitals, the UHF parametrization used below reduces to a
single-parameter problem.
Figure~\ref{fig:h2-two-centers} defines the two-center geometry and
the associated atomic basis functions.
\begin{center}
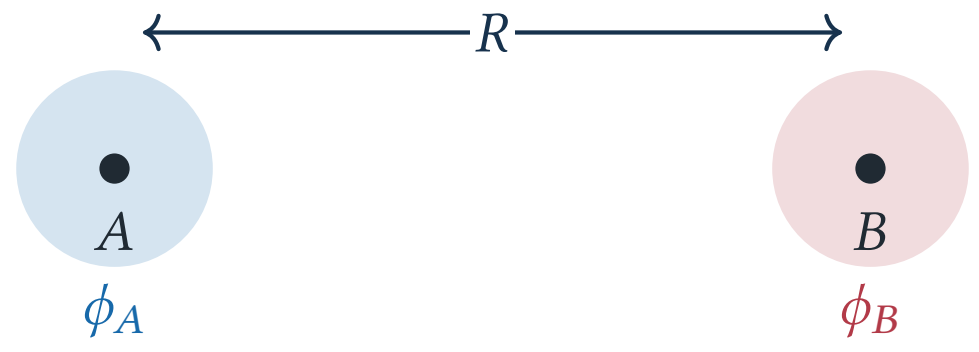

\begin{tikzpicture}[x=1.25cm,y=1.25cm]
  \fill[hfBlue!18] (-2.0,0) circle (0.52);
  \fill[hfRed!18] (2.0,0) circle (0.52);
  \fill[hfInk] (-2.0,0) circle (0.08) node[below=0.15cm] {$A$};
  \fill[hfInk] (2.0,0) circle (0.08) node[below=0.15cm] {$B$};
  \draw[hfNavy,thick,<->] (-1.85,0.72)--(1.85,0.72)
    node[midway,fill=white,inner sep=1pt] {$R$};
  \node[hfBlue] at (-2.0,-0.75) {$\phi_A$};
  \node[hfRed] at (2.0,-0.75) {$\phi_B$};
\end{tikzpicture}
\captionof{figure}{Minimal two-center description of $\ce{H2}$: atomic functions
$\phi_A$ and $\phi_B$ are centered on nuclei $A$ and $B$, separated
by the internuclear distance $R$.}
\label{fig:h2-two-centers}
\end{center}

\subsubsection{The restricted dissociation limit}

In the restricted description both electrons occupy the gerade bonding
orbital. The RHF wavefunction and energy are
\begin{align}
  \ket{\Psi_\text{RHF}}&=\ket{\psi_g\Bar{\psi}_g}
  &
  E_\text{RHF}&=2h_{gg}+J_{gg}+\frac{1}{R}
\end{align}
At large $R$, where $S\rightarrow0$,
\begin{equation}
  \psi_g(1)\psi_g(2)\longrightarrow\frac{1}{2}\qty[
    \underbrace{\phi_A(1)\phi_B(2)+\phi_B(1)\phi_A(2)}
      _{\text{covalent}}
    +\underbrace{\phi_A(1)\phi_A(2)+\phi_B(1)\phi_B(2)}
      _{\text{ionic}}
  ]
\end{equation}
RHF retains \red{$50\%$ ionic character} as $R\rightarrow\infty$. Both
electrons are forced to share the same spatial orbital, so its energy lies
above that of two separated H atoms. Hence, RHF is qualitatively correct
near equilibrium but not at dissociation.
The defect is not caused by an inadequate optimization of the bonding orbital
within RHF. It follows from the restricted ansatz itself: placing both
electrons in the same spatial orbital inevitably retains the ionic
configurations together with the covalent ones.

\subsubsection{Unrestricted symmetry breaking and the Coulson--Fischer point}

\hfBioCoulsonFischer

The two-dimensional orbital space spanned by $\psi_g$ and $\psi_u$ permits
the $\alpha$ and $\beta$ orbitals to rotate in opposite directions. The UHF
orbitals can be parametrized as
\begin{align}
    \psi^\alpha(\theta)&=\cos\theta\,\psi_g+\sin\theta\,\psi_u
    &
    \psi^\beta(\theta)&=\cos\theta\,\psi_g-\sin\theta\,\psi_u
\end{align}
with $\braket{\psi^\alpha}{\psi^\beta}=\cos(2\theta)$.
The two choices $\pm\theta$ are degenerate and exchange the spin
localization.
The parameter $\theta$ therefore provides a continuous path from the
restricted determinant at $\theta=0$ to orthogonal spatial orbitals at
$\theta=\pi/4$. In the dissociation limit, these endpoint orbitals become
localized on separate atoms. The orbital overlap $\cos(2\theta)$ decreases
along this path and vanishes at its endpoint; at finite atomic overlap,
orthogonality alone does not imply complete localization on individual atoms.
Figure~\ref{fig:h2-orbital-rotation} illustrates the progressive
localization of the two spin channels along this path.
\begin{center}
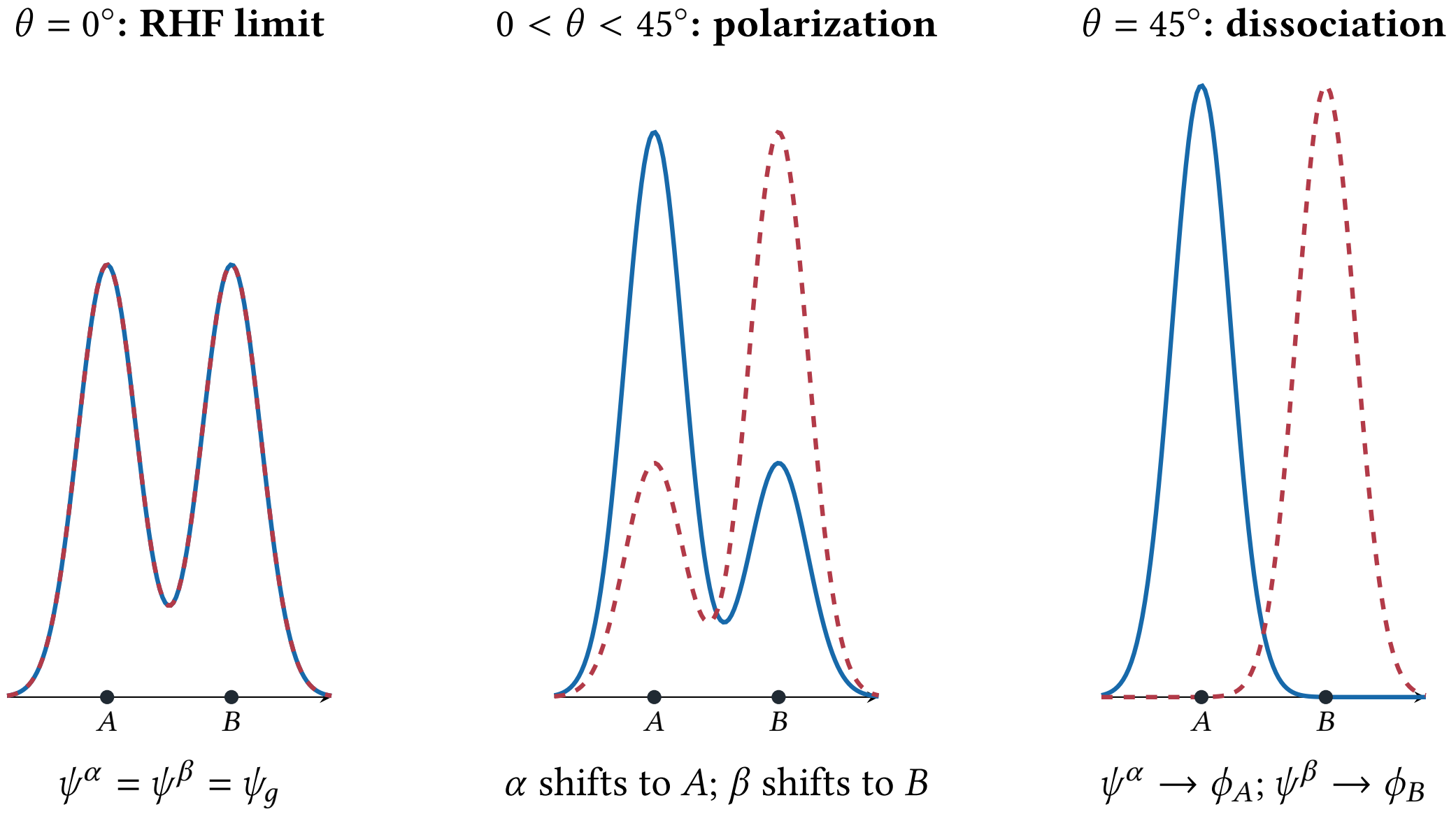

\begin{minipage}{\textwidth}
\centering
\begin{minipage}[t]{0.31\textwidth}
  \centering\strut\textbf{$\theta=0^\circ$: RHF limit}

  \HtwoOrbitalPlot{0}

  \strut$\psi^\alpha=\psi^\beta=\psi_g$
\end{minipage}\hfill
\begin{minipage}[t]{0.31\textwidth}
  \centering\strut\textbf{$0<\theta<45^\circ$: polarization}

  \HtwoOrbitalPlot{22.5}

  \strut$\alpha$ shifts to $A$; $\beta$ shifts to $B$
\end{minipage}\hfill
\begin{minipage}[t]{0.31\textwidth}
  \centering\strut\textbf{$\theta=45^\circ$: dissociation}

  \HtwoOrbitalPlot{45}

  \strut$\psi^\alpha\to\phi_A$; $\psi^\beta\to\phi_B$
\end{minipage}
\captionof{figure}{Schematic spatial profiles of the $\alpha$ (solid blue) and
$\beta$ (dashed red) UHF orbitals under opposite rotations by $\theta$.
The three panels show the restricted limit, partial spin polarization,
and localization on separate atoms at dissociation.}
\label{fig:h2-orbital-rotation}
\end{minipage}
\end{center}

With $J_{pq}=\rbraket{pp}{qq}$ and $K_{pq}=\rbraket{pq}{pq}$, the
angle-dependent electronic energy is given below. Unlike the preceding RHF
total-energy expression, this polynomial omits the nuclear constant $1/R$.
That constant does not affect minimization with respect to $\theta$ at
fixed $R$, but must be added when comparing total energies as $R$ varies.
\begin{equation}
  E_\text{UHF}(\theta)
  =2\cos^2\theta\,h_{gg}
  +2\sin^2\theta\,h_{uu}
  +\cos^4\theta\,J_{gg}
  +\sin^4\theta\,J_{uu}
  \\
  +2\sin^2\theta\cos^2\theta\qty(J_{gu}-2K_{gu})
\end{equation}
Introducing
\begin{align}
  x&=\cos^2\theta
  &
  A&=J_{gu}-2K_{gu}
\end{align}
allows the same energy to be written as
\begin{equation}
  E_\text{UHF}(x)
  =
  2xh_{gg}+2(1-x)h_{uu}
  +x^2J_{gg}+(1-x)^2J_{uu}
  +2x(1-x)A
\end{equation}
Its derivative is
\begin{equation}
  \dv{E_\text{UHF}}{x}
  =
  2\qty[
  h_{gg}-h_{uu}-J_{uu}+A
  +x\qty(J_{gg}+J_{uu}-2A)
  ]
\end{equation}
Because $\dv*{x}{\theta}=-\sin(2\theta)$, stationarity requires either
$\sin(2\theta)=0$ or $\dv*{E_\text{UHF}}{x}=0$. The first condition contains
the restricted solution $\theta=0$; the second gives the broken-symmetry
solution.
Besides the restricted stationary point $\theta=0$, a
broken-symmetry stationary solution satisfies
\begin{equation}
    \cos^2\theta^\star
    =\frac{h_{uu}-h_{gg}+J_{uu}-J_{gu}+2K_{gu}}
      {J_{gg}+J_{uu}-2J_{gu}+4K_{gu}}
\end{equation}
At short bond lengths, only the minimum at $\theta=0$ is physically
accessible and UHF reduces to RHF. On the dissociating branch, the
coefficient $J_{gg}+J_{uu}-2J_{gu}+4K_{gu}$ is positive. When
$0<\cos^2\theta^\star<1$, two minima appear at $\pm\theta^\star$ and
RHF becomes unstable to spin polarization.
The endpoint $\cos^2\theta^\star=1$ is the point at which these solutions
meet RHF; positivity of the quadratic coefficient distinguishes a minimum
from a maximum of $E_\text{UHF}(x)$.

\begin{figure}[htbp]
\centering
\begin{minipage}[t]{0.31\textwidth}
\centering\textbf{$R<R_\text{CF}$}
\begin{tikzpicture}
\begin{axis}[width=\linewidth,height=0.65\linewidth,xmin=-45,xmax=45,
ymin=-0.06,ymax=1.08,axis x line=bottom,axis y line=left,
xtick={-45,0,45},xticklabels={$-45^\circ$,$0$,$45^\circ$},ytick=\empty,
xlabel={$\theta$},ylabel={$E(\theta)$},samples=120,domain=-45:45]
\addplot[hfNavy,very thick] {(x/45)^2};
\addplot[only marks,mark=*,mark size=2.2pt,hfTeal] coordinates {(0,0)};
\end{axis}
\end{tikzpicture}

\green{one minimum at $\theta=0$}
\end{minipage}\hfill
\begin{minipage}[t]{0.31\textwidth}
\centering\textbf{$R=R_\text{CF}$}
\begin{tikzpicture}
\begin{axis}[width=\linewidth,height=0.65\linewidth,xmin=-45,xmax=45,
ymin=-0.06,ymax=1.08,axis x line=bottom,axis y line=left,
xtick={-45,0,45},xticklabels={$-45^\circ$,$0$,$45^\circ$},ytick=\empty,
xlabel={$\theta$},samples=120,domain=-45:45]
\addplot[hfViolet,very thick] {(x/45)^4};
\addplot[only marks,mark=*,mark size=2.2pt,hfRed] coordinates {(0,0)};
\end{axis}
\end{tikzpicture}

the minimum becomes flat
\end{minipage}\hfill
\begin{minipage}[t]{0.31\textwidth}
\centering\textbf{$R>R_\text{CF}$}
\begin{tikzpicture}
\begin{axis}[width=\linewidth,height=0.65\linewidth,xmin=-45,xmax=45,
ymin=-0.06,ymax=1.08,axis x line=bottom,axis y line=left,
xtick={-45,0,45},xticklabels={$-45^\circ$,$0$,$45^\circ$},ytick=\empty,
xlabel={$\theta$},samples=120,domain=-45:45]
\addplot[hfRed,very thick] {0.65*((x/30)^2-1)^2};
\addplot[only marks,mark=*,mark size=2.2pt,hfTeal]
coordinates {(-30,0) (30,0)};
\addplot[only marks,mark=o,mark size=2.5pt,hfRed] coordinates {(0,0.65)};
\end{axis}
\end{tikzpicture}

\green{two UHF minima at $\pm\theta^\star$}
\end{minipage}
\caption{Evolution of the stationary points at the Coulson--Fischer point.}
\label{fig:coulson-fischer}
\end{figure}
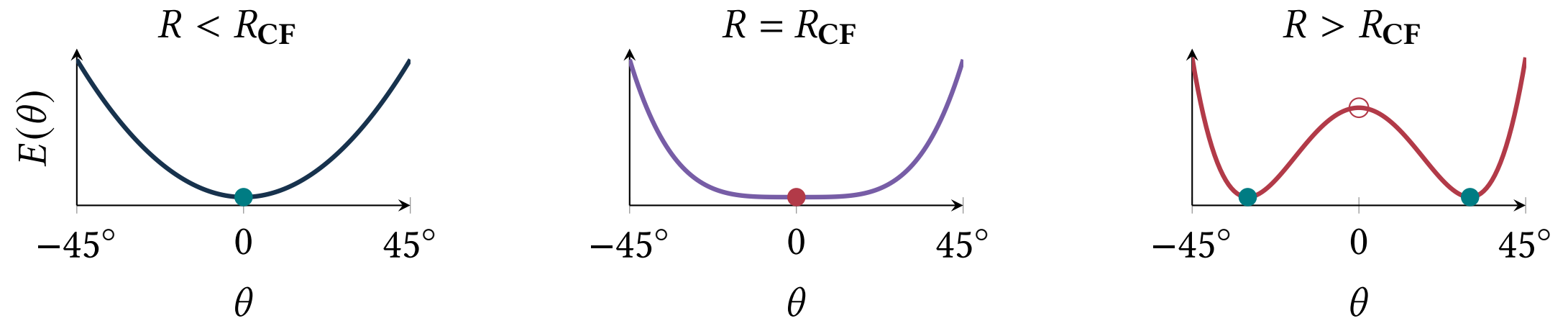

The RHF curvature is
\begin{equation}
  \eval{\dv[2]{E}{\theta}}_{\theta=0}
  =4\qty(h_{uu}-h_{gg}-J_{gg}+J_{gu}-2K_{gu})
\end{equation}
and the Coulson--Fischer point \citep{CoulsonFischer1949} is defined by
\begin{equation}
    R=R_\text{CF}
    \quad\Longleftrightarrow\quad
    \eval{\dv[2]{E}{\theta}}_{\theta=0}=0
\end{equation}
For $R<R_\text{CF}$, the RHF curvature is positive and there is one minimum
at $\theta=0$. At $R=R_\text{CF}$, it vanishes and the minimum becomes flat.
For $R>R_\text{CF}$ it is negative and two UHF minima occur at
$\pm\theta^\star$. The RHF determinant does not disappear: it changes from a
minimum to an unstable stationary point while two degenerate UHF minima
emerge.
This is a bifurcation of stationary solutions. The vanishing curvature marks
the geometry at which an infinitesimal spin-polarizing orbital rotation ceases
to raise the energy and begins to lower it.
The three regimes are illustrated in Fig.~\ref{fig:coulson-fischer}.

\begin{figure}[htbp]
\centering
\begin{tikzpicture}
\begin{axis}[width=0.72\textwidth,height=0.42\textwidth,
xmin=0.45,xmax=6.15,ymin=-1.02,ymax=0.80,xlabel={$R$},
ylabel={$E(R)-2E_\text{H}$},xtick={2.30},xticklabels={$R_\text{CF}$},
ytick={0,0.55},yticklabels={$0$,$\tfrac12U$},
legend style={draw=none,fill=white,font=\scriptsize,
at={(0.98,0.03)},anchor=south east},clip=false]
\addplot[hfNavy,very thick,smooth] coordinates {
(0.55,0.72) (0.80,-0.18) (1.15,-0.68) (1.45,-0.79)
(1.75,-0.73) (2.05,-0.60) (2.30,-0.48) (2.65,-0.27)
(3.05,-0.02) (3.50,0.19) (4.00,0.35) (4.60,0.46)
(5.30,0.52) (6.05,0.55)};
\addlegendentry{RHF}
\addplot[hfRed,very thick,smooth] coordinates {
(0.55,0.72) (0.80,-0.18) (1.15,-0.68) (1.45,-0.79)
(1.75,-0.73) (2.05,-0.60) (2.30,-0.48) (2.55,-0.34)
(2.85,-0.22) (3.25,-0.13) (3.75,-0.065) (4.35,-0.027)
(5.05,-0.008) (6.05,0.00)};
\addlegendentry{UHF}
\addplot[hfTeal,thick,dashed,smooth] coordinates {
(0.55,0.67) (0.80,-0.24) (1.15,-0.76) (1.45,-0.88)
(1.75,-0.82) (2.10,-0.67) (2.50,-0.48) (2.95,-0.31)
(3.50,-0.17) (4.10,-0.080) (4.80,-0.030) (5.50,-0.008)
(6.05,0.00)};
\addlegendentry{near-exact}
\end{axis}
\end{tikzpicture}
\caption{Schematic RHF, UHF, and near-exact dissociation curves for
minimal-basis \ce{H2}.}
\label{fig:h2-dissociation}
\end{figure}
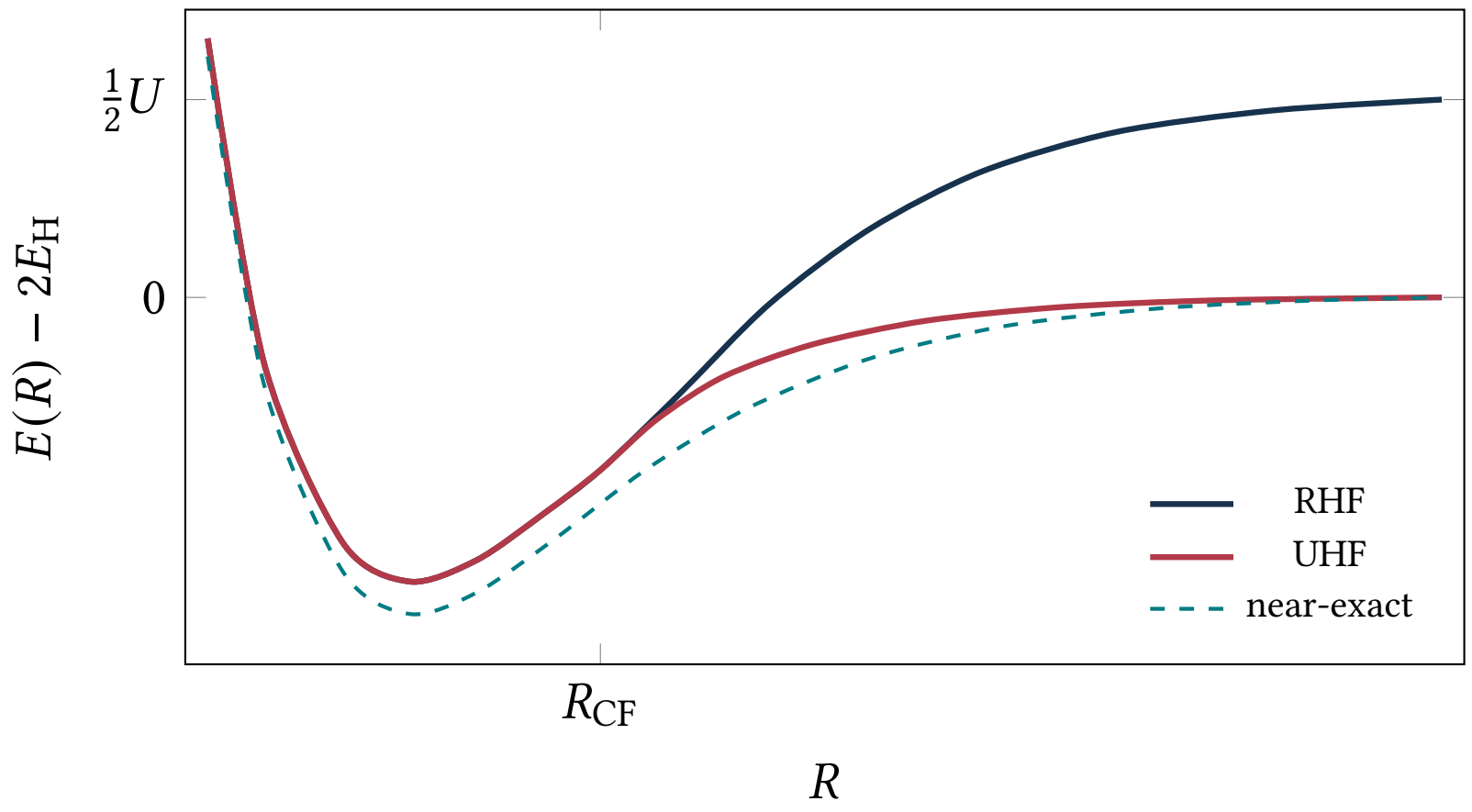

At dissociation,
\begin{align}
  S&\rightarrow0
  &
  \psi_g&\rightarrow\frac{\phi_A+\phi_B}{\sqrt2}
  &
  U&=(AA|AA)>0
\end{align}
Writing $E_\text{H}$ for the isolated-atom energy represented in the same
basis, the restricted and unrestricted total-energy limits are
\begin{align}
  E_\text{RHF}^\infty&=2E_\text{H}+\frac12U
  &
  E_\text{UHF}^\infty&=2E_\text{H}
\end{align}
The ionic terms therefore leave a finite RHF error, whereas
$\psi^\alpha\to\phi_A$ and $\psi^\beta\to\phi_B$ give the correct UHF limit.
Figure~\ref{fig:h2-dissociation} compares the resulting schematic dissociation
curves with a near-exact reference.

For the two-electron $M_S=0$ determinant,
\begin{equation}
    \expval{\purple{\chS^2}}_\text{UHF}
    =1-\abs{\braket{\psi^\alpha}{\psi^\beta}}^2
    =\sin^2(2\theta)
\end{equation}
Near equilibrium, $\theta=0$ and $\expval*{\purple{\chS^2}}=0$, so the
determinant is a pure singlet. At dissociation, $\theta\rightarrow\pi/4$ and
$\expval*{\purple{\chS^2}}\rightarrow1$, so it is an equal mixture of a
singlet and an $M_S=0$ triplet:
\begin{equation}
  \ket{A\alpha\,B\beta}
  =\frac{1}{\sqrt2}\qty(
    \ket{\text{covalent singlet}}
    +\ket{\text{triplet},M_S=0})
  \qquad\text{as }R\to\infty
\end{equation}
Thus, the UHF energy is physically useful, but the UHF wavefunction is
spin-contaminated. Three conclusions follow. A stationary SCF solution need
not be stable, the Coulson--Fischer point marks a continuous RHF-to-UHF
transition, and symmetry breaking mimics static correlation at the price of a
contaminated wavefunction.
The example exposes the characteristic compromise of broken-symmetry
mean-field theory. The determinant reproduces the correct separated-atom
energy through spin localization, but represents the singlet character only
through a mixture of spin eigenfunctions.

\subsection{Generalized Hartree--Fock theory}

Unrestricted Hartree--Fock remains collinear: every occupied orbital is either
purely $\alpha$ or purely $\beta$ with respect to one common quantization
axis. Generalized Hartree--Fock removes this remaining restriction. The key
idea is that GHF allows every orbital to choose its spin direction.
A GHF orbital is a two-component spinor:
\begin{equation}
  \boxed{
    \chi_p^\text{GHF}(\bx)
    =\alpha(\omega)\blue{\psi_p^\alpha(\br)}
    +\beta(\omega)\red{\psi_p^\beta(\br)}
  }
\end{equation}
Both components are expanded in the same spatial AO basis,
\begin{align}
  \psi_i^\sigma(\br)&=\sum_{\mu=1}^{K}C_{\mu i}^\sigma\phi_\mu(\br)
  &
  \boldsymbol{\mathcal C}&=\mqty(\bC^\alpha\\\bC^\beta)
  &
  \boldsymbol{\mathcal C}&\in\mathbb C^{2K\times N}
\end{align}
RHF pairs electrons in one spatial orbital and preserves both
$\purple{\chS^2}$ and $\violet{\chS_z}$. UHF uses orbitals that are purely
$\alpha$ or $\beta$, has a collinear spin density, and preserves
$\violet{\chS_z}$. GHF permits each orbital to mix $\alpha$ and $\beta$, so
in general neither $\purple{\chS^2}$ nor $\violet{\chS_z}$ is preserved.
GHF therefore does not require a common spin-quantization axis or fixed values
of $N^\alpha$ and $N^\beta$.
Figure~\ref{fig:ghf-spin-directions} distinguishes these possibilities from
a mere global rotation of the common UHF spin axis.
RHF and UHF are consequently contained as special cases of GHF. A GHF
coefficient matrix has twice as many rows because every spatial basis function
can appear in both spin components of each occupied spinor.

\begin{figure}[htbp]
\centering
\begin{tikzpicture}[>=stealth,scale=1]
\foreach \x/\name/\desc in {
 -5.7/RHF/{same spatial orbital},
 -1.9/UHF/{different spatial orbitals},
  1.9/{Rotated UHF}/{a common tilted axis},
  5.7/GHF/{different tilted axes}}{
 \begin{scope}[shift={(\x,1.0)}]
  \draw[hfInk,thick] (0,0) circle (1.00);
  \draw[hfInk!50] (-1.00,0) arc (180:360:1.00 and 0.28);
  \draw[hfInk!50,dashed] (-1.00,0) arc (180:0:1.00 and 0.28);
  \draw[->,hfInk!40] (0,-1.18)--(0,1.18);
  \node[right,hfInk!60] at (0,1.2) {$z$};
 \end{scope}
 \node[above] at (\x,2.20) {\textbf{\name}};
 \node[align=center,text width=3.2cm] at (\x,-0.60) {\desc};
}
\draw[->,very thick,hfBlue] (-5.7,1.0)--(-5.7,1.80);
\draw[->,very thick,hfRed] (-5.7,1.0)--(-5.7,0.20);
\node[right,hfBlue] at (-5.65,1.62) {$\alpha$};
\node[right,hfRed] at (-5.65,0.38) {$\beta$};
\draw[->,very thick,hfBlue] (-1.9,1.0)--(-1.9,1.80);
\draw[->,very thick,hfRed] (-1.9,1.0)--(-1.9,0.20);
\node[right,hfBlue] at (-1.85,1.62) {$\alpha$};
\node[right,hfRed] at (-1.85,0.38) {$\beta$};
\draw[->,very thick,hfViolet] (1.9,1.0)--(2.56,1.58);
\draw[->,very thick,hfOrange] (1.9,1.0)--(1.24,0.42);
\node[above right,hfViolet] at (2.25,1.28) {$\boldsymbol{s}$};
\node[below left,hfOrange] at (1.55,0.72) {$\boldsymbol{s}'$};
\draw[->,very thick,hfViolet] (5.7,1.0)--(6.40,1.56);
\draw[->,very thick,hfOrange] (5.7,1.0)--(5.22,1.72);
\node[above right,hfViolet] at (6.05,1.28) {$\boldsymbol{s}$};
\node[above left,hfOrange] at (5.46,1.36) {$\boldsymbol{s}'$};
\end{tikzpicture}
\begin{equation}
 \underbrace{\text{UHF}\xrightarrow{\text{global spin rotation}}
 \text{rotated UHF}}_{\text{physically equivalent collinear solutions}}
 \qquad\neq\qquad
 \underbrace{\text{genuine GHF}}_{\text{non-collinear}}
\end{equation}
\caption{RHF, UHF, globally rotated UHF, and genuinely noncollinear GHF spin
directions. A tilted spin axis alone does not imply a genuinely noncollinear
GHF solution.}
\label{fig:ghf-spin-directions}
\end{figure}

The GHF density matrix is
\begin{align}
    \boldsymbol{\mathcal P}&=\mqty(
      \bP^{\alpha\alpha}&\bP^{\alpha\beta}\\
      \bP^{\beta\alpha}&\bP^{\beta\beta})
  &
  P_{\mu\nu}^{\sigma\sigma'}
  &=\sum_{i=1}^{N}C_{\mu i}^\sigma C_{\nu i}^{\sigma' *}
\end{align}
The diagonal blocks describe the two collinear spin populations, whereas the
off-diagonal blocks encode coherence between the $\alpha$ and $\beta$
components. These off-diagonal blocks vanish for an ordinary UHF determinant.
Its charge and magnetization blocks are
\begin{align}
  \bP&=\frac12\qty(\bP^{\alpha\alpha}+\bP^{\beta\beta})
  &
  \bM_z&=\frac12\qty(\bP^{\alpha\alpha}-\bP^{\beta\beta})
  \\
  \bM_y&=\frac{1}{2\ii}\qty(\bP^{\beta\alpha}-\bP^{\alpha\beta})
  &
  \bM_x&=\frac12\qty(\bP^{\beta\alpha}+\bP^{\alpha\beta})
\end{align}
Equivalently, using the Pauli spin matrices $\boldsymbol{\sigma}_k$ and
the Kronecker product $\otimes$,
\begin{equation}
  \boldsymbol{\mathcal P}
  =\mqty(
    \bP+\bM_z&\bM_x-\ii\bM_y\\
    \bM_x+\ii\bM_y&\bP-\bM_z)
  =\bI_\text{spin}\otimes\bP
   +\sum_{k=x,y,z}\boldsymbol{\sigma}_k\otimes\bM_k
\end{equation}
with
\begin{align}
  \bM&=\mqty(\bM_x\\\bM_y\\\bM_z)
  &
  \boldsymbol{\sigma}
  &=\mqty(\boldsymbol{\sigma}_x\\\boldsymbol{\sigma}_y\\
    \boldsymbol{\sigma}_z)
\end{align}
Here $\bI_\text{spin}$ is the $2\times2$ identity matrix in the spin basis.
In this decomposition,
$\bP$ is half of the total charge-density matrix, whereas
$\boldsymbol{\mathcal P}$ is the full spinor density matrix.

In spinor space, $\boldsymbol{\mathcal C}$ stacks the two spin components
of the orbital coefficient matrix, $\boldsymbol{\mathcal F}$ is the
spinor Fock matrix, $\beps$ contains the orbital energies,
and $\boldsymbol{\mathcal S}$ is the block-diagonal overlap metric.
The Roothaan--Hall equation is
\begin{equation}
    \boldsymbol{\mathcal F}\cdot\boldsymbol{\mathcal C}
    =\boldsymbol{\mathcal S}\cdot\boldsymbol{\mathcal C}
     \cdot\beps
\end{equation}
with
\begin{equation}
  \boldsymbol{\mathcal S}=\mqty(\bS&\bO\\\bO&\bS)
\end{equation}
and normalization
\begin{equation}
  \boldsymbol{\mathcal C}^\dagger\cdot\boldsymbol{\mathcal S}
  \cdot\boldsymbol{\mathcal C}=\bI
\end{equation}
The GHF Fock matrix is given below, using the same AO core-Hamiltonian
matrix $\bH$ as in the finite-basis derivation.
\begin{equation}
\boldsymbol{\mathcal F}=\mqty(
  \bH+\bJ[\bP^{\alpha\alpha}]+\bJ[\bP^{\beta\beta}]
    -\bK[\bP^{\alpha\alpha}]
  &-\bK[\bP^{\alpha\beta}]\\
  -\bK[\bP^{\beta\alpha}]
  &\bH+\bJ[\bP^{\alpha\alpha}]+\bJ[\bP^{\beta\beta}]
    -\bK[\bP^{\beta\beta}])
\end{equation}
where
\begin{align}
  J_{\mu\nu}[\bP]
  &=\sum_{\lambda\sigma}P_{\lambda\sigma}
    \rbraket{\mu\nu}{\lambda\sigma}
  &
  K_{\mu\nu}[\bP]
  &=\sum_{\lambda\sigma}P_{\lambda\sigma}
    \rbraket{\mu\lambda}{\sigma\nu}
\end{align}
The AO functions in these contractions are real, but the spinor coefficients
and density blocks may be complex. The exchange indices follow the definition
$P_{\lambda\sigma}=\sum_i C_{\lambda i}C_{\sigma i}^*$; exchanging
$\lambda$ and $\sigma$ in the integral without also transposing the density
would be incorrect for complex orbitals.
The off-diagonal exchange blocks couple the $\alpha$ and $\beta$ components
self-consistently. The SCF procedure builds all four density blocks, builds
all four Fock blocks, solves a $2K\times2K$ generalized eigenvalue problem,
occupies the $N$ lowest spinors, and iterates.

The mathematical structure is the same nonlinear generalized eigenvalue
problem as before, now expressed in a doubled spinor space. The off-diagonal
exchange blocks are precisely the new terms that permit the spin direction to
rotate away from the common UHF axis.

\subsubsection{Variational hierarchy and stability}

For the corresponding global minima in the same one-electron basis,
\begin{equation}
    \{\Phi_\text{RHF}\}\subset\{\Phi_\text{UHF}\}
    \subset\{\Phi_\text{GHF}\}
  \qquad\Longrightarrow\qquad
  \boxed{\red{E_\text{GHF}\leq E_\text{UHF}\leq E_\text{RHF}}}
\end{equation}
A converged UHF determinant may remain unstable to occupied-virtual rotations
that mix $\alpha$ and $\beta$ orbitals; following a negative-curvature
direction gives a lower-energy GHF solution. Not every GHF solution is
noncollinear: RHF and UHF are special cases, and a global spin rotation of a
UHF determinant changes its apparent quantization axis while remaining
physically collinear.

The energy inequalities compare minima over nested variational spaces. They do
not imply that every numerical GHF stationary point lies below every UHF or
RHF stationary point. Establishing whether a converged solution is a local
minimum requires a stability analysis with respect to the orbital rotations
allowed in the larger space \citep{Thouless1960,SeegerPople1977}.

Lowering the variational energy may require breaking exact symmetries. UHF
may break $\chS^2$; GHF may break both $\chS^2$ and $\chS_z$; and complex GHF
may also break time-reversal symmetry. GHF remains a single determinant, so
correlation is not included, and $S$ and $M_S$ are generally not good quantum
numbers. The many stationary solutions make the initial guess and
stability analysis important. A putative GHF solution may be a globally
rotated UHF solution, requiring a collinearity test. Equilateral
\red{\ce{H3}} and triangular magnetic clusters are typical genuine
noncollinear examples. GHF is the most general number-conserving
single-determinant HF ansatz; particle-number symmetry can instead be broken
within Hartree--Fock--Bogoliubov (HFB) theory \citep{RingSchuck1980,Scuseria2011}.

\paragraph{Practical interpretation.}

The charge and magnetization matrices provide a direct diagnostic of the
solution. Collinear determinants admit one global spin axis along which only
one magnetization component remains. A genuinely noncollinear GHF solution
cannot be reduced to that form by a single global spin rotation. This is why a
tilted axis, by itself, is insufficient evidence of noncollinearity.

\subsection{Classification of Hartree--Fock solutions}

The labels RHF, UHF, and GHF describe broad variational classes, but each class
contains further possibilities depending on whether the orbitals are real or
complex and which spin and time-reversal symmetries are retained. The
Stuber--Paldus notation \citep{StuberPaldus2003} organizes these stationary solutions by symmetry and
by the block structure of their coefficient matrices.
For the Fukutome designations used in the table, see \citet{Fukutome1981}.
The Stuber--Paldus classification is summarized in
Table~\ref{tab:stuber-paldus}. In this table, $\chK$ denotes complex
conjugation, rather than the exchange operator used earlier, and
$\hat\Theta$ denotes time reversal.
\hfBioPaldus

\begin{table}[htbp]
\centering
\hftablestyle
\caption{Stuber--Paldus classification of Hartree--Fock solutions.
Reproduced in part from Table~I of Henderson, Jim\'enez-Hoyos, and Scuseria
\citep{Henderson2018}, with the density-matrix columns omitted and notation
adjusted to the present chapter.}
\label{tab:stuber-paldus}
\setlength{\tabcolsep}{2.5pt}
\begin{tabular}{
  @{}>{\centering\arraybackslash}m{0.14\textwidth}
  >{\centering\arraybackslash}m{0.19\textwidth}
  >{\centering\arraybackslash}m{0.17\textwidth}
  >{\centering\arraybackslash}m{0.39\textwidth}@{}}
\toprule
\hftablehead{\makecell{Fukutome\\designation}}&
\hftablehead{\makecell{Stuber--Paldus\\designation}}&
\hftablehead{\makecell{Symmetries\\preserved}}&
\hftablehead{\makecell{Structure of the orbital\\
          coefficient matrix $\bC$}}\\
\midrule
\hftablesection{4}{Restricted Hartree--Fock}
$\text{TICS}^{a}$&real RHF&$\chS^2,\chS_z,\chK,\hat\Theta$&
$\mqty(\bC_{\sigma\sigma}&\bO\\\bO&\bC_{\sigma\sigma}),
\ \bC\in\mathbb R$\\
\addlinespace[3pt]
$\text{CCW}^{b}$&complex RHF&$\chS^2,\chS_z$&
$\mqty(\bC_{\sigma\sigma}&\bO\\\bO&\bC_{\sigma\sigma})$\\
\hftablesection{4}{Unrestricted Hartree--Fock}
$\text{ASCW}^{c}$&paired UHF&$\chS_z,\hat\Theta$&
$\mqty(\bC_{\sigma\sigma}&\bO\\\bO&\bC_{\sigma\sigma}^*)$\\
\addlinespace[3pt]
$\text{ASDW}^{d}$&real UHF&$\chS_z,\chK$&
$\mqty(\bC_{\sigma\sigma}&\bO\\\bO&\bC_{\sigma'\sigma'}),
\ \bC\in\mathbb R$\\
\addlinespace[3pt]
$\text{ASW}^{e}$&complex UHF&$\chS_z$&
$\mqty(\bC_{\sigma\sigma}&\bO\\\bO&\bC_{\sigma'\sigma'})$\\
\hftablesection{4}{Generalized Hartree--Fock}
$\text{TSCW}^{f}$&paired GHF&$\hat\Theta$&
$\mqty(\bC_{\sigma\sigma}&\bC_{\sigma\sigma'}\\
-\bC_{\sigma\sigma'}^*&\bC_{\sigma\sigma}^*)$\\
\addlinespace[3pt]
$\text{TSDW}^{g}$&real GHF&$\chK$&
$\mqty(\bC_{\sigma\sigma}&\bC_{\sigma\sigma'}\\
\bC_{\sigma'\sigma}&\bC_{\sigma'\sigma'}),\ \bC\in\mathbb R$\\
\addlinespace[3pt]
$\text{TSW}^{h}$&complex GHF&---&
$\mqty(\bC_{\sigma\sigma}&\bC_{\sigma\sigma'}\\
\bC_{\sigma'\sigma}&\bC_{\sigma'\sigma'})$\\
\bottomrule
\end{tabular}
\end{table}

Here TICS denotes time-reversal-invariant closed-shell; CCW, charge current
wave; ASCW, axial spin current wave; ASDW, axial spin density wave; ASW,
axial spin wave; TSCW, torsional spin current wave; TSDW, torsional spin
density wave; and TSW, torsional spin wave.
Figure~\ref{fig:stuber-paldus} displays the connections between these classes
as successive constraints are released.

\begin{figure}[htbp]
  \centering

\begingroup
{\renewcommand{\arraystretch}{1.6}
\begin{tabularx}{\linewidth}{>{\bfseries}lXXX}
\toprule
Space & Real coefficients & Time-reversal paired & Complex coefficients\\
\midrule
\rowcolor{hfGreen!7}
RHF & RHF: $\chS^2,\chS_z,\chK,\hat\Theta$&
Not a separate class here&
c-RHF: $\chS^2,\chS_z$\\
\rowcolor{hfOrange!8}
UHF & UHF: $\chS_z,\chK$&
p-UHF: $\chS_z,\hat\Theta$&
c-UHF: $\chS_z$\\
\rowcolor{hfRed!6}
GHF & GHF: $\chK$&
p-GHF: $\hat\Theta$&
c-GHF: none of these\\
\bottomrule
\end{tabularx}}
\par\smallskip
{\small Cells name the retained symmetries. Real and paired classes are
alternative constraints.\par}
\par\medskip
{\small\bfseries\color{hfGreenDark}Constraint-release connections\par}\smallskip
{\small\renewcommand{\arraystretch}{1.35}
\begin{tabularx}{\linewidth}{XX}
RHF $\to$ c-RHF: release $\hat\Theta,\chK$&
RHF $\to$ UHF: release $\chS^2,\hat\Theta$\\
RHF $\to$ p-UHF: release $\chS^2,\chK$&
c-RHF $\to$ c-UHF: release $\chS^2$\\
UHF $\to$ GHF: release $\chS_z$&
UHF $\to$ c-UHF: release $\chK$\\
p-UHF $\to$ p-GHF: release $\chS_z$&
p-UHF $\to$ c-UHF: release $\hat\Theta$\\
c-UHF $\to$ c-GHF: release $\chS_z$&
GHF $\to$ c-GHF: release $\chK$\\
p-GHF $\to$ c-GHF: release $\hat\Theta$&
\end{tabularx}}
\endgroup


  \caption{Stuber--Paldus classification of Hartree--Fock
  solutions, retaining all eight classes and eleven constraint-release
  connections. New comparison layout based on Fig.~1 of Burton, Thom, and Loos,
  \emph{Parity-Time Symmetry in Hartree--Fock Theory}
  \citep{BurtonThomLoos2019}.}
  \label{fig:stuber-paldus}
\end{figure}

The table makes the variational hierarchy concrete. Moving from restricted to
unrestricted and then generalized orbitals removes constraints from the
coefficient matrix; allowing complex coefficients removes additional
antiunitary symmetries. These larger spaces can contain lower stationary
energies, but identifying their physical symmetry requires examining the
solution rather than relying only on its numerical energy.


\clearpage
\section{One-electron basis sets}
\label{sec:basis-sets}

The Hartree--Fock formalism determines the optimal orbitals only within the
one-electron space made available to it. A practical calculation therefore
requires a second set of choices: the functions used to represent the
orbitals and the algorithms used to evaluate their matrix elements. The
remainder of the chapter develops these ingredients, beginning with basis-set
construction and continuing with Gaussian molecular integrals and their
low-rank representations.

\subsection{Basis-set expansions and design principles}

The matrix equations in Section~\ref{sec:finite-basis-hf} assumed that the
orbital basis had already been chosen. We now examine that choice: which
functions can represent the radial and angular variations of molecular
orbitals accurately with a manageable number of coefficients?
In practical molecular calculations, each molecular orbital is expanded in a
finite set of known one-electron functions,
\begin{equation}
  \psi_p(\br)=\sum_{\mu=1}^{K}C_{\mu p}\phi_\mu(\br)
\end{equation}
The basis functions $\phi_\mu$ are fixed, whereas the coefficients $C_{\mu p}$
are determined variationally. Enlarging a nested sequence of basis spaces
increases the variational freedom and lowers, or leaves unchanged, the minimum
Hartree--Fock energy as the complete-basis-set (CBS) limit is approached. Basis exponents
and contraction coefficients are fixed numerical parameters, often obtained
from atomic electronic-structure calculations. Their presence does not by
itself make the method empirical; they specify the finite representation of
the otherwise ab initio problem.
Orbital optimization therefore changes the molecular combination of the
building blocks, not their shapes or nonlinear parameters. At a fixed
geometry these are selected before the SCF calculation begins; the centers
move with the nuclei when atom-centered calculations are repeated at new
geometries.

A useful basis must be accurate, compact, numerically stable, and practical.
These requirements compete: a larger and more flexible basis improves the
description of the orbitals, but increases both the computational cost and the
risk of linear dependence.

\subsubsection{The LEGO principle}

\begin{figure}[htbp]
  \centering
  \includegraphics[width=0.62\textwidth]{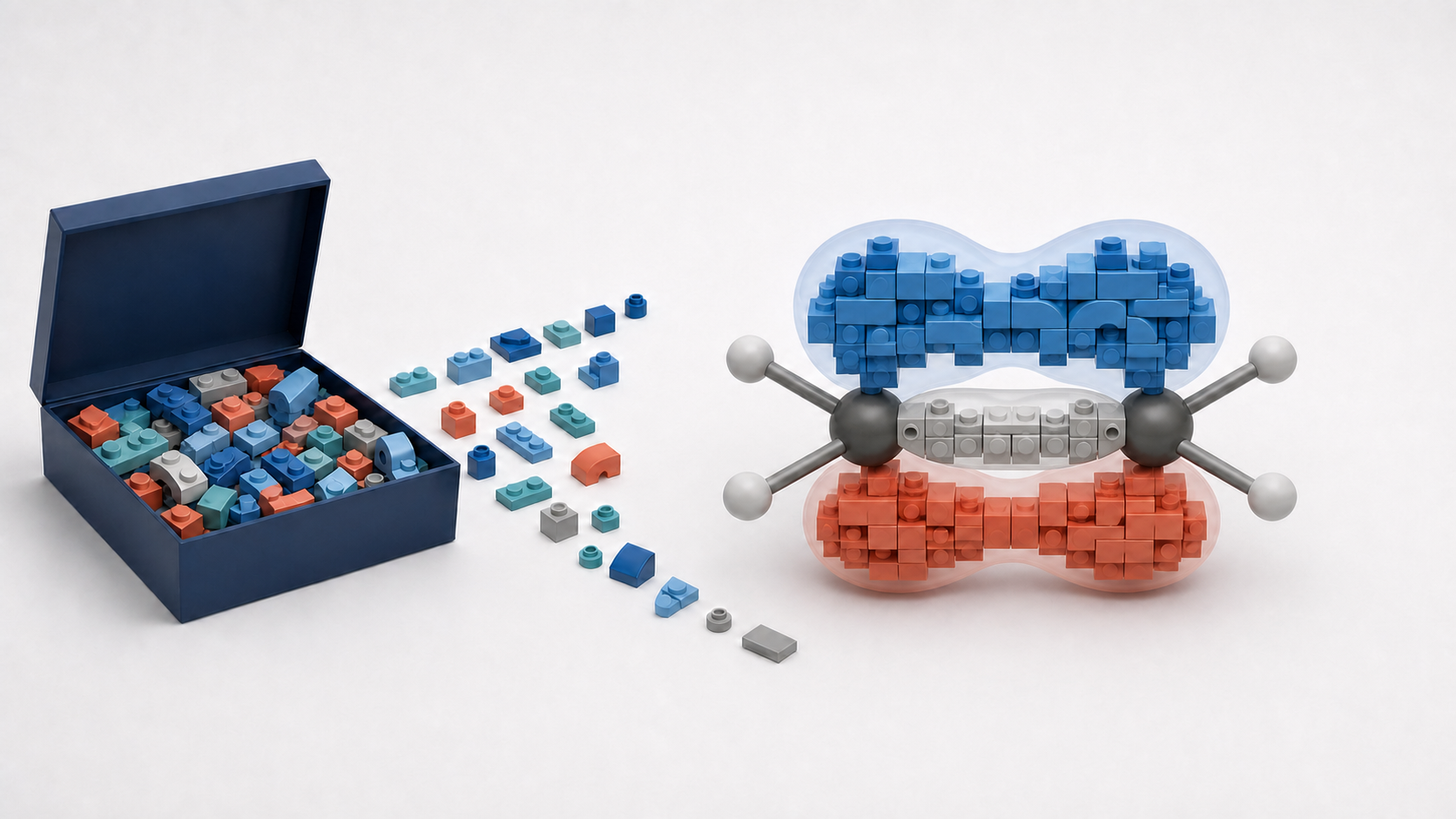}
  \caption{The basis functions are the elementary building blocks from which
  molecular orbitals are assembled.}
  \label{fig:basis-building-blocks}
\end{figure}

Basis-set construction follows a LEGO-like philosophy: a small collection of
well-designed building blocks is combined to describe chemically diverse
systems, as illustrated in Fig.~\ref{fig:basis-building-blocks}.

The analogy also captures transferability. Atom-centered functions designed
for individual elements can be assembled in different combinations without
redesigning the basis for every molecule. The molecular environment enters
through the variational coefficients that combine these fixed building blocks.

\subsubsection{Principal families of basis functions}

Several families of one-electron functions are routinely used.
Gaussian-type orbitals (GTOs), which are proportional to
$\exp(-\alpha r^2)$, are atom centered and are the standard choice for
molecules because their multicenter integrals can be evaluated efficiently.
Here $r$ is the distance from the function's center, and $\alpha$ is a
positive Gaussian exponent.
They are used, for example, in \textsc{Orca} \citep{ORCA2022},
\textsc{Q-Chem} \citep{QChem2021}, \textsc{Molcas} \citep{Molcas2016},
\textsc{Molpro} \citep{Molpro2020}, \textsc{PySCF} \citep{PySCF2020},
\textsc{Psi4} \citep{Psi42020}, \textsc{CFOUR} \citep{CFOUR2020},
\textsc{MRCC} \citep{MRCC2020}, and \textsc{GAMESS} \citep{GAMESS1993}.

Slater-type orbitals (STOs) are also atom centered, but are
proportional to $\exp(-\zeta r)$. Their radial form can reproduce a nuclear
cusp and an exponential long-range tail, and they are used, for example, in
\textsc{ADF} \citep{ADF2001}. The positive exponent $\zeta$ controls their radial extent.
By contrast,
plane waves (PWs), proportional to $\exp(\ii\bk\cdot\br)$, are
delocalized, mutually orthonormal, and naturally adapted to periodic boundary
conditions. Here $\bk$ is the wavevector and $\ii$ the imaginary unit.
Their basis size is controlled by an energy cutoff, and they are
used in \textsc{Quantum Espresso} \citep{QuantumEspresso2009},
\textsc{VASP} \citep{VASP1996}, \textsc{ABINIT} \citep{ABINIT2020}, and
\textsc{CASTEP} \citep{CASTEP2005}. Finally, numerical atomic orbitals (NAOs) have
tabulated radial parts. Their compactness and flexibility make them useful in
programs such as \textsc{FHI-aims} \citep{FHIAims2009} and
\textsc{SIESTA} \citep{SIESTA2002}.

\subsection{Plane-wave basis sets}

The four families balance radial accuracy, integral efficiency, periodicity,
and numerical flexibility differently. We first examine the delocalized
plane-wave representation, then return to atom-centered functions.

Plane waves are the natural one-electron functions of a translationally
invariant system. Under periodic boundary conditions, only reciprocal-lattice
wavevectors are compatible with the periodicity of the simulation cell.
In a unit cell of volume $\Omega$, a normalized plane wave is
\begin{align}
  \phi_{\bk}(\br)&=\boxed{\frac{1}{\sqrt{\Omega}}\exp(\ii\bk\cdot\br)}
  &
  \abs{\bk}&\leq k_\text{cut}
  &
  E_\text{cut}&=\frac{k_\text{cut}^2}{2}
\end{align}
The reciprocal-lattice vector is indexed by three integers:
\begin{equation}
  \bk=k_1\bb_1+k_2\bb_2+k_3\bb_3
\end{equation}
where, for direct-lattice vectors $\ba_1,\ba_2,\ba_3$,
\begin{align}
  \Omega&=\ba_1\cdot(\ba_2\times\ba_3)
  &\bA&=(\ba_1,\ba_2,\ba_3)\\
  \bb_1&=2\pi\frac{\ba_2\times\ba_3}{\Omega}
  &\bb_2&=2\pi\frac{\ba_3\times\ba_1}{\Omega}
  &\bb_3&=2\pi\frac{\ba_1\times\ba_2}{\Omega}
\end{align}
The reciprocal vectors satisfy $\ba_i\cdot\bb_j=2\pi\delta_{ij}$. The energy
cutoff selects reciprocal-lattice vectors inside a sphere of radius
$k_\text{cut}$ and therefore supplies a single systematic parameter for
enlarging the basis.
The overlap integrals are diagonal:
\begin{equation}
\begin{split}
  \braket{\bk_1}{\bk_2}
  &=\frac{1}{\Omega}\int_\Omega
    \exp[-\ii\bk_1\cdot\br]\exp[\ii\bk_2\cdot\br]\dd\br\\
  &=\frac{1}{\Omega}\int_\Omega
    \exp[\ii(\bk_2-\bk_1)\cdot\br]\dd\br
  =\delta_{\bk_1\bk_2}
\end{split}
\end{equation}
The kinetic-energy integrals are diagonal as well:
\begin{equation}
\begin{split}
  \mel{\bk_1}{-\frac12\laplacian}{\bk_2}
  &=\frac{1}{\Omega}\int_\Omega
    \exp[-\ii\bk_1\cdot\br]
    \qty(-\frac12\laplacian)\exp[\ii\bk_2\cdot\br]\dd\br\\
  &=\frac{k_2^2}{2\Omega}\int_\Omega
    \exp[\ii(\bk_2-\bk_1)\cdot\br]\dd\br
  =\frac{k_2^2}{2}\delta_{\bk_1\bk_2}
\end{split}
\end{equation}
For the plane-wave ERI, we use the physicists' index grouping:
the first and third indices belong to electron 1, and the second
and fourth to electron 2. The integral begins as
\begin{equation}
\begin{split}
  \braket{\bk_1\bk_2}{\bk_3\bk_4}
  &=\frac{1}{\Omega^2}\iint_\Omega
  \frac{\exp[\ii(\bk_3-\bk_1)\cdot\br_1]
        \exp[\ii(\bk_4-\bk_2)\cdot\br_2]}
       {r_{12}}\dd\br_1\dd\br_2
\end{split}
\end{equation}
Under periodic boundary conditions, the Coulomb kernel is understood as its
periodic counterpart. With its zero Fourier component omitted, the
reciprocal-space resolution is
\begin{equation}
  v_\Omega(\br_1-\br_2)=\frac{1}{\Omega}\sum_{\bk\ne\bO}
  \frac{4\pi}{k^2}\exp[\ii\bk\cdot(\br_1-\br_2)]
\end{equation}
The $\bk=\bO$ term requires a specified electrostatic convention, for
example a neutralizing background. The symbol $r_{12}^{-1}$ in the preceding
integral represents this periodic kernel in the following calculation.
Thus, we have
\begin{equation}
\begin{split}
  \braket{\bk_1\bk_2}{\bk_3\bk_4}
  &=\frac{1}{\Omega^3}\sum_{\bk\ne\bO}\frac{4\pi}{k^2}
  \iint_\Omega\exp[\ii(\bk_3-\bk_1+\bk)\cdot\br_1]
  \exp[\ii(\bk_4-\bk_2-\bk)\cdot\br_2]\dd\br_1\dd\br_2\\
  &=\frac{1}{\Omega}\sum_{\bk\ne\bO}\frac{4\pi}{k^2}
  \delta_{\bk_3-\bk_1+\bk,\bO}
  \delta_{\bk_4-\bk_2-\bk,\bO}\\
  &=\frac{4\pi}{\Omega\abs{\bk_1-\bk_3}^2}
  \delta_{\bk_1+\bk_2,\bk_3+\bk_4}
\end{split}
\end{equation}
The Kronecker delta expresses conservation of crystal momentum.
The last expression applies when $\bk_1\ne\bk_3$. For zero momentum
transfer, the omitted zero Fourier component gives zero under the convention
used here; the fraction must not be evaluated with a zero denominator.
The simplicity of these expressions follows because differentiation and
convolution become multiplication in reciprocal space. Overlap and kinetic
energy are diagonal, while a two-electron integral survives only when the
total incoming and outgoing wavevectors are equal.

The free-electron solutions provide the building blocks; their use as a
basis does not require the molecular or crystalline potential to be a weak
perturbation. Plane waves are orthonormal and independent of
the atomic positions; naturally compatible with periodic boundary conditions;
systematically convergent through the single parameter $E_\text{cut}$; and
diagonalize the kinetic and Coulomb operators in reciprocal space. They also
permit efficient fast-Fourier-transform evaluation and have neither basis-set
superposition error nor atom-centered Pulay forces. Their limitations are
large basis sizes,
\begin{equation}
  N_\text{PW}\propto\Omega E_\text{cut}^{3/2}
\end{equation}
Core orbitals require prohibitively large cutoffs, motivating
pseudopotentials; the cost increases with supercell size; delocalized functions
provide little real-space sparsity; and the large virtual spaces make
correlated methods expensive.
Plane waves are therefore especially effective for periodic valence-electron
calculations, usually in conjunction with pseudopotentials \citep{Payne1992}.

\subsection{Slater-type and Gaussian-type orbitals}

For isolated atoms and molecules, atom-centered functions provide a compact
description by placing radial flexibility where the electronic density is
concentrated. The central choice is between radial forms that reproduce the
exact local behavior and forms that permit efficient multicenter integration.
In the following Slater- and Gaussian-type orbitals, $n$ labels the radial
form, $\ell$ and $m$ are the angular-momentum and magnetic quantum numbers,
and $Y_{\ell m}(\hat{\br})$ is a spherical harmonic evaluated in the direction
of $\br$. The Slater radial form is associated historically with screened atomic
orbitals \citep{Slater1930Orbitals}. The general forms used in the comparison are
\begin{equation}
\begin{split}
  \phi_{\zeta n\ell m}^\text{STO}(\br)
 &\propto r^{n-1}e^{-\zeta r}Y_{\ell m}(\hat{\br})
  \\
  \phi_{\alpha n\ell m}^\text{GTO}(\br)
 &\propto r^{2(n-1)-\ell}e^{-\alpha r^2}Y_{\ell m}(\hat{\br})
\end{split}
\end{equation}
For their $1s$ representatives, the STO has a nonzero radial slope at the nucleus,
\begin{align}
  \phi_\text{1s}^\text{STO}(r)&\propto e^{-\zeta r}
  &
  \eval{\dv{\phi_\text{1s}^\text{STO}(r)}{r}}_{r=0}
 &=-\zeta\phi_\text{1s}^\text{STO}(0)
\end{align}
The logarithmic slope is $-\zeta$; the electron--nucleus cusp condition
requires $-Z$ for a point nucleus of charge $Z$ \citep{Kato1957}, and is satisfied by this
representative when $\zeta=Z$. The factor $\phi(0)$ accounts for the
normalization dependence of the radial derivative.
By contrast, the GTO is flat at the nucleus,
\begin{align}
  \phi_\text{1s}^\text{GTO}(r)&\propto e^{-\alpha r^2}
  &
  \eval{\dv{\phi_\text{1s}^\text{GTO}(r)}{r}}_{r=0}&=0
\end{align}
The STO also has the correct exponential long-range behavior of a molecular
orbital. For the HOMO, its leading radial exponential factor in the
frozen-orbital interpretation is
\begin{align}
  \psi_\text{HOMO}^\text{STO}(r)
 &\propto e^{-\sqrt{-2\epsilon_\text{HOMO}}r}
  &
  \mathrm{IP}&=-\epsilon_\text{HOMO}
\end{align}
A Gaussian instead decays too rapidly. Multicenter STO integrals are
substantially more difficult to evaluate than their Gaussian counterparts.
This computational advantage usually favors GTOs, although more Gaussian
functions are required to reproduce comparable radial flexibility.
The practical success of Gaussian basis sets is therefore computational. A
linear combination of several Gaussians is used to recover the radial
flexibility that a single Slater function supplies more naturally. The
Gaussian strategy was established in Boys's early molecular calculations
and developed further through systematic atomic expansions
\citep{Boys1950,Huzinaga1965}.

\hfBioBoys

\hfBioHuzinaga

\subsubsection{Primitive and contracted Gaussian functions}

A Gaussian basis function has two layers of structure. A primitive specifies
one exponent and one Cartesian polynomial. A contracted function combines
several primitives of the same center and angular momentum with fixed
coefficients, so that many primitive shapes share a single variational MO
coefficient \citep{Raffenetti1973}.
A primitive Cartesian Gaussian centered at $\bA$ is labeled by
$\ba=(a_x,a_y,a_z)$, a vector of nonnegative integer powers. Its total angular
momentum is $a=a_x+a_y+a_z$. In the integral discussion, these angular-momentum
labels replace the earlier use of $a,b,c,d$ for virtual orbital indices.
The square ket denotes a primitive function:
\begin{equation}
  \boxed{
  \sket{\ba}=(x-A_x)^{a_x}(y-A_y)^{a_y}(z-A_z)^{a_z}
  \exp[-\alpha\abs{\br-\bA}^2]
  }
\end{equation}
An ordinary ket denotes a contracted Gaussian, a fixed linear combination
of primitives with coefficients $D_k$:
\begin{equation}
  \boxed{
  \ket{\ba}\equiv\phi_{\ba}^{\bA}(\br)
  =\sum_{k=1}^{K}D_k\sket{\ba_k}
  }
\end{equation}
The square-ket definition is unnormalized. Its normalization factor can be
absorbed into $D_k$; coefficients tabulated for normalized primitives must
be combined with those factors when this unnormalized convention is used.
Here $\alpha$ is the primitive exponent. All primitives in a contraction have
the same center and polynomial powers but may have different exponents.
The upper limit $K$ in this local contraction sum counts primitives;
elsewhere $K$ denotes the number of orbital basis functions.

An $s$ shell contains $(0,0,0)$; a $p$ shell contains $(1,0,0)$,
$(0,1,0)$, and $(0,0,1)$; a Cartesian $d$ shell contains
$(2,0,0)$, $(0,2,0)$, $(0,0,2)$, $(1,1,0)$, $(1,0,1)$, and $(0,1,1)$;
and a Cartesian $f$ shell contains $(3,0,0)$, $(0,3,0)$, $(0,0,3)$, $(2,1,0)$, $(2,0,1)$,
$(1,2,0)$, $(0,2,1)$, $(0,1,2)$, $(1,0,2)$, and $(1,1,1)$.
In general, a Cartesian shell of total angular momentum $\ell$ contains
\begin{equation}
  N_\ell^\text{cart}=\frac{(\ell+1)(\ell+2)}{2}
\end{equation}
which gives $1,3,6,10$ functions for $s,p,d,f$. Spherical harmonics instead
give $N_\ell^\text{sph}=2\ell+1$, namely $1,3,5,7$. Thus
\begin{align}
  6d&=5d+1s
  &
  10f&=7f+3p
\end{align}
because, for example, $x^2+y^2+z^2=r^2$. Most molecular calculations use
pure spherical functions.
The distinction appears from $d$ functions onward. Cartesian shells are
convenient for integral generation, whereas transformation to pure spherical
functions removes the lower-angular-momentum combinations and restores the
expected $2\ell+1$ components.

Contraction reduces the number of variational coefficients, but it does not
remove the primitive-integral work. For four contracted functions,
\begin{equation}
  \rbraket{\ba\bb}{\bc\bd}
  =\sum_{i=1}^{K_a}\sum_{j=1}^{K_b}\sum_{k=1}^{K_c}\sum_{l=1}^{K_d}
  D_i^aD_j^bD_k^cD_l^d
  \sbraket{\ba_i\bb_j}{\bc_k\bd_l}
\end{equation}
For example, a cc-pVTZ carbon $1s$ function contracted from ten primitives
requires $10^4$ primitive $ssss$ integrals for a single four-center integral.
The complete contraction in this example is
\begin{equation}
 \rbraket{1s1s}{1s1s}
 =\sum_{i=1}^{10}\sum_{j=1}^{10}\sum_{k=1}^{10}\sum_{l=1}^{10}
 D_iD_jD_kD_l
 \sbraket{s_i^{\alpha_i}s_j^{\alpha_j}}{s_k^{\alpha_k}s_l^{\alpha_l}}
\end{equation}
The corresponding exponents and contraction coefficients are listed in
Table~\ref{tab:carbon-contraction}.
Basis-set definitions can be inspected and downloaded from the
\href{https://www.basissetexchange.org}{Basis Set Exchange}.
The contraction thus reduces the dimensions of the Fock and density
matrices, while integral algorithms must still reorganize the underlying
primitive sums efficiently.

\begin{table}[htbp]
  \centering
  \hftablestyle
  \caption{Primitive exponents and coefficients of one ten-primitive $s$
  contraction for carbon in cc-pVTZ. The shell scale is $1.00$. Values are
  extracted from the Basis Set Exchange; the basis family is due to Dunning
  \citep{Dunning1989,Pritchard2019}.}
  \label{tab:carbon-contraction}

\begin{tabular}{@{}crr@{}}
\toprule
\hftablehead{Primitive $i$} & \hftablehead{Exponent $\alpha_i$} &
\hftablehead{Coefficient $D_i$}\\
\midrule
1  & $8.236000\times10^{3}$  & $5.310000\times10^{-4}$\\
2  & $1.235000\times10^{3}$  & $4.108000\times10^{-3}$\\
3  & $2.808000\times10^{2}$  & $2.108700\times10^{-2}$\\
4  & $7.927000\times10^{1}$  & $8.185300\times10^{-2}$\\
5  & $2.559000\times10^{1}$  & $2.348170\times10^{-1}$\\
6  & $8.997000\times10^{0}$  & $4.344010\times10^{-1}$\\
7  & $3.319000\times10^{0}$  & $3.461290\times10^{-1}$\\
8  & $9.059000\times10^{-1}$ & $3.937800\times10^{-2}$\\
9  & $3.643000\times10^{-1}$ & $-8.983000\times10^{-3}$\\
10 & $1.285000\times10^{-1}$ & $2.385000\times10^{-3}$\\
\bottomrule
\end{tabular}


\end{table}

\subsection{Gaussian basis-set design and selection}

No single enlargement of a Gaussian basis improves every physical feature in
the same way. Different physical effects demand different forms of
flexibility.
Radial flexibility is provided by multiple-$\zeta$ functions, as in
double-$\zeta$, triple-$\zeta$, and quadruple-$\zeta$ bases.
Angular flexibility is supplied by polarization functions, such as
$d$ functions on second-row atoms and $p$ functions on hydrogen.
Long-range flexibility is supplied by diffuse functions with small
exponents. Such functions are important for anions, Rydberg states,
noncovalent interactions, multipoles, and (hyper)polarizabilities.
Core flexibility, in turn, is supplied by tight functions and is
needed for core correlation, nuclear magnetic shielding, and spin-spin
coupling constants.

\begin{figure}[htbp]
  \centering
  \includegraphics[width=\textwidth]{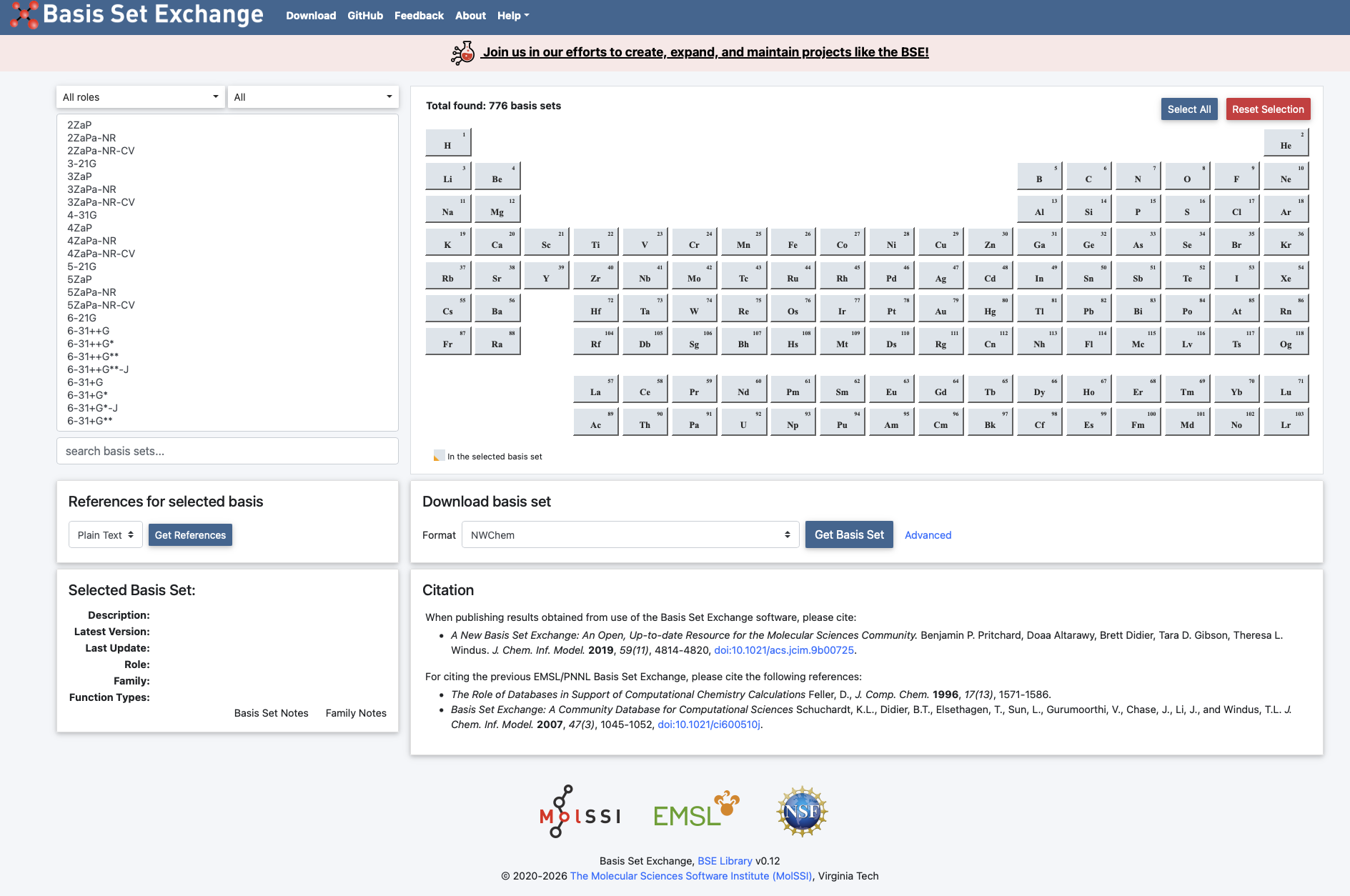}
  \caption{The Basis Set Exchange provides standardized definitions for a
  large number of orbital and auxiliary basis sets \citep{Pritchard2019}.}
  \label{fig:basis-set-exchange}
\end{figure}

The phrase \emph{larger basis} therefore needs a physical qualification:
the added functions must address the radial, angular, diffuse, or core
flexibility that controls the target energy or property. Increasing the
function count alone does not identify which of these deficiencies is being
reduced.
Figure~\ref{fig:basis-set-exchange} shows the Basis Set Exchange interface
through which the definitions of these basis families can be accessed.

\subsubsection{STO-$n$G minimal basis sets}

\begin{figure}[htbp]
  \centering
  \includegraphics[width=0.80\textwidth]{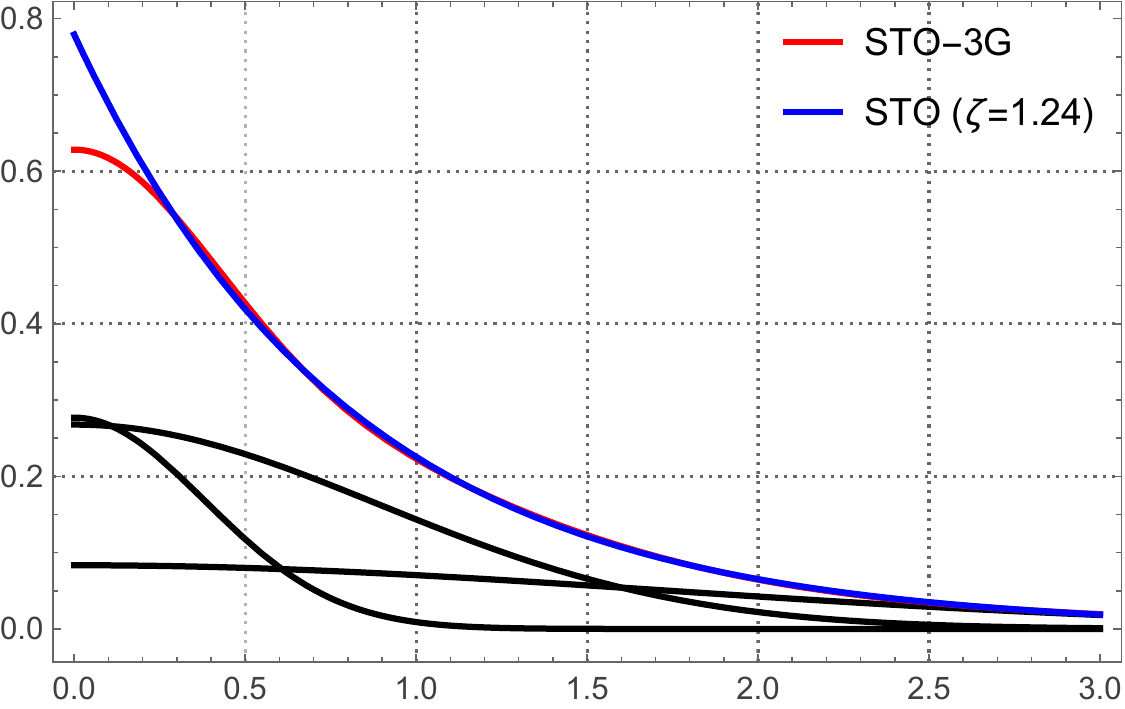}
  \caption{Radial dependence of the hydrogen $1s$ STO with
  $\zeta=1.24\,a_0^{-1}$ (blue) and its STO-3G approximation (red).
  The black curves are the three coefficient-weighted normalized Gaussian
  primitives whose sum forms the contraction, using the parameters in
  Table~\ref{tab:hydrogen-sto3g}. The horizontal coordinate is the distance
  $r$ from the nucleus in Bohr radii; the curves show spatial-orbital
  amplitudes, not radial probability densities. The plot is retained from
  the original lecture material.}
  \label{fig:hydrogen-sto3g}
\end{figure}

In an STO-$n$G basis, there is one contracted basis function for every occupied
orbital in the minimal atomic description. Each contracted GTO contains $n$
primitive GTOs, whose contraction coefficients and exponents are selected so
that the contraction approximates an STO. STO-$n$G bases are therefore minimal,
single-$\zeta$ basis sets; STO-3G is the most familiar member
\citep{Hehre1969}.
For hydrogen, Table~\ref{tab:hydrogen-sto3g} lists the three primitives in
the contraction, and Fig.~\ref{fig:hydrogen-sto3g} shows how their sum
approximates the Slater-type $1s$ orbital.

\begin{table}[htbp]
  \centering
  \hftablestyle
  \caption{Primitive exponents and contraction coefficients for the hydrogen
  $1s$ function in STO-3G, with shell scale $1.00$. The dimensionless
  coefficients $D_i$ multiply normalized primitive Gaussians; $a_0$ is the
  Bohr radius. Values are extracted from the Basis Set
  Exchange \citep{Hehre1969,Pritchard2019}.}
  \label{tab:hydrogen-sto3g}

\begin{tabular}{@{}crr@{}}
\toprule
\hftablehead{Primitive $i$} & \hftablehead{Exponent $\alpha_i$ ($a_0^{-2}$)} &
\hftablehead{Coefficient $D_i$}\\
\midrule
1 & $3.425250914$ & $0.1543289673$\\
2 & $0.6239137298$ & $0.5353281423$\\
3 & $0.1688554040$ & $0.4446345422$\\
\bottomrule
\end{tabular}


\end{table}

The quality of the contraction is controlled by $n$, but the molecular
orbital space still contains only the minimal number of functions. Such a
basis is useful for illustrating the construction, while its lack of extra
radial and angular flexibility limits quantitative accuracy.

\subsubsection{Pople split-valence basis sets}

Pople basis sets \citep{Hehre1972} extend the minimal description by splitting the valence space
and by adding optional diffuse and polarization shells. Their notation
explicitly records the contractions and augmentations. In
6-31+G*, the core function is contracted from six primitives; the valence
space is double-$\zeta$, with contractions of three and one primitives; the
``+'' adds diffuse functions to non-hydrogen atoms \citep{Clark1983}; and ``*'' adds polarization
functions to non-hydrogen atoms, conventionally a $d$ shell. In 6-311++G**,
the valence space is triple-$\zeta$ with contractions $3,1,1$ \citep{Krishnan1980}; the second
``+'' adds diffuse functions to hydrogen; and the second ``*'' adds
polarization functions to hydrogen. Finally, 6-311G(2df,2pd) adds two $d$ and
one $f$ polarization shells to non-hydrogen atoms, and two $p$ and one $d$
shell to hydrogen \citep{Frisch1984}.

The notation describes how the basis is assembled: the contraction pattern is
followed by symbols or parentheses identifying diffuse and polarization
functions. It is therefore more detailed than a simple label of basis size.

\subsubsection{Dunning correlation-consistent basis sets}

\hfBioDunning

The correlation-consistent construction \citep{Dunning1989} organizes functions into successive
balanced layers designed to recover angular correlation systematically. The
main sequence is
\begin{align}
  \text{cc-pVXZ},&& X&=\mathrm{D,T,Q,5,6}
\end{align}
In the basis names, D, T, and Q denote double-, triple-, and quadruple-$\zeta$
quality. The numerical cardinal number $X$ takes the corresponding values
$2,3,4,5,6$. A standard correlation-energy extrapolation
\citep{Helgaker1997} is
\begin{equation}
  \boxed{
  \Ec(X)\approx\Ec^\text{CBS}+AX^{-3}
  }
\end{equation}
Here $\Ec^\text{CBS}$ is the complete-basis-set correlation energy and $A$
is an extrapolation coefficient.
Diffuse augmentation gives aug-cc-pVXZ, d-aug-cc-pVXZ, and t-aug-cc-pVXZ
\citep{Kendall1992}.
Core correlation is described by cc-pCVXZ, optimized for core-core and
core-valence correlation \citep{Woon1995}, or by cc-pwCVXZ, which weights the core-valence
contribution \citep{Peterson2002}.

Increasing the cardinal number adds both radial flexibility and a new
layer of higher angular momentum, as the explicit shell patterns below
demonstrate. The inverse-cubic expression concerns correlation energy, not
the Hartree--Fock contribution; the polarization-consistent discussion below uses a
different Hartree--Fock extrapolation form.
For a second-row atom (the second period, Li--Ne),
Table~\ref{tab:cc-shell-content} lists the valence
correlation-consistent sequence and its diffuse augmentations.

\begin{table}[htbp]
\centering
\hftablestyle
\caption{Shell contents of correlation-consistent basis sets for
second-row atoms, with successive diffuse augmentations. The entries count
contracted radial functions in each angular-momentum shell
\citep{Dunning1989,Kendall1992}.}
\label{tab:cc-shell-content}
\begin{tabular}{@{}ll@{}}
\toprule
\hftablehead{Basis} & \hftablehead{Shell content}\\
\midrule
\hftablesection{2}{Unaugmented valence sets}
cc-pVDZ & $3s2p1d$\\
cc-pVTZ & $4s3p2d1f$\\
cc-pVQZ & $5s4p3d2f1g$\\
\hftablesection{2}{Single diffuse augmentation}
aug-cc-pVDZ & $4s3p2d$\\
aug-cc-pVTZ & $5s4p3d2f$\\
aug-cc-pVQZ & $6s5p4d3f2g$\\
\hftablesection{2}{Double diffuse augmentation}
d-aug-cc-pVDZ & $5s4p3d$\\
d-aug-cc-pVTZ & $6s5p4d3f$\\
d-aug-cc-pVQZ & $7s6p5d4f3g$\\
\bottomrule
\end{tabular}
\end{table}

The successive entries make the layered construction explicit. For the
core-valence sequence, the corresponding decomposition is
\begin{equation}
\begin{split}
 \text{cc-pVDZ: }&3s2p1d=1s+2s2p1d\\
 \text{cc-pVTZ: }&4s3p2d1f=1s+3s3p2d1f\\
 \text{cc-pVQZ: }&5s4p3d2f1g=1s+4s4p3d2f1g\\
 \text{cc-pCVDZ: }&4s3p1d=2s1p+2s2p1d\\
 \text{cc-pCVTZ: }&6s5p3d1f=3s2p1d+3s3p2d1f\\
 \text{cc-pCVQZ: }&8s7p5d3f1g=4s3p2d1f+4s4p3d2f1g
\end{split}
\end{equation}

\subsubsection{Jensen polarization-consistent basis sets}

Jensen's polarization-consistent sets \citep{Jensen2001} apply systematic improvement to Hartree--Fock and
Kohn--Sham calculations, with functions selected according to their energetic
importance. The sequence has shell contents
\begin{align}
  \text{pc-0}&:3s2p &\text{pc-1}&:3s2p1d\\
  \text{pc-2}&:4s3p2d1f &\text{pc-3}&:6s5p4d2f1g
\end{align}
Thus pc-0 is unpolarized, pc-1 adds one polarization layer, and the later
members add systematically more radial and angular flexibility. Diffuse variants
are denoted aug-pc-$n$. Their Hartree--Fock convergence is described by the
extrapolation form \citep{Jensen2005}
\begin{equation}
  E_\text{HF}(L)=E_\text{HF}^\text{CBS}
  +A(L+1)\exp(-B\sqrt L)
\end{equation}
Here $L$ is the highest angular momentum present in the basis, while $A$ and
$B$ are extrapolation coefficients. This Hartree--Fock extrapolation is distinct from
the inverse-cubic correlation-energy extrapolation above.

These hierarchies illustrate why Gaussian bases are widely used: they are
compact, atom centered, chemically transferable, supported
by analytic integral technology, and available in systematic hierarchies.
They make all-electron calculations practical for light and medium elements,
and their locality supports integral screening and sparse algorithms. The
physical picture is that a molecule is assembled from atoms whose electron
distributions are distorted by their environment. Because the basis functions
move with the nuclei, nuclear derivatives contain Pulay terms in addition to
the derivatives of the electronic Hamiltonian.
Their principal limitations are the absent nuclear cusp, the incorrect
long-range Gaussian decay, basis-set superposition error, and possible linear
dependence when many diffuse functions are used. There is consequently no
single best Gaussian basis: the appropriate choice depends on the property,
the electronic-structure method, and the desired accuracy.

\subsubsection{Basis-set superposition error and counterpoise correction}

\hfBioBernardi

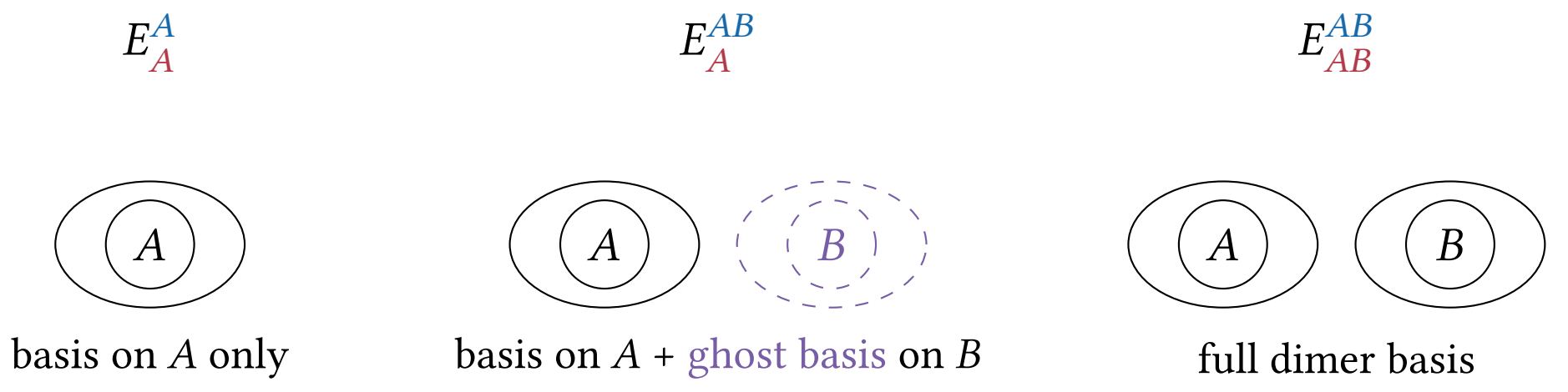
\begin{figure}[htbp]
 \centering
 \begin{tikzpicture}[
  atom/.style={circle,draw,minimum size=0.7cm,inner sep=0pt},
  ghost/.style={circle,draw,dashed,minimum size=0.7cm,inner sep=0pt},
  basis/.style={ellipse,draw,minimum width=1.5cm,minimum height=1.0cm},
  ghostbasis/.style={ellipse,draw,dashed,minimum width=1.5cm,minimum height=1.0cm}]
  \node at (-4.5,1.6) {$E_{\red A}^{\blue A}$};
  \node[basis] at (-4.5,0) {};
  \node[atom] at (-4.5,0) {$A$};
  \node at (-4.5,-0.9) {\small basis on $A$ only};
  \node at (0,1.6) {$E_{\red A}^{\blue{AB}}$};
  \node[basis] at (-0.9,0) {};
  \node[atom] at (-0.9,0) {$A$};
  \node[ghostbasis,hfViolet] at (0.9,0) {};
  \node[ghost,hfViolet] at (0.9,0) {$B$};
  \node at (0,-0.9) {\small basis on $A$ + \violet{ghost basis} on $B$};
  \node at (4.9,1.6) {$E_{\red{AB}}^{\blue{AB}}$};
  \node[basis] at (4.0,0) {};
  \node[atom] at (4.0,0) {$A$};
  \node[basis] at (5.8,0) {};
  \node[atom] at (5.8,0) {$B$};
  \node at (4.9,-0.9) {\small full dimer basis};
 \end{tikzpicture}
 \caption{A fragment in its own basis, the same fragment supplemented by
 ghost functions, and the full dimer. Dashed violet shapes retain basis
 functions without adding nuclear charge or electrons.}
 \label{fig:ghost-basis}
\end{figure}

The locality that makes atom-centered bases compact also creates a possible
imbalance when fragment energies are compared. Atom-centered basis functions
belong formally to the complete calculation
rather than to immutable fragments. When two fragments approach one another,
each gains access to functions centered on the other, and the dimer and
isolated-fragment energies no longer possess equivalent variational
flexibility.
Let $E_X^Y$ denote the energy of system $X$ evaluated in basis $Y$. In a
dimer $AB$, each fragment can borrow the basis functions of the other fragment
as ghost functions. The variational principle therefore gives
\begin{align}
  E_A^{AB}&\leq E_A^A
  &
  E_B^{AB}&\leq E_B^B
\end{align}
The corresponding uncorrected interaction energy is
\begin{equation}
  \Delta E_\text{int}=E_{AB}^{AB}-E_A^A-E_B^B
\end{equation}
which is too attractive relative to a comparison in the common dimer basis.
This basis imbalance is distinct from other basis-incompleteness or
electronic-structure errors; counterpoise does not establish an exact
interaction energy. The Boys--Bernardi counterpoise correction
\citep{BoysBernardi1970} uses
the full dimer basis for every term,
\begin{equation}
  \boxed{
  \Delta E_\text{int}^\text{CP}
  =E_{AB}^{AB}-E_A^{AB}-E_B^{AB}
  }
\end{equation}
and the basis-set superposition error (BSSE) is
\begin{equation}
  \mathrm{BSSE}
  =\Delta E_\text{int}^\text{CP}-\Delta E_\text{int}
  =(E_A^A-E_A^{AB})+(E_B^B-E_B^{AB})\geq0
\end{equation}
The partner's basis functions are retained as ghost functions: they provide
variational flexibility but carry neither electrons nor nuclear charge. The
counterpoise expression thereby compares the dimer and both fragments in the
same one-electron space.
Figure~\ref{fig:ghost-basis} illustrates the difference between an isolated
fragment basis and the full dimer basis retained in a counterpoise calculation.

\subsection{Effective core potentials and pseudopotentials}

Core electrons are numerous and spatially compact, yet they often participate
only indirectly in chemical bonding. Effective core potentials (ECPs)
exploit this
separation by replacing selected core electrons and their nuclear interaction
with an operator acting on the remaining explicit electrons.
The key idea of an effective core potential \citep{WadtHay1985} is to treat only the chemically
relevant electrons explicitly. It replaces the nuclear attraction and the
chemically inactive core electrons
\begin{equation}
 \hh(\br)=-\frac{\nabla^2}{2}
 -\sum_A\frac{Z_A}{\abs{\br-\bR_A}}
\end{equation}
by an effective one-electron operator
\begin{align}
 \hh^\text{ECP}(\br)&=-\frac{\nabla^2}{2}
 -\sum_A\frac{\green{Q_A}}{\abs{\br-\bR_A}}
 +\sum_A\Hat{U}_A^\text{ECP}(\br)
 &
 \green{Q_A}&=Z_A-N_A^\text{core}
\end{align}
Here $N_A^\text{core}$ is the number of electrons removed by the ECP on atom
$A$, and $Q_A$ is the charge of the remaining ionic core. It is not the
population-analysis charge $q_A$ introduced earlier.
This lowers the cost by reducing the number of explicit electrons and removing
the basis functions needed for core orbitals. It can incorporate
scalar-relativistic effects and, with a suitable operator, spin-orbit
coupling, and it can be used in both molecular and periodic calculations with
Gaussian or plane-wave bases. The price is a new class of one-electron
integrals that can be difficult to evaluate; core properties are not described
explicitly, and the accuracy depends on the ECP quality.

For one center, with $r$ measured from that center and $L$ labeling the local
channel, a common semilocal form is
\begin{equation}
  \Hat{U}^\text{ECP}=U_L(r)+
  \sum_{\ell=0}^{L-1}\sum_{m=-\ell}^{\ell}
  \ket{Y_{\ell m}}U_\ell(r)\bra{Y_{\ell m}}
\end{equation}
The first term is local, or unprojected; the remaining angular-momentum
channels are nonlocal, or projected. In this convention, each projected
$U_\ell(r)$ is a correction relative to the local potential $U_L(r)$, not
the full potential in channel $\ell$. Their radial parts are generally expanded
as
\begin{align}
  U_\ell(r)&=\sum_kD_{\ell k}r^{n_{\ell k}}
  \exp(-\eta_{\ell k}r^2)
  &
  n_{\ell k}&\in\{-2,-1,0\}
\end{align}
The $D_{\ell k}$ and $\eta_{\ell k}$ are fitted parameters. Although these are
one-electron operators, their angular projectors make the integrals nontrivial.

For silver, a large-core ECP removes 36 electrons and treats explicitly the
$4d^{10}5s^1$ shells, leaving 11 electrons. A small-core ECP removes 28
electrons and retains $4s^24p^64d^{10}5s^1$, leaving 19 explicit electrons;
the semicore $4s$ and $4p$ shells are therefore retained, providing more
flexibility at higher cost. Several Gaussian-basis ECP families are commonly
used. LANL \citep{WadtHay1985} and SBK/SBKJC \citep{Stevens1992} are established families; in particular, LANL2DZ
combines an ECP with a double-$\zeta$ valence basis. The
\href{https://www.tc.uni-koeln.de/PP/clickpse.en.html}
{Stuttgart/Cologne energy-consistent pseudopotentials} are offered with
different core sizes and dedicated bases \citep{Bergner1993}, while the def2-ECPs are paired with
the def2 bases of Weigend and Ahlrichs \citep{WeigendAhlrichs2005}. The Burkatzki--Filippi--Dolg (BFD)
potentials were developed for quantum Monte Carlo calculations \citep{Burkatzki2007}, whereas the
\href{https://pseudopotentiallibrary.org}{correlation-consistent ECPs
(ccECPs)} were developed for high-accuracy correlated many-body methods \citep{Bennett2017}.
Most ECPs have dedicated basis sets.

The core size is consequently a controlled compromise. A large-core potential
is cheaper, whereas a small-core potential retains more semicore response and
flexibility. Consistency requires the ECP and its accompanying valence basis
to be treated as a matched pair.

\section{Molecular integrals over Gaussian functions}
\label{sec:gaussian-integrals}

Once the orbital basis and any effective core potential have been selected,
the abstract matrix elements of the finite-basis Hartree--Fock equations must
be evaluated. Gaussian functions are especially valuable because their
products and derivatives can be reduced to a small set of fundamental
integrals. The discussion below develops that reduction from overlap and
one-electron terms to four-center electron-repulsion integrals, and then shows
how recurrence relations and screening turn the resulting formulas into a
practical algorithm \citep{Gill1994}.

The Cartesian Gaussian functions in this section are real. Primitive
formulas use the unnormalized square-ket convention introduced above;
normalization and contraction factors are applied separately. This scope is
important when using the permutation symmetries and screening bounds below:
the more general complex-orbital symmetries were stated earlier.

\subsection{The Gaussian product theorem and overlap integrals}
\label{sec:gaussian-product}

The Gaussian product rule reduces multicenter products to single-center
quantities: the product of two $s$-type Gaussian factors is another Gaussian
times a constant. Polynomial factors in higher-angular-momentum functions
are handled by the recurrence relations developed later.
This closure property underlies the classical Gaussian integral formulation
of Boys \citep{Boys1950}.
Writing the $s$-type Gaussian factors as
\begin{align}
  G_{\red{\alpha},\red{\bA}}(\br)
 &=\exp[-\red{\alpha}\abs{\br-\red{\bA}}^2]
  &
  G_{\blue{\beta},\blue{\bB}}(\br)
 &=\exp[-\blue{\beta}\abs{\br-\blue{\bB}}^2]
\end{align}
then
\begin{hftheorem}{Gaussian product theorem}
For positive exponents $\alpha$ and $\beta$, the product of the two $s$-type
Gaussian factors is a Gaussian centered at the exponent-weighted point $\bP$,
multiplied by a separation-dependent prefactor:
\begin{equation}
 G_{\red{\alpha},\red{\bA}}(\br)
 G_{\blue{\beta},\blue{\bB}}(\br)
 =K_{\violet{\zeta},\violet{\bP}}
 G_{\violet{\zeta},\violet{\bP}}(\br)
 \label{eq:gaussian-product}
\end{equation}
with
\begin{align}
  \violet{\zeta}&=\red{\alpha}+\blue{\beta}
  &
  \violet{\bP}&=\frac{\red{\alpha\bA}+\blue{\beta\bB}}
 {\red{\alpha}+\blue{\beta}}
\end{align}
and
\begin{equation}
 K_{\violet{\zeta},\violet{\bP}}
 =\exp[-\frac{\red{\alpha}\blue{\beta}}
 {\red{\alpha}+\blue{\beta}}\abs{\red{\bA}-\blue{\bB}}^2]
\end{equation}
\end{hftheorem}
The product center $\bP$ lies on the line joining $\bA$ and $\bB$ and is
weighted toward the Gaussian with the larger exponent. The prefactor contains
the dependence on the separation of the original centers: its Gaussian decay
makes distant pairs rapidly negligible. The bold $\bP$ here denotes a
coordinate, not the density matrix used in the Hartree--Fock equations.
For two $s$ Gaussians, the overlap illustrates Cartesian factorization:
\begin{equation}
\begin{split}
 \int G_{\red{\alpha},\red{\bA}}(\br)
 G_{\blue{\beta},\blue{\bB}}(\br)\dd\br
 &=K_{\violet{\zeta},\violet{\bP}}
   \int G_{\violet{\zeta},\violet{\bP}}(\br)\dd\br\\
 &=K_{\violet{\zeta},\violet{\bP}}
 \qty(\int_{-\infty}^{+\infty}e^{-\violet{\zeta}x^2}\dd x)
 \qty(\int_{-\infty}^{+\infty}e^{-\violet{\zeta}y^2}\dd y)
 \qty(\int_{-\infty}^{+\infty}e^{-\violet{\zeta}z^2}\dd z)\\
 &=K_{\violet{\zeta},\violet{\bP}}
 \qty(\frac{\pi}{\violet{\zeta}})^{3/2}
\end{split}
\end{equation}

The product formula above uses unnormalized Gaussian factors.
Exercise~\ref{ex:normalized-gaussian-overlap} shows how normalization changes the overlap and provides useful
limiting cases for checking an integral routine.

\begin{hfexercise}[phantom={\label{ex:normalized-gaussian-overlap}}]{Normalized Gaussian overlap}
\violet{\textit{Normalize two $s$-type Gaussian primitives and derive their
overlap for arbitrary exponents and centers.}}
\end{hfexercise}
\begin{hfsolution}{Normalized Gaussian overlap}
For $\alpha>0$, define a normalized $s$-type primitive centered at $\bA$ by
\begin{equation}
  \phi_{\alpha,\bA}(\br)
  =\mathcal{N}_\alpha\exp[-\alpha\abs{\br-\bA}^2]
\end{equation}
Normalization requires the integral of the squared function to be one.
Translating the integration variable to the center removes any dependence
on $\bA$. The three Cartesian integrals then factorize:
\begin{equation}
\begin{split}
  1
  &=\mathcal{N}_\alpha^2
    \int\exp[-2\alpha\abs{\br-\bA}^2]\dd\br\\
  &=\mathcal{N}_\alpha^2
    \qty(\int_{-\infty}^{+\infty}e^{-2\alpha x^2}\dd x)^3\\
  &=\mathcal{N}_\alpha^2\qty(\frac{\pi}{2\alpha})^{3/2}
\end{split}
\end{equation}
Choosing the positive root gives
\begin{equation}
  \boxed{\mathcal{N}_\alpha=\qty(\frac{2\alpha}{\pi})^{3/4}}
\end{equation}
The exponent has dimensions of inverse length squared, so the normalized
three-dimensional function has dimensions of inverse length to the power
$3/2$, as required by its normalization integral.

For two such real functions, write
\begin{equation}
\begin{split}
  S_{\alpha\bA,\beta\bB}
  &\equiv\braket{\phi_{\alpha,\bA}}{\phi_{\beta,\bB}}\\
  &=\mathcal{N}_\alpha\mathcal{N}_\beta
    \int\exp[-\alpha\abs{\br-\bA}^2
             -\beta\abs{\br-\bB}^2]\dd\br
\end{split}
\end{equation}
Using $\zeta=\alpha+\beta$ and
$\bP=(\alpha\bA+\beta\bB)/\zeta$, the exponent can be completed to a
square:
\begin{equation}
  \alpha\abs{\br-\bA}^2+\beta\abs{\br-\bB}^2
  =\zeta\abs{\br-\bP}^2
   +\frac{\alpha\beta}{\zeta}\abs{\bA-\bB}^2
\end{equation}
Thus the separation-dependent factor leaves the integral:
\begin{equation}
\begin{split}
  S_{\alpha\bA,\beta\bB}
  &=\mathcal{N}_\alpha\mathcal{N}_\beta
    \exp[-\frac{\alpha\beta}{\zeta}\abs{\bA-\bB}^2]
    \int\exp[-\zeta\abs{\br-\bP}^2]\dd\br\\
  &=\mathcal{N}_\alpha\mathcal{N}_\beta
    \qty(\frac{\pi}{\zeta})^{3/2}
    \exp[-\frac{\alpha\beta}{\zeta}\abs{\bA-\bB}^2]
\end{split}
\end{equation}
Substituting the normalization factors yields
\begin{equation}
  \boxed{
  S_{\alpha\bA,\beta\bB}
  =\qty(\frac{2\sqrt{\alpha\beta}}{\alpha+\beta})^{3/2}
    \exp[-\frac{\alpha\beta}{\alpha+\beta}\abs{\bA-\bB}^2]
  }
\end{equation}
The result is dimensionless and symmetric under exchange of the two
functions. Both factors are at most one: the first follows from
$2\sqrt{\alpha\beta}\leq\alpha+\beta$, and the second from the
nonpositive exponent. The positive $s$ primitives therefore have an
overlap between zero and one.

For coincident centers the exponential is one, but the overlap is
\begin{equation}
  \eval{S_{\alpha\bA,\beta\bB}}_{\bB=\bA}
  =\qty(\frac{2\sqrt{\alpha\beta}}{\alpha+\beta})^{3/2}
\end{equation}
It is strictly less than one when $\alpha\ne\beta$: sharing a center does
not make two different radial functions identical. If the exponents also
coincide, the functions are identical, and the overlap is exactly one,
recovering the normalization condition.
For equal exponents at arbitrary separation, the prefactor is one and
\begin{equation}
  \eval{S_{\alpha\bA,\beta\bB}}_{\beta=\alpha}
  =\exp[-\frac{\alpha}{2}\abs{\bA-\bB}^2]
\end{equation}
For any fixed positive exponents, the overlap tends to zero as the centers
are separated without bound. This decay is Gaussian in the separation;
more diffuse primitives retain appreciable overlap over longer distances.

As a numerical illustration, take $\alpha=1\,a_0^{-2}$,
$\beta=2\,a_0^{-2}$, $\bA=(0,0,0)\,a_0$, and
$\bB=(1,0,0)\,a_0$, where $a_0$ is the Bohr radius.
The combined exponent and center are
\begin{align}
  \zeta&=3\,a_0^{-2}\\
  \bP&=\qty(\frac23,0,0)\,a_0
\end{align}
The center lies closer to $\bB$, which carries the tighter primitive.
The overlap is
\begin{equation}
  S_{\alpha\bA,\beta\bB}
  =\qty(\frac{2\sqrt2}{3})^{3/2}e^{-2/3}
  \simeq0.4700
\end{equation}
The product theorem determines the shape and center of the product, not
its normalization. Even when both original functions are normalized, their
product is not generally a normalized primitive; its prefactor must be
retained when evaluating the overlap.
\end{hfsolution}

\subsection{One-electron Gaussian integrals}

Once the product theorem has combined two basis functions, overlap, kinetic,
and nuclear-attraction integrals reduce to Gaussian moments about the product
center. The $s$-type formulas expose the essential structure before angular
momentum is introduced.

\subsubsection{Kinetic-energy integrals}

The Laplacian of an $s$ Gaussian is
\begin{equation}
 -\frac12\laplacian G_{\blue{\beta},\blue{\bB}}(\br)
 =\qty[3\blue{\beta}-2\blue{\beta}^2
 \abs{\br-\blue{\bB}}^2]G_{\blue{\beta},\blue{\bB}}(\br)
\end{equation}
Consequently, using the Gaussian product theorem
\eqref{eq:gaussian-product}, its kinetic-energy integral becomes
\begin{equation}
\begin{split}
 -\frac12\int G_{\red{\alpha},\red{\bA}}(\br)\laplacian
 G_{\blue{\beta},\blue{\bB}}(\br)\dd\br
 &=\int G_{\red{\alpha},\red{\bA}}(\br)
 \qty[3\blue{\beta}-2\blue{\beta}^2
 \abs{\br-\blue{\bB}}^2]
 G_{\blue{\beta},\blue{\bB}}(\br)\dd\br\\
 &=\frac{\red{\alpha}\blue{\beta}}{\violet{\zeta}}
 \qty[3-2\frac{\red{\alpha}\blue{\beta}}{\violet{\zeta}}
 \abs{\red{\bA}-\blue{\bB}}^2]
 K_{\violet{\zeta},\violet{\bP}}
 \qty(\frac{\pi}{\violet{\zeta}})^{3/2}
\end{split}
\end{equation}
The result has the same overlap prefactor as the preceding integral. The
additional polynomial factor is the consequence of differentiating the
Gaussian centered at $\blue{\bB}$ twice. It contains both a local curvature
term, $3\blue{\beta}$, and a term that depends on the separation of the two
centers.

\subsubsection{Nuclear-attraction integrals}

For a nucleus of charge $Z_C$ located at $\bC$, the nuclear-attraction
integral introduces the function $F_0$, the zeroth-order Boys function defined
in Eq.~\eqref{eq:boys-functions}:
\begin{equation}
\begin{split}
 -Z_C\int G_{\red{\alpha},\red{\bA}}(\br)
 \frac{1}{\abs{\br-\bC}}
 G_{\blue{\beta},\blue{\bB}}(\br)\dd\br
 &=-Z_CK_{\violet{\zeta},\violet{\bP}}
 \int\frac{e^{-\violet{\zeta}\abs{\br-\violet{\bP}}^2}}
 {\abs{\br-\bC}}\dd\br\\
 &=-\frac{2Z_C}{\sqrt\pi}K_{\violet{\zeta},\violet{\bP}}
 \int_0^\infty\dd t\int
 e^{-\violet{\zeta}\abs{\br-\violet{\bP}}^2}
 e^{-t^2\abs{\br-\bC}^2}\dd\br\\
 &=-\frac{2\pi Z_C}{\violet{\zeta}}
 K_{\violet{\zeta},\violet{\bP}}
 F_0\qty(\violet{\zeta}\abs{\violet{\bP}-\bC}^2)
\end{split}
\end{equation}

\subsection{Four-center electron-repulsion integrals}

Electron-repulsion integrals are the most numerous quantities in conventional
molecular Hartree--Fock theory. Four basis functions and the nonseparable
Coulomb kernel appear, but the product theorem reduces the four original
centers to two Gaussian product centers.

\subsubsection{Gaussian-pair reduction and shell structure}

For four unnormalized primitive $s$ Gaussians, applying the Gaussian product
rule on both electron coordinates gives the following reduction, with
$\bO=(0,0,0)$ denoting zero angular momentum. Higher-angular-momentum
primitives also contain polynomial factors, which are generated through the
recurrences in Section~\ref{sec:os-recurrences}.
\begin{equation}
\begin{split}
 \sbraket{\red{\bO}\blue{\bO}}{\orange{\bO}\green{\bO}}
 &=\iint G_{\red{\alpha},\red{\bA}}(\br_1)
 G_{\blue{\beta},\blue{\bB}}(\br_1)\frac{1}{r_{12}}
 G_{\orange{\gamma},\orange{\bC}}(\br_2)
 G_{\green{\delta},\green{\bD}}(\br_2)\dd\br_1\dd\br_2\\
 &=K_{\violet{\zeta},\violet{\bP}}
 K_{\purple{\eta},\purple{\bQ}}
 \iint G_{\violet{\zeta},\violet{\bP}}(\br_1)
 \frac{1}{r_{12}}G_{\purple{\eta},\purple{\bQ}}(\br_2)
 \dd\br_1\dd\br_2
\end{split}
\end{equation}
The second pair has combined exponent $\eta=\gamma+\delta$ and product
center $\bQ=(\gamma\bC+\delta\bD)/\eta$, with a prefactor
$K_{\eta,\bQ}$ defined in the same way as for the first pair.
There are formally $\order{K^4}$ ERIs. At a fixed numerical threshold and
with a fixed local basis per atom, the significant subset can approach
$\order{K^2}$ in sufficiently large, spatially extended systems. This is a
locality-based screening estimate, not an exact reduction in the number of
nonzero Coulomb integrals.

\paragraph{Functions, shell pairs, and shell quartets.}

The notation distinguishes individual functions from shells and primitives
from contractions. Thus $\sbra{\red{\ba}\blue{\bb}}$ is one primitive bra,
whereas $\sbra{\red a\blue b}$ is a complete primitive shell pair;
$\sbraket{\red{\ba}\blue{\bb}}{\orange{\bc}\green{\bd}}$ is one primitive
integral, whereas $\sbraket{\red a\blue b}{\orange c\green d}$ is a primitive
shell quartet. Likewise, $\rbra{\red{\ba}\blue{\bb}}$ and
$\rbra{\red a\blue b}$ denote a contracted bra and contracted shell pair,
and $\rbraket{\red{\ba}\blue{\bb}}{\orange{\bc}\green{\bd}}$ and
$\rbraket{\red a\blue b}{\orange c\green d}$ denote a contracted integral
and a contracted shell quartet.

\subsubsection{The fundamental $ssss$ integral}

Higher-angular-momentum ERIs can be generated from a hierarchy of fundamental
$s$-type quantities. The derivation begins by replacing the Coulomb operator
with its Laplace representation, after which every Cartesian coordinate enters
through an ordinary Gaussian integral.
For four primitive $s$ functions,
\begin{equation}
\begin{split}
 \sbraket{\bs\bs}{\bs\bs}
 &=K_{\violet{\zeta},\violet{\bP}}
 K_{\purple{\eta},\purple{\bQ}}
 \iint G_{\violet{\zeta},\violet{\bP}}(\br_1)
 \frac{1}{r_{12}}G_{\purple{\eta},\purple{\bQ}}(\br_2)
 \dd\br_1\dd\br_2\\
 &=K_{\violet{\zeta},\violet{\bP}}
 K_{\purple{\eta},\purple{\bQ}}
 \iint e^{-\violet{\zeta}\abs{\br_1-\violet{\bP}}^2}
 \frac{1}{r_{12}}
 e^{-\purple{\eta}\abs{\br_2-\purple{\bQ}}^2}
 \dd\br_1\dd\br_2
\end{split}
\end{equation}
The Laplace transform
\begin{equation}
 \frac{1}{r_{12}}=\frac{2}{\sqrt\pi}\int_0^\infty
 e^{-t^2r_{12}^2}\dd t
 =\frac{2}{\sqrt\pi}\int_0^\infty
 e^{-t^2(x_1-x_2)^2}e^{-t^2(y_1-y_2)^2}
 e^{-t^2(z_1-z_2)^2}\dd t
\end{equation}
turns the six-dimensional integral into three Cartesian factors:
\begin{equation}
\begin{split}
 \sbraket{\bs\bs}{\bs\bs}
 ={}&\frac{2}{\sqrt\pi}K_{\violet{\zeta},\violet{\bP}}
 K_{\purple{\eta},\purple{\bQ}}\int_0^\infty\dd t
 \qty[\iint e^{-\violet{\zeta}(x_1-\violet P_x)^2}
 e^{-t^2(x_1-x_2)^2}e^{-\purple{\eta}(x_2-\purple Q_x)^2}
 \dd x_1\dd x_2]\\
 &\cdot\qty[\iint e^{-\violet{\zeta}(y_1-\violet P_y)^2}
 e^{-t^2(y_1-y_2)^2}e^{-\purple{\eta}(y_2-\purple Q_y)^2}
 \dd y_1\dd y_2]\\
 &\cdot\qty[\iint e^{-\violet{\zeta}(z_1-\violet P_z)^2}
 e^{-t^2(z_1-z_2)^2}e^{-\purple{\eta}(z_2-\purple Q_z)^2}
 \dd z_1\dd z_2]
\end{split}
\end{equation}
Defining
\begin{equation}
 \vartheta=\frac{\violet{\zeta}\purple{\eta}}
 {\violet{\zeta}+\purple{\eta}}
\end{equation}
the Cartesian integrations yield, step by step,
\begin{equation}
\begin{split}
 \sbraket{\bs\bs}{\bs\bs}
 ={}&\frac{2}{\sqrt\pi}K_{\violet{\zeta},\violet{\bP}}
 K_{\purple{\eta},\purple{\bQ}}\int_0^\infty
 \prod_{i=x,y,z}\qty[
 \frac{\pi}{\sqrt{\violet{\zeta}\purple{\eta}}}
 \frac{\exp[-\vartheta(\violet P_i-\purple Q_i)^2
 t^2/(t^2+\vartheta)]}
 {\sqrt{1+t^2/\vartheta}}]\dd t\\
 ={}&\frac{2\pi^{5/2}}
 {(\violet{\zeta}\purple{\eta})^{3/2}}
 K_{\violet{\zeta},\violet{\bP}}K_{\purple{\eta},\purple{\bQ}}
 \int_0^\infty
 \frac{\exp[-\vartheta\abs{\violet{\bP}-\purple{\bQ}}^2
 t^2/(t^2+\vartheta)]}{(1+t^2/\vartheta)^{3/2}}\dd t
\end{split}
\end{equation}
With the substitution $t^2=\vartheta u^2/(1-u^2)$,
\begin{equation}
\begin{split}
 \sbraket{\bs\bs}{\bs\bs}
 ={}&\sqrt{\frac{4\vartheta}{\pi}}
 \qty(\frac{\pi}{\violet{\zeta}})^{3/2}
 K_{\violet{\zeta},\violet{\bP}}
 \qty(\frac{\pi}{\purple{\eta}})^{3/2}
 K_{\purple{\eta},\purple{\bQ}}
 \underbrace{\int_0^1e^{-\vartheta
 \abs{\violet{\bP}-\purple{\bQ}}^2u^2}\dd u}
 _{F_0(\vartheta\abs{\violet{\bP}-\purple{\bQ}}^2)}
\end{split}
\end{equation}
The zero angular-momentum vector $\bO=(0,0,0)$ denotes an $s$ function in
this integral notation. The fundamental auxiliary notation is therefore
\begin{equation}
  \boxed{
 \sbraket{\bO\bO}{\bO\bO}^{(0)}
 =\sqrt{\frac{4\vartheta}{\pi}}
 \qty(\frac{\pi}{\violet{\zeta}})^{3/2}K_{\violet{\zeta},\violet{\bP}}
 \qty(\frac{\pi}{\purple{\eta}})^{3/2}K_{\purple{\eta},\purple{\bQ}}
 F_0\qty(\vartheta\abs{\violet{\bP}-\purple{\bQ}}^2)
  }
\end{equation}
The original six-dimensional Coulomb integral has thus been reduced to
elementary Gaussian prefactors and a one-dimensional function. This reduction
is the foundation of the recurrence construction below.

\subsubsection{Boys functions and auxiliary integrals}
\label{sec:boys-functions}

The remaining one-dimensional integral is collected into the generalized Boys
function. Here $m$ is a nonnegative integer order and $x$ a nonnegative,
dimensionless argument, not a Cartesian coordinate:
\begin{align}
  F_m(x)&=\boxed{\int_0^1t^{2m}\exp(-xt^2)\dd t}
  &
  F_0(x)&=\sqrt{\frac{\pi}{4x}}\erf(\sqrt x)
  \label{eq:boys-functions}
\end{align}
At the origin, the exponential becomes unity and the limiting value follows
directly:
\begin{equation}
 F_m(0)=\int_0^1t^{2m}\dd t=\frac{1}{2m+1}
\end{equation}
This limit is particularly useful because the expression for $F_0(x)$ in
terms of the error function has an apparent singularity at $x=0$, although
the Boys function itself is perfectly finite there.

In implementations, $F_m(x)$ is evaluated from pretabulated grid values and
a Taylor expansion, often with a Chebyshev interpolation table \citep{Gill1994}. Lower orders
are then obtained by the downward recurrence
\begin{equation}
 F_m(x)=\frac{2xF_{m+1}(x)+\exp(-x)}{2m+1}
\end{equation}
This downward recurrence is numerically stable, whereas upward recursion can
become unstable for small $x$. At large $x$,
\begin{equation}
 F_m(x)\sim\frac{(2m-1)!!}{2^{m+1}}
 \sqrt{\frac{\pi}{x^{2m+1}}}
\end{equation}

The physical $ssss$ integral contains only $F_0$, but differentiation to
generate angular momentum produces higher Boys orders. It is therefore useful
to define the complete hierarchy
\begin{equation}
 \sbraket{\bO\bO}{\bO\bO}^{(\red m)}
 =\sqrt{\frac{4\vartheta}{\pi}}
 \qty(\frac{\pi}{\violet{\zeta}})^{3/2}K_{\violet{\zeta},\violet{\bP}}
 \qty(\frac{\pi}{\purple{\eta}})^{3/2}K_{\purple{\eta},\purple{\bQ}}
 F_{\red m}\qty(\vartheta\abs{\violet{\bP}-\purple{\bQ}}^2)
\end{equation}
The $m=0$ member is the physical $ssss$ ERI; the $m>0$ members are auxiliary
intermediates. Angular-momentum recurrences couple orders $m$ and $m+1$.
For total angular momentum $L=a+b+c+d$, fundamental auxiliary integrals are
needed through $m=L$.
The recurrence construction must therefore retain the necessary Boys orders
alongside the angular-momentum labels.

\subsubsection{Obara--Saika recurrence relations}
\label{sec:os-recurrences}

Deriving a separate closed expression for every angular-momentum combination
would be inefficient. Obara--Saika instead uses differentiation identities to
generate the required quantities recursively from lower-angular-momentum
integrals \citep{ObaraSaika1986}.

\paragraph{Generating angular momentum vertically.}

Here $i$ labels a Cartesian direction, $x$, $y$, or $z$, rather than an
occupied orbital. The vector $\bI_i$ has unit component in direction $i$ and
zero components in the other two directions. For an unnormalized primitive
Cartesian Gaussian, the key identity is
\begin{align}
  \pdv{\sket{\red{\ba}}}{A_i}
 &=2\alpha\sket{\red{\ba}+\bI_i}
 -a_i\sket{\red{\ba}-\bI_i}
  &
  \sket{\red{\ba}+\bI_i}
 &=\frac{1}{2\alpha}\qty[
 \pdv{\sket{\red{\ba}}}{A_i}
 +a_i\sket{\red{\ba}-\bI_i}]
\end{align}
Applying this identity to auxiliary ERIs creates the Obara--Saika vertical
recurrence relations (VRRs), so that integrals are built recursively rather
than independently:
\begin{equation}
 \sbraket{00}{00}^{(m)}\longrightarrow
 \sbraket{\red a0}{00}^{(m)}\longrightarrow
 \sbraket{\red a0}{\orange c0}^{(m)}
\end{equation}

On the bra pair, angular momentum is built on $\bA$ according to
\begin{equation}
\begin{split}
 \sbraket{(\red{\ba}+\bI_i)\bO}{\bO\bO}^{(m)}
 &=(P_i-A_i)\sbraket{\red{\ba}\bO}{\bO\bO}^{(m)}
 -\frac{\eta}{\zeta+\eta}(P_i-Q_i)
 \sbraket{\red{\ba}\bO}{\bO\bO}^{(m+1)}\\
 &+\frac{a_i}{2\zeta}\qty[
 \sbraket{(\red{\ba}-\bI_i)\bO}{\bO\bO}^{(m)}
 -\frac{\eta}{\zeta+\eta}
 \sbraket{(\red{\ba}-\bI_i)\bO}{\bO\bO}^{(m+1)}]
\end{split}
\end{equation}
On the ket pair, angular momentum is built on $\bC$ according to
\begin{equation}
\begin{split}
 \sbraket{\red{\ba}\bO}{(\orange{\bc}+\bI_i)\bO}^{(m)}
 &=(Q_i-C_i)\sbraket{\red{\ba}\bO}{\orange{\bc}\bO}^{(m)}
 -\frac{\zeta}{\zeta+\eta}(Q_i-P_i)
 \sbraket{\red{\ba}\bO}{\orange{\bc}\bO}^{(m+1)}\\
 &+\frac{c_i}{2\eta}\qty[
 \sbraket{\red{\ba}\bO}{(\orange{\bc}-\bI_i)\bO}^{(m)}
 -\frac{\zeta}{\zeta+\eta}
 \sbraket{\red{\ba}\bO}{(\orange{\bc}-\bI_i)\bO}^{(m+1)}]\\
 &+\frac{a_i}{2(\zeta+\eta)}
 \sbraket{(\red{\ba}-\bI_i)\bO}{\orange{\bc}\bO}^{(m+1)}
\end{split}
\end{equation}

The signs of the product-center terms can be checked directly from the
fundamental integral. With $T=\vartheta\abs{\bP-\bQ}^2$, the definition of
the Boys function gives $\dv*{F_m(T)}{T}=-F_{m+1}(T)$. Since
$\pdv*{P_i}{A_i}=\alpha/\zeta$, differentiating the fundamental integral
and dividing by $2\alpha$ yields
\begin{equation}
 \sbraket{\bI_i\bO}{\bO\bO}^{(m)}
 =(P_i-A_i)\sbraket{\bO\bO}{\bO\bO}^{(m)}
 -\frac{\vartheta}{\zeta}(P_i-Q_i)
 \sbraket{\bO\bO}{\bO\bO}^{(m+1)}
\end{equation}
Here $\vartheta/\zeta=\eta/(\zeta+\eta)$. The ket-pair result follows
by exchanging the two pairs. The derivative argument fixes both
product-center terms to have negative signs.

The required angular-momentum build and contraction are organized as
\begin{equation}
 \sbraket{00}{00}^{(m)}
 \xrightarrow{\red{\mathrm{VRR}}}
 \sbraket{\red a0}{\orange c0}^{(m)}
 \xrightarrow{\orange{\mathrm{contraction}}}
 \rbraket{\red a0}{\orange c0}
 \xrightarrow{\blue{\mathrm{HRR}}}
 \rbraket{\red a\blue b}{\orange c\green d}
\end{equation}
The VRRs create angular momentum on one center of each Gaussian pair;
contraction combines primitive integrals; and the horizontal recurrence
relations (HRRs) redistribute angular momentum within each pair.

\paragraph{Redistributing angular momentum horizontally.}

\begin{figure}[htbp]
  \centering

\resizebox{\linewidth}{!}{%
\begin{tikzpicture}
  \node[draw=hfGreenDark,fill=hfGreenLight,rounded corners=2mm,
    line width=0.6pt,inner sep=3mm] {%
    $\displaystyle
    \underbrace{\sbraket{00}{00}^{(m)}}_{
      \substack{\text{fundamental}\\\text{auxiliary integrals}}}
    \xrightarrow{\;\red{\mathrm{VRR}}\;}
    \underbrace{\sbraket{\red{a}0}{\orange{c}0}}_{
      \text{primitive integrals}}
    \xrightarrow{\;\orange{\mathrm{contract}}\;}
    \underbrace{\rbraket{\red{a}0}{\orange{c}0}}_{
      \text{contracted integrals}}
    \xrightarrow{\;\blue{\mathrm{HRR}}\;}
    \underbrace{\rbraket{\red{a}\blue{b}}{\orange{c}\green{d}}}_{
      \text{target ERIs}}
    $};
\end{tikzpicture}%
}


  \caption{Late-contraction path for ERI evaluation, adapted from slide~38
  of the basis-set lecture. Square brackets denote primitive integrals and
  round brackets contracted integrals; $0$ denotes an $s$-type shell and
  $m$ the auxiliary Boys-function order. The labels $a,b,c,d$ specify shell
  angular momenta. The intermediate sets include all angular momenta
  required to reach the target shell quartet $\rbraket{ab}{cd}$.
  The exponent-dependent VRRs precede contraction of the required $m=0$
  primitive integrals, while the exponent-independent HRRs act afterwards
  \citep{HeadGordonPople1988}.}
  \label{fig:late-contraction}
\end{figure}
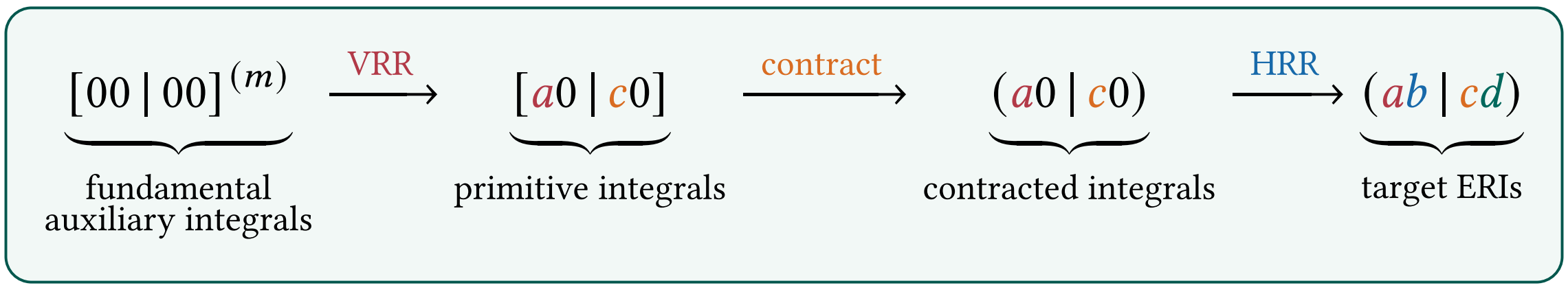

Once angular momentum has been generated on one center, it is shifted to the
other through
\begin{equation}
 \sbraket{\red{\ba}(\blue{\bb}+\bI_i)}
 {\orange{\bc}\green{\bd}}
 =\sbraket{(\red{\ba}+\bI_i)\blue{\bb}}
 {\orange{\bc}\green{\bd}}
 +(A_i-B_i)\sbraket{\red{\ba}\blue{\bb}}
 {\orange{\bc}\green{\bd}}
\end{equation}
An analogous relation transfers angular momentum from $\bC$ to $\bD$.
Because the HRR depends only on $A_i-B_i$, it is independent of the Gaussian
exponents and applies directly to contracted integrals:
\begin{equation}
  \boxed{
 \rbraket{\red{\ba}(\blue{\bb}+\bI_i)}
 {\orange{\bc}\green{\bd}}
 =\rbraket{(\red{\ba}+\bI_i)\blue{\bb}}
 {\orange{\bc}\green{\bd}}
 +(A_i-B_i)\rbraket{\red{\ba}\blue{\bb}}
 {\orange{\bc}\green{\bd}}
  }
\end{equation}
The late-contraction algorithm begins by
computing the required Boys functions $F_m(T)$ from
Section~\ref{sec:boys-functions} and forming the fundamental
auxiliary integrals $\sbraket{00}{00}^{(m)}$. The VRRs are then used to build
angular momentum on $\bA$ and $\bC$, after which the primitive integrals are
contracted. Finally, the exponent-independent HRRs transfer angular momentum
according to $\bA\rightarrow\bB$ and $\bC\rightarrow\bD$.
This combination of vertical and horizontal recurrences is the
Head--Gordon--Pople late-contraction strategy \citep{HeadGordonPople1988}; the vertical recurrences are
those of Obara--Saika. Thus the expensive exponent-dependent recurrences are performed before
contraction, whereas the cheap exponent-independent HRRs are performed after
contraction.
The resulting reduction in the number of quantities handled by the HRRs
complements the screening savings considered next.
Figure~\ref{fig:late-contraction} summarizes this path, distinguishing the
primitive and contracted stages.

\subsection{Integral bounds, screening, and direct SCF}
\label{sec:integral-screening}

A formal count of $K^4$ ERIs greatly overestimates the useful work for large
local systems because most products of distant basis functions are negligible.
Screening converts this locality into a computational saving by testing a
cheap upper bound before evaluating an integral or shell quartet.

\subsubsection{Cauchy--Schwarz bounds}

A useful ERI upper bound must be simple and inexpensive, strong enough to be a
good estimate, and scaling consistent, meaning
$K_\text{sig}=\order{N_\text{upper}}$. Here $K_\text{sig}$ counts significant
integrals and $N_\text{upper}$ counts those retained by the bound at the
chosen threshold. The Cauchy--Schwarz inequality gives
\begin{equation}
  \boxed{
 \abs{\rbraket{\red{\ba}\blue{\bb}}{\orange{\bc}\green{\bd}}}
 \leq
 \sqrt{\rbraket{\red{\ba}\blue{\bb}}{\red{\ba}\blue{\bb}}}
 \sqrt{\rbraket{\orange{\bc}\green{\bd}}{\orange{\bc}\green{\bd}}}
  }
\end{equation}
or, in terms of the Gaussian product centers,
\begin{equation}
 \abs{\rbraket{\violet{\bP}}{\purple{\bQ}}}
 \leq\sqrt{\rbraket{\violet{\bP}}{\violet{\bP}}}
 \sqrt{\rbraket{\purple{\bQ}}{\purple{\bQ}}}
\end{equation}
The integral $\rbraket{\violet{\bP}}{\purple{\bQ}}$ should decay as
$1/\abs{\violet{\bP}-\purple{\bQ}}$ or faster. The product on the right has
no information about the separation between the two charge distributions, so
the Cauchy--Schwarz bound is rigorous and cheap but not tight for well-separated
shell pairs. Even so, a cheap upper bound avoids evaluating negligible ERIs.
The practical use of integral bounds in direct-SCF screening is discussed by
\citet{HaserAhlrichs1989}.

\subsubsection{Shell-quartet screening}

At the individual-integral level, only $\order{K^2}$ diagonal factors
$\rbraket{\red{\ba}\blue{\bb}}{\red{\ba}\blue{\bb}}$ are required to bound
$\order{K^4}$ four-center integrals. More efficiently, define a shell-pair
bound
\begin{equation}
 Q_{\red a\blue b}
 =\max_{\substack{\red{\ba}\in a\\\blue{\bb}\in b}}
 \sqrt{\rbraket{\red{\ba}\blue{\bb}}{\red{\ba}\blue{\bb}}}
\end{equation}
Then an entire shell class obeys
\begin{equation}
 \abs{\rbraket{\red a\blue b}{\orange c\green d}}
 \leq Q_{\red a\blue b}Q_{\orange c\green d}
\end{equation}
For a positive screening threshold $\tau$, if
\begin{equation}
  \boxed{
 Q_{\red a\blue b}Q_{\orange c\green d}<\tau
  }
\end{equation}
the entire shell quartet can be discarded. This is consequential: even the
simple $\rbraket{dd}{dd}$ class contains 1296 Cartesian integrals.

The number of significant integrals may be represented by
\begin{equation}
  \rbraket{\red{\ba}\blue{\bb}}{\orange{\bc}\green{\bd}}
 \equiv\rmel{\red{\ba}\blue{\bb}}{\chO_2}
 {\orange{\bc}\green{\bd}}
\end{equation}
and $K_\text{sig}=cK^\alpha$.
Here $K$ is the number of basis functions, $c$ is a proportionality factor,
and $\alpha$ is the effective scaling exponent.
The numerical data in Table~\ref{tab:screening-scaling} contrast the long-range Coulomb operator with a
short-range Gaussian operator:

\begin{table}[htbp]
\centering
\hftablestyle
\caption{Numbers of significant integrals and effective scaling exponents
for Coulomb and short-range Gaussian operators. Numerical data reproduced
from the lecture slides.}
\label{tab:screening-scaling}
\begin{tabular}{@{}lrrrrr@{}}
\toprule
\rowcolor{hfGreenLight}
& &\multicolumn{2}{c}{\red{$\chO=r_{12}^{-1}$}}
&\multicolumn{2}{c}{\orange{$\chO=e^{-r_{12}^2}$}}\\
\cmidrule(lr){3-4}\cmidrule(lr){5-6}
\hftablehead{Molecule}&\hftablehead{$K$}&\hftablehead{$K_\text{sig}$}&
\hftablehead{$\alpha$}&\hftablehead{$K_\text{sig}$}&\hftablehead{$\alpha$}\\
\midrule
propene&12&1,625&---&1,650&---\\
butadiene&16&5,020&3.9&5,020&3.9\\
hexatriene&24&24,034&3.9&23,670&3.8\\
octatetraene&32&63,818&3.4&52,808&2.8\\
decapentaene&40&119,948&2.8&81,404&1.9\\
dodecahexaene&48&192,059&2.6&109,965&1.6\\
\bottomrule
\end{tabular}
\end{table}

The Coulomb interaction remains long ranged, so its effective exponent falls
more slowly than that of the short-range Gaussian operator. Nevertheless, the
data show that the number of numerically significant Coulomb integrals grows
much more slowly than the formal quartic count in sufficiently large systems.

\begin{figure}[htbp]
 \centering
 \begin{tikzpicture}[>=stealth',font=\small]
  \node[draw,fill=hfViolet!15,align=center] (target) at (0,2.5)
   {Two-electron integrals\\$\rbraket{\ba\bb}{\bc\bd}$};
  \node[draw,fill=hfRed!15,align=center,text width=3.7cm] (fundamental) at (-4.2,0)
   {Fundamental integrals\\$\sbraket{\bO\bO}{\bO\bO}^{(m)}$};
  \node[draw,fill=hfOrange!15,align=center,text width=3.7cm] (recurrence) at (0,0)
   {Recurrence relations\\VRRs and HRRs};
  \node[draw,fill=hfGreen!15,align=center,text width=3.7cm] (bound) at (4.2,0)
   {Upper bounds\\$\abs{\rbraket{\ba\bb}{\bc\bd}}\leq B$};
  \draw[very thick,->] (fundamental)--(target);
  \draw[very thick,->] (recurrence)--(target);
  \draw[very thick,->] (bound)--(target);
 \end{tikzpicture}
 \caption{The three ingredients of integral evaluation: fundamental
 integrals, recurrence relations, and screening bounds.}
 \label{fig:integral-recipe}
\end{figure}
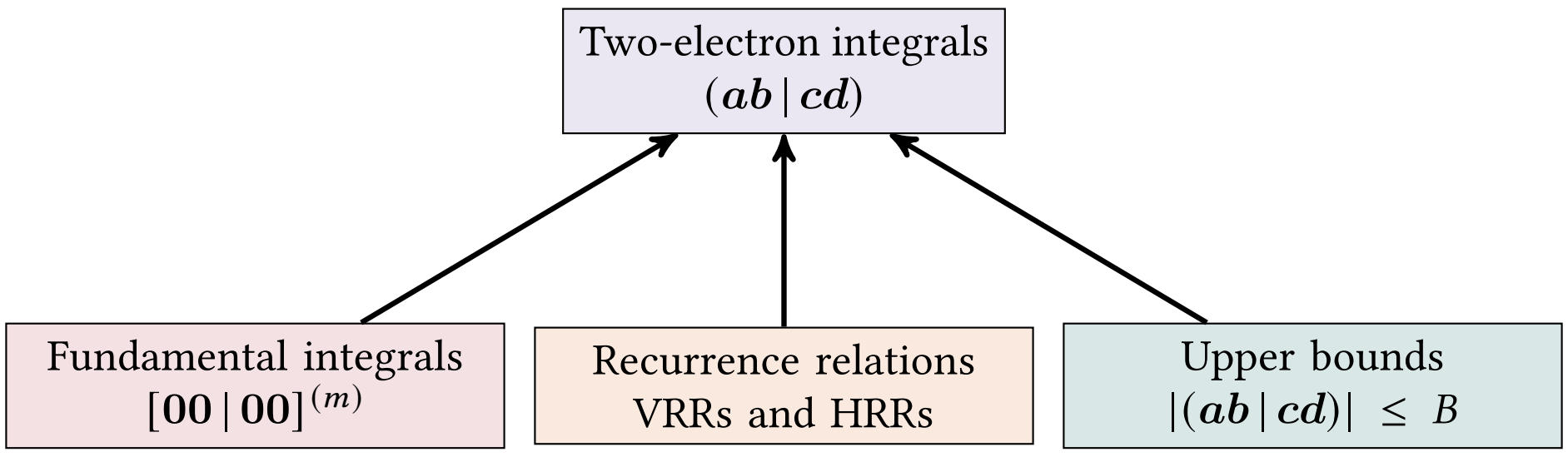

The complete integral-evaluation strategy has three components, summarized in
Fig.~\ref{fig:integral-recipe}. The target
quantities are the two-electron integrals
$\rbraket{\ba\bb}{\bc\bd}$. Their construction requires the fundamental
integrals $\rbraket{\bO\bO}{\bO\bO}^{(m)}$, the recurrence relations,
represented schematically by
$\rbraket{(\ba+\bI_i)\bO}{\bO\bO}
=\rbraket{\ba\bO}{\bO\bO}
+\rbraket{(\ba-\bI_i)\bO}{\bO\bO}$, and an upper bound
$\abs{\rbraket{\ba\bb}{\bc\bd}}\leq B$ that permits insignificant
integrals to be discarded before evaluation.

\subsubsection{Direct SCF evaluation}

Once screening has identified the significant shell quartets, the resulting
integrals need not be stored. Direct SCF contracts each batch with the current
density immediately after it is generated.
For the real AO functions and real symmetric density used in the following
direct-SCF contractions, the Coulomb and exchange matrices are
\begin{align}
 J_{\mu\nu}&=\sum_{\lambda\sigma}P_{\lambda\sigma}
 \rbraket{\mu\nu}{\lambda\sigma}
 &
 K_{\mu\nu}&=\sum_{\lambda\sigma}P_{\lambda\sigma}
 \rbraket{\mu\lambda}{\nu\sigma}
\end{align}
Direct SCF \citep{Almlof1982} trades disk storage for repeated integral evaluation. Instead of
storing all $\order{K^4}$ ERIs, it loops over the shell quartets
$\rbraket{\mu\nu}{\lambda\sigma}$ and uses an upper bound to discard
negligible quartets before their integrals are evaluated. The significant
ERIs are computed, their contributions are accumulated immediately in $\bJ$
and $\bK$, and the integrals are then discarded.
The density-matrix elements can themselves be included in the screening
criterion. In compact form, the algorithm is
\begin{equation}
 \text{screen}\longrightarrow\text{compute}\longrightarrow
 \text{accumulate}\longrightarrow\text{discard}
\end{equation}

Because the Fock build contains products of density-matrix elements and ERIs,
screening can be tightened by bounding the contribution to the matrix rather
than the integral alone. A shell quartet whose bound is below the integral
threshold is a candidate for omission, while a density-weighted criterion can also discard a
formally non-negligible quartet when the density elements multiplying it are
small at the current SCF iteration. An integral threshold alone does not
bound the accumulated error in the Fock matrix or energy: the density
weights and the number of discarded contributions also matter. The threshold
must therefore be chosen consistently with the target accuracy of the SCF
energy and residual.

\section{Resolution of the identity and low-rank ERI representations}
\label{sec:ri}

Screening avoids work associated with small individual integrals, but it does
not alter the four-index structure of the ERI tensor. Resolution of the
identity (RI), also called density fitting (DF), addresses this complementary
problem by replacing four-center quantities with contractions over a smaller
auxiliary index. Its starting point is that products of orbital basis
functions span a large, highly redundant space. Representing that space
economically connects the integral technology developed above to less costly
Fock-matrix construction.
In this section, orbital and auxiliary functions are taken to be real, as in
the Gaussian integral formulas above. An ERI is the Coulomb interaction
between two pair densities,
\begin{equation}
 \rbraket{\mu\nu}{\lambda\sigma}
 =\iint
 \underbrace{\phi_\mu(\br_1)\phi_\nu(\br_1)}_{\rho_{\mu\nu}(\br_1)}
 \frac{1}{r_{12}}
 \underbrace{\phi_\lambda(\br_2)\phi_\sigma(\br_2)}
 _{\rho_{\lambda\sigma}(\br_2)}\dd\br_1\dd\br_2
\end{equation}
With $K$ orbital basis functions, there are formally $\order{K^2}$ different
pair densities. RI/DF replaces this
large product space by a compact auxiliary basis
$\{\varphi_A^\text{aux}\}$ \citep{Whitten1973,Dunlap1979,Vahtras1993}:
\begin{equation}
  \rho_{\mu\nu}(\br)\approx
 \widetilde\rho_{\mu\nu}(\br)
 =\sum_AC_{\mu\nu}^A\varphi_A^\text{aux}(\br)
\end{equation}
with $K_\text{aux}=\order{K}$.
Here $\widetilde\rho_{\mu\nu}$ is the fitted pair density and
$K_\text{aux}$ the number of auxiliary functions. The coefficients
$C_{\mu\nu}^A$ are fitting coefficients, not the MO coefficients
$C_{\mu p}$. The auxiliary basis is therefore not a second orbital basis.
It is designed
to represent products of orbital functions, and its index $A$ labels
functions in this approximate product space. The reduction from
$\order{K^2}$ pair densities to $\order{K}$ auxiliary functions is the source
of the lower-rank representation.

\subsection{Fitting pair densities in the Coulomb metric}

The fitted coefficients depend on the norm used to measure the residual pair
density. For electron-repulsion integrals, the natural choice weights fitting
errors by their Coulomb self-interaction. Define the Coulomb inner product
\begin{equation}
 \rbraket{f}{g}=\iint f(\br_1)\frac{1}{r_{12}}g(\br_2)
 \dd\br_1\dd\br_2
\end{equation}
The optimal fitting coefficients minimize the Coulomb norm of the error,
\begin{equation}
 \Delta_{\mu\nu}=\rbraket{\rho_{\mu\nu}-\widetilde\rho_{\mu\nu}}
 {\rho_{\mu\nu}-\widetilde\rho_{\mu\nu}}
\end{equation}
We abbreviate an auxiliary function by its index inside an integral:
$\rbraket{A}{B}$ is the Coulomb interaction between two auxiliary functions,
and $\rbraket{A}{\mu\nu}$ couples an auxiliary function to an AO pair density.
Substituting the auxiliary expansion and using the symmetry of the Coulomb
metric gives
\begin{equation}
\begin{split}
 \Delta_{\mu\nu}
 ={}&\rbraket{\rho_{\mu\nu}}{\rho_{\mu\nu}}
 -2\sum_A C_{\mu\nu}^A\rbraket{A}{\mu\nu}
 +\sum_{AB}C_{\mu\nu}^A
 \underbrace{\rbraket{A}{B}}_{V_{AB}}C_{\mu\nu}^B
\end{split}
\end{equation}
The first term is independent of the fitting coefficients. Differentiating
the remaining two terms with respect to one coefficient gives the
stationarity condition,
\begin{equation}
 \pdv{\Delta_{\mu\nu}}{C_{\mu\nu}^A}
 =-2\rbraket{A}{\mu\nu}
 +2\sum_BV_{AB}C_{\mu\nu}^B=0
\end{equation}
Hence the least-squares equations are
\begin{align}
  \sum_B\underbrace{\rbraket{A}{B}}_{V_{AB}}C_{\mu\nu}^B
 &=\underbrace{\rbraket{A}{\mu\nu}}_{\text{three-center integral}}
  &
  C_{\mu\nu}^A&=\sum_B(\bV^{-1})_{AB}\rbraket{B}{\mu\nu}
\end{align}
Thus the Coulomb metric determines the optimal auxiliary expansion. The
stationarity condition also states that the residual is Coulomb-orthogonal
to every auxiliary function:
\begin{equation}
  \boxed{
 \rbraket{A}{\rho_{\mu\nu}-\widetilde\rho_{\mu\nu}}=0
  }
 \label{eq:ri-projection-orthogonality}
\end{equation}
The inverse metric accounts for the nonorthogonality of the auxiliary
functions. These expressions assume that the retained auxiliary functions
are linearly independent in the Coulomb metric. If the metric has
near-zero eigenvalues, the inverse is formed in the retained eigenspace,
using the same principle as canonical orthogonalization of the orbital
overlap matrix. The resulting coefficients are optimal within that
auxiliary space under the Coulomb norm.

Because a fitted density belongs to the auxiliary space,
Eq.~\eqref{eq:ri-projection-orthogonality} gives
\begin{equation}
 \rbraket{\widetilde\rho_{\mu\nu}}{\widetilde\rho_{\lambda\sigma}}
 =\rbraket{\rho_{\mu\nu}}{\widetilde\rho_{\lambda\sigma}}
 =\rbraket{\widetilde\rho_{\mu\nu}}{\rho_{\lambda\sigma}}
\end{equation}
Thus fitting both densities is equivalent to fitting either one in this
Coulomb projection. Inserting the fitted pair densities converts the
four-center ERI to three-center quantities:
\begin{equation}
\begin{split}
 \rbraket{\mu\nu}{\lambda\sigma}
 &\approx\sum_A C_{\lambda\sigma}^A\rbraket{\mu\nu}{A}\\
 &=\sum_{AB}\rbraket{\mu\nu}{A}(\bV^{-1})_{AB}
 \rbraket{B}{\lambda\sigma}
\end{split}
\end{equation}
Equivalently,
\begin{equation}
  \boxed{
 \rbraket{\mu\nu}{\lambda\sigma}
 \approx\sum_{AB}\rbraket{\mu\nu}{A}(\bV^{-1})_{AB}
 \rbraket{B}{\lambda\sigma}
  }
\end{equation}
With the symmetrically orthonormalized quantities, where $P$ labels an
orthonormalized auxiliary direction,
\begin{equation}
 B_{\mu\nu}^P=\sum_A\rbraket{\mu\nu}{A}(\bV^{-1/2})_{AP}
\end{equation}
the factorization becomes
\begin{equation}
  \boxed{
 \rbraket{\mu\nu}{\lambda\sigma}
 \approx\sum_PB_{\mu\nu}^PB_{\lambda\sigma}^P
  }
\end{equation}
The four-index tensor has thereby been replaced by products of three-index
tensors.
This last form follows by splitting the inverse metric symmetrically,
$\bV^{-1}=\bV^{-1/2}\cdot\bV^{-1/2}$. The explicit matrix multiplication
symbol emphasizes that the auxiliary indices are contracted through the
metric rather than multiplied element by element.

\subsection{RI, Cholesky decomposition, and practical contractions}

Both density fitting and Cholesky decomposition expose the low numerical rank
of the ERI matrix in compound pair indices. They differ in how the
three-index representation is constructed.
In RI/DF, the AO products are expanded in a predefined auxiliary basis,
\begin{equation}
 \phi_\mu(\br)\phi_\nu(\br)
 \approx\sum_AC_{\mu\nu}^A\varphi_A^\text{aux}(\br)
 \quad\Longrightarrow\quad
 \rbraket{\mu\nu}{\lambda\sigma}
 \approx\sum_PB_{\mu\nu}^PB_{\lambda\sigma}^P
\end{equation}
The quality of the approximation therefore depends on the auxiliary basis.
Alternatively, the ERI tensor
$\bV_{(\mu\nu),(\lambda\sigma)}=\rbraket{\mu\nu}{\lambda\sigma}$ is a
positive-semidefinite matrix in the compound indices. Here $\bV$ denotes the
full AO-pair ERI matrix, rather than the auxiliary Coulomb metric
$V_{AB}$ of the preceding derivation. A pivoted Cholesky
decomposition constructs an auxiliary space adaptively
\citep{BeebeLinderberg1977},
\begin{equation}
 \rbraket{\mu\nu}{\lambda\sigma}
 \approx\sum_{P=1}^{N_\text{CD}}L_{\mu\nu}^PL_{\lambda\sigma}^P
\end{equation}
until the residual is below a prescribed threshold $\tau_\text{CD}$. RI/DF
therefore uses a predefined auxiliary space, whereas Cholesky decomposition
constructs one adaptively.

For the Coulomb matrix,
\begin{equation}
\begin{split}
 J_{\mu\nu}
 &=\sum_{\lambda\sigma}P_{\lambda\sigma}
 \rbraket{\mu\nu}{\lambda\sigma}\\
 &\approx\sum_{AB}\rbraket{\mu\nu}{A}(\bV^{-1})_{AB}
 \underbrace{\sum_{\lambda\sigma}P_{\lambda\sigma}
 \rbraket{B}{\lambda\sigma}}_{d_B}
\end{split}
\end{equation}
The practical sequence is
\begin{equation}
 \bP\longrightarrow\{d_A\}\longrightarrow
 \{c_A\}=\bV^{-1}\cdot\{d_A\}\longrightarrow\bJ
\end{equation}
The full $\order{K^4}$ ERI tensor need not be stored; only two-center
$\rbraket{A}{B}$ and three-center $\rbraket{\mu\nu}{A}$ integrals are needed;
and optimized auxiliary bases make the fitting error very small
\citep{Weigend2006}. RI/DF thus greatly reduces the cost and memory
requirements of Coulomb contractions.
The computational gain is that contractions can be ordered through the
auxiliary index instead of four orbital indices \citep{Weigend2002}. The approximation is
controlled by the auxiliary basis in RI/DF or by the decomposition threshold
in Cholesky decomposition.

Taken together, the basis-set and integral constructions complete the
practical formulation of Hartree--Fock theory. The basis determines the
variational one-electron space, Gaussian integral algorithms provide the
matrix elements required within that space, and screening or low-rank
factorization controls the cost of assembling the mean-field operators. These
components lead back to the Roothaan--Hall and SCF equations developed in
Section~\ref{sec:finite-basis-hf}.


\section{Concluding perspective}
\label{sec:conclusion}

The developments in this chapter connect a variational approximation to a
practical molecular calculation. Hartree--Fock theory restricts the
electronic wavefunction to a single determinant while retaining
antisymmetry and the Coulomb and exchange contributions to its energy.
Orbital optimization gives a Fock operator that depends on the occupied
orbitals themselves, and hence a self-consistent problem. Restricted,
unrestricted, and generalized formulations impose different constraints on
that optimization (Section~\ref{sec:open-shell-hf}). A converged solution is
not, by convergence alone, guaranteed to be the lowest-energy determinant
or to possess the desired spin and spatial symmetries.

The basis-set approximation is a separate restriction. Expanding the
orbitals in a finite one-electron space converts the orbital equations into
the Roothaan--Hall matrix problem of Section~\ref{sec:finite-basis-hf}.
The basis families discussed in Section~\ref{sec:basis-sets} offer different
ways of approaching the required radial and angular flexibility. Improving
that space reduces basis-set incompleteness, but does not remove the
single-determinant restriction or recover the correlation missing from HF.

Integral evaluation makes this finite-basis formulation operational. The
Gaussian product theorem and recurrence relations provide the one- and
two-electron matrix elements needed to build the Fock matrix
(Section~\ref{sec:gaussian-integrals}). Screening avoids negligible
contributions, while density fitting and Cholesky decomposition replace the
four-index ERI representation by more economical contractions
(Section~\ref{sec:ri}). Their thresholds or auxiliary spaces control
numerical approximations distinct from the choice of orbital basis.

Interpreting a calculation therefore requires keeping these levels
separate: the chosen HF variational space, the orbital basis, the integral
representation, and the convergence of the SCF solution. The resulting
total energy, orbital energies, and molecular properties must then be read
with their respective assumptions in mind. In particular, a stationary HF
reference supplies a useful starting point for correlated methods, not a
substitute for them. The link between these theoretical assumptions and
their numerical realization is what makes Hartree--Fock theory both a
foundational model and a practical framework for electronic structure.

\section*{Further reading}

Useful general references include Jensen's \emph{Introduction to
Computational Chemistry} \citep{Jensen2017}, Cramer's \emph{Essentials of Computational
Chemistry} \citep{Cramer2004}, Szabo and Ostlund's \emph{Modern Quantum Chemistry}
\citep{SzaboOstlund1996}, and
Helgaker, J{\o}rgensen, and Olsen's \emph{Molecular Electronic Structure
Theory} \citep{Helgaker2000}.

\section*{ESQC 2026}

\begin{center}
  \includegraphics[width=\linewidth]{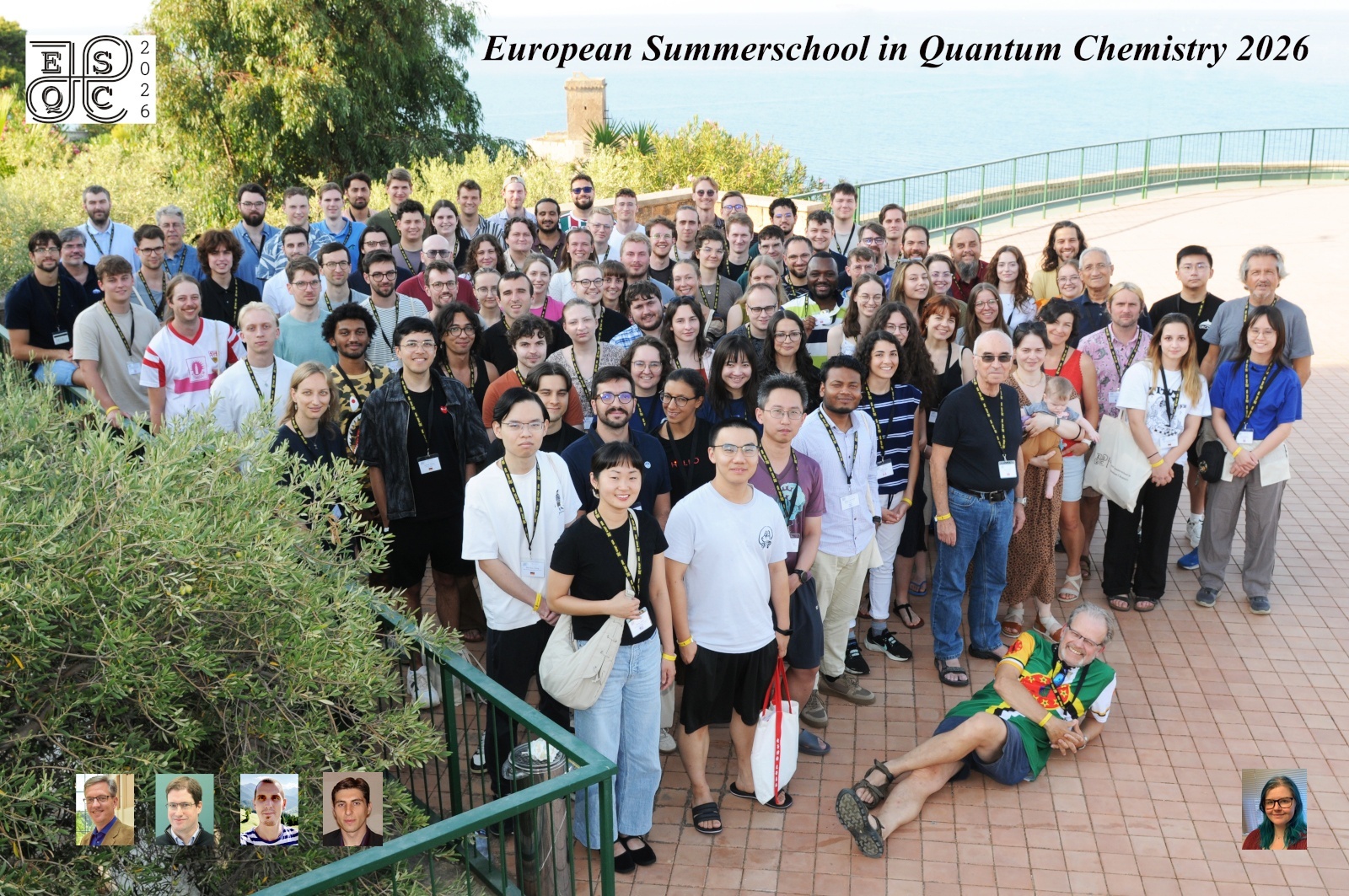}
\end{center}

\section*{Acknowledgements}
I would like to thank all my supervisors, collaborators, and group members over the years, with whom I have had the opportunity to discuss many of the topics presented here.
I am particularly grateful to Xavier Assfeld, Peter Gill, and Hugh Burton for the many stimulating discussions we have had about Hartree--Fock theory over the years.
I would also like to thank Trond Saue for inviting me to teach at ESQC 2026 and for his constructive comments and suggestions during the lectures.
A special thanks also goes to all the ESQC tutors, who do an incredible job behind the scenes and contribute greatly to the success of the summerschool.

\renewcommand{\bibsection}{\section*{References}}
\bibliographystyle{chapter-unsrtnat}
\bibliography{ESQC_Loos_HF_BF_chapter}

\end{document}